\PassOptionsToPackage{dvipsnames}{xcolor}
\documentclass[11pt,a4paper]{article}
\usepackage[utf8]{inputenc}
\usepackage{jheppub}
\usepackage[english]{babel}
\usepackage{amsmath}
\usepackage{amsfonts}
\usepackage{placeins}
\usepackage{amssymb}
\usepackage{graphicx}
\usepackage{lmodern}
\usepackage{enumitem}
\usepackage{braket}
\usepackage{extarrows}
\usepackage{subcaption}
\usepackage{float}
\usepackage{pdfpages}
\usepackage{slashed}
\usepackage{booktabs}

\usepackage{microtype}

\newcommand{\DC}{\Delta C}
\newcommand{\Fcal}{\mathcal{F}}
\newcommand{\Gcal}{\mathcal{G}}
\newcommand{\Ccal}{\mathcal{C}}

\newcommand{\qb}{\bar{q}}
\newcommand{\qp}{{q^\prime}}
\newcommand{\qbp}{{\bar{q}^\prime}}

\newcommand{\dd}{\mathrm{d}}
\newcommand{\NC}{\mathrm{NC}}
\newcommand{\CC}{\mathrm{CC}}
\newcommand{\modd}{\mathrm{odd}}
\newcommand{\meven}{\mathrm{even}}
\newcommand{\vh}{\hat{v}}
\newcommand{\ah}{\hat{a}}
\newcommand{\ih}{\hat{i}}
\newcommand{\xh}{\hat{x}}
\newcommand{\zh}{\hat{z}}
\newcommand\numberthis{\addtocounter{equation}{1}\tag{\theequation}}

\newcommand{\NoW}{\mathrm{NoW}}
\newcommand{\Fcon}{\mathrm{Fcon}}
\newcommand{\vSa}{\mathrm{A}}
\newcommand{\aSa}{\mathrm{A}}
\newcommand{\SaSa}{\mathrm{AA}}
\newcommand{\GeV}{\mathrm{GeV}}

\newcommand{\CF}{C_\mathrm{F}}
\newcommand{\CA}{C_\mathrm{A}}
\newcommand{\TR}{T_\mathrm{R}}
\newcommand{\nf}{n_f}

\newcommand{\MSbar}{\overline{\textrm{MS}}}
\newcommand{\Larin}{\mathrm{Larin}}

\newcommand{\tr}{\operatorname{Tr}}

\def\code#1{\texttt{#1}}

\allowdisplaybreaks

\title{Polarized Neutral and Charged Current Semi-Inclusive Deep-Inelastic Scattering at NNLO in QCD}
\author[a]{Leonardo Bonino,}
\author[a]{Thomas Gehrmann,}
\author[a]{Markus Löchner,}
\author[a]{Kay Schönwald,}
\author[b,c]{and Giovanni Stagnitto}

\affiliation[a]{Universität Zürich, Winterthurerstrasse 190, CH-8057 Zürich, Switzerland}
\affiliation[b]{Università degli Studi di Milano-Bicocca \& INFN, Piazza della Scienza 3, I-20126 Milano, Italy}
\affiliation[c]{Università di Genova \& INFN, Via Dodecaneso 33, I-16146 Genova, Italy}

\emailAdd{leonardo.bonino@physik.uzh.ch}
\emailAdd{thomas.gehrmann@uzh.ch}
\emailAdd{markus.loechner@physik.uzh.ch}
\emailAdd{kay.schoenwald@physik.uzh.ch}
\emailAdd{giovanni.stagnitto@unimib.it}

\abstract{The semi-inclusive
production of identified hadrons in deeply inelastic lepton scattering on polarized nucleons
allows to probe the spin structure of the nucleon target in a more detailed level than through fully inclusive
processes.
We compute the NNLO QCD corrections to longitudinally polarized semi-inclusive deep inelastic processes
mediated by electroweak neutral and charged currents.
We present the first calculation of polarized electroweak structure functions in the Larin scheme up to NNLO.
Additionally, we reformulate the finite scheme transformation
to the $\MSbar$ scheme  in a quark flavour basis
and identify issues in its assignment to partonic channels in previous calculations.
Using our results we perform a detailed phenomenological study of
polarized cross sections and of single and double spin asymmetries.
We observe large electroweak effects, which need to be included in precision studies of polarized observables.
}

\begin{document}

\begin{flushright}
 ZU-TH 64/25
\end{flushright}

\maketitle

\section{Introduction}

The precision study of the spin structure of the proton~\cite{Aidala:2012mv} is
among the main physics objectives of the planned Electron-Ion Collider (EIC) at BNL, which
will be able to measure a large variety of different final states in polarized electron-proton
collisions~\cite{Accardi:2012qut}.
Semi-inclusive production of identified hadrons in deep inelastic scattering (SIDIS)
will play a major role~\cite{AbdulKhalek:2021gbh,Aschenauer:2019kzf} in the EIC spin physics program.

Compared to
fully inclusive deep inelastic  scattering (DIS), SIDIS offers more detailed information on the
quark flavour decomposition of the proton's partonic content, described in terms of unpolarized and polarized
parton distribution functions (PDFs). This information is particularly valuable in the determination of
polarized PDFs,
which have to rely on a much more restricted and less diverse data set than their
unpolarized counterparts.
The determination of polarized PDFs from a global fit to experimental data mainly relied on a
next-to-leading order (NLO) QCD framework until very recently. The
availability of  next-to-next-to-leading order (NNLO)
QCD corrections to polarized Altarelli-Parisi splitting
functions~\cite{Moch:2014sna,Blumlein:2021enk,Blumlein:2021ryt}, polarized inclusive
DIS~\cite{Zijlstra:1993sh,Blumlein:2022gpp} and SIDIS~\cite{Bonino:2024wgg,Goyal:2024tmo,Goyal:2024emo}
 coefficient functions is now enabling a leap in theoretical accuracy in polarized PDF fits towards
 NNLO~\cite{Bertone:2024taw,Borsa:2024mss}.

Compared to previous generations of fixed-target experiments, the EIC will probe SIDIS at considerably larger
virtualities $Q^2$ of the exchanged gauge boson, thereby enhancing contributions that are mediated
by the massive electroweak gauge bosons $Z,W^\pm$ over the dominant photon exchange. The $Z$ exchange
modifies the chirality structure in the neutral current (giving rise for example to
single lepton
spin asymmetries on an unpolarized nucleon target), while the $W^\pm$ gives rise to entirely new observables in
charged-current SIDIS. In a previous study, we have investigated these electroweak tree-level effects
on SIDIS observables on unpolarized nucleons~\cite{Bonino:2025qta}, including NNLO QCD corrections
that complemented earlier NNLO QCD results
on photon-mediated SIDIS~\cite{Bonino:2024qbh,Goyal:2023zdi}.

In the present work, we compute the NNLO QCD corrections to neutral and charged current SIDIS on
longitudinally polarized targets and perform detailed phenomenological studies of the polarized SIDIS cross sections
and on single and double spin asymmetries derived from them. The paper is structured as
follows. In Section~\ref{sec:kin}, we review the kinematics of SIDIS and introduce the decomposition of
the unpolarized and polarized SIDIS cross sections in terms of structure functions. These structure
functions are expressed in terms of perturbatively calculable parton-level coefficient functions, which
are systematically decomposed according to the incoming and outgoing parton types in Section~\ref{sec:coeff}.
We describe the computation of the NNLO QCD corrections to these coefficient functions in
Section~\ref{sec:calculation}. Extensive phenomenological studies of SIDIS on polarized targets for neutral current
and charged current processes are performed in Section~\ref{sec:num} and our findings are summarised
in Section~\ref{sec:conc}.
Appendix~\ref{sec:Scheme_transform} documents the mass factorization and polarized
scheme transformation of the coefficient functions with particular emphasis on
the structure of the quark flavour basis.
Appendix~\ref{sec:files} guides the reader through the content and the notation in the ancillary file containing the analytic expressions of the coefficient functions up to NNLO.

\section{Kinematics of SIDIS}
\label{sec:kin}
\begin{figure}[tb]
\ifx\nodiags\undefined
	\centering
	\includegraphics{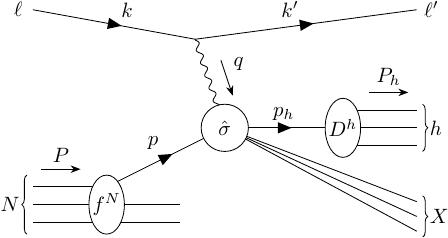}
\fi
\caption{Kinematics of semi-inclusive deep-inelastic scattering}
\label{fig:SIDIS_had}
\end{figure}

We consider the production of a final-state hadron $h$ from the scattering of a highly energetic lepton $\ell$ on a nucleon $N$, $\ell(k) N(P) \to \ell'(k') h(P_h) X$, with $X$ denoting any additional final-state radiation, Figure~\ref{fig:SIDIS_had},
and employ the commonly used kinematic variables
\begin{align*}
&q=k-k' &&\text{four-momentum of the exchanged virtual vector boson,} \\
&Q^2=-q^2 > 0 && \text{virtuality of the exchanged virtual vector boson,} \\
&x=\frac{Q^2}{2P\cdot q} &&\text{Bjorken variable,}  \\
&z=\frac{P\cdot P_h}{P\cdot q}  && \begin{tabular}{@{}p{0.7\textwidth}}\text{scaling variable related to the momentum fraction
picked up} \\ \text{by the identified hadron,} \end{tabular}  \\
&y=\frac{P\cdot q}{P\cdot k} &&\text{energy transfer to the virtual vector boson,}  \\
&W^2=(P+q)^2 &&\text{invariant mass of the hadronic final-state.} \numberthis
\label{eq:hadronic_kinematics}
\end{align*}

The above variables are related to the squared centre-of-mass energy of the lepton-nucleon system, $s = (P+k)^2$, through $Q^2=x y s$. Semi-inclusive ($z$-differential) deep-inelastic scattering describes the kinematic regime $Q^2, W^2 \gg M_N^2$, with $M_N$ being the mass of the incoming nucleon.

\subsection{Hadronic tensor decomposition}
At leading order in the electroweak theory a single gauge boson is exchanged.
The leptonic and hadronic parts
consequently factorise, and the cross section is given by the general form
\begin{align}
\frac{\dd^3\sigma^h}{\dd x \, \dd y \, \dd z} = \frac{2\pi y \alpha^2}{Q^4} \sum_k \eta_k L_{k}^{\mu\nu} W_{\mu\nu}^{h,k},
\label{eq:lept_had}
\end{align}
with $\alpha$ being the QED coupling constant.
For neutral-current ($\NC$) interactions, the initial and final state leptons have equal flavour. The neutral current is mediated by $\gamma$ or $Z$ exchange at amplitude level, which can interfere at cross section level.
The index $k \in \{\gamma\gamma, ZZ, \gamma Z\}$ denotes the resulting different contributions.
Depending on the exchanged current, the cross section is modified by
\begin{align}
\eta_{\gamma\gamma} &= 1 \, , &
\eta_{\gamma Z} & = \left(\frac{G_FM_Z^2}{2\sqrt{2}\pi \alpha}\right) \left(\frac{Q^2}{Q^2+M_Z^2}\right)\, ,&
\eta_{ZZ} = \eta_{\gamma Z}^2 \,  ,
\end{align}
with the Fermi coupling $G_F$ and the $Z$ boson mass $M_Z$.
In charged-current ($\CC$) interactions the flavours of the initial and final state leptons are different.
There are four different leptonic transitions, $\ell^-\to \nu_\ell$ and $\bar{\nu}_\ell\to \ell^+$ with a mediating $W^-$, and $\ell^+\to \bar{\nu}_\ell$ and $\nu_\ell \to \ell^-$ with a mediating $W^+$.
Our convention for the charge of the mediating $W$ boson is given by the electric charge flow from the lepton to the quark line.
Again, the total cross section takes the form of~\eqref{eq:lept_had} with
\begin{align}
\eta_W = \frac{1}{2}\left( \frac{G_F M_W^2}{4\pi \alpha}\frac{Q^2}{Q^2+M_W^2}\right)^2 \, ,
\end{align}
with $M_W$ denoting the $W$ boson mass.

The leptonic tensor $L^{\mu\nu}$ for an initial state lepton $\ell$ of lepton number $l_{\ell}=\pm 1$ and helicity $\lambda_{\ell}=\pm 1$ is given by
\begin{align}
& L_{\mu\nu}^{\gamma\gamma} = 2\left( k_\mu^{\phantom{\prime}} k_\nu^\prime + k_\mu^\prime k_\nu^{\phantom{\prime}} - (k\cdot k^\prime - m_\ell^2) g_{\mu\nu}^{\phantom{\prime}} - i \lambda_{\ell} \varepsilon_{\mu\nu\rho\sigma}^{\phantom{\prime}}k^\rho k^{\prime \sigma}\right)\, , \nonumber \\
& L_{\mu\nu}^{\gamma Z} = \left( g_V^\ell - l_{\ell} \lambda_{\ell} g_A^\ell \right) L_{\mu\nu}^{\gamma\gamma} \,
 , \quad L_{\mu\nu}^{ZZ} = (g_V^\ell - l_{\ell} \lambda_{\ell} g_A^\ell)^2 L_{\mu\nu}^{\gamma\gamma} \, , \quad
L_{\mu\nu}^{W} =& (1 - l_{\ell}\lambda_{\ell})^2 L_{\mu\nu}^{\gamma\gamma} \, .
\end{align}
with
\begin{align}
g_V^{e,\mu,\tau} =& -\frac{1}{2} +2\sin^2\theta_\mathrm{W}\, , &
g_A^{e,\mu,\tau} =& -\frac{1}{2}\, , &
g_V^{\nu,\bar{\nu}} &= g_A^{\nu,\bar{\nu}} = \frac{1}{2} \, .
\end{align}

The hadronic tensor $W_{\mu\nu}^{h,k}$ for a resolved final state hadron $h$ is defined by the current insertion of the electromagnetic or electroweak current belonging to $k$,
\begin{align}
W^{h,k}_{\mu\nu}
= \frac{1}{4\pi} \int \dd^4r \, e^{iq\cdot r} \sum\limits_{X} \Braket{P,S | J_{\mu,k}^\dagger (r) | h,X } \Braket{ h,X | J_{\nu,k}^{\phantom{\dagger}}(0) | P,S}
\end{align}
for a nucleon with momentum $P$ and spin vector $S$ with $P^2=M^2$, $S^2=-M^2$, and $S\cdot P = 0$.

The hadronic tensor is decomposed into structure functions by~\cite{ParticleDataGroup:2024cfk,deFlorian:2012wk}
\begin{align}
W_{\mu \nu}^{h,k} =& \bigg(-g_{\mu \nu} + \frac{q_\mu q_\nu}{q^2}\bigg) \Fcal_1^{h,k}(x,z,Q^2) + \frac{\hat{P}_\mu \hat{P}_\nu}{P\cdot q} \Fcal_2^{h,k}(x,z,Q^2) - i \varepsilon_{\mu\nu\rho\sigma} \frac{q^{\rho} P^\sigma}{2P\cdot q} \Fcal_3^{h,k}(x,z,Q^2) \nonumber \\
& + i \varepsilon_{\mu\nu\rho\sigma} \frac{q^\rho }{P\cdot q} \left[ S^\sigma g_1^{h,k}(x,z,Q^2) + \left(S^\sigma -\frac{S\cdot q}{P\cdot q} \hat{P}^\sigma \right) g_2^{h,k}(x,z,Q^2)\right] \nonumber \\
&+ \frac{1}{P\cdot q} \left[ \frac{1}{2} \left(\hat{P}_\mu \hat{S}_\nu + \hat{S}_\mu \hat{P}_\nu\right) - \frac{S\cdot q}{P\cdot q} \hat{P}_\mu \hat{P}_\nu \right] g_3^{h,k}(x,z,Q^2) \nonumber \\
&+ \frac{S\cdot q}{P\cdot q} \left[ \frac{\hat{P}_\mu \hat{P}_\nu}{P\cdot q} g_4^{h,k}(x,z,Q^2) + \left( -g_{\mu\nu}+ \frac{q_\mu q_\nu}{q^2} \right) g_5^{h,k}(x,z,Q^2) \right]
\label{eq:had_tens}
\end{align}
with
\begin{align}
\hat{P}_\mu=& P_\mu - \frac{P\cdot q}{q^2}q_\mu \, , &
\hat{S}_\mu=& S_\mu - \frac{S\cdot q}{q^2}q_\mu\, .
\end{align}
In the following we assume that the nucleon spin vector $S$ is aligned to the momentum $k$ of the incident lepton in the nucleon rest frame.
In the massless limit $M^2/Q^2 \to 0$ the structure functions $g_2^h$ and $g_3^h$ then vanish at leading twist.

The spin-independent structure functions $\Fcal_1^{h,k},\Fcal_2^{h,k},\Fcal_3^{h,k}$ are accessible in spin-averaged collisions, whereas the spin-dependent structure functions $g_1^{h,k},\dots,g_5^{h,k}$ vanish when averaging over the nucleon spin.
Instead, these structure functions are only accessible in polarized DIS experiments within spin differences.

\subsection{Spin-independent cross section}

Instead of $\Fcal_1^{h,k}$ and $\Fcal_2^{h,k}$ we express our results in terms of the transverse and longitudinal structure functions $\Fcal_T^{h,k} = 2x\, \Fcal_1^{h,k}$ and $\Fcal_L^{h,k} = \Fcal_2^{h,k} - 2x \, \Fcal_1^{h,k} \,$.
The structure functions associated with an \mbox{(anti-)symmetric} tensor structure will be referred to as \mbox{(anti-)symmetric} in the following.

The structure functions depend on $x$, $z$, and $Q^2$.
In the NC case they apply to scattering processes with conserved lepton flavour.
For a charged lepton, they are obtained by summing $\gamma$, $Z$ and interference contributions:
\begin{align} \label{eq:SFNC}
\mathcal{F}_{T,L}^{h,\NC}&= \mathcal{F}_{T,L}^{h,\gamma\gamma}-\left(g^{\ell}_V - l_{\ell} \lambda_{\ell} g^{\ell}_A\right)\eta_{\gamma Z}\,\mathcal{F}_{T,L}^{h,\gamma Z}+\left(g^{\ell \, 2}_V+g^{\ell \, 2}_A - 2 l_{\ell} \lambda_{\ell} g^{\ell}_Vg^{\ell}_A\right)\eta_{ZZ}\,\mathcal{F}_{T,L}^{h,ZZ} \, , \nonumber \\
\mathcal{F}_{3}^{h,\NC}&= -\left(g^{\ell}_A - l_{\ell} \lambda_{\ell} g^{\ell}_V\right)\eta_{\gamma Z}\,\mathcal{F}_{3}^{h,\gamma Z} + \left[2 g^{\ell}_V g^{\ell}_A - l_{\ell} \lambda_{\ell} \left(g^{\ell \, 2}_V+g^{\ell \, 2}_A \right)\right]\eta_{ZZ}\, \mathcal{F}_{3}^{h,ZZ}\,  .
\end{align}
The lepton-flavour changing $\CC$ structure functions are given by
\begin{align}
\mathcal{F}_{T,L}^{h,\CC}&= \mathcal{F}_{T,L}^{h,W^{\pm}} \, , &
\mathcal{F}_{3}^{h,\CC}&= \mathcal{F}_{3}^{h,W^{\pm}} \,  .
\end{align}
The neutral current/charged current SIDIS cross section in terms of these structure functions is then given by
\begin{align}\label{eq:xsNCCC}
\frac{\dd^3 \sigma^{h,\NC / \CC}}{\dd x\,\dd y\,\dd z} = \frac{4\pi \alpha^2}{Q^2} \eta^{\NC/\CC} \Bigg[  \frac{1+(1-y)^2}{2y}\Fcal_T^{h,\NC / \CC}+\frac{1-y}{y}\Fcal_L^{h,\NC / \CC} & \nonumber \\
+ l_{\ell} \frac{1-(1-y)^2}{2y}\Fcal_3^{h,\NC / \CC}&  \Bigg] \,  .
\end{align}
Furthermore $\eta^{\NC}=1$, whereas $\eta^{\CC}=(1 - l_{\ell}\lambda_{\ell})^2\eta_W$, with the helicity of the (anti-)lepton $\lambda_{\ell}=\pm 1$.

\subsection{Spin-dependent cross section}

Instead of expressing our results in terms of $g^h_1$, $g^h_4$, and $g^h_5$ as in~\cite{deFlorian:2012wk} we introduce the structure functions $\Gcal^h_1=2g^h_1$, $\Gcal^h_T=g^h_5$, and $\Gcal^h_L=(g^h_4-2x g_5^h)/x$, to align with the unpolarized structure functions in~\eqref{eq:xsNCCC}.

The polarized cross section difference $\Delta \sigma = \sigma(\lambda_N=-1, \lambda_\ell) - \sigma(\lambda_N=1,\lambda_\ell)$, where $\lambda_N$ is the helicity of the nucleon, is then expressed by 
\begin{align} \label{eq:SFNCpol}
\frac{\dd^3 \Delta \sigma^{h,\NC/\CC}}{\dd x \, \dd y \, \dd z} =&  \frac{4\pi \alpha^2}{Q^2} \eta^{\NC/\CC} \Big[ l_{\ell}\frac{1-(1-y)^2}{y} \Gcal_1^{h,\NC/\CC}(x,z,Q^2) \nonumber \\
&-\frac{1+(1-y)^2}{y} 2 \Gcal_T^{h,\NC/\CC}(x,z,Q^2) - \frac{1-y}{y}2 \Gcal_L^{h,\NC/\CC}(x,z,Q^2) \Big] \, .
\end{align}
For NC SIDIS the structure functions consist of photon and $Z$ contributions and their interference
\begin{align*}
\Gcal_1^{h,\NC}&=-l_{\ell}\lambda_{\ell} \Gcal_1^{h,\gamma\gamma}-\left(g_A^{\ell} - l_{\ell}\lambda_{\ell} g_V^{\ell}\right)\eta_{\gamma Z}\, \Gcal_1^{h,\gamma Z}+\left(2g_V^{\ell}g^{\ell}_A - l_{\ell}\lambda_{\ell} \left(g^{\ell \, 2}_V+g^{\ell \, 2}_A\right)\right)\eta_{ZZ}\, \Gcal_1^{h,ZZ} \, , \\
\Gcal_{T,L}^{h,\NC}&=-\left(g_V^{\ell} - l_{\ell}\lambda_{\ell} g_A^{\ell}\right)\eta_{\gamma Z} \, \Gcal_{T,L}^{h,\gamma Z}+\left(g^{\ell \, 2}_V+g^{\ell \, 2}_A - 2 l_{\ell} \lambda_{\ell} g_V^{\ell}g^{\ell}_A\right)\eta_{ZZ}\, \Gcal_{T,L}^{h,ZZ} \, . \numberthis
\end{align*}
The lepton flavour changing $\CC$ structure functions are given by
\begin{align}
\Gcal_{1}^{h,\CC}&= \Gcal_{1}^{h,W^{\pm}} \, , &
\Gcal_{T,L}^{h,\CC}&= \Gcal_{T,L}^{h,W^{\pm}} \,  .
\end{align}

\subsection{Parton-model expressions}

Factorization in SIDIS separates the cross section into three pieces:
\begin{enumerate}[label=(\roman*)]
\item Long-range initial-state interactions, with a parton $p$ extracted from the incoming nucleon $N$ according to a collinear parton distribution function (PDF) $f_p^N$ evaluated at the initial state factorization scale $\mu_F$.
\item Long-range final-state interactions, with a parton $p'$ fragmenting into the detected hadron $h$ through the fragmentation function (FF) $D_{p'}^h$ evaluated at the final state factorization scale $\mu_A$.
\item The short-distance cross section $\hat{\sigma}$ at the hard scale $Q$, which is calculated at the partonic level in perturbative QCD. The renormalization of $\hat{\sigma}$ introduces a third scale, the renormalization scale $\mu_R$.
\end{enumerate}
This factorization is expressed at the level of the structure functions:
\begin{align}\label{eq:Fcal_Gcal}
\Fcal_i^{h,k}(x,z,Q^2) =& \sum_{p,p'}\int_x^1 \frac{\dd \xh}{\xh} \int_z^1 \frac{\dd \zh}{\zh} \, f_p^N\bigg(\frac{x}{\xh},\mu_F^2\bigg)\, D^h_{p'}\bigg(\frac{z}{\zh},\mu_A^2\bigg) \, \Ccal^{i,k}_{p'p}(\xh,\zh,Q^2,\mu_R^2,\mu_F^2,\mu_A^2) \, , \nonumber \\
\Gcal_j^{h,k}(x,z,Q^2) =& \sum_{p,p'}\int_x^1 \frac{\dd \xh}{\xh} \int_z^1 \frac{\dd \zh}{\zh} \, \Delta f_p^N\bigg(\frac{x}{\xh},\mu_F^2\bigg)\, D^h_{p'}\bigg(\frac{z}{\zh},\mu_A^2\bigg) \, \Delta \Ccal^{j,k}_{p'p}(\xh,\zh,Q^2,\mu_R^2,\mu_F^2,\mu_A^2) \,
\end{align}
where $(\Delta)\Ccal^{i,k}_{p'p}$ denotes the (polarized) parton-level coefficient function.
The indices $i\in\{T,L,3\}$ refer to the respective unpolarized coefficient function, $j\in\{1,T,L\}$ to the respective polarized coefficient function, and $k\in\{\gamma\gamma, \gamma Z, ZZ, W^\pm \}$ refers to the electroweak gauge bosons exchanged in the process.
For this purpose, we introduce the partonic kinematics in analogy with~\eqref{eq:hadronic_kinematics} as
\begin{align*}
&\xh=\frac{Q^2}{2p\cdot q} &&\text{momentum fraction of the initial state parton,}  \\
&\zh=\frac{p\cdot p_h}{p\cdot q}  &&\text{momentum fraction
picked up by the identified parton.} \numberthis
\label{eq:partonic_kinematics}
\end{align*}
The coefficient functions describing the hard scattering process can be calculated in perturbation theory by expanding in the strong coupling constant $\alpha_s$,
\begin{align}
(\Delta)\Ccal^{i,k}_{p'p}=(\Delta)\Ccal^{i,k, (0)}_{p'p}+\frac{\alpha_s(\mu_R^2)}{2\pi} (\Delta)\Ccal^{i,k, (1)}_{p'p}+\left(\frac{\alpha_s(\mu_R^2)}{2\pi}\right)^2 (\Delta)\Ccal^{i,k, (2)}_{p'p} +\mathcal{O}(\alpha_s^3)\,  .
\end{align}

In a previous publication \cite{Bonino:2025qta} we described the unpolarized electroweak coefficient functions for unpolarized SIDIS up to NNLO.
In the following sections we present the individual contributions to the \emph{longitudinally polarized} SIDIS coefficient functions up to NNLO.

\section{SIDIS Coefficient Functions up to NNLO}
\label{sec:coeff}
\subsection{Electroweak Channel Decomposition}

For the polarized process we employ the same channel decomposition as for the unpolarized case,
\begin{alignat*}{3}
& (\Delta)\Ccal^{i,k,(2)}_{qq} &&= (\Delta)C_{qq}^{i,k,(2)} + (\Delta)C^{i,k,(2)}_{qq, \Fcon} + (\Delta)\Ccal^{i,k,(2)}_{qq,\NoW} \hspace{-.52pt} + \left. (\Delta)C^{i,k,(2)}_{\qp q} \right|_{\qp=q} \hspace{-1pt} + \left. (\Delta)C^{i,k,(2)}_{\qp q, \NoW} \right|_{\qp=q} \, \hspace{-2pt}  , \\
& (\Delta)\Ccal^{i,k,(2)}_{\qb q} &&= (\Delta)C^{i,k,(2)}_{\qb q} + (\Delta)C^{i,k,(2)}_{\qb q, \Fcon} + \left. (\Delta)C^{i,k,(2)}_{\qbp q} \right|_{\qbp = \qb} + \left. (\Delta)C^{i,k,(2)}_{\qbp q, \NoW} \right|_{\qbp = \qb} \,  , \\
& (\Delta)\Ccal^{i,k,(2)}_{\qp q} &&=  (\Delta)C^{i,k,(2)}_{\qp q} + (\Delta)C^{i,k,(2)}_{\qp q, \NoW} && \hspace{-8cm} (\text{for }\qp \neq q) \,  , \\
& (\Delta)\Ccal^{i,k,(2)}_{\qbp q} &&=  (\Delta)C^{i,k,(2)}_{\qbp q} + (\Delta)C^{i,k,(2)}_{\qbp q, \NoW} && \hspace{-8cm} (\text{for }\qbp \neq \qb) \, , \\
& (\Delta)\Ccal^{i,k,(2)}_{gq} &&=  (\Delta)C^{i,k,(2)}_{gq} + (\Delta)C^{i,k,(2)}_{gq,\NoW} \, , \\
& (\Delta)\Ccal^{i,k,(2)}_{qg} &&=  (\Delta)C^{i,k,(2)}_{qg} + (\Delta)C^{i,k,(2)}_{qg,\NoW} \, , \\
& (\Delta)\Ccal^{i,k,(2)}_{gg} &&= (\Delta)C^{i,k,(2)}_{gg} + (\Delta)C^{i,k,(2)}_{gg,\NoW}  \numberthis \, ,
 \label{eq:channel_decomp}
\end{alignat*}
where $q,\qp$ denote quarks and $\qb,\qbp$ denote anti-quarks.
A detailed discussion of this channel decomposition can be found in~\cite{Bonino:2025qta}.

In Section~\ref{sec:pol_NC} the precise composition of the $\Delta \Ccal^{i,k,(2)}_{p p'}$ for $\NC$ exchange in terms of unique building blocks is discussed.
In Section~\ref{sec:pol_CC} we use these building blocks to assemble the $\CC$ coefficient functions.

\subsection{Neutral Current SIDIS}
\label{sec:pol_NC}

In the Standard Model the $\NC$ SIDIS coefficient functions describe the partonic hard scattering in the processes
\begin{align}
& \ell^- N \to \ell^- h X\, , & & \ell^+ N \to \ell^+ h X \, , & &  \nu_\ell N \to \nu_\ell h X \, , & & \bar{\nu}_\ell N \to \bar{\nu}_\ell h X\,.
\end{align}
In the following we establish the notation and present the coefficient functions for NC SIDIS up to NNLO in QCD.

The $\NC$ structure functions are present in photon and $Z$ boson exchange, and in their interference ($k\in\{\gamma \gamma , ZZ ,\gamma Z\}$).
We introduce generic vector or axial vector couplings
\begin{align}
v^{\gamma}_q &=e_q^{\phantom{\gamma}} \,, & v^{Z}_q &= g^q_V \,, &
a^{\gamma}_q &=0\,, & a^{Z}_q &= g^q_A\,,
\end{align}
with
\begin{align*}
e_u &= \phantom{+}\frac{2}{3}\, , &
g^u_V &= \phantom{+}\frac{1}{2} - \frac{4}{3}\sin^2\theta_\mathrm{W} \, , &
g^u_A &= \phantom{+}\frac{1}{2} \, , \\
e_d &= -\frac{1}{3} \, , &
g^d_V &= -\frac{1}{2} + \frac{2}{3}\sin^2\theta_\mathrm{W} \, ,&
g^d_A &= -\frac{1}{2} \numberthis
\end{align*}
for up- and down-like quarks with the weak mixing angle $\sin^2 \theta_\mathrm{W}$.
To write the coefficient functions for $\NC$ SIDIS in uniform notation, we define
\begin{align*}
\vh^{\gamma\gamma}_{q\qp} &=e_q e_\qp \,, & \vh^{\gamma Z}_{q\qp} &= e_q g_V^{\qp} + e_\qp g_V^q \, , & \vh^{ZZ}_{q\qp} &= g^q_V\, g^{\qp}_V\,, \\
\ah^{\gamma\gamma}_{q\qp} &=0\,, & \ah_{q\qp}^{\gamma Z} &= 0 \, , & \ah^{ZZ}_{q\qp} &= g^q_A \, g^{\qp}_A \,, \\
\ih^{\gamma\gamma}_{q\qp} &=0\,, & \ih^{\gamma Z}_{q\qp} &= 2 e_q^{\phantom{q}} g^{\qp}_A\,,  & \ih^{ZZ}_{q\qp} &= g^q_V g^{\qp}_A + g^{\qp}_V g^q_A  \numberthis
\label{eq:couplings}
\end{align*}
for two arbitrary quarks $q,\qp$.
Here, $\vh$ denotes combinations of vector couplings, $\ah$ denotes combinations of axial vector couplings, and $\ih$ denotes interferences between vector and axial vector couplings.
Note that $\ih_{q\qp}^{\gamma Z}$ is not symmetric in the quark indices: $\ih_{q\qp}^{\gamma Z}\neq \ih_{\qp q}^{\gamma Z}$.

\subsubsection{Antisymmetric Polarized Coefficient Functions}\label{sec:polNCAntisymmCF}
In this section we describe the neutral current (NC) polarized coefficient functions for the $\Gcal_1^h$ structure function up to NNLO.
We follow the notation of the unpolarized symmetric NC coefficient functions in~\cite{Bonino:2025qta}.

At LO only the $q\to q$ channel contributes,
\begin{align*}
\Delta\Ccal^{1,k,(0)}_{qq} = \left[ \vh_{qq}^k + \ah_{qq}^k \right] \DC^{1,(0)}_{qq} \,  , \numberthis
\end{align*}
with $\DC^{1,(0)}_{qq} = \delta(1-\zh)\delta(1-\xh)$.
The $\Gcal^h_1$ structure function at LO is therefore given by
\begin{alignat*}{2}
& \Big[ \Gcal_{1}^{h,\gamma\gamma},~ \Gcal_{1}^{h,\gamma Z},~ \Gcal_{1}^{h,ZZ} \Big] &&= \sum_{q} \Big[ e_q^2,~ 2 \, e_q \, g_V^q ,~ (g_V^q)^2 + (g_A^q)^2 \Big] \big( \Delta f_q^N \, D_q^h + \Delta f_{\qb}^N \, D_{\qb}^h \big) \, . \numberthis
\end{alignat*}

The polarized SIDIS coefficient functions at NLO were first computed in \cite{deFlorian:2012wk}.
At NLO the $q\to g$ and $g\to q$ channels open,
\begin{align*}
\Delta\Ccal^{1,k,(1)}_{qq} &=  \left[ \vh_{qq}^k + \ah_{qq}^k \right] \DC^{1,(1)}_{qq} \,  , \\
\Delta\Ccal^{1,k,(1)}_{gq} &= \left[ \vh_{qq}^k + \ah_{qq}^k \right] \DC^{1,(1)}_{gq} \,  , \\
\Delta\Ccal^{1,k,(1)}_{qg} &=  \left[ \vh_{qq}^k + \ah_{qq}^k \right] \DC^{1,(1)}_{qg} \,  . \numberthis
\end{align*}

The contributions entering the right hand side of~\eqref{eq:channel_decomp} are given in terms of the electroweak couplings from~\eqref{eq:couplings},
multiplying building blocks of different diagrammatic origin:
\begin{alignat*}{2}
&\DC^{1,k,(2)}_{qq} && = \left[ \vh_{qq}^k + \ah_{qq}^k \right]\DC^{1,(2)}_{qq}  \,  , \\
&\DC^{1,k,(2)}_{qq, \Fcon} &&= \left[ \vh_{qq}^k + \ah_{qq}^k \right] \DC^{1,(2)}_{qq,\Fcon,1} + \sum_{\qp} \left[ \vh_{\qp\qp}^k + \ah_{\qp\qp}^k \right] \DC^{1,(2)}_{qq,{\Fcon},2}  \,  , \\
&\DC^{1,k,(2)}_{qq,\NoW} &&= a^k_q \sum_{\qp} \left(a^k_{\qp}\right) \DC^{1,(2)}_{qq, \aSa,\NoW}  \,  , \\
&\DC^{1,k,(2)}_{\qb q} && = \left[ \vh_{qq}^k + \ah_{qq}^k \right]\DC^{1,(2)}_{\qb q}  \,  , \\
&\DC^{1,k,(2)}_{\qb q, \Fcon} &&= \left[ \vh_{qq}^k + \ah_{qq}^k \right] \DC^{1,(2)}_{\qb q, \Fcon}  \,  , \\
&\DC^{1,k,(2)}_{\qp q} && =  \left[ \vh_{qq}^k + \ah_{qq}^k \right]\DC^{1,(2)}_{\qp q, 1} + \left[ \vh_{\qp\qp}^k + \ah_{\qp\qp}^k \right] \DC^{1,(2)}_{\qp q, 2} \,  , \\
&\DC^{1,k,(2)}_{\qp q, \NoW} &&= \vh_{q\qp}^k \, \DC^{1,(2)}_{\qp q, \NoW, 3} + \ah_{q\qp}^k \, \DC^{1,(2)}_{\qp q, \NoW, 4}  \,  , \\
&\DC^{1,k,(2)}_{\qbp q} && = \left[ \vh_{qq}^k + \ah_{qq}^k \right]\DC^{1,(2)}_{\qp q, 1} + \left[ \vh_{\qp\qp}^k + \ah_{\qp\qp}^k \right]\DC^{1,(2)}_{\qp q, 2} \,  , \\
&\DC^{1,k,(2)}_{\qbp q, \NoW} &&= - \vh_{q\qp}^k \, \DC^{1,(2)}_{\qp q,\NoW, 3} + \ah_{q\qp}^k \, \DC^{1,(2)}_{\qp q,\NoW, 4}  \,  , \\
&\DC^{1,k,(2)}_{gq} &&= \left[ \vh_{qq}^k + \ah_{qq}^k \right] \DC^{1,(2)}_{gq} \, ,  \\
&\DC^{1,k,(2)}_{gq,\NoW} &&= a^k_q \sum_{q'} \left( a^k_{q'} \right) \DC^{1,(2)}_{gq, \aSa,\NoW}  \,  , \\
&\DC^{1,k,(2)}_{qg} &&= \left[ \vh_{qq}^k + \ah_{qq}^k \right] \DC^{1,(2)}_{qg} \, ,  \\
&\DC^{1,k,(2)}_{qg,\NoW} &&= a^k_q \sum_{q'} \left( a^k_{q'} \right) \DC^{1,(2)}_{q g, \aSa,\NoW}  \,  , \\
&\DC^{1,k,(2)}_{gg} &&=\sum_{q'} \left[ \vh_{\qp\qp}^k + \ah_{\qp\qp}^k \right] \DC^{1,(2)}_{gg} \,, \\
&\DC^{1,k,(2)}_{gg,\NoW} &&= \left[ \sum_{\qp} a_{\qp}^k \right]^2 \DC^{1,(2)}_{gg,\SaSa,\NoW} \, . \numberthis
\label{eq:polNC_CoeffF_Full}
\end{alignat*}
Note that $\DC^{1,(2)}_{gg,\SaSa,\NoW}$ in~\eqref{eq:polNC_CoeffF_Full} is inherently symmetric and vanishes when contracted with the antisymmetric projector onto $\Gcal^h_1$.
As no contribution arises from the vector-axial vector interferences, the entire channel $\DC^{i,k,(2)}_{gg,\NoW}$ in~\eqref{eq:channel_decomp} is zero.

The missing coefficient functions with anti-quarks in the initial state are obtained by charge conjugation,
\begin{equation}
\Delta\Ccal^{1,k,(j)}_{p'p} = \Delta\Ccal^{1,k,(j)}_{\bar{p}'\bar{p}}\quad \mathrm{for} \quad j \in \{0,1,2\} \,  .
\end{equation}

The contributions with both electroweak couplings connecting to the same quark line coincide for vector-vector and axial-vector-axial-vector coupling as is expected from na\"ively anticommuting $\gamma_5$.

If the axial vector vertices appear on different fermion lines, such an equality is no longer expected and also not realised, cf.\ $\Delta C_{\qp q,\NoW,3}^{1,(2)} \neq \Delta C_{\qp q,\NoW,4}^{1,(2)}$.
The building blocks $\DC^{1,(2)}_{\qp q, 1}$, $\DC^{1,(2)}_{\qp q, 2}$, $\DC^{1,(2)}_{\qp q,\NoW, 3}$, and $\DC^{1,(2)}_{\qp q,\NoW, 4}$ from the $q\to \qp$ channel reemerge in the $q\to \qbp$ channel, which is related by charge conjugation applied to the secondary quark line.


We note that unlike in inclusive DIS, the pure-singlet contributions are  not restricted to a single channel.
The appearance of pure-singlet splitting functions in the mass factorization of $\Delta C^{1,(2)}_{\qp q,2}$ associates the corresponding coefficient functions to the pure-singlet of SIDIS.
The contribution $\Delta C^{1,(2)}_{qq,\Fcon,2}$, which also constitutes a part of the pure-singlet, is finite and does not require mass factorization.

\subsubsection{Symmetric Polarized Coefficient Functions}\label{sec:polNCSymmCF}

In this section we present the first calculation of the NNLO NC coefficient functions for the structure functions $\Gcal_T^h$ and $\Gcal_L^h$.
These structure functions describe the interference contributions between vector and axial vector coupling.
In $\NC$ SIDIS they contribute for $k\in \{\gamma Z, ZZ\}$.

At LO only the $q\to q$ channel contributes,
\begin{align*}
\Delta\Ccal^{\{T,L\},k,(0)}_{qq} = \ih_{qq}^k \DC^{\{T,L\},(0)}_{qq} \,  , \numberthis
\end{align*}
with $\DC^{T,(0)}_{qq} = -\frac{1}{2} \delta(1-\zh)\delta(1-\xh)$ and $\DC^{L,(0)}_{qq} = 0$.
The polarized symmetric LO structure functions are given by
\begin{alignat*}{2}
& \Big[ \Gcal_{T}^{h,\gamma\gamma},~ \Gcal_{T}^{h,\gamma Z},~ \Gcal_{T}^{h,ZZ} \Big] &&= \frac{1}{2} \sum_{q} \Big[ 0,~ 2 \, e_q \, g_A^q ,~ 2 g_V^q g_A^q \Big] \big(  \Delta f_{\qb}^N \, D_{\qb}^h - \Delta f_q^N \, D_q^h \big) \, , \\
& \Big[ \Gcal_{L}^{h,\gamma\gamma},~ \Gcal_{L}^{h,\gamma Z},~ \Gcal_{L}^{h,ZZ} \Big] &&= 0 \, . \numberthis
\end{alignat*}

At NLO the $qg$ and $gq$ channels open and the longitudinal components do not vanish any more,
\begin{align*}
\Delta\Ccal^{\{T,L\},k,(1)}_{qq} &=  \ih_{qq}^k \DC^{\{T,L\},(1)}_{qq} \,  , \\
\Delta\Ccal^{\{T,L\},k,(1)}_{gq} &=  \ih_{qq}^k \DC^{\{T,L\},(1)}_{gq} \,  , \\
\Delta\Ccal^{\{T,L\},k,(1)}_{qg} &=  \ih_{qq}^k \DC^{\{T,L\},(1)}_{qg} \,  . \numberthis
\end{align*}
The $\Delta\Ccal^{\{T,L\},(1)}_{p'p}$ coefficient functions above were computed up to NLO in~\cite{deFlorian:2012wk}.
Note that $\Delta\Ccal^{L,k,(1)}_{qg} = 0$.

The missing coefficient functions with anti-quarks in the initial state are obtained by charge conjugation through
\begin{equation}
\Delta\Ccal^{\{T,L\},k,(j)}_{\bar{p}'\bar{p}} = - \Delta\Ccal^{\{T,L\},k,(j)}_{p'p} \quad \mathrm{for} \quad j \in \{0,1,2\} \,  .
\end{equation}

At NNLO, the contributions appearing in~\eqref{eq:channel_decomp} in terms of elementary building blocks read as follow:
\begin{alignat*}{2}
&\DC^{\{T,L\},k,(2)}_{qq} && = \ih_{qq}^k \DC^{\{T,L\},(2)}_{qq}  \,  , \\
&\DC^{\{T,L\},k,(2)}_{qq, \Fcon} &&= \ih_{qq}^k \DC^{\{T,L\},(2)}_{qq, \Fcon, 1}  \,  , \\
&\DC^{\{T,L\},k,(2)}_{qq,\NoW} &&= 2 v^k_q \sum_{q'} \left( a_{q'}^k \right) \DC^{\{T,L\},(2)}_{qq, \vSa,\NoW}  \,  , \\
&\DC^{\{T,L\},k,(2)}_{\qb q} && = \ih_{qq}^k \DC^{\{T,L\},(2)}_{\qb q}  \,  , \\
&\DC^{\{T,L\},k,(2)}_{\qb q,\Fcon} &&= \ih_{qq}^k \DC^{\{T,L\},(2)}_{\qb q, \Fcon}  \,  , \\
&\DC^{\{T,L\},k,(2)}_{\qp q} && = \ih_{qq}^k \DC^{\{T,L\},(2)}_{\qp q, 1} + \ih_{\qp\qp}^k \, \DC^{\{T,L\},(2)}_{\qp q, 2} \,  , \\
&\DC^{\{T,L\},k,(2)}_{\qp q, \NoW} &&= \ih_{q\qp}^k \DC^{\{T,L\},(2)}_{\qp q, \NoW , 3} + \ih_{\qp q}^k \DC^{\{T,L\},(2)}_{\qp q, \NoW, 4}  \,  , \\
&\DC^{\{T,L\},k,(2)}_{\qbp q} && = \ih_{qq}^k \DC^{\{T,L\},(2)}_{\qp q, 1} - \ih_{\qp\qp}^k \, \DC^{\{T,L\},(2)}_{\qp q, 2} \,  , \\
&\DC^{\{T,L\},k,(2)}_{\qbp q, \NoW} &&= \ih_{q\qp}^k \DC^{\{T,L\},(2)}_{\qp q, \NoW , 3} - \ih_{\qp q}^k \DC^{\{T,L\},(2)}_{\qbp q, \NoW , 4}  \,  , \\
&\DC^{\{T,L\},k,(2)}_{gq} &&= \ih_{qq}^k \DC^{\{T,L\},(2)}_{gq} \, , \\
&\DC^{\{T,L\},k,(2)}_{gq,\NoW} &&= 2 v^k_q \sum_{q'} \left( a^k_{q'} \right) \DC^{\{T,L\},(2)}_{gq, \vSa,\NoW}  \,  , \\
&\DC^{\{T,L\},k,(2)}_{qg} &&= \ih_{qq}^k \DC^{\{T,L\},(2)}_{qg} \, ,  \\
&\DC^{\{T,L\},k,(2)}_{qg,\NoW} &&= 2 v^k_q \sum_{q'} \left( a^k_{q'} \right) \DC^{\{T,L\},(2)}_{q g, \vSa\NoW}  \, . \numberthis
\label{eq:NC_CoeffF_gTgL_Full}
\end{alignat*}
The purely gluonic channels $\DC^{\{T,L\},k}_{gg}$ vanish to all orders with a line reversal argument like in~$\Fcal_3^h$.
The contribution $\DC^{\{T,L\},k,(2)}_{qq,\Fcon,2}$ vanishes likewise.

The contributions $\DC^{\{T,L\},(2)}_{\qp q,1}$, $\DC^{\{T,L\},(2)}_{\qp q, 2}$, $\DC^{\{T,L\},(2)}_{\qp q, \NoW, 3}$, and $\DC^{\{T,L\},(2)}_{\qp q, \NoW, 4}$ appear in two channels due to a charge conjugation symmetry on the secondary quark line ($\qp\to\qbp$).


\subsection{Charged Current SIDIS}
\label{sec:pol_CC}

We proceed with describing the production of one identified hadron in polarized charged current SIDIS, i.e.\ the processes
\begin{align}
& \ell^- N \to \nu_\ell h X \, , && \ell^+ N \to \bar{\nu}_\ell h X \, ,& & \nu_{\ell}  N \to \ell^{-} h X \,  & & \bar{\nu}_{\ell}  N\to \ell^{+} h X\, .
\end{align}
In the following we present the coefficient functions in polarized charged current SIDIS at NNLO in QCD corresponding to the symmetric and antisymmetric structure functions.

\subsubsection{Antisymmetric Polarized Coefficient Functions}\label{sec:polCCAntisymmCF}

In this section we describe the polarized CC coefficient functions contributing to the $\Gcal_1$ structure function, up to NNLO.
In $\CC$ SIDIS the allowed quark transitions are restricted by the charge of the current entering the hadronic line.
In the following we explicitly give the contributions for $W^-$ exchange.
The contributions for $W^+$ exchange can be inferred.

Following the notation of~\cite{Bonino:2025qta},
we denote the set of up-type quarks with $\mathfrak{u}=\{u,c,(t)\}$ and the set of down-type quarks with $\mathfrak{d}=\{d,s,(b)\}$.
At LO we have
\begin{align*}
\Delta\Ccal^{1,W^{-},(0)}_{\beta\alpha} = 2\left|V_{\alpha\beta}\right|^2 \DC^{1,(0)}_{qq} \,  ,  \\
\Delta\Ccal^{1,W^{-},(0)}_{\bar{\alpha}\bar{\beta}} = 2\left|V_{\alpha\beta}\right|^2 \DC^{1,(0)}_{qq} \,  , \numberthis
\end{align*}
where $\alpha \in \mathfrak{u}$ and  $\beta \in \mathfrak{d}$.
The factors of $2$ arise from summing the vector and axial vector contributions.
The coefficients $V_{\alpha\beta}$ denote the entries of the Cabibbo-Kobayashi-Maskawa (CKM) matrix
\begin{align}
V_\mathrm{CKM}=\begin{pmatrix}
V_{ud} & V_{us} & V_{ub}\\
V_{cd} & V_{cs} & V_{cb}\\
V_{td} & V_{ts} & V_{tb}
\end{pmatrix} \, ,
\end{align}
and describe the size of the flavour mixing.

At LO the antisymmetric polarized SIDIS $\CC$ structure functions are therefore given by
\begin{alignat*}{2}
& \Big[ \Gcal_{1}^{h,W^-},~ \Gcal_{1}^{h,W^+} \Big] &&= 2 \sum_{\phantom| \alpha\in \mathfrak{u} \phantom|} \sum_{\phantom| \beta  \in \mathfrak{d} \phantom| } |V_{\alpha\beta}|^2 \Big[ \Delta f_\alpha^N \, D_\beta^h + \Delta f_{\bar{\beta}}^N \, D_{\bar{\alpha}}^h ,~ \Delta f_\beta^N \, D_\alpha^h + \Delta f_{\bar{\alpha}}^N \, D_{\bar{\beta}}^h \Big] \, . \numberthis
\end{alignat*}

At NLO there exist the quark-initiated contributions
\begin{align*}
\Delta\Ccal^{1,W^{-},(1)}_{\beta\alpha} &= 2\left|V_{\alpha\beta}\right|^2 \DC^{1,(1)}_{qq} \,  , \\
\Delta\Ccal^{1,W^{-},(1)}_{g\alpha} &= 2\sum_\beta \left|V_{\alpha\beta}\right|^2 \DC^{1,(1)}_{gq} \,  , \numberthis
\end{align*}
the anti-quark-initiated contributions
\begin{align*}
\Delta\Ccal^{1,W^{-},(1)}_{\bar{\alpha}\bar{\beta}} &= \DC^{1,W^{-},(1)}_{\beta\alpha}  \,  , \\
\Delta\Ccal^{1,W^{-},(1)}_{g\bar{\beta}} & = 2 \sum_\alpha \left|V_{\alpha\beta}\right|^2 \DC^{1,(1)}_{gq} \, ,\numberthis
\end{align*}
and the gluon-initiated contributions
\begin{align*}
\Delta\Ccal^{1,W^{-},(1)}_{\beta g} &=  2\sum_\alpha \left|V_{\alpha\beta}\right|^2 \DC^{1,(1)}_{qg} \,  , \\
\Delta\Ccal^{1,W^{-},(1)}_{\bar{\alpha}g}  & =  2\sum_\beta \left|V_{\alpha\beta}\right|^2 \DC^{1,(1)}_{qg} \, . \numberthis
\end{align*}

At NNLO the quark-initiated contributions are
\begin{align*}
\DC^{1,W^-,(2)}_{\beta\alpha} & =2\left|V_{\alpha\beta} \right|^2 \DC^{1,(2)}_{qq}  \,  , \\
\DC^{1,W^-,(2)}_{qq,\Fcon} &= 2\sum_{\alpha,\beta}  \left|V_{\alpha\beta} \right|^2  \DC^{1,(2)}_{qq, \Fcon, 1} \,  , \\
\DC^{1,W^-,(2)}_{\alpha\alpha,\Fcon} &= 2\sum_\beta \left|V_{\alpha\beta} \right|^2 \DC^{1,(2)}_{qq, \Fcon}   \,  , \\
\DC^{1,W^-,(2)}_{\bar{\beta} \alpha} & = 2\left|V_{\alpha\beta} \right|^2 \DC^{1,(2)}_{\qb q}  \,  , \\
\DC^{1,W^-,(2)}_{\bar{\alpha} \alpha, \Fcon} &= 2\sum_\beta \left|V_{\alpha\beta} \right|^2 \DC^{1,(2)}_{\qb q, \Fcon}  \,  , \\
\DC^{1,W^-,(2)}_{\qp \alpha} & = 2\sum_\beta \left|V_{\alpha\beta}\right|^2 \DC^{1,(2)}_{\qp q, 1}  \,  , \\
\DC^{1,W^-,(2)}_{\beta' q} & =  2\sum_\alpha \left|V_{\alpha\beta}\right|^2 \DC^{1,(2)}_{\qp q, 2} \,  , \\
\DC^{1,W^-,(2)}_{\qbp \alpha} & = 2\sum_\beta \left|V_{\alpha\beta}\right|^2  \DC^{1,(2)}_{\qp q, 1}  \,  , \\
\DC^{1,W^-,(2)}_{\bar{\alpha}' q} & = 2\sum_\beta \left|V_{\alpha\beta}\right|^2 \DC^{1,(2)}_{\qp q, 2} \,  , \\
\DC^{1,W^-,(2)}_{g\alpha} &=  2\sum_\beta \left|V_{\alpha\beta}\right|^2 \DC^{1,(2)}_{gq} \, , \numberthis
\end{align*}
where $q\in\mathfrak{u}\cup\mathfrak{d}$ denotes a quark of any active flavour; the anti-quark-initiated contributions are given by
 \begin{align*}
\DC^{1,W^-,(2)}_{\bar{\alpha}\bar{\beta}} & = \DC^{1,W^-,(2)}_{\beta\alpha}  \,  , \\
\DC^{1,W^-,(2)}_{\bar{q}\bar{q},\Fcon} &= 2\sum_{\alpha,\beta}  \left|V_{\alpha\beta} \right|^2  \DC^{1,(2)}_{qq, \Fcon, 2}   \,  , \\
\DC^{1,W^-,(2)}_{\bar{\beta}\bar{\beta},\Fcon} &= 2\sum_\alpha \left|V_{\alpha\beta} \right|^2 \DC^{1,(2)}_{qq, \Fcon, 1}  \,  , \\
\DC^{1,W^-,(2)}_{\alpha\bar{\beta}} & = \DC^{1,W^-,(2)}_{\bar{\beta} \alpha} \,  , \\
\DC^{1,W^-,(2)}_{\beta\bar{\beta}, \Fcon} &= 2\sum_\alpha \left|V_{\alpha\beta} \right|^2 \DC^{1,(2)}_{\qb q, \Fcon}  \,  , \\
\DC^{1,W^-,(2)}_{\qp \bar{\beta}} & = 2\sum_\alpha \left|V_{\alpha\beta}\right|^2 \DC^{1,(2)}_{\qp q, 1}  \,  , \\
\DC^{1,W^-,(2)}_{\beta \bar{q}} & =  2\sum_\alpha \left|V_{\alpha\beta}\right|^2 \DC^{1,(2)}_{\qp q, 2} \,  , \\
\DC^{1,W^-,(2)}_{\qbp \bar{\beta}} & = 2\sum_\alpha \left|V_{\alpha\beta}\right|^2  \DC^{1,(2)}_{\qp q, 1}  \,  , \\
\DC^{1,W^-,(2)}_{\bar{\alpha}' \bar{q}} & = 2\sum_\beta \left|V_{\alpha\beta}\right|^2 \DC^{1,(2)}_{\qp q, 2} \,  , \\
\DC^{1,W^-,(2)}_{g\bar{\beta}} &=  2\sum_\alpha \left|V_{\alpha\beta}\right|^2 \DC^{1,(2)}_{gq} \,  , \numberthis
\end{align*}
and the gluon-initiated contributions read
\begin{align*}
\DC^{1,W^-,(2)}_{\beta g} &= 2\sum_\alpha \left|V_{\alpha\beta}\right|^2 \DC^{1,(2)}_{qg}  \, ,  \\
\DC^{1,W^-,(2)}_{\bar{\alpha}g} &= 2\sum_\beta \left|V_{\alpha\beta}\right|^2 \DC^{1,(2)}_{qg} \, ,  \\
\DC^{1,W^-,(2)}_{gg} &= 2\sum_{\alpha,\beta} \left|V_{\alpha\beta}\right|^2 \DC^{1,(2)}_{gg} \, . \numberthis
\end{align*}
In order to obtain a coefficient function from the channel decomposition, all the above contributions which admit the initial and final state particle need to be summed.

The coefficient functions $\Delta\Ccal^{1,W^+,(j)}_{p'p}$ describing the exchange of a $W^+$ boson are obtained for $\alpha\in \mathfrak{d}$ and $\beta\in \mathfrak{u}$.

\subsubsection{Symmetric Polarized Coefficient Functions}\label{sec:polCCSymmCF}
The symmetric polarized CC structure functions at LO are
\begin{align*}
\Delta\Ccal^{T,W^{-},(0)}_{\beta\alpha} &= 2 \left|V_{\alpha\beta}\right|^2 \DC^{T,(0)}_{qq} \,  ,  \\
\Delta\Ccal^{T,W^{-},(0)}_{\bar{\alpha}\bar{\beta}} &= -2 \left|V_{\alpha\beta}\right|^2 \DC^{T,(0)}_{qq} \,  , \numberthis
\end{align*}
where $\alpha\in \mathfrak{u}$ and  $\beta\in \mathfrak{d}$.
The factors of $2$ arise from our choice of normalization of $\DC^{T,(0)}_{qq} = -\frac{1}{2}\delta(1-\xh) \delta(1-\zh)$, like for NC.
The polarized longitudinal coefficient functions at LO vanish.

At NLO the quark-initiated coefficient functions are given by
\begin{align*}
\Delta\Ccal^{\{T,L\},W^{-},(1)}_{\beta\alpha} &= 2 \left|V_{\alpha\beta}\right|^2 \DC^{\{T,L\},(1)}_{qq} \,  , \\
\Delta\Ccal^{\{T,L\},W^{-},(1)}_{g\alpha} &= 2 \sum_\beta \left|V_{\alpha\beta}\right|^2 \DC^{\{T,L\},(1)}_{gq}  \numberthis \,  ,
\end{align*}
the anti-quark-initiated coefficient functions by
\begin{align*}
\Delta\Ccal^{\{T,L\},W^{-},(1)}_{\bar{\alpha}\bar{\beta}} &= -\DC^{\{T,L\},W^{-},(1)}_{\beta\alpha}  \,  , \\
\Delta\Ccal^{\{T,L\},W^{-},(1)}_{g\bar{\beta}} & = -2\sum_\alpha \left|V_{\alpha\beta}\right|^2 \DC^{\{T,L\},(1)}_{gq} \, , \numberthis
\end{align*}
and the gluon-initiated coefficient functions by
\begin{align*}
\Delta\Ccal^{T,W^{-},(1)}_{\beta g} &= 2 \sum_\alpha \left|V_{\alpha\beta}\right|^2 \DC^{T,(1)}_{qg} \,  ,  \\
\Delta\Ccal^{T,W^{-},(1)}_{\bar{\alpha}g}  & =  -2 \sum_\beta \left|V_{\alpha\beta}\right|^2 \DC^{T,(1)}_{qg} \numberthis \, .
\end{align*}
As in the NC case the gluon-initiated longitudinal coefficient functions $\DC^{L,W^\pm,(1)}_{qg}$ and $\DC^{L,W^\pm,(1)}_{\bar qg}$ are zero.

Therefore the leading order $\Gcal_T^h$, $\Gcal_L^h$ structure functions are given by
\begin{alignat*}{3}
& \Big[ \Gcal_{T}^{h,W^-},~ \Gcal_{T}^{h,W^+} \Big] &&= \sum_{\phantom| \alpha\in \mathfrak{u} \phantom|} \sum_{\phantom| \beta\in \mathfrak{d} \phantom| } |V_{\alpha\beta}|^2 \Big[ \Delta f_{\bar{\beta}}^N \, D_{\bar{\alpha}}^h - \Delta f_\alpha^N \, D_\beta^h ,~ \Delta f_{\bar{\alpha}}^N \, D_{\bar{\beta}}^h - \Delta f_\beta^N \, D_\alpha^h \Big] \, , \\
& \Big[ \Gcal_{L}^{h,W^-},~ \Gcal_{L}^{h,W^+} \Big] &&= 0 \, . \numberthis
\end{alignat*}

At NNLO we the quark-initiated contributions read
\begin{align*}
\DC^{\{T,L\},W^-,(2)}_{\beta\alpha} & = 2 \left|V_{\alpha\beta} \right|^2 \DC^{\{T,L\},(2)}_{qq}  \,  , \\
\DC^{\{T,L\},W^-,(2)}_{\alpha\alpha,\Fcon} &= 2 \sum_\beta \left|V_{\alpha\beta} \right|^2 \DC^{\{T,L\},(2)}_{qq, \Fcon}  \,  , \\
\DC^{\{T,L\},W^-,(2)}_{\bar{\beta} \alpha} & = 2 \left|V_{\alpha\beta} \right|^2 \DC^{\{T,L\},(2)}_{\qb q}  \,  , \\
\DC^{\{T,L\},W^-,(2)}_{\bar{\alpha} \alpha, \Fcon} &= 2 \sum_\beta \left|V_{\alpha\beta} \right|^2 \DC^{\{T,L\},(2)}_{\qb q, \Fcon}  \,  , \\
\DC^{\{T,L\},W^-,(2)}_{\qp \alpha} & = 2 \sum_\beta \left|V_{\alpha\beta}\right|^2 \DC^{\{T,L\},(2)}_{\qp q, 1}  \,  , \\
\DC^{\{T,L\},W^-,(2)}_{\beta' q} & = 2 \sum_\alpha \left|V_{\alpha\beta}\right|^2 \DC^{\{T,L\},(2)}_{\qp q, 2} \,  , \\
\DC^{\{T,L\},W^-,(2)}_{\qbp \alpha} & = 2 \sum_\beta \left|V_{\alpha\beta}\right|^2  \DC^{\{T,L\},(2)}_{\qp q, 1}  \,  , \\
\DC^{\{T,L\},W^-,(2)}_{\bar{\alpha}' q} & = -2 \sum_\beta \left|V_{\alpha\beta}\right|^2 \DC^{\{T,L\},(2)}_{\qp q, 2} \,  , \\
\DC^{\{T,L\},W^-,(2)}_{g\alpha} &= 2 \sum_\beta \left|V_{\alpha\beta}\right|^2 \DC^{\{T,L\},(2)}_{gq} \, , \numberthis
\end{align*}
with $q\in\mathfrak{u}\cup\mathfrak{d}$; the anti-quark-initiated contributions are
 \begin{align*}
\DC^{\{T,L\},W^-,(2)}_{\bar{\alpha}\bar{\beta}} & = -\DC^{\{T,L\},W^-,(2)}_{\beta\alpha}  \,  , \\
\DC^{\{T,L\},W^-,(2)}_{\bar{\beta}\bar{\beta},\Fcon} &= -2 \sum_\alpha \left|V_{\alpha\beta} \right|^2 \DC^{\{T,L\},(2)}_{qq, \Fcon} \,  , \\
\DC^{\{T,L\},W^-,(2)}_{\bar{\alpha}\beta} & = -\DC^{\{T,L\},W^-,(2)}_{\bar{\beta} \alpha} \,  , \\
\DC^{\{T,L\},W^-,(2)}_{\beta\bar{\beta}, \Fcon} &=-2 \sum_\alpha \left|V_{\alpha\beta} \right|^2 \DC^{\{T,L\},(2)}_{\qb q, \Fcon}  \,  , \\
\DC^{\{T,L\},W^-,(2)}_{\qp \bar{\beta}} & = -2 \sum_\alpha \left|V_{\alpha\beta}\right|^2 C^{\{T,L\},(2)}_{\qp q, 1}  \,  , \\
\DC^{\{T,L\},W^-,(2)}_{\beta \bar{q}} & =  2 \sum_\alpha \left|V_{\alpha\beta}\right|^2 \DC^{\{T,L\},(2)}_{\qp q, 2} \,  , \\
\DC^{\{T,L\},W^-,(2)}_{\qbp \bar{\beta}} & = -2 \sum_\alpha \left|V_{\alpha\beta}\right|^2  \DC^{\{T,L\},(2)}_{\qp q, 1}  \,  , \\
\DC^{\{T,L\},W^-,(2)}_{\bar{\alpha}' \bar{q}} & = -2 \sum_\beta \left|V_{\alpha\beta}\right|^2 \DC^{\{T,L\},(2)}_{\qp q,2} \,  , \\
\DC^{\{T,L\},W^-,(2)}_{g\bar{\beta}} &=  -2 \sum_\alpha \left|V_{\alpha\beta}\right|^2 \DC^{3,(2)}_{gq}  \numberthis \, ,
\end{align*}
and the gluon initiated contributions read
\begin{align*}
\DC^{\{T,L\},W^-,(2)}_{\beta g} &= 2\sum_\alpha \left|V_{\alpha\beta}\right|^2 \DC^{\{T,L\},(2)}_{qg} \, ,  \\
\DC^{\{T,L\},W^-,(2)}_{\bar{\alpha}g} &= -2 \sum_\beta \left|V_{\alpha\beta}\right|^2 \DC^{\{T,L\},(2)}_{qg} \, , \\
\DC^{\{T,L\},W^-,(2)}_{gg} &=0 \,  . \numberthis
\end{align*}
The NNLO coefficient functions in the channel decomposition are obtained by summing all the contributions above admitting the appropriate identified partons.
For $W^+$-exchange one instead has to consider $a\in \mathfrak{d}$ and $b\in \mathfrak{u}$.

\section{Analytical Calculation of the Polarized Coefficient Functions}
\label{sec:calculation}
Like in the unpolarized case~\cite{Bonino:2025qta} we use our established toolchain \cite{Bonino:2024qbh,Bonino:2024wgg} to calculate the NNLO corrections to neutral- and charged-current SIDIS.
The Feynman diagrams are generated with \code{QGRAF}~\cite{Nogueira:1991ex} and converted into amplitudes with in-house routines in \code{FORM}~\cite{Vermaseren:2000nd} and \code{Mathematica}.

To obtain the structure functions $\Gcal_1^h$, $\Gcal_T^h$, and $\Gcal_L^h$ from~\eqref{eq:had_tens} we apply the same projectors as for $\mathcal{F}_3^h$, $\mathcal{F}_T^h$, and $\mathcal{F}_L^h$ respectively.
With respect to the unpolarized case the projection onto the polarization difference for a quark or gluon introduces an additional antisymmetric structure.
During the calculation of the polarized coefficient functions we encountered several subtleties in
treating $\gamma_5$, which are discussed next, before turning to technical details of the calculation
and checks we performed on our results.

\subsection[\texorpdfstring {Treatment of $\gamma_5$}{Treatment of \textbackslash{}gamma\_5}]{\boldmath Treatment of $\gamma_5$}
\label{sec:g5}
Since the treatment of $\gamma_5$ in $d\neq 4$ dimensions is not unique, we have to specify the scheme we are using for our calculation.
We choose to work in the Larin scheme \cite{Larin:1993tq}, which, in our case, is defined by replacing
\begin{align}
\gamma_\mu \gamma_5 \to \frac{i}{3!} \varepsilon_{\mu\nu\rho\sigma} \gamma^\nu \gamma^\rho \gamma^\sigma,
\label{eq:Larin_g5_pol}
\end{align}
and subsequently contracting pairs of Levi-Civita tensors into $d$-dimensional scalar products via the identity
\begin{align}
\varepsilon_{\mu\nu\rho\sigma} \, \varepsilon_{\alpha\beta\gamma\delta} = \operatorname{det}
\begin{pmatrix}
g_{\mu \alpha} & g_{\mu \beta} & g_{\mu \gamma} & g_{\mu \delta} \\
g_{\nu \alpha} & g_{\nu \beta} & g_{\nu \gamma} & g_{\nu \delta} \\
g_{\rho \alpha} & g_{\rho \beta} & g_{\rho \gamma} & g_{\rho \delta} \\
g_{\sigma \alpha} & g_{\sigma \beta} & g_{\sigma \gamma} & g_{\sigma \delta}
\end{pmatrix}.
\label{eq:Larin_epstens_pol}
\end{align}

However, with the standard Larin scheme replacement the known NLO results for the $\Gcal_1^h$, $\Gcal^h_T$ and $\Gcal^h_L$ structure functions~\cite{deFlorian:1994wp,Forte:2001ph}
obtained in the 't Hooft-Veltman-Breitenlohner-Maison (HVBM) scheme~\cite{tHooft:1972tcz,Breitenlohner:1977hr} are not necessarily reproduced.
The discrepancies can be traced back to two problems which have to be addressed:
\begin{enumerate}
	\item How to project correctly onto the polarization of an external quark.
	\item How to deal with contractions of four Levi-Civita tensors.
\end{enumerate}

\paragraph{Projector of polarized external quarks} \ \\
A quark-induced Wilson coefficient in polarized DIS can be expressed as
\begin{equation}
\DC_{q}(x,Q^2) = \tr [ \Pi(p,n) \, \Gamma(p,n) ] \, ,
\label{eq:Wilson_coeff_proj}
\end{equation}
where $\Pi$ denotes the polarized quark projector, $\Gamma$ is the sum of amputated Feynman diagrams contributing to the Wilson coefficient $\DC_q$, and $n$ is the light-like reference vector $n = q + x p$, which satisfies $n^2 =0$, $n \cdot q = Q^2/2 $ and $n \cdot p = Q^2/2x$.
In $d=4$ dimensions the general tensor structure of $\Gamma$ can be written as
\begin{align}
\Gamma (p,n) \xlongrightarrow{d\to 4} A \, \slashed{p} \gamma_5 + B \, \slashed{n} \gamma_5 \,.
\end{align}
The projectors onto these two structures are given in four dimensions by
\begin{align}
\Pi_A^{(4)} =& -\frac{x}{2 Q^2} \slashed{n} \gamma_5 \,, &\Pi_B^{(4)} =& -\frac{x}{2 Q^2} \slashed{p} \gamma_5 \,,
\end{align}
where $\Pi_B^{(4)}$ is the projector onto the relevant  coefficient $B$.

We observe that choosing to apply the usual Larin prescription for the polarized quark projector \cite{Zijlstra:1993sh},
which amounts to the replacement
\begin{align}
	\Pi_B^{(4)} \to & -\frac{i x}{12 Q^2} \varepsilon^{\mu_1\mu_2\mu_3\mu_4}p_{\mu_4} \gamma_{\mu_1} \gamma_{\mu_2} \gamma_{\mu_3}
	\label{eq:naive_projector}
\end{align}
does not give rise to the appropriate non-singlet splitting functions~\cite{Vogelsang:1996im,Mertig:1995ny} appearing in the coefficient functions belonging to $\Gcal^h_T$ and $\Gcal^h_L$ starting from NNLO.
Likewise, the finite parts of the $\MSbar$ NLO coefficient functions~\cite{deFlorian:1994wp,Forte:2001ph}, which were computed in the HVBM $\gamma_5$ scheme
are not correctly reproduced with the usual (process independent) scheme transformation~\cite{Matiounine:1998re}.
The non-singlet coefficient functions, however, have to coincide with their unpolarized counter parts belonging to $\Fcal_T^h$ and $\Fcal_L^h$ by na\"ive anticommutation of $\gamma_5$ to the vertex.
In the following we show why the na\"ive Larin prescription for $\slashed{p} \gamma_5$ fails
and extend the prescription to correctly deal with the projection of the external polarization in this scheme.
The same problem was encountered before in the calculation of polarized operator matrix elements~\cite{Klein:2009ig}
and solved in that context~\cite{Schonwald:2019gmn,Behring:2019tus}.

We notice that in the Larin scheme the analytic continuation of the tensor structures $\slashed{p} \gamma_5$ and $\slashed{n} \gamma_5$ to $d$ dimensions in the presence of an additional external vector is not unique.
For off-shell quarks we can define four independent tensors which reduce to the two 4-dimensional structures $\slashed{p} \gamma_5$ and $\slashed{n} \gamma_5$,
\begin{alignat}{2}
	T_1 &= \frac{i}{6} \varepsilon^{\mu_1\mu_2\mu_3 \mu_4} p_{\mu_4}\gamma_{\mu_1} \gamma_{\mu_2} \gamma_{\mu_3} &&\xlongrightarrow{d\to 4} \slashed{p} \gamma_5 \,,
	\\
	T_2 &= \frac{i}{6} \varepsilon^{\mu_1\mu_2\mu_3 \mu_4} n_{\mu_4} \gamma_{\mu_1} \gamma_{\mu_2} \gamma_{\mu_3} &&\xlongrightarrow{d\to 4} \slashed{n} \gamma_5 \,,
        \\
	T_3 &= -i\frac{x}{Q^2} \varepsilon^{\mu_1\mu_2 \mu_3 \mu_4} p_{\mu_3} n_{\mu_4} \gamma_{\mu_1} \gamma_{\mu_2} \slashed{p} \, &&\xlongrightarrow{d\to 4} \slashed{p} \gamma_5 \,,
        \\
	T_4 &= i\frac{x}{Q^2} \varepsilon^{\mu_1\mu_2 \mu_3 \mu_4} p_{\mu_3} n_{\mu_4} \gamma_{\mu_1} \gamma_{\mu_2} \slashed{n} &&\xlongrightarrow{d\to 4} \slashed{n} \gamma_5 \,.
\end{alignat}
In a $d$-dimensional Larin scheme the Dirac structure therefore decomposes into
\begin{alignat}{2}
	\Gamma &&=~& ~ A_1 T_1 + A_2 T_2 + A_3 T_3 + A_4 T_4  \nonumber
	\\
	&&\xlongrightarrow{d\to 4} & \left( A_1 + A_3 \right) \slashed{p} \gamma_5 + \left( A_2 + A_4 \right) \slashed{n} \gamma_5 \,.
\end{alignat}
For on-shell quarks in $4$ dimensions the projector $\slashed{p} \gamma_5$ yields
\begin{align}
\tr [\Pi_B^{(4)} \,\Gamma] \xlongrightarrow{d\to 4} A_2+A_4 \, ,
\end{align}
which is proportional to the coefficient of the tensor structure $\slashed n \gamma_5$.
However, in the $d$ dimensional Larin scheme all tensor structures which reduce to $\slashed{n} \gamma_5$ in the $4$-dimensional limit have to be taken into account simultaneously.
This implies that we need to construct a projector $\Pi_2$ in $d=4-2\epsilon$ dimensions fulfilling
\begin{align}
&\tr [\Pi_B \, \Gamma ] = f(\epsilon) \, (A_2 + A_4) \,, \label{eq:projector_matching}
\end{align}
with $f(0)=1$.
Without loss of generality we set $f(\epsilon) = 1$.
The corresponding projector then reads
\begin{align}
	\Pi_B &= \frac{i}{2(1-\epsilon)(1-2\epsilon)} \frac{x^2}{Q^4} \varepsilon^{\mu_1\mu_2 \mu_3 \mu_4} p_{\mu_3} n_{\mu_4} \gamma_{\mu_1} \gamma_{\mu_2} \slashed{p} \, .
\label{eq:quark_projector}
\end{align}
Conversely, the na\"ive prescription of~\eqref{eq:naive_projector} introduces a non-trivial $\epsilon$ dependence between $A_2$ and $A_4$,
\begin{align}
	\tr[\Pi_B^{(4)} \,\Gamma] \to &-\frac{i x}{12 Q^2} \varepsilon^{\mu_1\mu_2\mu_3 \mu_4} p_{\mu_4} \tr[\gamma_{\mu_1}\gamma_{\mu_2}\gamma_{\mu_3} \Gamma] \nonumber\\
	= &
	   (1-\epsilon)(1-2\epsilon) \left[
	  		\frac{3-2\epsilon}{3} A_2
	  		+  A_4
	  \right] \,,
\end{align}
explaining why the $\MSbar$ results for the coefficient functions of $\Gcal^h_T$ and $\Gcal^h_L$ are not reproduced with the standard renormalization constants.

In four dimensions the projector onto the spin difference of the external quark state is given by $\slashed{p} \gamma_5$,
which in the Larin scheme translates into
\begin{align}
	\slashed{p} \gamma_5 \to \frac{-i}{(1-\epsilon)(1-2\epsilon)} \frac{x}{Q^2} \varepsilon^{\mu_1\mu_2 \mu_3 \mu_4} p_{\mu_3} q_{\mu_4} \gamma_{\mu_1} \gamma_{\mu_2} \slashed{p}\,.
\end{align}
Note that we replaced $n_{\mu_4}$ with $q_{\mu_4}$, which is equivalent due to the antisymmetry of the Levi-Civita tensor.

Interestingly, the projection onto the structure function $\Gcal_1^h$ is unaffected by the change
of projector, since the na\"ive prescription given by $\Pi_B^{(4)}$ and $\Pi_B$ yields the same result:
\begin{align}
\left( \varepsilon^{\mu \nu \rho \sigma} p_\rho q_\sigma \right) \left( \frac{i}{6}\varepsilon^{\mu_1\mu_2\mu_3 \mu_4} p_{\mu_4} \gamma_{\mu_1} \gamma_{\mu_2} \gamma_{\mu_3} \right) = \left( \varepsilon^{\mu \nu \rho \sigma} p_\rho q_\sigma \right) \left( - i\frac{x}{Q^2}  \varepsilon^{\mu_1\mu_2 \mu_3 \mu_4}  p_{\mu_3}  q_{\mu_4} \gamma_{\mu_1} \gamma_{\mu_2} \slashed{p}\right)\,,
\end{align}
which can be seen by explicitly contracting the two Levi-Civita tensors on both sides according to~\eqref{eq:Larin_epstens_pol}.
This shows that any observable related to the $\Gcal_1^h$ structure function is unaffected by this modification of the quark projector.

\paragraph{Contraction of four Levi-Civita tensors} \phantom* \\
When calculating the coefficient functions for neutral- and charged-current interactions of the structure function
$\Gcal^h_1$, we encounter the contraction of up to four Levi-Civita tensors.
Starting from the contraction of three Levi-Civita tensors, i.e.
$\displaystyle \varepsilon_{\mu\nu\rho\sigma}\varepsilon^{\mu\nu\rho\sigma} \varepsilon_{\alpha\beta\gamma\delta}$
the contraction in the Larin scheme is ambiguous,
since there are two ways of associating the Levi-Civita tensors for pairwise contraction~\cite{Belusca-Maito:2023wah},
\begin{align*}
(\varepsilon_{\mu\nu\rho\sigma}\varepsilon^{\mu\nu\rho\sigma})\, \varepsilon_{\alpha\beta\gamma\delta} &= (24 - 100\varepsilon + 140 \varepsilon^2 - 80 \varepsilon^3 + 16 \varepsilon^4)\,  \varepsilon_{\alpha\beta\gamma\delta}
\\
\varepsilon_{\mu\nu\rho\sigma} \, (\varepsilon^{\mu\nu\rho\sigma} \varepsilon_{\alpha\beta\gamma\delta}) &= 24 \,  \varepsilon_{\alpha\beta\gamma\delta} \,,
\end{align*}
which yield different results for $\epsilon \neq 0$.
Hence there is an ambiguity in $d\neq 4$ depending on the order of contracting the Levi-Civita tensors.

In the case of SIDIS the Levi-Civita tensors either stem from couplings to axial vector currents or the projection onto external polarizations.
To obtain consistent results, a unique contraction order for the Levi-Civita tensors must be established.
We therefore apply the following prescription for $\Gcal^h_1$, which can be applied at any order in perturbation theory:
\begin{enumerate}
\item Replace axial vector vertices $\gamma_\mu \gamma_5$ according to~\eqref{eq:Larin_g5_pol} and the polarized quark projector according to~\eqref{eq:quark_projector}.
\item Contract Levi-Civita tensors from the projectors onto $\Gcal^h_1$ and the polarized parton state through~\eqref{eq:Larin_epstens_pol}.
\item Contract Levi-Civita tensors from the axial vector couplings (if present) through~\eqref{eq:Larin_epstens_pol}.
\end{enumerate}
The divergent and finite axial vector coupling renormalizations~\cite{Larin:1993tq} and the finite renormalization of the polarized quark current~\cite{Matiounine:1998re} are then carried out in the usual manner.

\begin{figure}[ht]
\centering
\includegraphics[width=.3\textwidth]{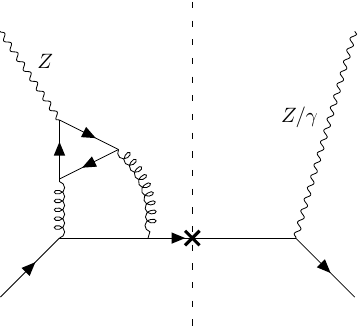}
\caption{UV-divergent 2-loop VV anomaly contribution to $C_{qq,A,\NoW}^{\{1,T\},(2)}$.}
\label{fig:div_anomaly}
\end{figure}

\begin{figure}[ht]
\centering
\includegraphics[width=.3\textwidth]{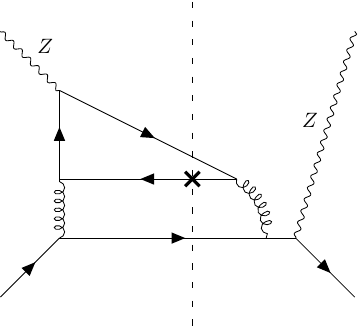}
\caption{Finite contribution with axial vector couplings attaching to different quark lines.}
\label{fig:B40_aap}
\end{figure}

With this prescription for the contraction of pairs of Levi-Civita tensors, the axial-vector coefficient functions and the vector coefficient functions coincide in the $\MSbar$ scheme whenever both electroweak couplings attach to the same quark line.
As a result, current conservation is restored in these contributions, as required by the chiral symmetry of massless QCD.
For contributions with axial vector couplings attaching to different quark lines in SIDIS, we distinguish two cases:
\begin{itemize}
\item Either (or both) axial vector couplings couple to a closed quark line, giving rise to the sum of axial vector couplings $\sum_{\qp} g^{\qp}_A$, which vanishes for an even number of flavours.
If an odd number of massless flavours is considered e.g.\ within a Variable Flavour Number Scheme, the anomaly contribution $\DC_{qq,A,\NoW}^{\{1,T\},(2)}$ arising in the 2-loop VV contribution, cf.\ Fig.~\ref{fig:div_anomaly}, is the only non-finite contribution.
While the contribution is UV divergent, it is also IR finite, and thus only requires the divergent and finite 2-loop pure-singlet UV renormalization of the axial vector vertex.
Therefore, no additional finite renormalization can be introduced from the polarized external current at NNLO.
\item The axial vector couplings couple to two different quark lines, which both contain an identified quark as shown in Fig.~\ref{fig:B40_aap}.
The corresponding coefficient functions $\Delta C^{\{i\},(2)}_{\qp q,\NoW,4}$ and $\Delta C^{\{i\},(2)}_{\qbp q,\NoW,4}$ are UV and IR finite after phase space integration and do not have to be renormalized.
This contribution may equivalently be calculated in $4$ dimensions and is therefore scheme independent.
\end{itemize}
Hence we are confident that we have applied a proper $\gamma_5$~prescription for the SIDIS coefficient functions.

\subsection{Technical details on the calculation}
\label{sec:technical}
Three classes of diagrams contribute at NNLO: the real-real (RR) contributions, describing the real radiation of two additional partons; the real-virtual (RV) contributions, describing the radiation of a single parton with an additional loop; and the virtual-virtual contributions with no extra radiation, but two additional loops, all with respect to the leading order.

With these prescriptions the calculation of the tree-level RR matrix elements becomes straightforward.
In order to reduce the loop integrals arising in the RV and VV contributions to a small set of master integrals, we use \code{REDUZE}~\cite{vonManteuffel:2012np} to generate integration-by-parts identities~\cite{Chetyrkin:1981qh,Laporta:2000dsw}.
The loop integrals in the RV matrix elements reduce to the 1-loop bubble and box integral~\cite{Gehrmann:1999as}.
In the VV we recover the four master integrals in~\cite{Gehrmann:2005pd}.

The coefficient functions are obtained by integrating the contributions over their corresponding phase spaces.
For the RV, the phase space integration is fully constrained by $\xh$ and $\zh$.
As a consequence it is sufficient to expand the unintegrated RV contributions in distributions in $(1-\xh)$ and $(1-\zh)$.
These expansions are described in detail in~\cite{Gehrmann:2022cih}, and
must be performed in the appropriate kinematical regions requiring analytical continuations of
the one-loop bubble and box master integrals~\cite{Gehrmann:2022cih,Haug:2022hkr}.
To integrate the RR phase space, we map the phase space integrals onto 14 integral families through reverse unitarity~\cite{Anastasiou:2002yz}.
These families contain a set of 8 independent propagators, 3 constraints from on-shellness, and the SIDIS phase space constraint introducing the fragmentation variable $z$.
They reduce to a set of 21 master integrals~\cite{Bonino:2024adk,Ahmed:2024owh}.

The resulting bare coefficient functions are initially expressed in terms of harmonic polylogarithms (HPLs)~\cite{Remiddi:1999ew} using the Mathematica packages \code{HPL}~\cite{Maitre:2005uu} and \code{PolyLogTools}~\cite{Duhr:2019tlz}, and subsequently converted into logarithms and polylogarithms.

The ultraviolet poles are removed by renormalizing the QCD coupling constant in the usual manner, and applying the divergent renormalization on the axial vector current~\cite{Larin:1993tq}.
Both renormalizations are multiplicative and independent of the kinematics, and yield the UV-finite coefficient functions in the Larin scheme.
Subsequently we apply mass factorization to cancel the IR poles associated with PDFs and FFs and obtain coefficient functions in the polarized and axial Larin scheme.
To convert our results from the Larin scheme into $\MSbar$, two steps have to be carried out.
\begin{itemize}
\item
The axial coupling is renormalized with the finite (multiplicative) renormalization~\cite{Larin:1993tq}, restoring current conservation at the axial vector vertex.
\item
The finite scheme transformation has to be applied to the polarized splitting functions~\cite{Matiounine:1998re,Moch:2014sna}.
Details about this scheme transformation are provided in Appendix~\ref{sec:Scheme_flavour_basis}.
\end{itemize}

\subsection{Checks on the Calculation}
\label{sec:checks}
In the $\MSbar$ scheme
coefficient functions for which the axial vector couplings couple to the initial state quark line or,
in absence of an axial vector coupling, if at least one vector coupling couples to the initial state quark line,
are related in the polarized and unpolarized case~\cite{Matiounine:1998re} by chirality conservation of massless fermions,
which can be enforced by the application of an anticommuting $\gamma_5$.
After performing the finite scheme transformation as described in~\cite{Bonino:2024wgg,Goyal:2024tmo} and in detail in \cite{Goyal:2024emo}, we recover the known results for photon exchange.
In the following we shall refer to the polarized results obtained in this way as ``na\"ive''.

Comparing these na\"ive polarized results to our unpolarized electroweak results presented in~\cite{Bonino:2025qta}
we observe discrepancies in
the SIDIS coefficient functions $\Delta C^{i,(2),\text{na\"ive}}_{qq}$ and $\Delta C^{i,(2),\text{na\"ive}}_{\qb q}$ for $i=1,T$.
For $\Delta C^{1,(2)}_{qq}$ the difference reads
\begin{align*}
C^{3,(2)}_{qq} - \Delta C^{1,(2),\text{na\"ive}}_{qq} = \frac{C_F}{N_C} \delta(1-z) \Bigg\{  (1+x) \bigg[ &  4  \ln(x)\ln(1+x) - \ln^2(x)   \\
&+ 4 \operatorname{Li}_2(-x) + \frac{\pi^2}{3} - 3 \ln(x) \bigg] - 7 (1-x) \Bigg\}\, . \numberthis
\label{eq:g1_Cqq_missing}
\end{align*}
The terms missing in $\Delta C^{1,(2),\text{na\"ive}}_{qq}$ appear as a surplus in $\Delta C^{1,(2),\text{na\"ive}}_{\qb q}$,
\begin{align}
C^{3,(2)}_{\qb q} - \Delta C^{1,(2),\text{na\"ive}}_{\qb q}  = - \Big( C^{3,(2)}_{qq} - \Delta C^{1,(2)}_{qq} \Big)\, ,
\end{align}
so that the sum of both contributions stays the same.
For $\Gcal^h_T$ we observe the same discrepancy.
However, the relation between the polarized and unpolarized results should hold at a diagrammatic level, and hence for the individual channels in SIDIS.

A closer inspection of the polarized scheme transformation offers insight.
The non-singlet polarized scheme transformation by definition corresponds to a finite renormalisation of the polarized non-singlet splitting functions.
At NNLO the non-singlet polarized splitting functions $\Delta P_{qq}^{\mathrm{NS},(1)}$ and $\Delta P_{\qb q}^{\mathrm{NS},(1)}$ however do not enter the same coefficient function in SIDIS, and hence may not be renormalized together.
This distinction will be relevant for any polarized process with sensitivity to a second parton, as is the case in SIDIS.
Applying the finite scheme transformation expressed in a basis of non-singlet and pure-singlet contributions na\"ively will therefore produce incorrect results in SIDIS and should be abandoned in favour of a scheme transformation in the flavour basis.
The details of this flavour basis decomposition and its application are discussed in Appendix~\ref{sec:Scheme_flavour_basis}.
The finite scheme transformation in the flavour basis resolves the discrepancy of~\eqref{eq:g1_Cqq_missing} and shifts the terms in the difference from $\Delta C_{\qb q}^{\{1,T\},(2)}$ into $\Delta C_{qq}^{\{1,T\},(2)}$.
The longitudinal polarized coefficient functions at NNLO are unaffected due to the vanishing structure function at leading order.
In this way we obtain the expected results in all polarized non-singlet channels.

Our results are expressed with the full scale dependence on $(\mu_A,\mu_F,\mu_R)$.
They are distributed inside an ancillary file with the \texttt{arXiv} submission.
A description of content and notation of the ancillary file is given in Appendix \ref{sec:files}.

To validate our results we have carried out the following checks:
Our results fulfill the scale dependence predicted by the renormalization group equations in $(\mu_F, \mu_A, \mu_R)$.
We have cross-checked our results for $\Gcal^h_1$ against our previous results for photon exchange~\cite{Bonino:2024wgg}, which were later confirmed by another group \cite{Goyal:2024tmo,Goyal:2024emo}.
We find full agreement for the coefficient functions with vector couplings upon undoing the polarized scheme transformation.
Furthermore we find agreement also for the axial vector coefficient functions with the axial vector couplings attaching to a common quark line.
An equality for axial vector coefficient functions, where couplings attach to different quark lines, is neither required by theory nor realized.

Finally, we have verified that the polarized non-singlet coefficient functions in the $\MSbar$ scheme presented in this paper agree with the results of our calculation of the unpolarized EW coefficient functions~\cite{Bonino:2025qta}.

\section{Phenomenological Results}
\label{sec:num}

In the following we provide phenomenological predictions for the electroweak effects in polarized neutral and charged current SIDIS at the BNL~EIC.
We study $\pi^{\pm}$ production at a centre-of-mass energy $\sqrt{s} = 140 \, \,\mathrm{GeV}$ in polarized electron-proton collisions.
We employ the same fiducial cuts as in the unpolarized phenomenological analysis presented in \cite{Bonino:2025qta}.
We impose
\begin{align}\label{eq:cutsxyz}
& 0.06 < y < 0.95 \, , && x < 0.8 \, , && 0.2 < z < 0.85
\end{align}
and we restrict ourselves to vector boson virtualities $25\,\GeV^2 < Q^2 < 14896\,\GeV^2$ divided in four logarithmically equispaced bins, which we denote as
\begin{alignat}{7}\label{eq:cutsQ2}
&\text{Low-}Q^2:~& \,  25\,\GeV^2 &< Q^2 < &~ 159 \,\GeV^2 & \, ,
&&\quad &&
\text{Mid-}Q^2:~& \, 159 \,\GeV^2 &< Q^2 < &~ 1007\,\GeV^2 & \, ,
\nonumber \\
&\text{High-}Q^2:~& \,1007\,\GeV^2 &< Q^2 < &~ 6400\,\GeV^2 & \, ,
&&\quad &&
\text{Extreme-}Q^2:~& \, 6400\,\GeV^2 &< Q^2 \, . \hspace{-.2em}
\end{alignat}

We employ the polarized PDF set \code{BDSSV24}~\cite{Borsa:2024mss} and the FF set \code{BDSSV22}~\cite{Borsa:2022vvp} at NNLO throughout.
The evolution of $\alpha_s$ is obtained from the unpolarized PDF set \code{NNPDF40\_nnlo\_as\_01180} \cite{NNPDF:2021njg} with the \code{LHAPDF} interface~\cite{Buckley:2014ana}, which is consistent with $\alpha_s$ used in~\cite{Borsa:2022vvp,Borsa:2024mss}.
The cross sections are evaluated for $\nf=5$ massless flavours,  which is sensible in the range of $Q^2$ considered.
To estimate the theory uncertainty stemming from missing higher orders in the cross section we perform a 7-point scale variation of $(\mu_R^2, \mu_F^2=\mu_A^2)$ around the central scale $Q^2$ and discard extreme variations.
We do not include additional uncertainties from the polarized PDFs and FFs.
The results for the fiducial SIDIS cross section are presented differentially in $Q^2$, $x$ or $z$.
All quantities displayed in plots are integrated with a Monte-Carlo technique over the respective bin.
Bin edges in $Q^2$ and $x$ are chosen logarithmically equispaced while bin edges in $z$ are linearly equispaced.
Numerical fluctuations, especially visible in plots of cross section ratios, are due to Monte-Carlo integration uncertainties.

As electroweak parameters we adopt the same values as in \cite{Bonino:2025qta}. In particular we use $G_F = 1.166379\times 10^{-5}~\mathrm{GeV}^{-2}$, $M_W = 80.369 \, \mathrm{GeV}$, $M_Z = 91.188~\mathrm{GeV}$, from which we derive $\alpha \simeq 1/132.10$ and $\sin^2 \theta_\mathrm{W} \simeq 0.2232$ with EW tree-level relations.
The full CKM matrix is used in predictions for the charged current cross section, with numerical values for the CKM entries equal to~\cite{ParticleDataGroup:2024cfk}:
\begin{align}
  |V_{ud}| &= 0.97435\,, & |V_{us}| &= 0.22500\,, & |V_{ub}| &= 0.00369\,, \nonumber \\
  |V_{cd}| &= 0.22486\,, & |V_{cs}| &= 0.97349\,, & |V_{cb}| &= 0.04182\,.
\end{align}

\subsection{Neutral Current}

In order to increase the sensitivity to electroweak effects we study cross section asymmetries, which isolate even and odd contributions in the lepton polarization  in the cross section difference~\eqref{eq:SFNCpol}.
The $\lambda_\ell$-odd component
\begin{equation} \label{eq:dDsodd}
\dd \Delta \sigma^h_{\modd} = \dd \Delta \sigma^h(\lambda_\ell=+1)-\dd \Delta \sigma^h(\lambda_\ell=-1) \,
\end{equation}
contains $\gamma\gamma$, $\gamma Z$ and $ZZ$ contributions.  The $\lambda_\ell$-even component instead
\begin{equation}\label{eq:dDseven}
\dd \Delta \sigma^h_{\meven} = \dd \Delta \sigma^h(\lambda_\ell=+1)+\dd \Delta \sigma^h(\lambda_\ell=-1) \equiv 2 \, \dd \Delta \sigma^h(\lambda_\ell=0) \,
\end{equation}
contains only $\gamma Z$ and $ZZ$ exchanges.
Note that due to the antisymmetric polarized quark projector, the role of $\lambda_\ell$-odd and $\lambda_\ell$-even contributions in the cross section difference is swapped with respect to unpolarized SIDIS.
While only the $\lambda_\ell$-even NC cross section receives contributions in unpolarized photon-induced SIDIS~\cite{Bonino:2025qta}, polarized photon-induced SIDIS contributes exclusively to the $\lambda_\ell$-odd NC cross section difference.

We also provide predictions for the structure functions decomposition of the cross section difference~\eqref{eq:dDsodd}
\begin{equation}
\dd \Delta \sigma^h_{\modd}=\dd \Delta \sigma^{h,\gamma\gamma,1}_{\modd}+\dd \Delta \sigma^{h,(\gamma Z+ZZ),1}_{\modd}+\dd \Delta \sigma^{h,(\gamma Z+ZZ),T}_{\modd}+\dd \Delta \sigma^{h,(\gamma Z+ZZ),L}_{\modd} \, ,
\end{equation}
with
\begin{equation}
\dd \Delta \sigma^{h,(\gamma Z+ZZ),i}_{\modd}=\dd \Delta \sigma^{h,\gamma Z,i}_{\modd}+\dd \Delta \sigma^{h,ZZ,i}_{\modd} \, , \quad \mathrm{for} \quad i =1,T,L
\end{equation}
in order to further isolate electroweak effects.

As already introduced in \cite{Bonino:2025qta} it is interesting to study asymmetries.
We present predictions for the \textit{polarized neutral current asymmetry}
\begin{equation}\label{eq:ANCm1}
\Delta A^h_\NC =- \frac{\dd \Delta \sigma^h_{\meven}}{\dd \Delta \sigma^h_{\modd}} \,  ,
\end{equation}
which is the inverse of the neutral current asymmetry $A^h_\NC$ of~\cite{Bonino:2025qta}, due to the presence of the photon-only exchange in the $\lambda_\ell$-odd component in polarized SIDIS.
This observable quantifies the size of the $\lambda_\ell$-even components in~\eqref{eq:SFNCpol} which are suppressed by $\eta_{\gamma Z}$ and $\eta_{Z Z}$ with respect to the photon exchange contribution in $\lambda_\ell$-odd.
All asymmetries shown in the following are subjected to the same kinematic cuts of~\eqref{eq:cutsxyz} and \eqref{eq:cutsQ2}.  The asymmetries are functions of the kinematic variables $Q^2$, $x$ or $z$, as a result of their construction from single-differential distributions. Theory uncertainties are obtained with a 31-point uncorrelated scale variation between numerator and denominator.

It is possible to estimate the size of the electroweak effects in the $n^\mathrm{th}$ order QCD corrections by dressing the photon-only $K$-factor with the LO NC exchange (see \cite{Bonino:2025qta}).
We estimate the precision that can be obtained following this procedure by computing the ratio
\begin{equation}\label{eq:Rratio}
R^{h,(n)} = \frac{\dd \Delta \sigma^{h,\NC,(n)}_{\modd}}{\dd \Delta \sigma^{h,\gamma\gamma,(n)}_{\modd}}\frac{\dd \Delta \sigma^{h,\NC,(0)}_{\modd}}{\dd \Delta \sigma^{h,\gamma\gamma,(0)}_{\modd}} \, ,
\end{equation}
with $n$ denoting the perturbative order in QCD.  The theory uncertainty for $R^{h,(n)}$ is given as a 7-point correlated scale variation between the order-dependent ratio $\dd \Delta \sigma^{h,\NC,(n)}_{\modd}/\dd \Delta \sigma^{h,\gamma\gamma,(n)}_{\modd}$.

Furthermore, we assess the importance of electroweak effects in neutral current exchanges on the \textit{double-longitudinal spin asymmetry}, described by
\begin{equation}\label{eq:AhLLNC}
A^{h,\NC}_{LL}=-\frac{1}{2}\frac{\dd\Delta\sigma^{h,\NC}_{\modd}}{\dd\sigma^{h,\NC}_{\meven}} \,  ,
\end{equation}
up to NNLO in QCD.
The factor of $-1/2$ accounts for our normalisation of the cross section difference \eqref{eq:SFNCpol} with respect to the unpolarized one \eqref{eq:xsNCCC}.
For the unpolarized cross section we use the unpolarized PDF set \code{NNPDF40\_nnlo\_as\_01180} \cite{NNPDF:2021njg}.
Both numerator and denominator contain $\gamma\gamma$, $\gamma Z$ and $ZZ$ exchanges.
The denominator is computed from~\eqref{eq:xsNCCC} as $\dd\sigma^{h,\NC}_{\meven}=2\,\dd\sigma^{h,\NC}(\lambda_{\ell}=0)$.
At lower values of virtuality, where to a good approximation only transverse photons are exchanged, the asymmetry approximates to
\begin{equation}
A_{LL}^{h,\gamma\gamma} \approx \frac{1}{2} A_1^h=\frac{2\Gcal^h_1}{\Fcal^h_T} \, .
\end{equation}
The $A_1^h$ asymmetry as a function of $x$ and $Q^2$ was previously measured by the COMPASS~\cite{COMPASS:2010hwr,COMPASS:2015mhb} and HERMES~\cite{HERMES:2004zsh,HERMES:2006jyl,HERMES:2018awh} collaborations and is a key ingredient in global fits of polarized PDFs~\cite{Borsa:2024mss,Bertone:2024taw}.
By comparing $A^{h,\gamma\gamma}_{LL}$ to $A^{h,\NC}_{LL}$ at high $Q^2$ we highlight the importance of a consistent treatment of electroweak effects in the double-longitudinal spin asymmetry~\eqref{eq:AhLLNC}.

Finally we study the neutral current \textit{single-longitudinal spin asymmetry}
\begin{equation}\label{eq:AhLNC}
A^{h,\NC}_{L}=\frac{1}{2}\frac{\dd\Delta\sigma^{h,\NC}_{\meven}}{\dd\sigma^{h,\NC}_{\meven}} \,  ,
\end{equation}
up to NNLO in QCD.
In this case in the numerator only electroweak effects are included with the $\lambda_\ell$-even component selecting only $\gamma Z$ and $ZZ$ exchanges.

\subsubsection[\texorpdfstring{$Q^2$-distributions}{Q\^{}2-distributions}]{\boldmath $Q^2$-distributions}

\begin{figure}[tb]
\centering
\includegraphics[width=\linewidth]{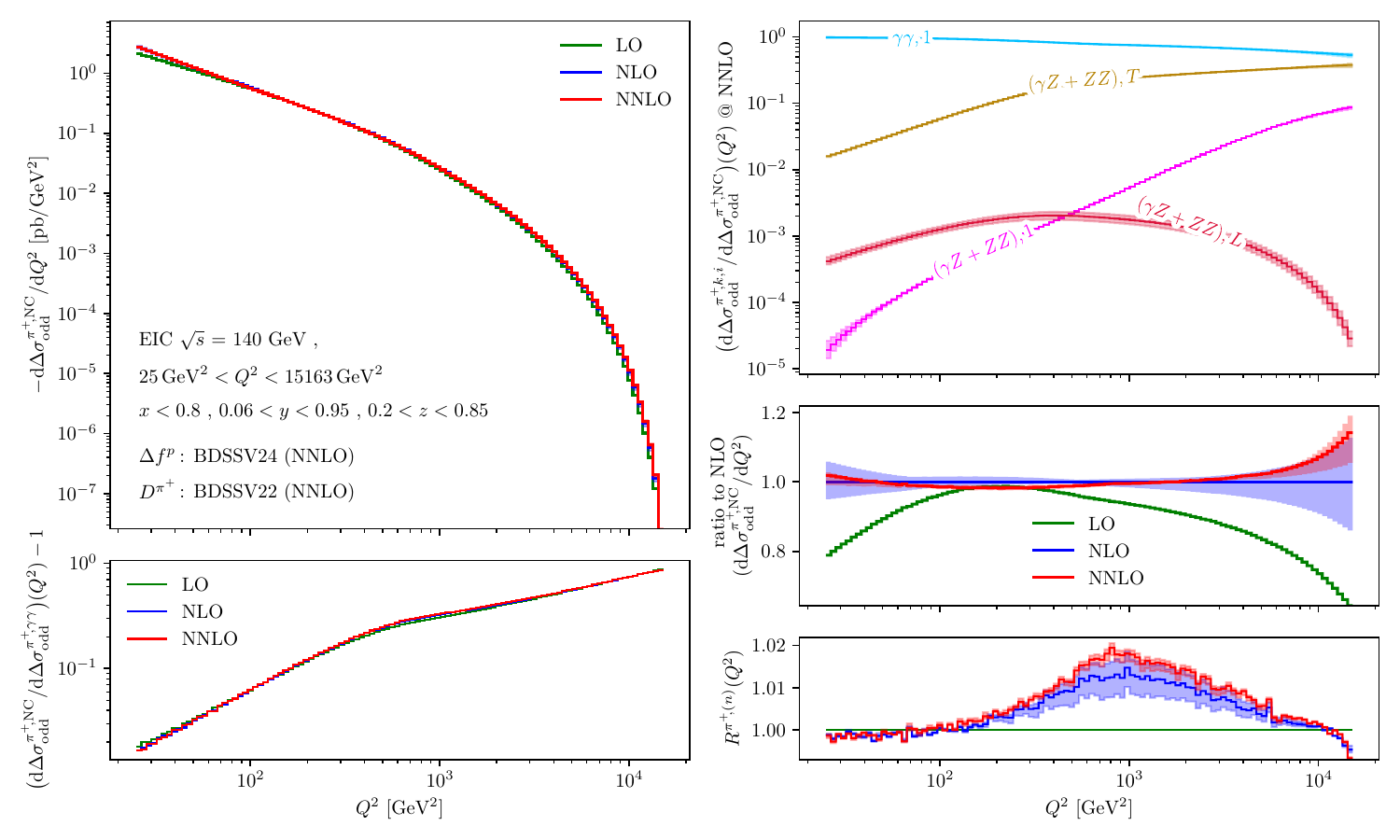}
\caption{
$\NC$ $\lambda_\ell$-odd $Q^2$-distribution for $\pi^+$ production.
Top left panel: total $\dd\Delta \sigma^{\NC}_{\modd}$ cross section difference.
Bottom left panel: ratio $(\dd\Delta \sigma^{\pi^+,\NC}_{\modd}/\dd\Delta \sigma^{\pi^+,\gamma\gamma}_{\modd})(Q^2)-1$.
Top right panel: structure function contributions to the total cross section.
Centre right panel: $\NC$ $\lambda_\ell$-odd ratio to NLO.
Bottom right panel: $R^{\pi^+,(n)} (Q^2)$.
}
\label{fig:Q2_NC_pip}
\end{figure}
\begin{figure}[tb]
\centering
\includegraphics[width=\linewidth]{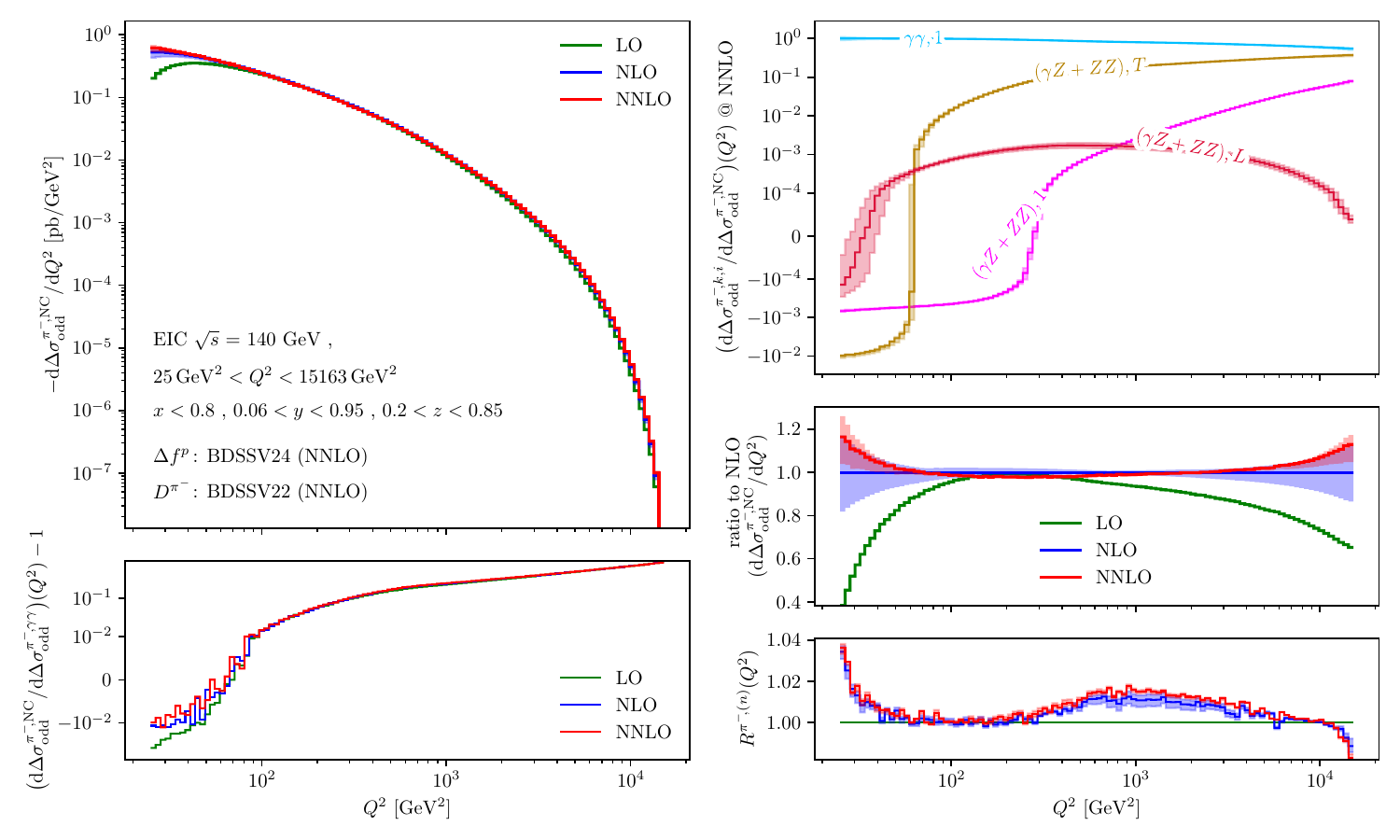}
\caption{
$\NC$ $\lambda_\ell$-odd $Q^2$-distribution for  $\pi^-$ production.
Top left panel: total $\dd\Delta\sigma^{\pi^-,\NC}_{\modd}$ cross section difference.
Bottom left panel: ratio $(\dd\Delta\sigma^{\pi^-,\NC}_{\modd}/\dd\Delta\sigma^{\pi^-,\gamma\gamma}_{\modd})(Q^2)-1$.
Top right panel: structure function contributions to the total cross section difference.
Centre right panel: $\dd\Delta\sigma^{\pi^-,\NC}_{\modd}$ Ratio to NLO.
Bottom right panel: $R^{\pi^-,(n)}(Q^2)$.
}
\label{fig:Q2_NC_pim}
\end{figure}

In this section we study the impact of electroweak effects in polarized NC SIDIS up to NNLO in QCD on the  $Q^2$-differential distributions.
Fig.~\ref{fig:Q2_NC_pip} shows the $Q^2$-distribution for the $\lambda_\ell$-odd cross section difference $\dd \Delta\sigma^{\pi^+,\NC}_{\modd}/\dd Q^2$.
In the top left panel the cross section difference is shown in the entire $Q^2$ kinematic range for all NC vector boson exchanges.
  A sharp change in slope can be observed for $Q^2>6000\,\GeV^2$; it is due to the corresponding high-$x$ region probed and resulting decrease of the polarized PDFs.
The electroweak effects are considerably larger than in  unpolarized SIDIS (see \cite{Bonino:2025qta}),
especially at Low-$Q^2$ and Mid-$Q^2$.
This can be seen from the ratio $(\dd \Delta^{\pi^+,\NC}_{\modd}/\dd \Delta^{\pi^+,\gamma\gamma}_{\modd})(Q^2)-1$ in the bottom left panel of Fig.~\ref{fig:Q2_NC_pip}. Electroweak effects are at $10\%$ level already at $Q^2\sim 200 \,\GeV^2$ and grow up to $90\%$ in the highest kinematic bins.
This is consistent with the enhancement of electroweak effects in polarized DIS observables already reported in \cite{Borsa:2022cap}.
The increase of the electroweak effects with $Q^2$ is connected to the growth of the suppression factors $\eta_{\gamma Z}$ and $\eta_{Z Z}$.

The ratio $(\dd \Delta^{\pi^+,\NC}_{\modd}/\dd \Delta^{\pi^+,\gamma\gamma}_{\modd})(Q^2)$ is affected by
perturbative corrections at the level of up to two per cent, as  can be seen in the bottom right panel of Fig.~\ref{fig:Q2_NC_pip} displaying the ratio $R^{\pi^+,(n)}$ from~\eqref{eq:Rratio}.
For $400 \,\GeV^2<Q^2<1300 \,\GeV^2$ estimates of the cross section obtained by dressing the $n^{\mathrm{th}}$ order photon exchange with the LO NC and photon exchange ratio are only accurate to low-percent level,
which stands in contrast with the sub-percent-level accuracy of this approximation in the unpolarized case \cite{Bonino:2025qta}.

In the top right panel of Fig.~\ref{fig:Q2_NC_pip} the structure function decomposition of the $\lambda_\ell$-odd NC cross section difference is shown.
The photon contribution is the most dominant at Low-$Q^2$ and Mid-$Q^2$.
The second most important contribution comes from $\Gcal^{\pi^+}_T$, starting from a few percent level at Low-$Q^2$ and Mid-$Q^2$.
For $Q^2>10^3\,\GeV^2$ $\Gcal^{\pi^+}_T$ becomes comparable in size to $\Gcal^{\pi^+,\gamma\gamma}_1$ despite including only electroweak effects.
Furthermore, $\Gcal^{\pi^+}_T$ is significantly larger than $\Gcal^{\pi^+,(\gamma Z+ZZ)}_1$ in the entire kinematic range.
Only in the Extreme-$Q^2$ regions electroweak effects in $\Gcal^{\pi^+}_1$ start to become sizeable.
The structure function decomposition highlights the importance of electroweak effects in polarized SIDIS: even at moderate $Q^2$ values the inclusion of the purely electroweak $\Gcal^{\pi^+}_T$ structure function is mandatory to achieve percent level precision.
We finally notice that the $\Gcal^{\pi^+}_L$ structure function is negligible in the entire kinematic region.
Its appearance starting from NLO results in enhanced theory uncertainty bands.
In unpolarized SIDIS the $\Fcal^{\pi^+,\gamma\gamma}_L$ structure function contributed significantly to the cross section at Low-$Q^2$ and Mid-$Q^2$, amplifying the dominant role played by the photon-only exchange \cite{Bonino:2025qta}.
The enhancement of electroweak effects in polarized with respect to unpolarized SIDIS is also due to the absence of the polarized longitudinal structure function $\Gcal^{\pi^+,\gamma\gamma}_L$ in photon exchange.

In the centre right panel of Fig.~\ref{fig:Q2_NC_pip} we show the size of the $\dd\Delta\sigma^{\pi^+,\NC,(n)}_{\modd}$ cross section  at $n^{\mathrm{th}}$ order in QCD relative to the NLO prediction.
We observe very large NLO corrections and sizeable NNLO corrections.
The LO prediction fails to describe the shape of the distribution in large parts of the $Q^2$ range.
NNLO corrections are rather small for most of the kinematical range, becoming sizeable only towards large $Q^2$,
where they reach up to 10\%.
Good perturbative convergence and stability is observed in the entire $Q^2$ range, with NLO and NNLO displaying overlapping theory uncertainty bands throughout.

In Fig.~\ref{fig:Q2_NC_pim} we show the NC $Q^2$-differential cross section for $\pi^-$ production in the same kinematical range of $Q^2$.
The shape of $\dd\Delta\sigma^{\pi^-,\NC}_{\modd}/\dd Q^2$ is comparable to  $\dd\Delta\sigma^{\pi^+,\NC}_{\modd}/\dd Q^2$, as can be seen in the top left panel of Fig.~\ref{fig:Q2_NC_pim}.
The size for $\dd\Delta\sigma^{\pi^-,\NC}_{\modd}/\dd Q^2$ is approximately a factor of 10 smaller than $\dd\Delta\sigma^{\pi^+,\NC}_{\modd}/\dd Q^2$ for the following two reasons.
Most prominently, we observe that the contribution from the negative $d\to d$ channel becomes more important, as it is flavour-favoured by both proton and $\pi^-$, see also Fig.~\ref{fig:z_NC_AS_channel_pim} below.
Moreover, we observe substantial contributions from sea quarks in $\pi^-$ production not only at Low-$Q^2$, where small values of $x$ are probed, but even at Mid-$Q^2$ and High-$Q^2$.
The $d$ and $\bar{d}$ initiated channels are negative due to the sign of $\Delta f_d^p$ and $\Delta f_{\bar{d}}^p$ in the BDSSV24 polarized PDF set~\cite{Borsa:2024mss}.

For $Q^2$ below $300 \, \GeV^2$ the three structure functions $\Gcal^{\pi^-,(\gamma Z+ZZ)}_T$, $\Gcal^{\pi^-,(\gamma Z+ZZ)}_1$ and $\Gcal^{\pi^-,(\gamma Z+ZZ)}_L$ cross zero and turn negative, as shown in the top right panel of Fig.~\ref{fig:Q2_NC_pim}.
This is due to large negative contributions from sizeable partonic channels, especially $d\to d$ and $\bar{d}\to\bar{d}$, and resulting cancellations.
For $Q^2 < 300 \, \GeV^2$ the $\Gcal^{\pi^-,\gamma\gamma}_1$ structure function is by far the largest contribution.
The most dominant electroweak contribution coming from $\Gcal^{\pi^-,(\gamma Z+ZZ)}_T$ grows quickly with $Q^2$ and is of the same order of magnitude as $\Gcal^{\pi^-,\gamma\gamma}_1$ already at a few hundred $\GeV^2$.
While the $\Gcal^{\pi^-,(\gamma Z+ZZ)}_L$ contribution is negligible in the entire kinematic range,  the contribution from $\Gcal^{\pi^-,(\gamma Z+ZZ)}_1$ is just one order of magnitude smaller than the one from $\Gcal^{\pi^-,(\gamma Z+ZZ)}_T$ in the highest kinematic bins.
Apart from the sign change at Low-$Q^2$, the overall trend of the structure functions is comparable between $\pi^+$ and $\pi^-$ production, resulting in large electroweak effects also for the latter.

The size of the electroweak effects can be inferred in more detail in the bottom left panel of Fig.~\ref{fig:Q2_NC_pim} with the ratio $(\dd\Delta\sigma^{\pi^-,\NC}_{\modd}/\dd\Delta\sigma^{\pi^-,\gamma\gamma}_{\modd})(Q^2)-1$.
Electroweak effects are at percent level for $Q^2>300\,\GeV^2$ and grow with $Q^2$.
Due to the depletion of the electroweak structure functions at Low-$Q^2$, the electroweak effects are overall smaller in $\pi^-$ production for $Q^2<10^3\,\GeV^2$ with respect to $\pi^+$.
In the highest $Q^2$ bins the electroweak effects reach the same size as the $\pi^+$ ones, due to the quick growth in $Q^2$ of the electroweak structure functions.

As shown in the bottom right panel of Fig.~\ref{fig:Q2_NC_pim} the ratio $R^{\pi^-,(n)}(Q^2)$ is comparable to $R^{\pi^+,(n)}(Q^2)$, apart for the Low-$Q^2$ region, for the same reasoning as above.
Overall it is possible to predict higher order QCD corrections to NC exchanges by dressing the photon-only cross section difference with the photon-only K-factor, at percent-level precision.
The perturbative stability and convergence of the QCD series is very similar in $\pi^+$ and $\pi^-$ production, as can be seen from the centre right panel in Fig.~\ref{fig:Q2_NC_pim}.
NNLO corrections are at most at percent level and always falling within the NLO theory uncertainty band.
The LO does not provide a satisfactory description of the distribution.

\begin{figure}[tb]
\centering
\includegraphics[width=0.49\textwidth]{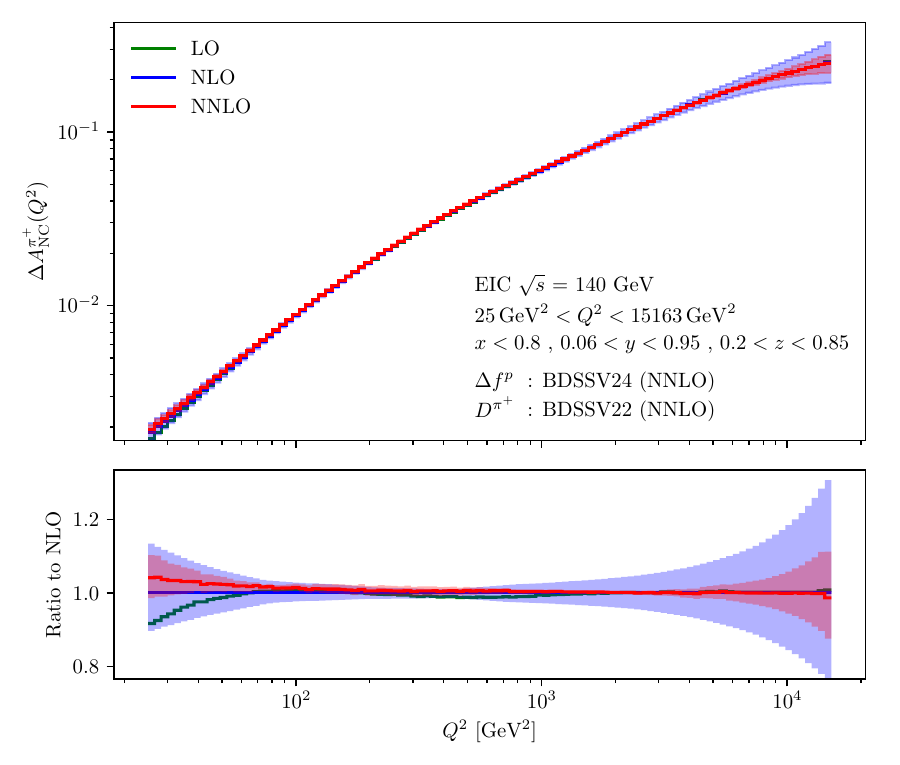}
\includegraphics[width=0.49\textwidth]{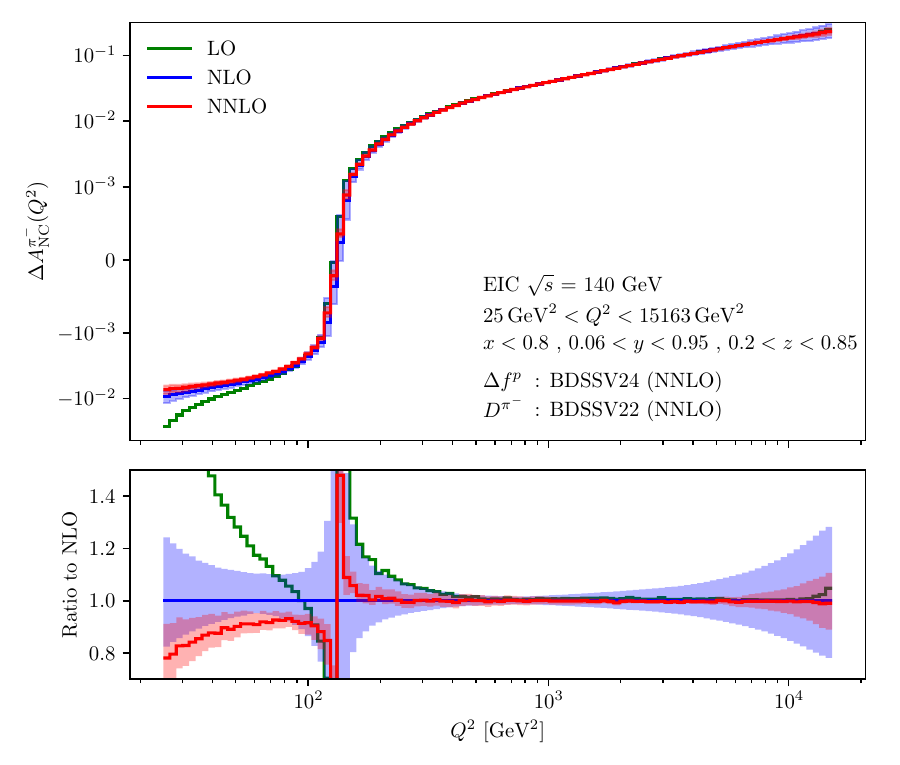}
\caption{Polarized NC
asymmetries
 $\Delta A_\mathrm{NC}^{\pi^+}$ (left) and $\Delta A_\mathrm{NC}^{\pi^-}$ (right) as function of $Q^2$.
Top panels: total asymmetries.
Bottom panels: Ratios to NLO.
}
\label{fig:Q2_A_NC_pi}
\end{figure}

Fig.~\ref{fig:Q2_A_NC_pi} displays the polarized neutral current asymmetry $\Delta A^{h}_\NC$, defined in~\eqref{eq:ANCm1},
 as a function of $Q^2$,
with $h=\pi^+$ and $h=\pi^-$ in the left and right panels respectively.
With this asymmetry it is possible to quantify the relative size of the $\lambda_{\ell}$-even contributions in~\eqref{eq:SFNCpol} with respect to the $\lambda_{\ell}$-odd ones.

$\Delta A^{\pi^+}_\NC(Q^2)$ is  small at low $Q^2$, reaching a percent level magnitude at $Q^2\sim 200\, \GeV^2$, and further increasing to $8\%$ at $Q^2\sim 1000\,\GeV^2$ to saturate at $30\%$ at the largest values of $Q^2$ probed.
The picture is very similar between unpolarized and polarized SIDIS, despite the fact that the roles of odd and even components are swapped.
The perturbative NNLO
corrections  to $\Delta A^{\pi^+}_\NC(Q^2)$, shown in the bottom panel, are sizeable at percent level only in the Low-$Q^2$ region.
For all other values of $Q^2$ they are at per-mille level.
We observe good perturbative convergence and stability.

The polarized NC asymmetry for $\pi^-$ production $\Delta A^{\pi^-}_\NC(Q^2)$ displays a similar
overall behaviour at large $Q^2$ compared to the $\pi^+$ case, while
significant differences appear for $Q^2<200\, \GeV^2$.
Since in the numerator of~\eqref{eq:ANCm1} the structure function $\Gcal^{\pi^-,\gamma\gamma}_1$ is absent, the $\lambda_\ell$-even asymmetry becomes zero and turns negative with the decrease of $Q^2$, following the trend of the electroweak structure functions (see Fig.~\ref{fig:Q2_NC_pim} top right panel).
For $Q^2<70\,\GeV^2$ the numerical value of the asymmetry becomes stable again and is about one order of magnitude larger than the $\pi^+$ asymmetry.
For $Q^2>500\,\GeV^2$ good perturbative convergence and stability is observed in the ratio to NLO.
Perturbative corrections are at per-mille level precision for $Q^2>500\,\GeV^2$.

\begin{figure}[tb]
\centering
\includegraphics[width=0.49\textwidth]{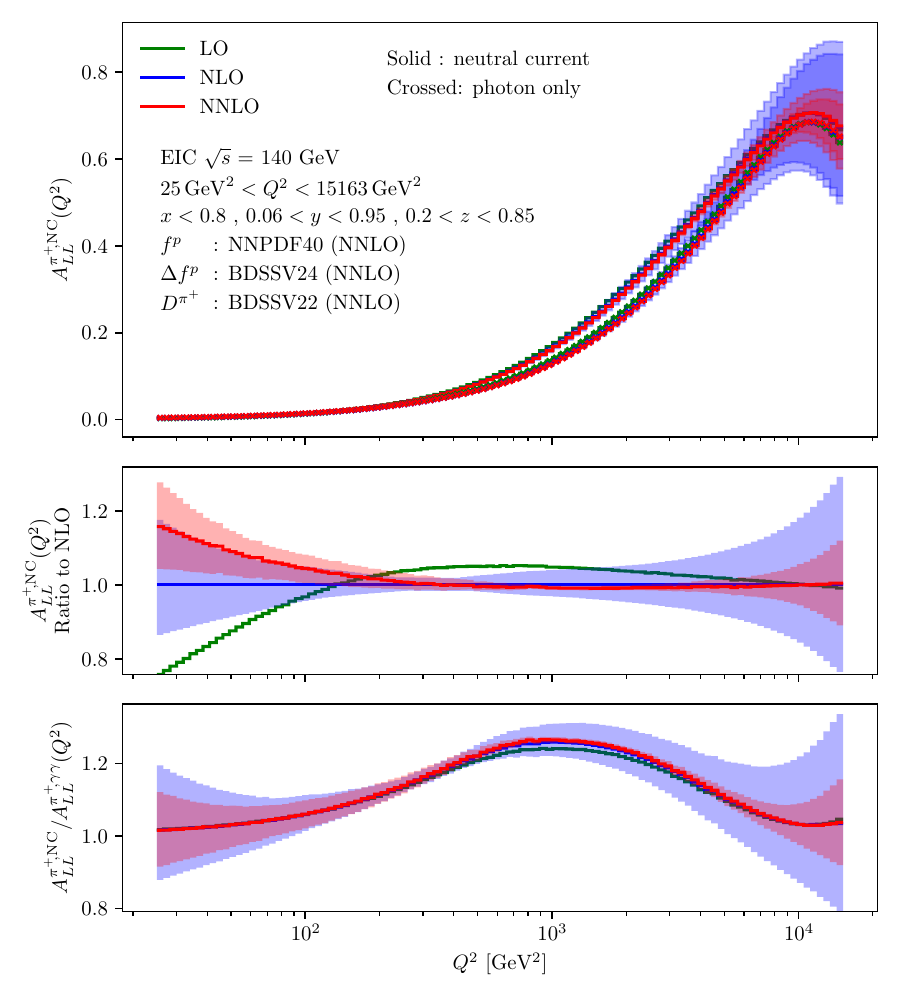}
\includegraphics[width=0.49\textwidth]{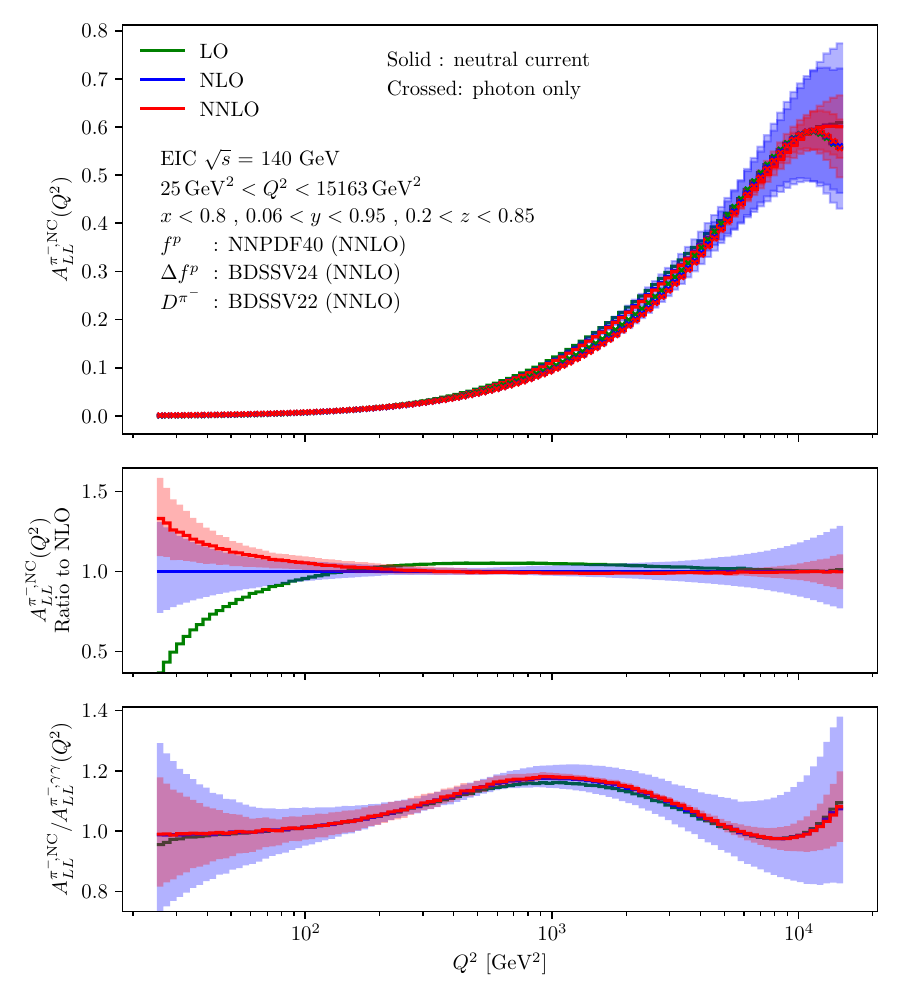}
\caption{Double-longitudinal spin asymmetries for NC exchange
 $A_{LL}^{\pi^+}$ (left) and $A_{LL}^{\pi^-}$ (right) as function of $Q^2$.
Top panels: total asymmetries for full NC (solid) and photon-only (crossed) exchanges.
Centre panels: NC Ratios to NLO.
Bottom panels: Ratio $A_{LL}^{h,\NC}/A_{LL}^{h,\gamma\gamma}$.
}\label{fig:Q2_A_LL_NC_pi}
\end{figure}

In Fig.~\ref{fig:Q2_A_LL_NC_pi}  we study the double-longitudinal spin asymmetry~\eqref{eq:AhLLNC} for $\pi^+$ and $\pi^-$ production respectively as a function of $Q^2$.
In the top panels the asymmetry is shown for full NC exchange (solid) and for photon-only exchange (crossed).

For $\pi^+$ production (left panels) in both current exchanges the asymmetry grows with $Q^2$ from close to zero up to about $0.7$.
The double-longitudinal spin asymmetry is larger for neutral current exchange ($A^{\pi^+,\NC}_{LL}$) than for photon-only exchange ($A^{\pi^+,\gamma\gamma}_{LL}$) for all $Q^2$-values, as visible in the ratio $(A^{\pi^+,\NC}_{LL}/A^{\pi^+,\gamma\gamma}_{LL})(Q^2)$ plotted in the bottom panel.
For $150 \, \GeV^2<Q^2<6000\,\GeV^2$
the enhancement by the neutral current is significant within theory uncertainties.
At $Q^2\sim 10^3\,\GeV^2$ the neutral current effects in $A^{\pi^+}_{LL}$ enhance the asymmetry by more than $20\%$ over $A^{\pi^+,\gamma\gamma}_{LL}$.
At Low-$Q^2$ and Extreme-$Q^2$ the ratio tends to unity.
This can be explained with an enhancement of electroweak effects observed for polarized SIDIS with respect to unpolarized SIDIS at Mid-$Q^2$ and High-$Q^2$, as already observed in Fig.~\ref{fig:Q2_NC_pip}.
This enhancement is aligned with a rapid growth of the $\Gcal^{\pi^+,(\gamma Z+ZZ)}_{T}$ structure function contribution to the total cross section difference (see top right panel in Fig.~\ref{fig:Q2_NC_pip}).
The asymmetry $A^{\pi^+,\NC}_{LL}$ displays large QCD corrections, with good perturbative convergence and stability.
The different orders in perturbation theory are in good agreement for the double-longitudinal spin asymmetry in a wide range of $Q^2$ ($Q^2 > 100\,\GeV^2$), with NLO and NNLO deviating
 from each other by at most $1\,\%$ for $Q^2 > 300\,\GeV^2$.
Larger corrections are visible only at low $Q^2$, with each order significantly enhancing the cross section.
The NLO and NNLO scale variation uncertainty bands are overlapping throughout.

The picture is similar for $\pi^-$ production (right panels), where
the asymmetries are about $20\%$ smaller than for $\pi^+$ production.
The electroweak effects display a similar shape, while being systematically smaller in the Mid-$Q^2$ and High-$Q^2$ regions in $\pi^-$ production.
At Low-$Q^2$ the perturbative corrections are enhanced in $\pi^-$ production, reaching multiples of the LO asymmetry, as visible in the centre panel.

\begin{figure}[tb]
\centering
\includegraphics[width=0.49\textwidth]{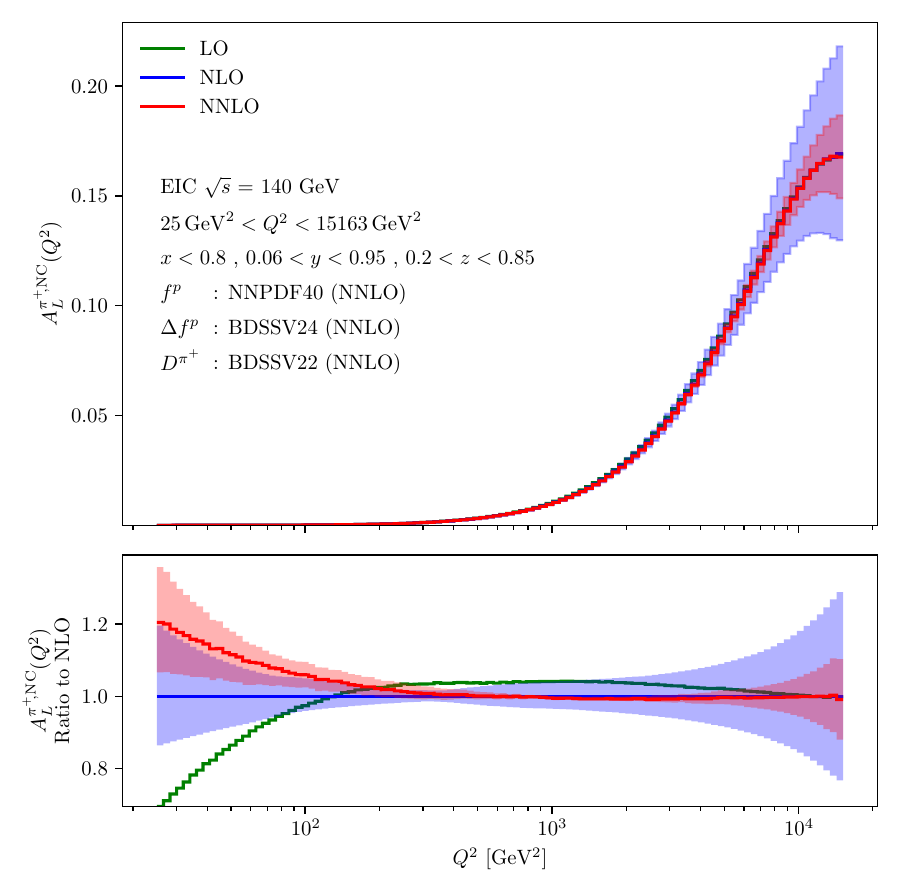}
\includegraphics[width=0.49\textwidth]{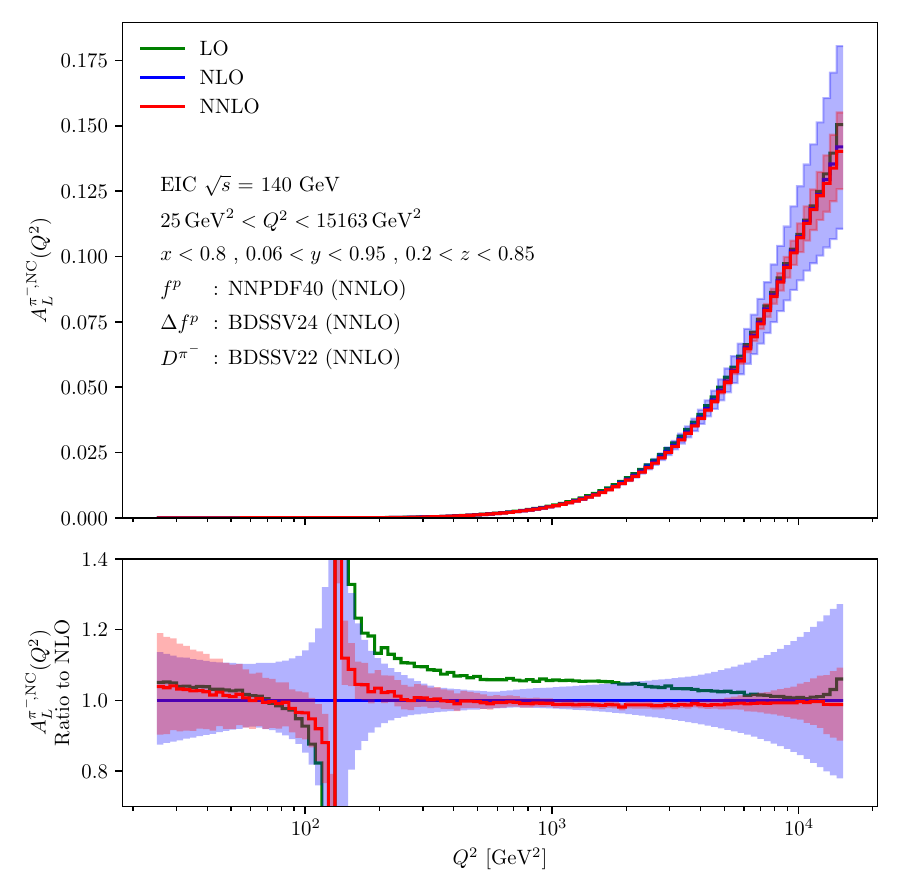}
\caption{Single-longitudinal spin asymmetries for NC exchange
 $A_{L}^{\pi^+,\NC}$ (left) and $A_{L}^{\pi^-,\NC}$ (right) as function of $Q^2$.
Top panels: total asymmetries.
Bottom panels: Ratios to NLO.
}\label{fig:Q2_A_L_pi}
\end{figure}

In Fig.~\ref{fig:Q2_A_L_pi} we study the NC single-longitudinal spin asymmetry~\eqref{eq:AhLNC} for $\pi^+$ and $\pi^-$ production as a function of $Q^2$.

For $\pi^+$ production the asymmetry slowly grows with increasing  $Q^2$ up to $Q^2\sim 10^3\,\GeV^2$, where
a  more rapid growth sets in.
At $Q^2\sim 15000\,\GeV^2$ the size of the electroweak effects in the polarized cross section difference is about $10\%$ the size of the full NC unpolarized cross section.
The asymmetry $A^{\pi^+,\NC}_{L}$ reaches a maximum of $\sim0.16$ in the highest $Q^2$ bin.
Perturbative corrections are large in the Low-$Q^2$ and Mid-$Q^2$ regions and they decrease with the increase of $Q^2$.
In the Low-$Q^2$ region perturbative corrections are above $10\%$ level for both NLO and NNLO.
 We observe overall good perturbative convergence and stability.

For $\pi^-$ the behaviour of the single-spin asymmetry is similar to the $\pi^+$ case, with
$A^{\pi^-,\NC}_{L}$ being about $20\%$ smaller than $A^{\pi^+,\NC}_{L}$ for all values of $Q^2$.
$A^{\pi^-,\NC}_{L}$ is negative up to $Q^2\sim 100\,\GeV^2$  and starts to becomes sizeable at $Q^2\sim 10^3\,\GeV^2$.
The behaviour of perturbative corrections is comparable between $\pi^+$ and $\pi^-$ production.

\subsubsection[\texorpdfstring{$z$-distributions}{z-distributions}]{\boldmath $z$-distributions}

\begin{figure}[p]
\centering
\begin{subfigure}{0.49\textwidth}
\centering
\hspace{-7pt}
\includegraphics[width=1.0\linewidth]{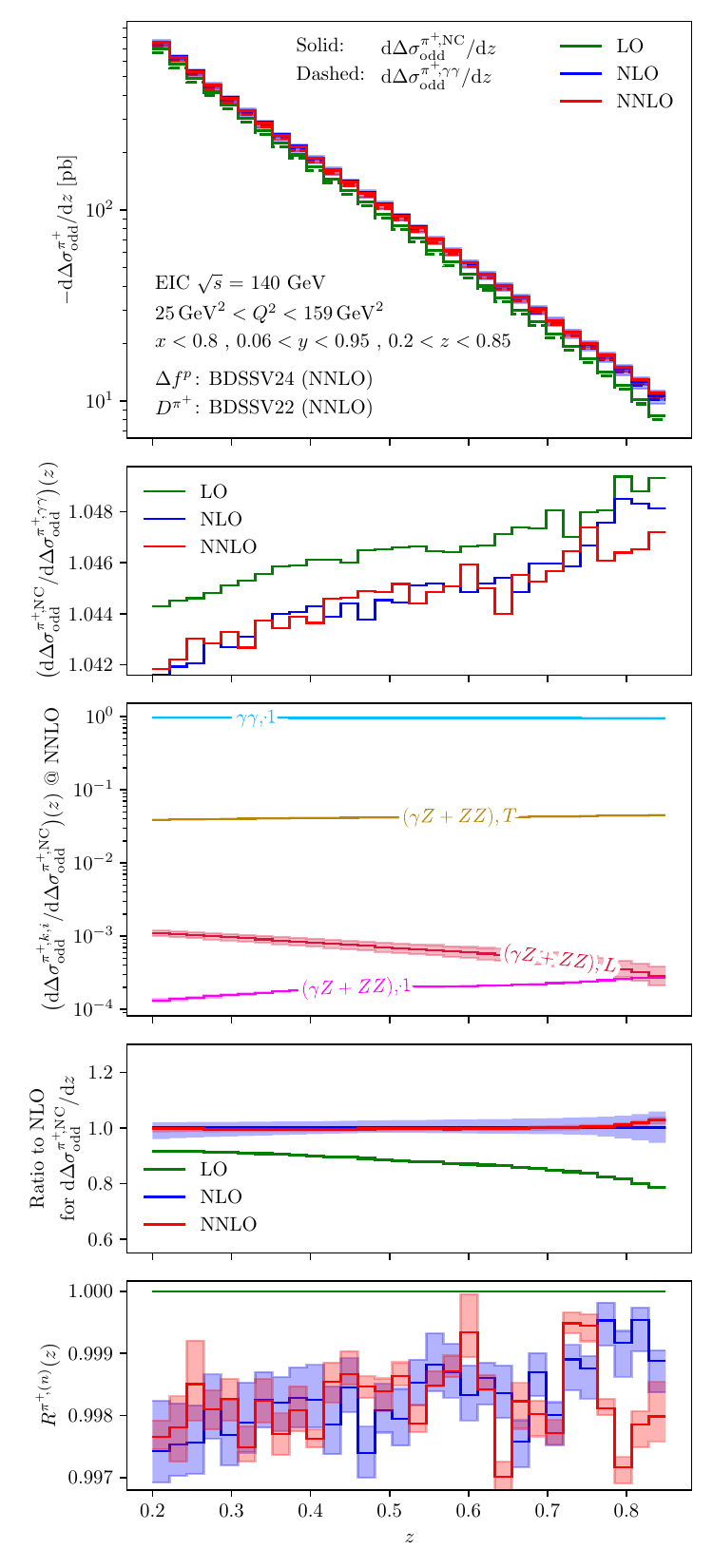}
\caption{Low-$Q^2$}
\end{subfigure}
\begin{subfigure}{0.49\textwidth}
\centering
\includegraphics[width=1.0\linewidth]{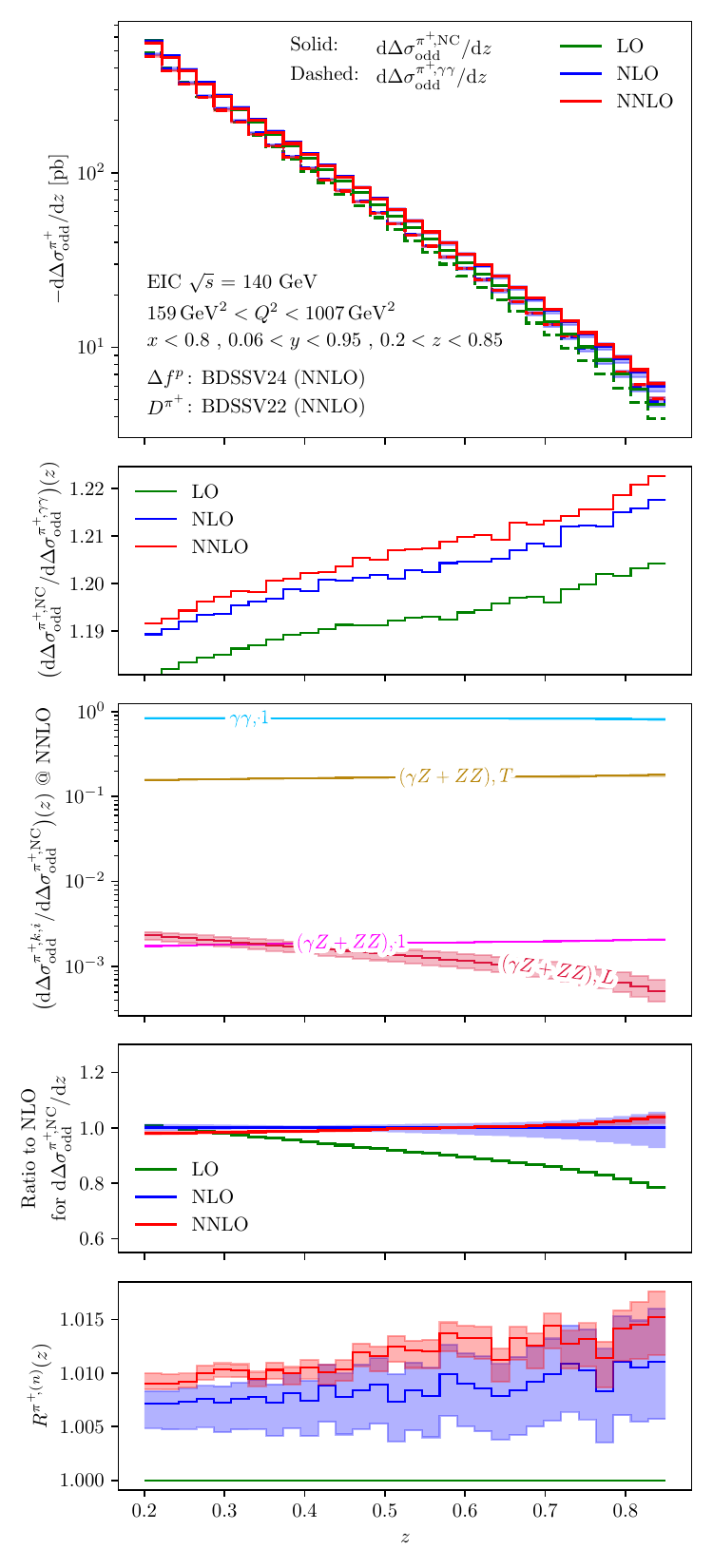}
\caption{Mid-$Q^2$}
\end{subfigure}

\caption{NC asymmetry $z$-distributions for $\pi^+$ production at low and intermediate $Q^2$.
First panels: total cross section difference for $\dd\Delta\sigma^{\pi^+,\NC}_{\modd}/\dd z$ and $\dd\Delta\sigma^{\pi^+,\gamma\gamma}_{\modd}/\dd z$.
Second panels: ratio $\left(\dd\Delta\sigma^{\pi^+,\NC}_{\modd}/\dd\Delta\sigma^{\pi^+,\gamma\gamma}_{\modd}\right)(z)$.
Third panels: structure function contributions to the total cross section asymmetry $\dd\Delta\sigma^{\pi^+,\NC}_{\modd}/\dd z$.
Fourth panels: $\dd\Delta\sigma^{\pi^+,\NC}_{\modd}$ Ratio to NLO.
Fifth Panels: $R^{\pi^+,(n)}$ ratio as function of $z$.
}
\label{fig:Z_0.0_NC_lowQ2_pip}

\end{figure}

\begin{figure}[p]
\centering
\begin{subfigure}{0.49\textwidth}
\centering
\includegraphics[width=\linewidth]{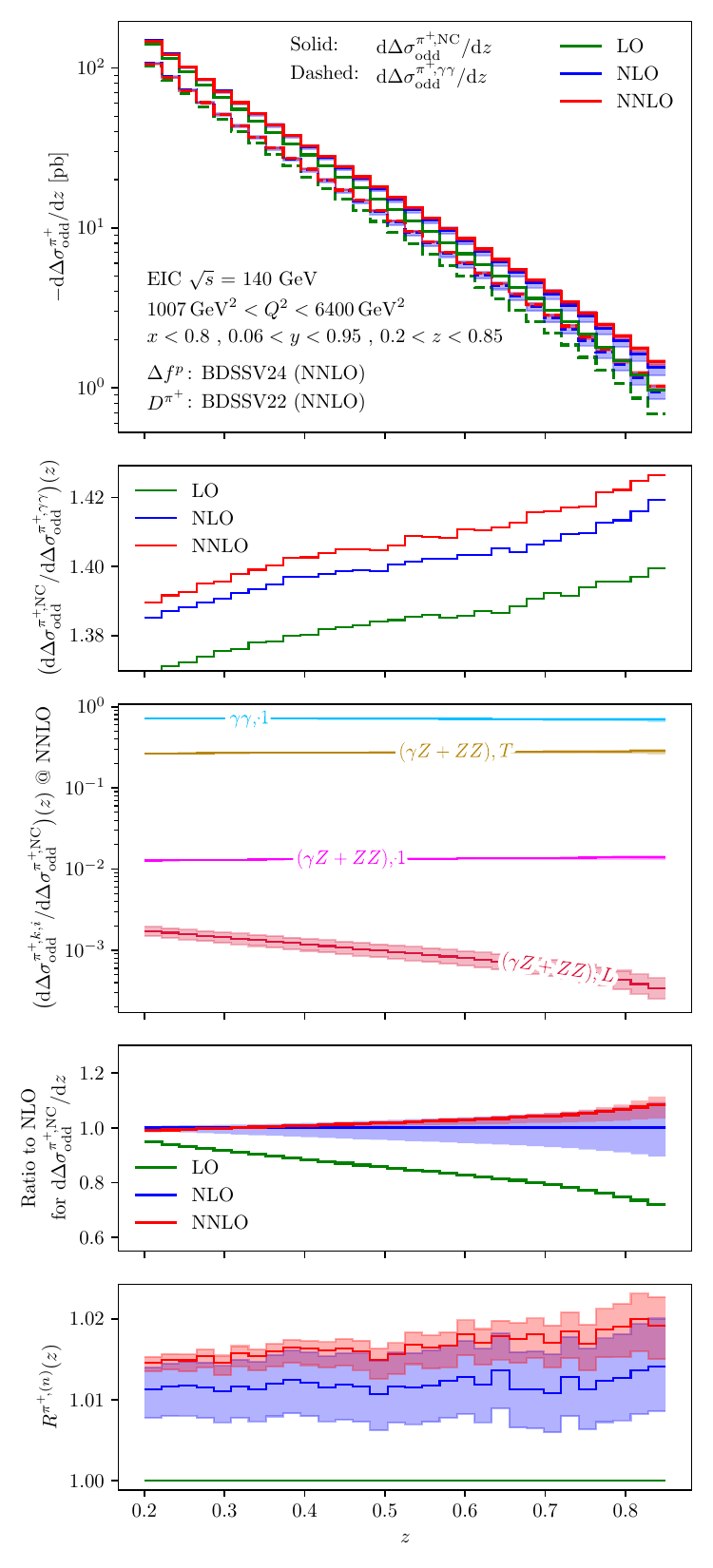}
\caption{High-$Q^2$}
\end{subfigure}
\begin{subfigure}{0.49\textwidth}
\centering
\includegraphics[width=\linewidth]{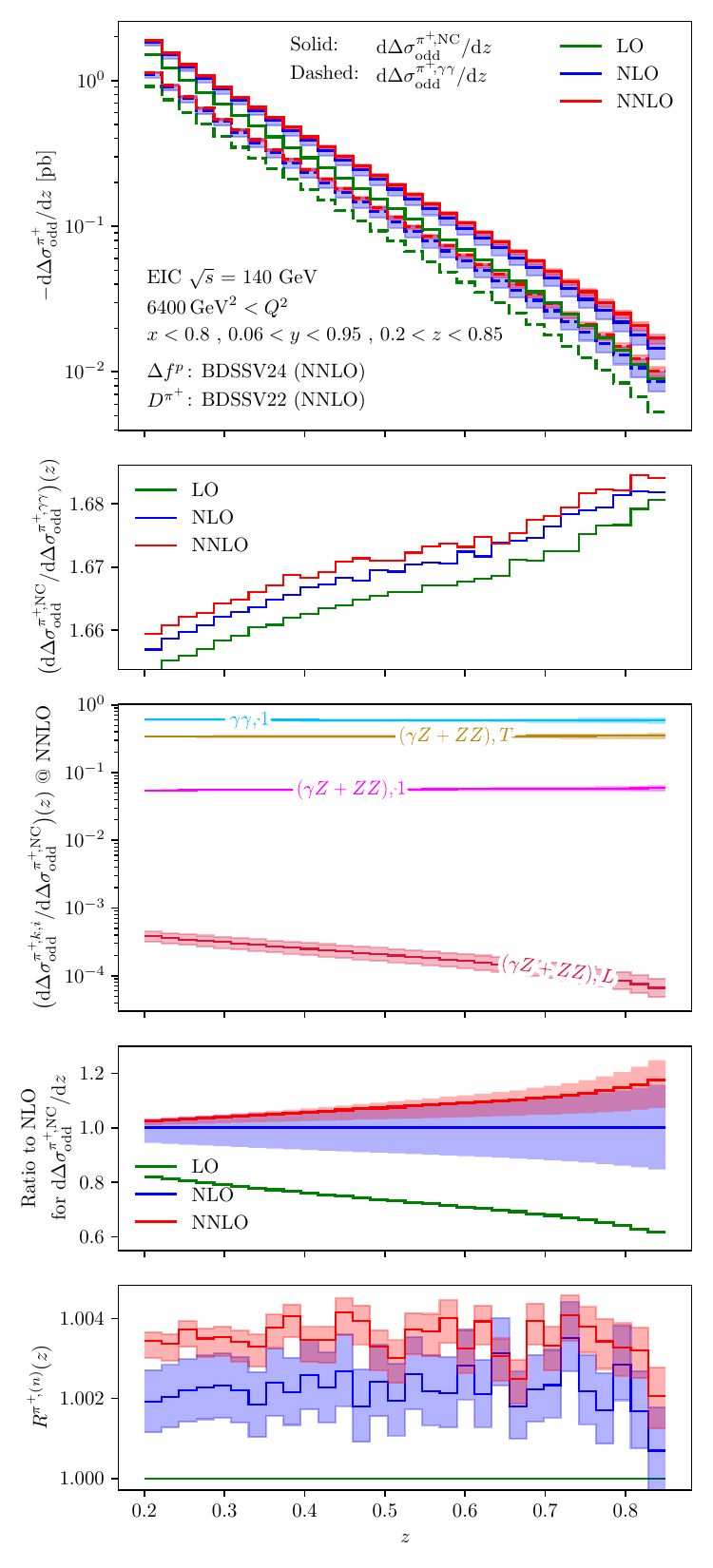}
\caption{Extreme-$Q^2$}
\end{subfigure}

\caption{
NC asymmetry $z$-distributions for $\pi^+$ production at high $Q^2$.
First panels: total cross section differences $\dd\Delta\sigma^{\pi^+,\NC}_{\modd}/\dd z$ and $\dd\Delta\sigma^{\pi^+,\gamma\gamma}_{\modd}/\dd z$.
Second panels: ratio $\left(\dd\Delta\sigma^{\pi^+,\NC}_{\modd}/\dd\Delta\sigma^{\pi^+,\gamma\gamma}_{\modd}\right)(z)$.
Third panels: structure function contributions to the total cross section asymmetry $\dd\Delta\sigma^{\pi^+,\NC}_{\modd}$.
Fourth panels: $\dd\Delta\sigma^{\pi^+,\NC}_{\modd}/\dd z$ Ratio to NLO.
Fifth Panels: $R^{\pi^+,(n)}$ ratio as function of $z$.
}
\label{fig:Z_0.0_NC_highQ2_pip}

\end{figure}

We now turn to the discussion of the $z$-distributions in $\pi^+$ production.
The cross section difference $\dd\Delta\sigma^{\pi^+,\NC}_{\modd}/\dd z$ is shown in Fig.~\ref{fig:Z_0.0_NC_lowQ2_pip} and~\ref{fig:Z_0.0_NC_highQ2_pip} for the four ranges in $Q^2$ defined in~\eqref{eq:cutsQ2}.
The differential cross section difference decreases with $Q^2$ following the DIS-like
scaling $1/Q^2$ as shown in the first panels.
The decrease in $z$ that is a common feature across the four distributions in~\ref{fig:Z_0.0_NC_lowQ2_pip} and~\ref{fig:Z_0.0_NC_highQ2_pip} is due to the behaviour of the fragmentation functions.
Only moderate changes in slope are observed across the four ranges.
In the first panel the $z$-differential distributions $\dd\Delta\sigma^{\pi^+}_{\modd}/\dd z$ for neutral current (solid) and photon exchange (dashed) are shown.
Their functional shape in $z$ is comparable, but electroweak effects visibly enhance the size of the cross section differences.
These electroweak effects are studied in more detail in the second panels of Fig.~\ref{fig:Z_0.0_NC_lowQ2_pip} and \ref{fig:Z_0.0_NC_highQ2_pip}, where the ratios $(\dd\Delta\sigma^{\pi^+,\NC}_{\modd}/\dd\Delta\sigma^{\pi^+,\gamma\gamma}_{\modd})(z)$ are plotted.
The size of the electroweak effects grows with $Q^2$.
Between different ranges in $Q^2$ it may be estimated by the propagator ratio $\eta_{\gamma Z}$ evaluated
at the lower bound in $Q^2$.
The relative size of the electroweak effects over photon exchange is thus comparable to unpolarized SIDIS,
see \cite{Bonino:2025qta},
ranging from a few percent level at Low-$Q^2$ up to more than $60\%$ in the Extreme-$Q^2$ region.

Unlike in unpolarized SIDIS, the electroweak effects display a non-trivial $z$-dependence and are further enhanced by higher orders by up to $3\,\%$, as visible in the second panels.
The $n^{\mathrm{th}}$-order QCD corrections and theory uncertainties to the cross section difference $\dd\Delta\sigma^{\pi^+,\NC}_{\modd}/\dd z$ can be predicted by rescaling the photon exchange cross section difference $\dd\Delta\sigma^{\pi^+,\gamma\gamma}_{\modd}$ with the LO ratio $\dd\Delta\sigma^{\pi^+,\NC,(0)}_{\modd}/\dd\Delta\sigma^{\pi^+,\gamma\gamma,(0)}_{\modd}$ at percent level precision.
This is shown with the ratio $R^{\pi^+,(n)}(z)$ in the fifth panel of each plot.
For unpolarized SIDIS instead per-mille level precision was observed~\cite{Bonino:2025qta}.

In the third panel of Fig.~\ref{fig:Z_0.0_NC_lowQ2_pip} and~\ref{fig:Z_0.0_NC_highQ2_pip} we display the decomposition in structure function contributions.
The photon-only structure function $\Gcal^{\pi^+,\gamma\gamma}_1$ is the dominant contribution in all $Q^2$ regions.
The second most relevant contributions come from the purely electroweak structure function $\Gcal^{\pi^+,(\gamma Z+ZZ)}_T$.
At Low-$Q^2$ its contribution is at the level of $5\,\%$.
Its size increases with $Q^2$, reaching close to $20\,\%$ of the contribution from $\Gcal^{\pi^+,\gamma\gamma}_1$ already in the Mid-$Q^2$ region.
From High-$Q^2$ to Extreme-$Q^2$ the contribution from $\Gcal^{\pi^+,(\gamma Z+ZZ)}_T$ grows from about $30\%$ to above $50\%$ of the contribution from $\Gcal^{\pi^+,\gamma\gamma}_1$.
The electroweak effects in $\Gcal^{\pi^+}_1$ display the strongest growth over several orders of magnitude between the different energy ranges, but still only reach a size of $1\,\%$ and $6\,\%$ of the total cross section at High-$Q^2$ and Extreme-$Q^2$ respectively.
The contribution from the longitudinal structure function is negligible in the entire kinematic range and appears with larger theory uncertainty bands due to its first emergence at NLO.
The $z$ dependence of the electroweak structure functions is dominated by the shape of the FFs as in the photon-only exchange.

In the fourth panels of Fig.~\ref{fig:Z_0.0_NC_lowQ2_pip} and \ref{fig:Z_0.0_NC_highQ2_pip} we show the ratio to NLO for $\dd\Delta\sigma^{\pi^+,\NC}_{\modd}/\dd z$.
The LO merely admits a rough estimate of the $z$-differential cross section in each kinematic range.
Higher order corrections are sizeable, especially at high $z$ where resummation should be taken into account~\cite{Goyal:2025bzf}.
We observe very good perturbative convergence and stability, with the NNLO prediction always within the NLO theory uncertainty.
Including NNLO corrections leads to a significant reduction of theory uncertainty.

\begin{figure}[p]
\centering
\begin{subfigure}{0.49\textwidth}
\centering
\hspace{-7pt}
\includegraphics[width=.995\linewidth]{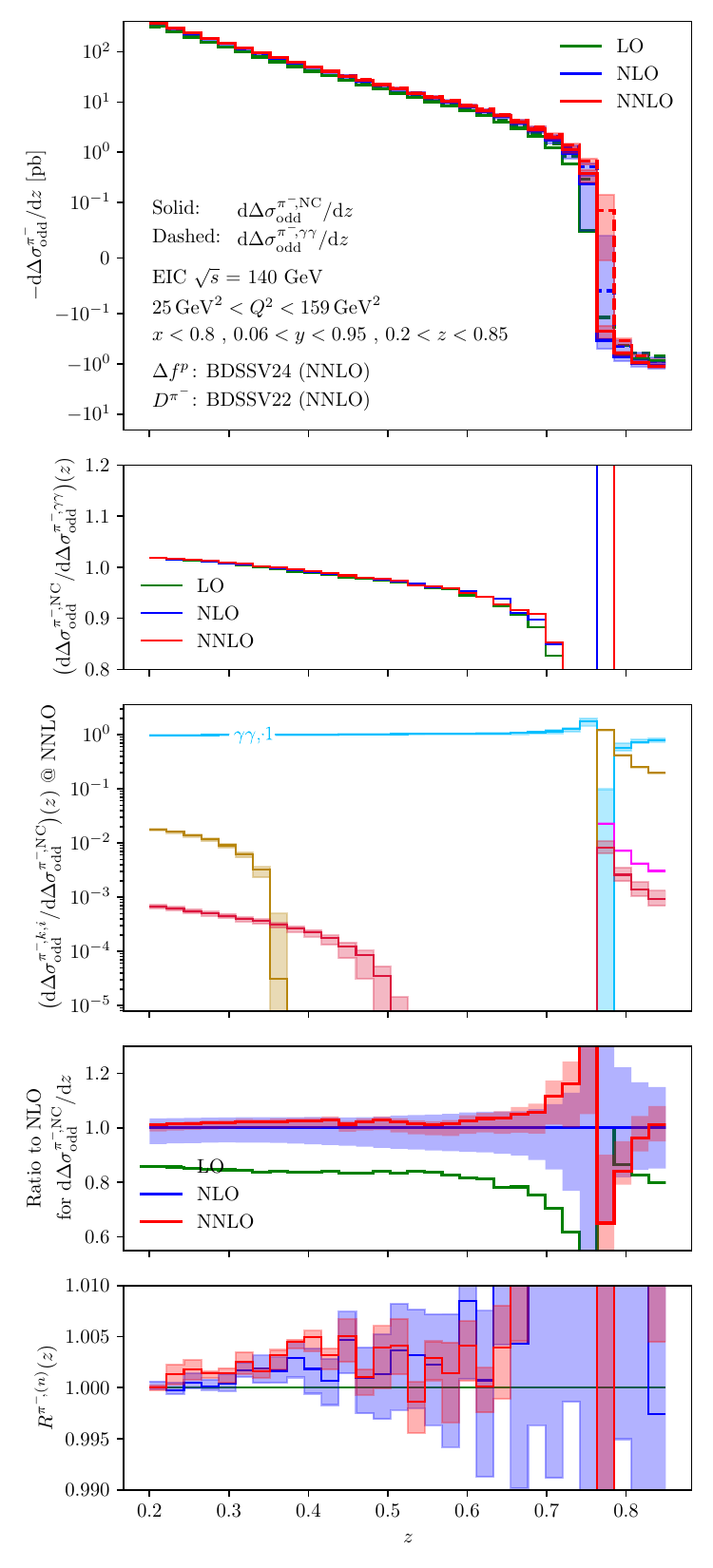}
\caption{Low-$Q^2$}
\end{subfigure}
\begin{subfigure}{0.49\textwidth}
\centering
\includegraphics[width=1.0\linewidth]{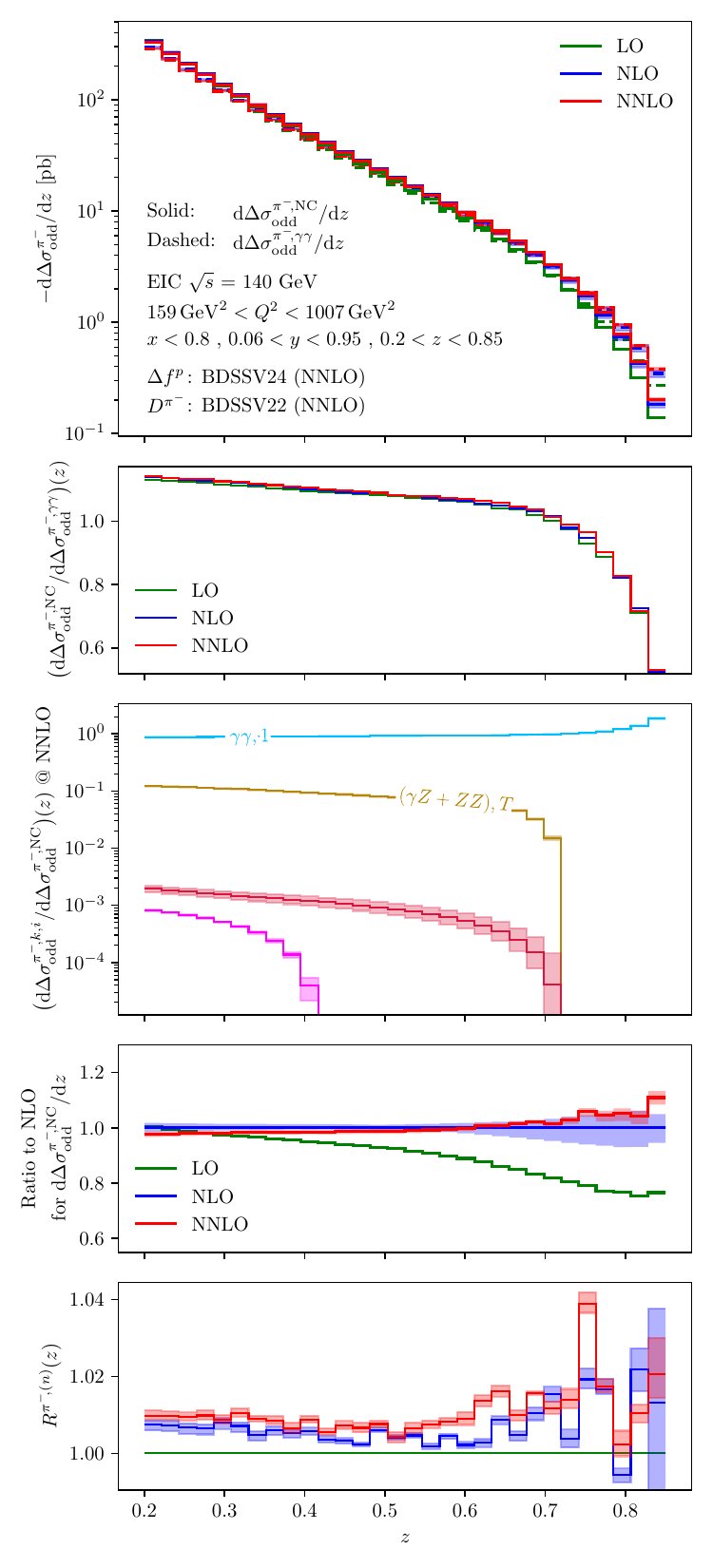}
\caption{Mid-$Q^2$}
\end{subfigure}

\caption{
NC cross section difference $z$-distributions for $\pi^-$ production at low and intermediate $Q^2$.
First panels: total cross section for $\dd\Delta\sigma^{\pi^-,\NC}_{\modd}/\dd z$ and $\dd\Delta\sigma^{\pi^-,\gamma\gamma}_{\modd}/\dd z$.
Second panels: ratio $\left(\dd\Delta\sigma^{\pi^-,\NC}_{\modd}/\dd\Delta\sigma^{\pi^-,\gamma\gamma}_{\modd}\right)(z)$.
Third panels: structure function contributions to the total cross section difference $\dd\Delta\sigma^{\pi^-,\NC}_{\modd}/\dd z$.
Fourth panels: $\dd\Delta\sigma^{\pi^-,\NC}_{\modd}/\dd z$ ratio to NLO.
Fifth Panels: $R^{\pi^-,(n)}$ ratio as function of $z$.
}
\label{fig:Z_0.0_NC_lowQ2_pim}

\end{figure}

\begin{figure}[p]
\centering
\begin{subfigure}{0.49\textwidth}
\centering
\includegraphics[width=\linewidth]{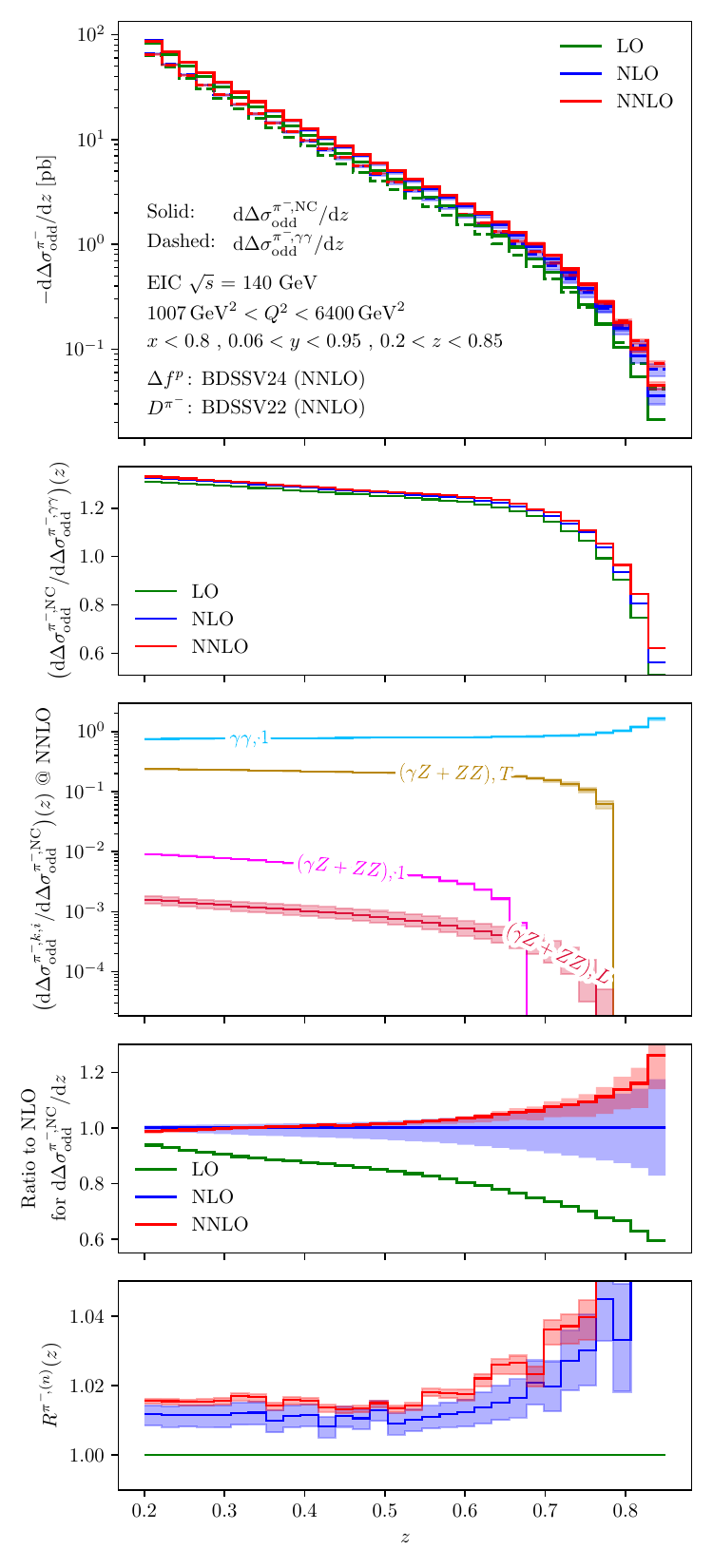}
\caption{High-$Q^2$}
\end{subfigure}
\begin{subfigure}{0.49\textwidth}
\centering
\includegraphics[width=\linewidth]{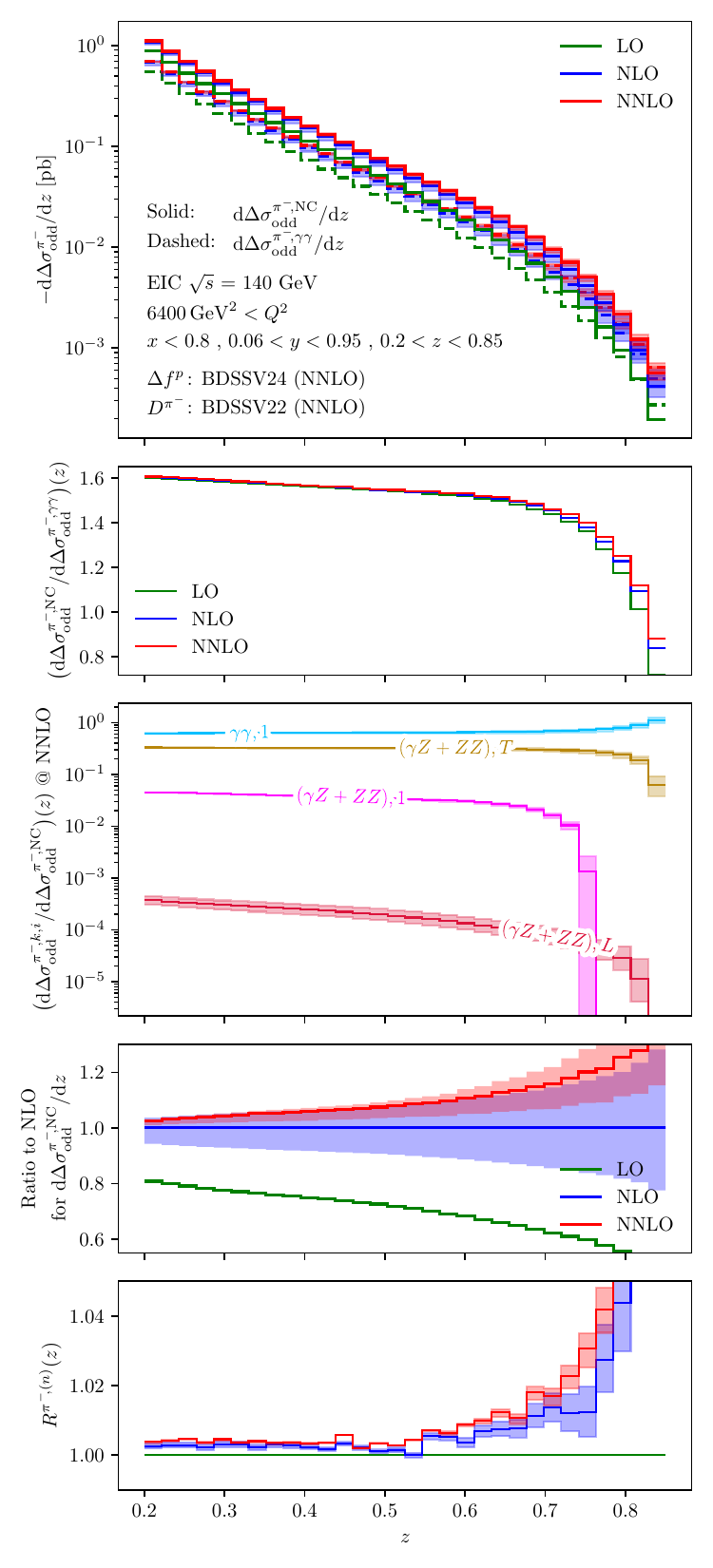}
\caption{Extreme-$Q^2$}
\end{subfigure}

\caption{
NC cross section difference $z$-distributions for $\pi^-$ production at high $Q^2$.
First panels: total cross sections $\dd\Delta\sigma^{\pi^-,\NC}_{\modd}/\dd z$ and $\dd\Delta\sigma^{\pi^-,\gamma\gamma}_{\modd}/\dd z$.
Second panels: ratio $\left(\dd\Delta\sigma^{\pi^-,\NC}_{\modd}/\dd\Delta\sigma^{\pi^-,\gamma\gamma}_{\modd}\right)(z)$.
Third panels: structure function contributions to the total cross section cross section difference $\dd\Delta\sigma^{\pi^-,\NC}_{\modd}/\dd z$.
Fourth panels: $\dd\Delta\sigma^{\pi^-,\NC}_{\modd}/\dd z$ ratio to NLO.
Fifth Panels: $R^{\pi^-,(n)}$ ratio as function of $z$.}
\label{fig:Z_0.0_NC_highQ2_pim}

\end{figure}

In Fig.~\ref{fig:Z_0.0_NC_lowQ2_pim} and \ref{fig:Z_0.0_NC_highQ2_pim} we show the $z$-differential cross section differences for $\pi^-$ production.
The first panels show that $\dd\Delta\sigma^{\pi^-,\NC}_{\modd}/\dd z$ is smaller than $\dd\Delta\sigma^{\pi^+,\NC}_{\modd}/\dd z$ for all $Q^2$ ranges.
This is due to the negative contribution from the flavour-favoured (and thus large in absolute value) $d\to d$ channel, see Fig.~\ref{fig:z_NC_AS_channel_pim} below.
The differential cross sections decrease with $Q^2$ following the DIS scaling in the same fashion as for $\pi^+$.
The $z$-dependence is given by the FFs with larger contributions from flavour-unfavoured quarks in the final state hadron.
The $\dd\Delta\sigma^{\pi^-}_{\modd}/\dd z$ distribution crosses zero at $z\sim 0.75$ for Low-$Q^2$.
This feature is already present in photon-only exchange at LO and is due to the growth of the negative flavour-favoured $d\to d$ contribution at large $z$.
The size of electroweak effects is studied in the ratio $(\dd\Delta\sigma^{\pi^-,\NC}_{\modd}/\dd\Delta\sigma^{\pi^-,\gamma\gamma}_{\modd})(z)$.
Like in unpolarized SIDIS and in polarized $\pi^+$ production, the ratio increases overall
 with increasing average value of $Q^2$ probed.
However, unlike the former two cases, the ratio decreases with increasing $z$ in polarized $\pi^-$ production.
At small $z$ in the Extreme-$Q^2$ range it reaches up to $60\%$.
The ratio is largely insensitive to the perturbative order as can be seen from the fifth panels.
For $z<0.7$ at per-mille level accuracy the $n^{\mathrm{th}}$-order QCD correction to $\dd\Delta\sigma^{\pi^-,\NC,(n)}_{\modd}/\dd z$ can be obtained by rescaling $\dd\Delta\sigma^{\pi^-,\gamma\gamma,(n)}_{\modd}/\dd z$ with the LO ratio $\dd\Delta\sigma^{\pi^-,\NC,(0)}_{\modd}/\dd\Delta\sigma^{\pi^-,\gamma\gamma,(0)}_{\modd}$.
This is analogous to what previously observed in unpolarized SIDIS~\cite{Bonino:2025qta}.

In the third panels of Fig.~\ref{fig:Z_0.0_NC_lowQ2_pim} and~\ref{fig:Z_0.0_NC_highQ2_pim} we show the structure function decomposition of $\dd\Delta\sigma^{\pi^-,\NC}_{\modd}/\dd z$.
The overall picture is identical to the $\pi^+$ case: the electroweak structure function $\Gcal^{\pi^-,(\gamma Z+ZZ)}_T$ quickly grows with $Q^2$ to reach the same size as the photon-only exchange.
At large $z$ the electroweak structure functions turn negative at a value of $z$ which depends on the average virtuality.
This is due to the negative flavour-favoured $d\to d$ channel reaching the same size as the positive $u\to u$ contribution, see Fig.~\ref{fig:z_NC_AS_channel_pim} below.
The electroweak structure functions are more sensitive to such large cancellations due to the smaller relative coupling suppression of the $d\to d$ channel.
Finally in the fourth panels we study the perturbative convergence of the QCD corrections.
We observe good perturbative convergence and stability with NNLO corrections falling in the NLO theory uncertainty envelope at small and moderate $z$.
At large $z$ the perturbative converges worsens, pointing at sizeable soft gluon resummation  effects \cite{Goyal:2025bzf}.

\begin{figure}[tb]
\centering
\begin{subfigure}{0.46\textwidth}
\centering
\includegraphics[width=1.0\linewidth]{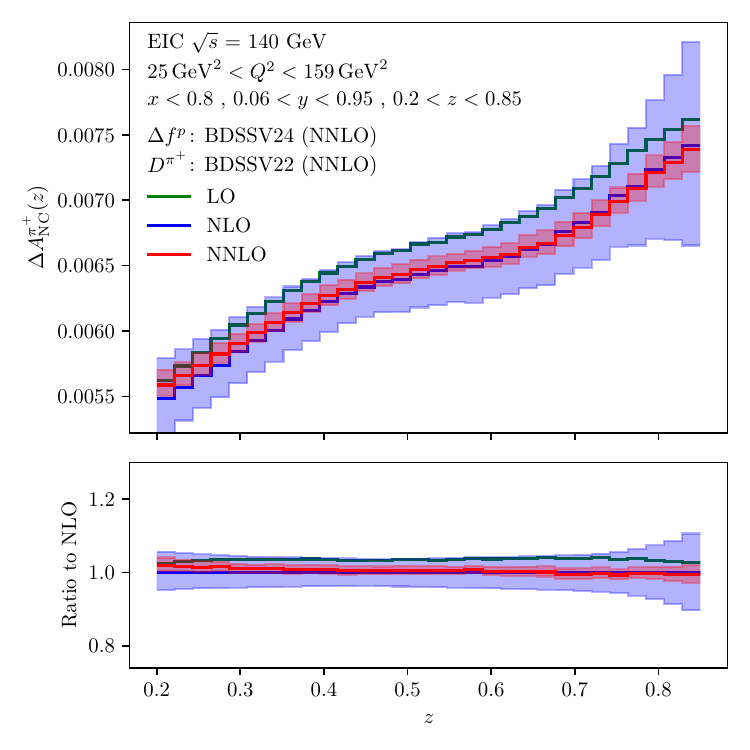}
\caption{Low-$Q^2$}
\end{subfigure}
\begin{subfigure}{0.46\textwidth}
\centering
\includegraphics[width=1.0\linewidth]{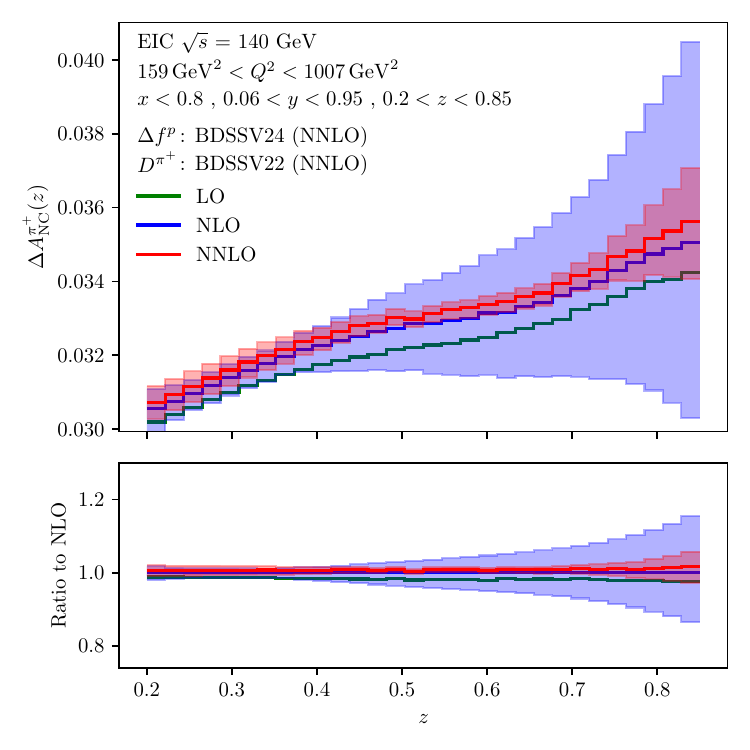}
\caption{Mid-$Q^2$}
\end{subfigure}
\begin{subfigure}{0.46\textwidth}
\centering
\includegraphics[width=1.0\linewidth]{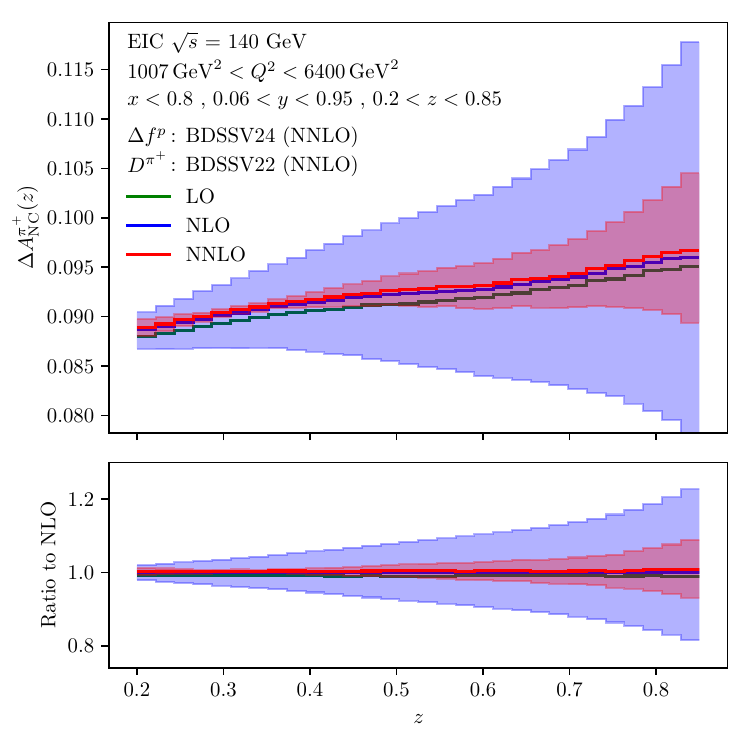}
\caption{High-$Q^2$}
\end{subfigure}
\begin{subfigure}{0.46\textwidth}
\centering
\includegraphics[width=\linewidth]{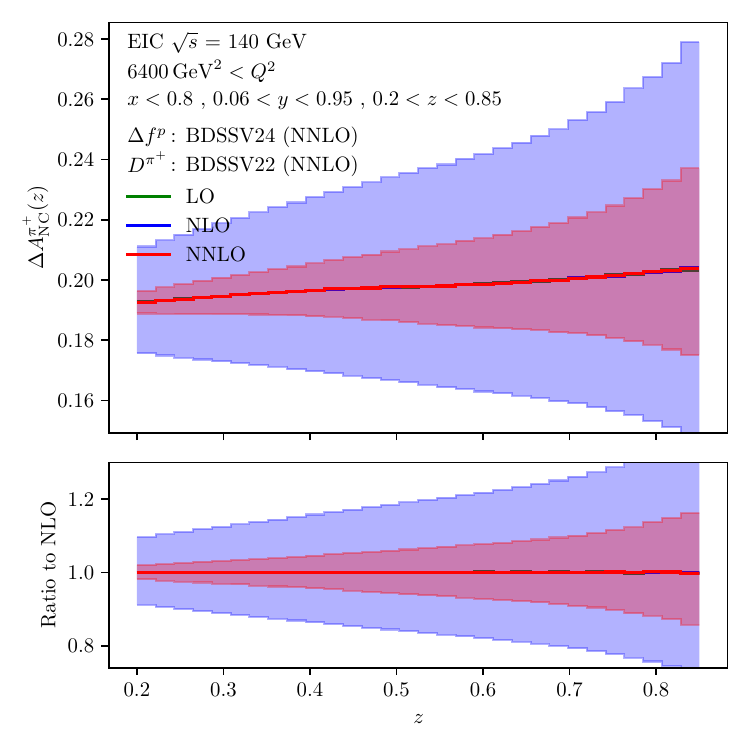}
\caption{Extreme-$Q^2$}
\end{subfigure}

\caption{
NC asymmetry $\Delta A_\NC^{\pi^+}$ as a function of $z$.
Top panels: Asymmetries.
Bottom panels: Ratio to NLO.
}
\label{fig:Z_A_NC_pip}

\end{figure}

\begin{figure}[tb]
\centering
\begin{subfigure}{0.46\textwidth}
\centering
\includegraphics[width=1.0\linewidth]{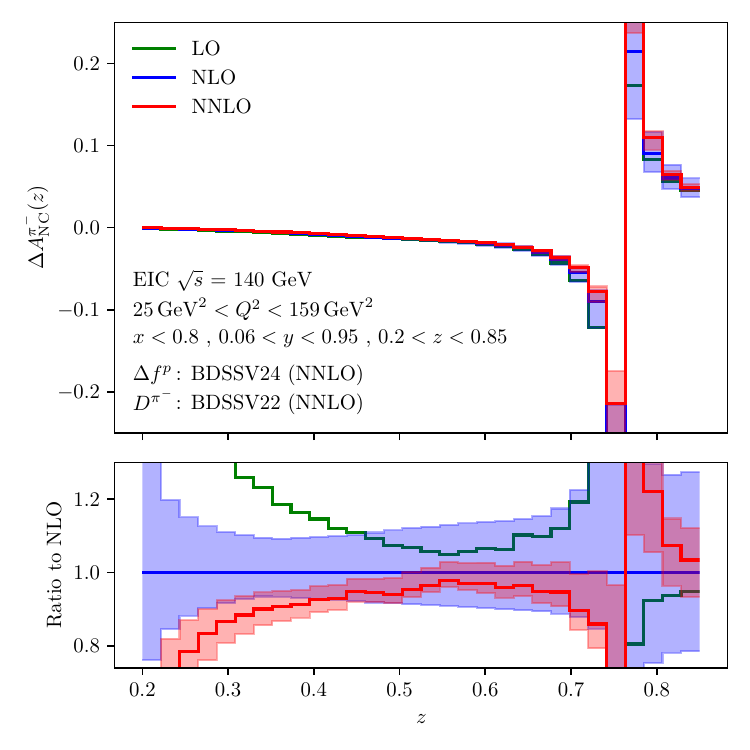}
\caption{Low-$Q^2$}
\end{subfigure}
\begin{subfigure}{0.46\textwidth}
\centering
\includegraphics[width=1.0\linewidth]{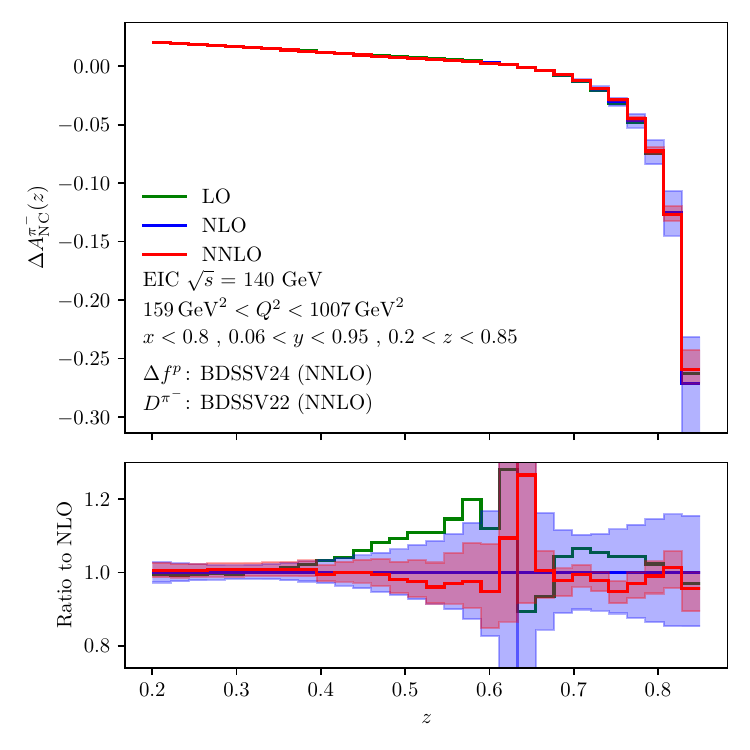}
\caption{Mid-$Q^2$}
\end{subfigure}
\begin{subfigure}{0.46\textwidth}
\centering
\includegraphics[width=1.0\linewidth]{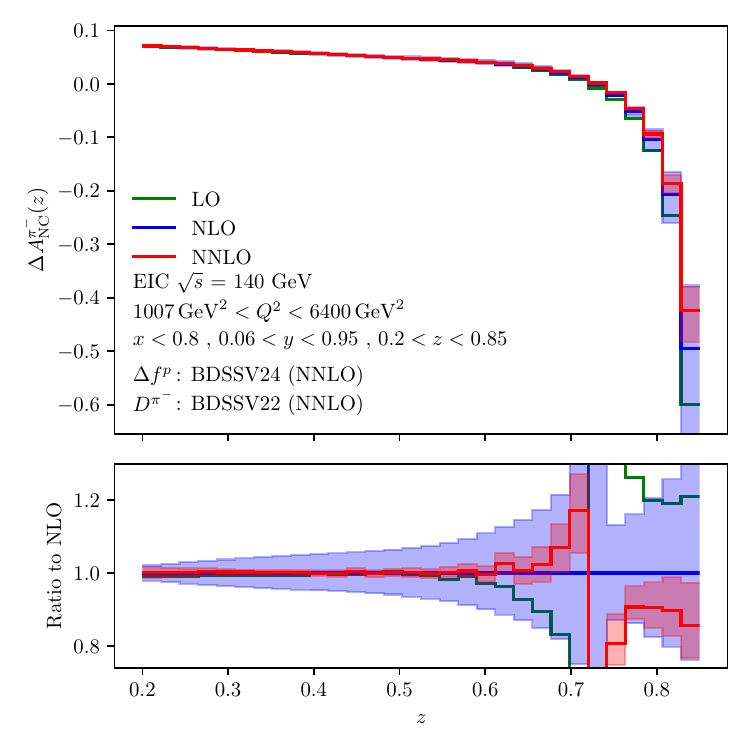}
\caption{High-$Q^2$}
\end{subfigure}
\begin{subfigure}{0.46\textwidth}
\centering
\includegraphics[width=1\linewidth]{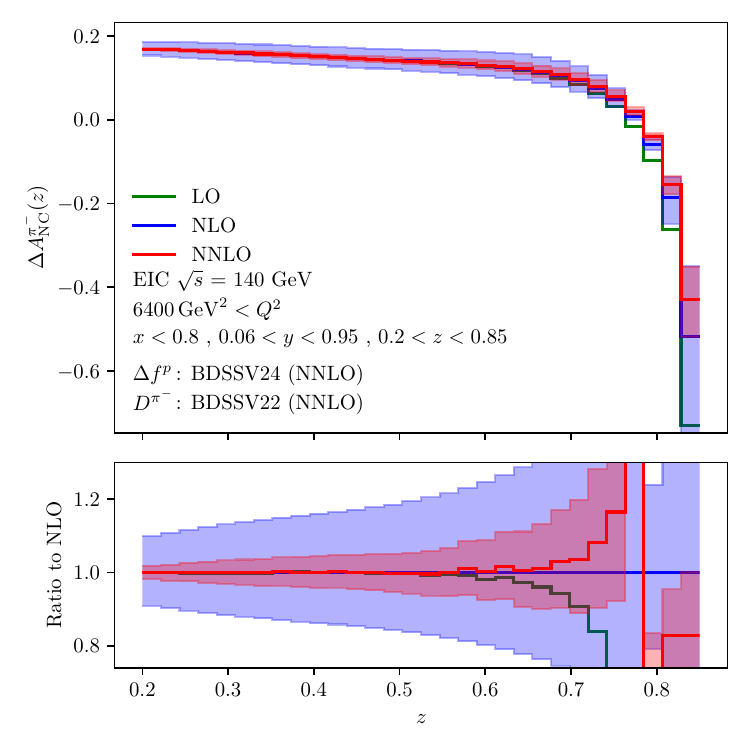}
\caption{Extreme-$Q^2$}
\end{subfigure}

\caption{
NC asymmetry $\Delta A_\NC^{\pi^-}$ as a function of $z$.
Top panels: asymmetry.
Bottom panels: Ratio to NLO.
}
\label{fig:Z_A_NC_pim}

\end{figure}

In Fig.~\ref{fig:Z_A_NC_pip} we study the polarized neutral current asymmetry~\eqref{eq:ANCm1} for $\pi^+$ production as a function of $z$ in the four kinematic ranges~\eqref{eq:cutsQ2}.
$\Delta A^{\pi^+}_{\NC}(z)$ carries only a moderate $z$ dependence but is sensitive to the average value of $Q^2$.
The absolute value increases with $Q^2$ from a sub-percent level at Low-$Q^2$ to just below $25\%$ at Extreme-$Q^2$.
At Extreme-$Q^2$ the $\lambda_\ell$-even electroweak contributions are about $20\%$ the size of the $\lambda_\ell$-odd ones.
The features of the asymmetry are very similar to the ones of the neutral current asymmetry in unpolarized SIDIS.
We observe a good perturbative convergence which improves with the increase of $Q^2$.

In Fig.~\ref{fig:Z_A_NC_pim} we show the polarized neutral current asymmetry~\eqref{eq:ANCm1} for $\pi^-$  production as a function of $z$.
Due to the behaviour of the purely electroweak structure functions the $\lambda_\ell$-even asymmetry crosses zero at a given value of $z$ which depends on the average $Q^2$.
$\Delta A^{\pi^-}_\NC$ shows a (moderate) downward trend in $z$.
While the polarized neutral current asymmetry in $\pi^+$ production is largely insensitive to the perturbative order in all energy ranges, the polarized neutral current asymmetry in $\pi^-$ production shows a large dependence on the respective orders at Low-$Q^2$.
At higher virtuality the different orders agree with each other away from the zero-crossing.

\begin{figure}[p]
\centering
\includegraphics[width=\textwidth]{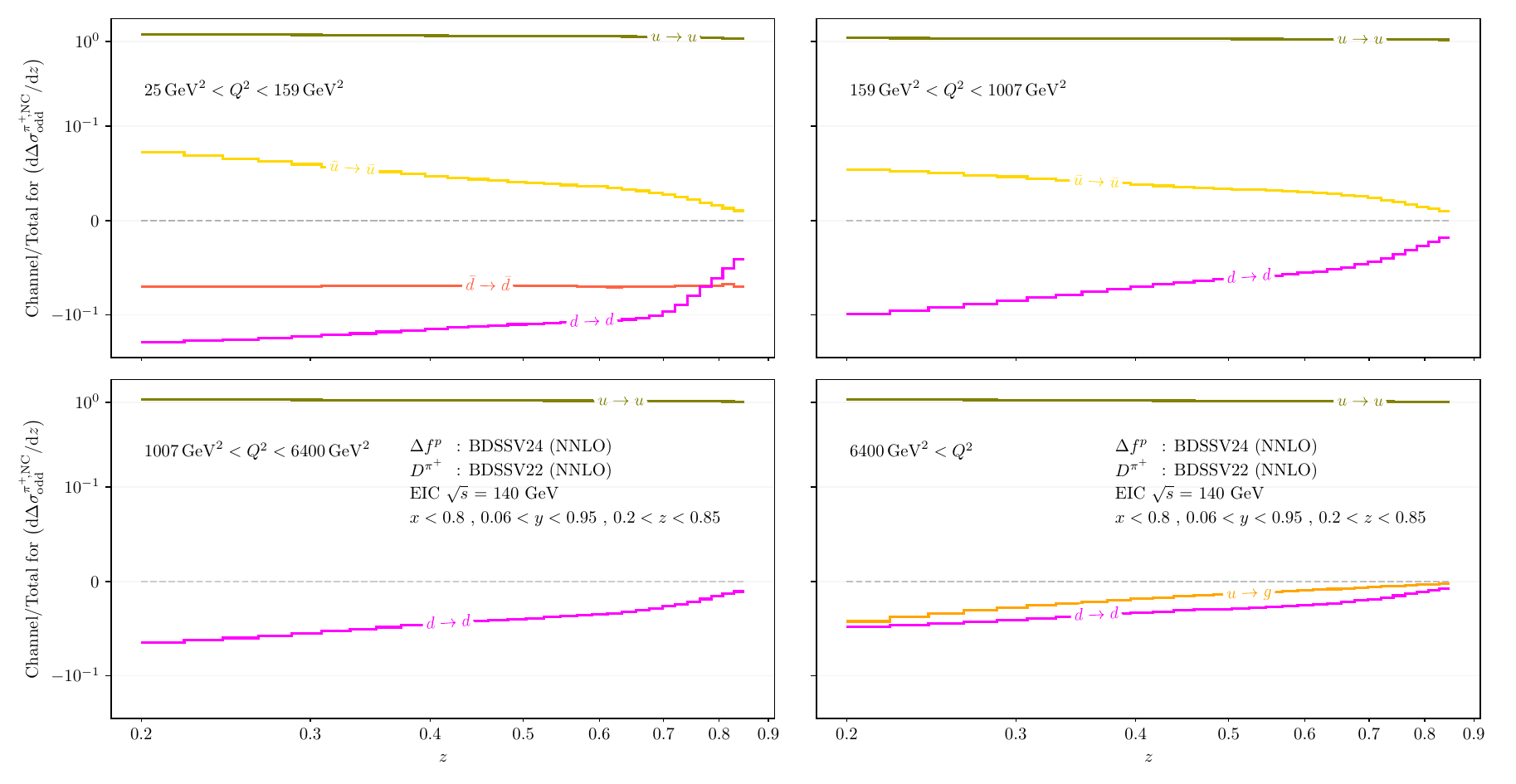}
\caption{
Channel decomposition for $\dd\Delta \sigma^{\pi^+,\NC}_{\modd}$ as a function of $z$.
Only channels contributing more than $4\,\%$ in any bin are displayed.
The range $(-0.1, 0.1)$ on the vertical axis is linear, the ranges above and below are plotted logarithmically.
}
\label{fig:z_NC_AS_channel_pip}

\bigskip

\includegraphics[width=\textwidth]{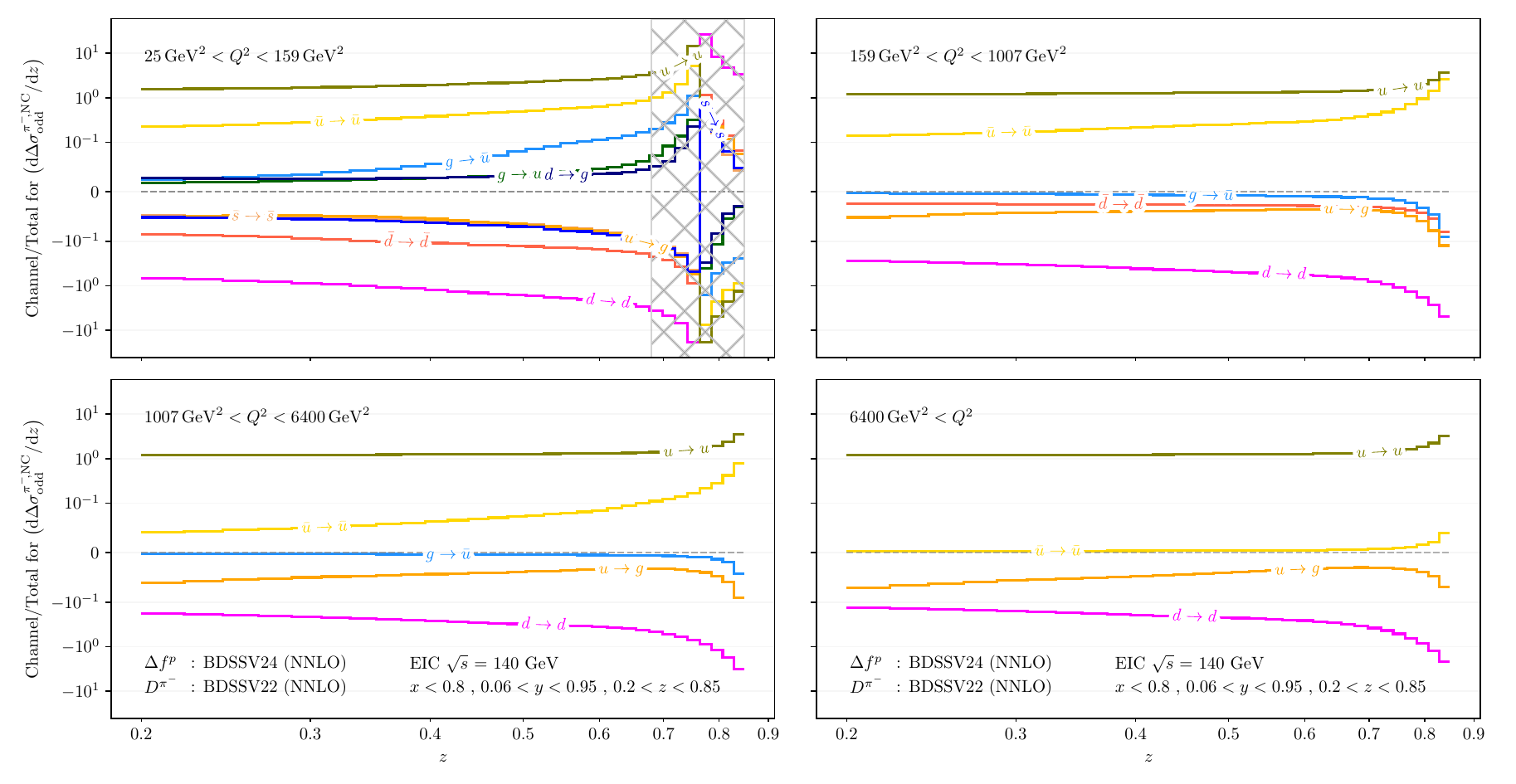}
\caption{
Channel decomposition for $\dd\Delta \sigma^{\pi^-,\NC}_{\modd}$ as a function of $z$.
Only channels contributing more than $4\,\%$ in any bin are displayed.
Bins entering in the dashed grey region where a division by zero occurs are not considered.
The range $(-0.1, 0.1)$ on the vertical axis is linear, the ranges above and below are plotted logarithmically.
}
\label{fig:z_NC_AS_channel_pim}
\end{figure}

In Fig.~\ref{fig:z_NC_AS_channel_pip} we show the partonic channel decomposition of $\dd\Delta\sigma^{\pi^+,\NC}_{\modd}/\dd z$ as a function of $z$ in the four $Q^2$ ranges.
Only channels that contribute more than $4\%$ to the total cross section asymmetry in a single bin are shown.
The flavour-favoured $u\to u$ channel ($u$ is valence of both initial and final state hadrons) is the largest contribution in each range of~$Q^2$.
It is the dominant production mode, accounting for a contribution of slightly above $100\,\%$ for all values of $z$.
The flavour-unfavoured channel $d\to d$ is the second most relevant and gives negative contributions of less than $10\,\%$ except for the lowest values of $Q^2$.
It is observed that the channels not producing final state valence content of $\pi^+$ are largest at low $z$ and fade into insignificance at the highest values of $z$.
Solely at Low-$Q^2$ the $\bar{d}\to\bar{d}$ production mode becomes relevant, and gives an almost $z$-independent relative contribution.

The channel decomposition of $\dd\Delta\sigma^{\pi^-,\NC}_{\modd}/\dd z$ for $\pi^-$ production,
Fig.~\ref{fig:z_NC_AS_channel_pim}, is significantly different from $\pi^+$.
The $u\to u$ channel is no longer flavour-favoured by the final state hadron and is therefore suppressed at large $z$ by the fragmentation mechanism.
Nevertheless it still remains the most dominant channel due to the abundance of $u$-type quarks in the proton and the stronger electromagnetic coupling.
The $d\to d$ channel is the second largest contribution in absolute value, while still being negative for all $z$, which is due to the negative polarized PDF $\Delta f_d^p(x)$.
At Low-$Q^2$ the proton's sea quarks and gluons are enhanced by the small values of $x$ probed.
Numerical cancellations between the increasingly preferred $d\to d$ channel and especially the $u\to u$ at large values of $z$ result in $\dd \Delta\sigma^{\pi^-,\NC}_{\modd}$ crossing zero at Low-$Q^2$.

\begin{figure}[p]
\centering
\includegraphics[width=\textwidth]{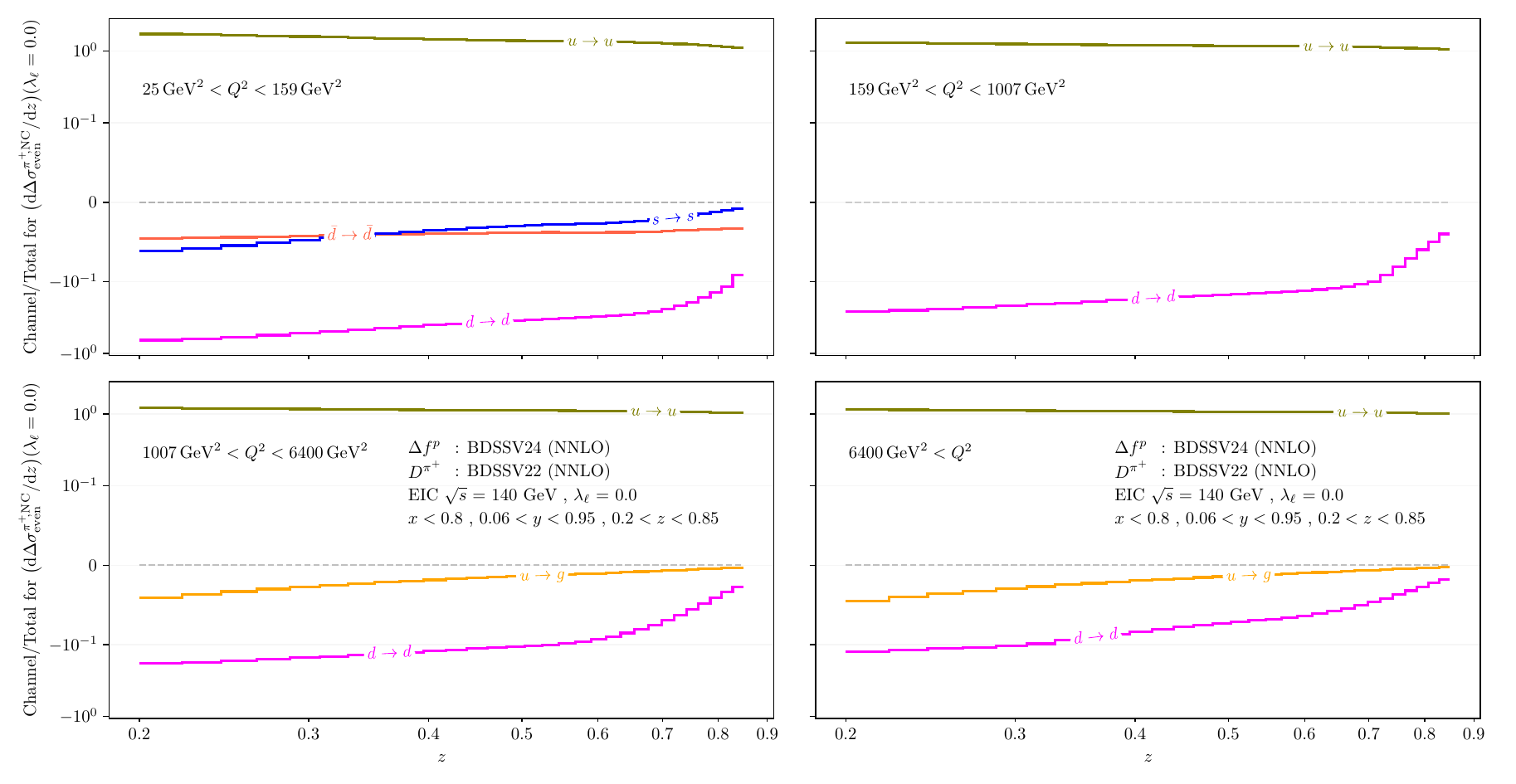}
\caption{
Channel decomposition for $\dd\Delta \sigma^{\pi^+,\NC}_{\meven}$ as a function of $z$.
Only channels contributing more than $4\,\%$ in any bin are displayed.
The range $(-0.1, 0.1)$ on the vertical axis is linear, the ranges above and below are plotted logarithmically.
}
\label{fig:z_NC_STD_channel_pip}

\bigskip

\includegraphics[width=\textwidth]{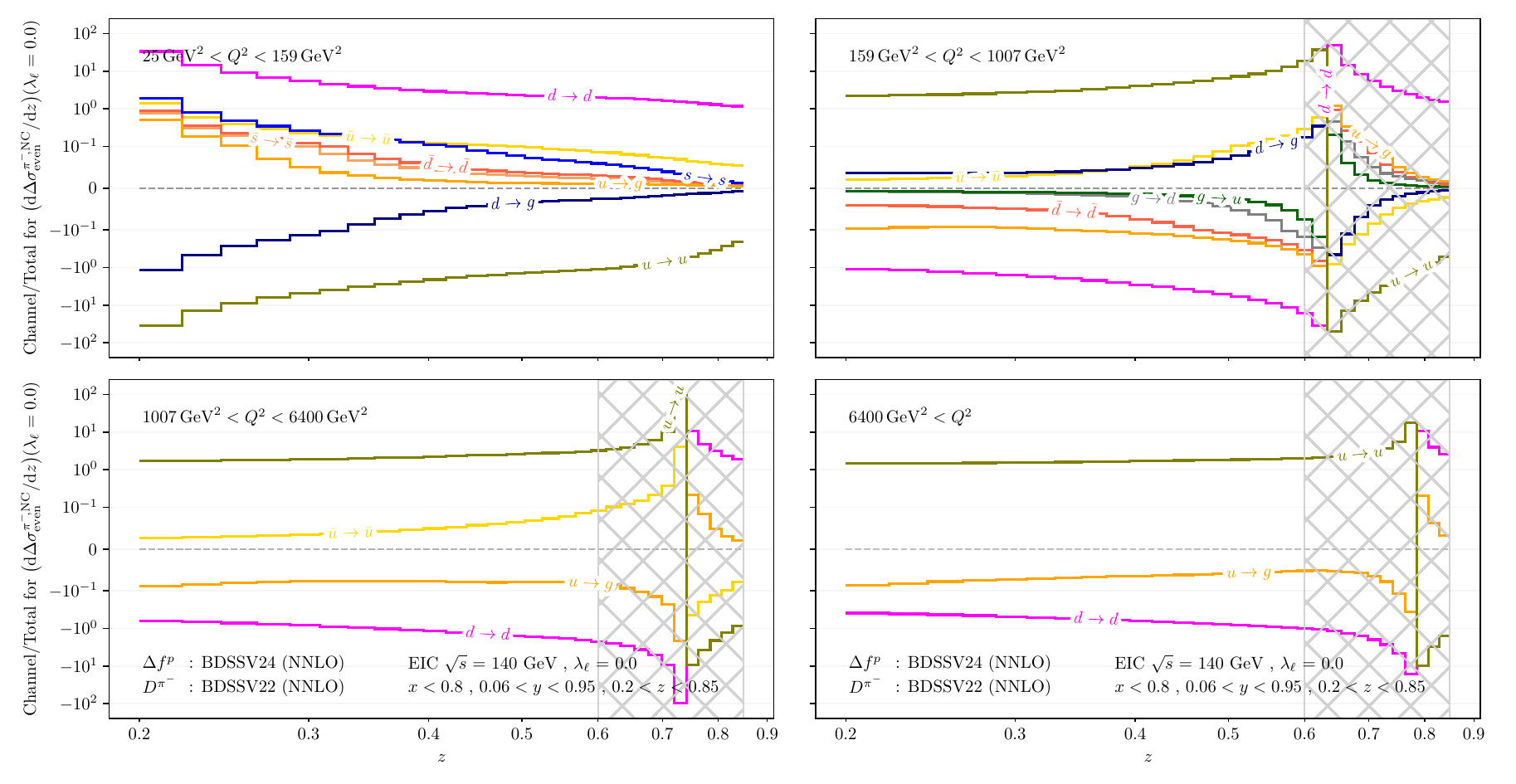}
\caption{
Channel decomposition for $\dd\Delta \sigma^{\pi^-,\NC}_{\meven}$ as a function of $z$.
Only channels contributing more than $4\,\%$ in any bin are displayed.
Bin entering in the dashed grey region where a division by zero occurs are not considered.
The range $(-0.1, 0.1)$ on the vertical axis is linear, the ranges above and below are plotted logarithmically.
}
\label{fig:z_NC_STD_channel_pim}
\end{figure}

In Fig.~\ref{fig:z_NC_STD_channel_pip} we show the channel decomposition for $\dd\Delta\sigma^{\pi^+,\NC}_{\meven}$ as a function of $z$.
Similar to the $\lambda_\ell$-odd case the $u\to u $ and $d\to d$ channels remain the two largest in all $Q^2$ ranges.
However they are both enhanced with respect to their $\lambda_\ell$-odd counterparts by a similar amount, which leads to a substantial enhancement especially for the smaller $d\to d$ channel.
The enhancement of the $d\to d$ transition is due to the smaller suppression of the $d$-channels with respect to the $u$-channels from the the electroweak couplings.

Similar conclusions can be drawn for $\pi^-$ production from a comparison between $\dd \Delta\sigma^{\pi^-,\NC}_{\meven}$ and $\dd \Delta\sigma^{\pi^-,\NC}_{\modd}$ for Mid-$Q^2$, High-$Q^2$, and Extreme-$Q^2$.
At Low-$Q^2$ the sign of the channels is swapped with respect to $\pi^+$ due to the normalisation to the negative $\dd \Delta\sigma^{\pi^-,\NC}_{\meven}$.
As shown in Fig.~\ref{fig:z_NC_STD_channel_pim} also in $\pi^-$ the $d\to d$ channel is enhanced by the coupling, leading to similarly sized contributions with opposite signs from $d\to d$ and $u\to u$.
At large $z$ the growth of two largest channels results in $\dd \Delta\sigma^{\pi^-,\NC}_{\meven}$ crossing zero and turning negative.

\begin{figure}[p]
\centering
\begin{subfigure}{0.45\textwidth}
\centering
\includegraphics[width=1.0\linewidth]{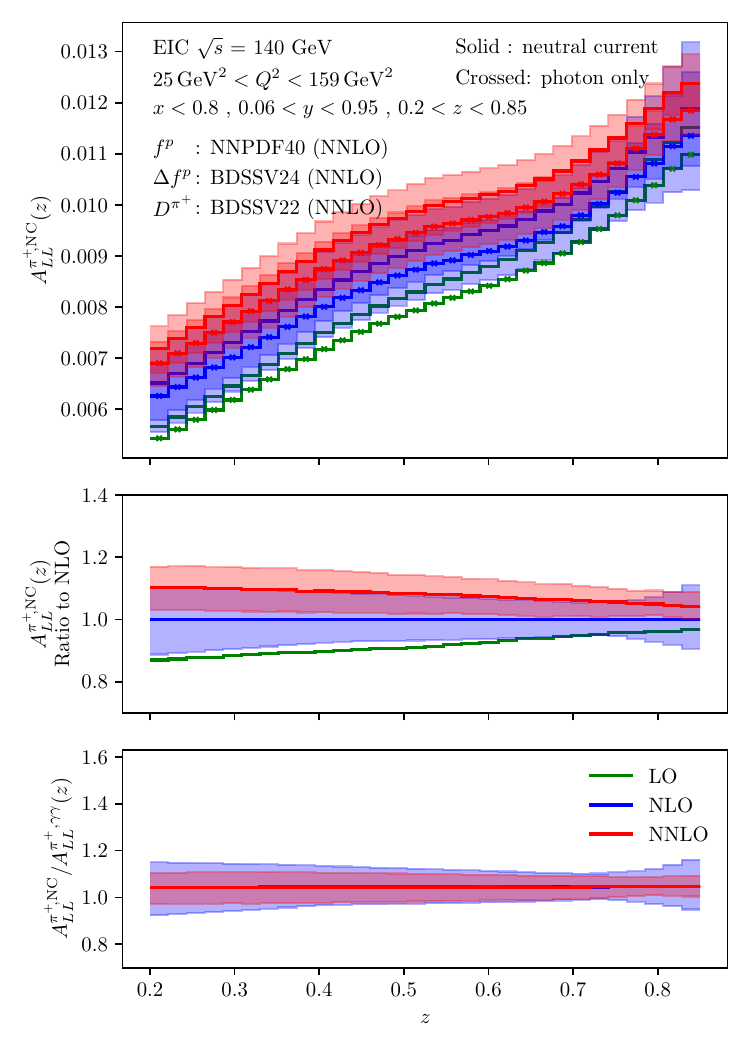}
\caption{Low-$Q^2$}
\end{subfigure}
\begin{subfigure}{0.45\textwidth}
\centering
\includegraphics[width=1.0\linewidth]{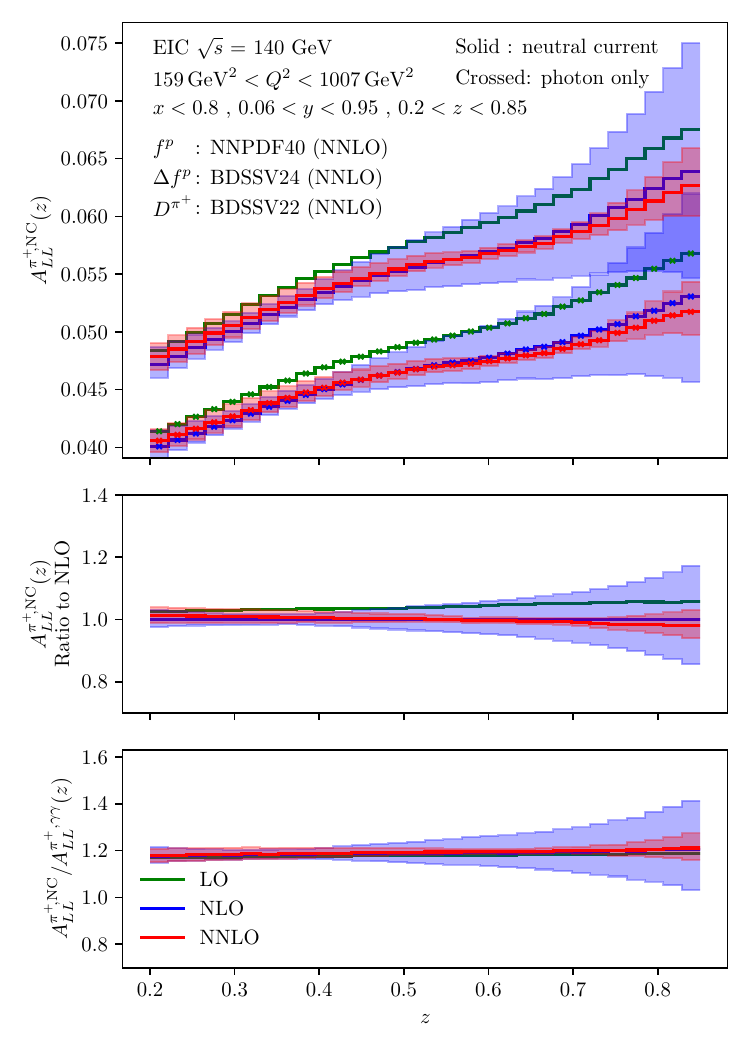}
\caption{Mid-$Q^2$}
\end{subfigure}
\begin{subfigure}{0.45\textwidth}
\centering
\includegraphics[width=1.0\linewidth]{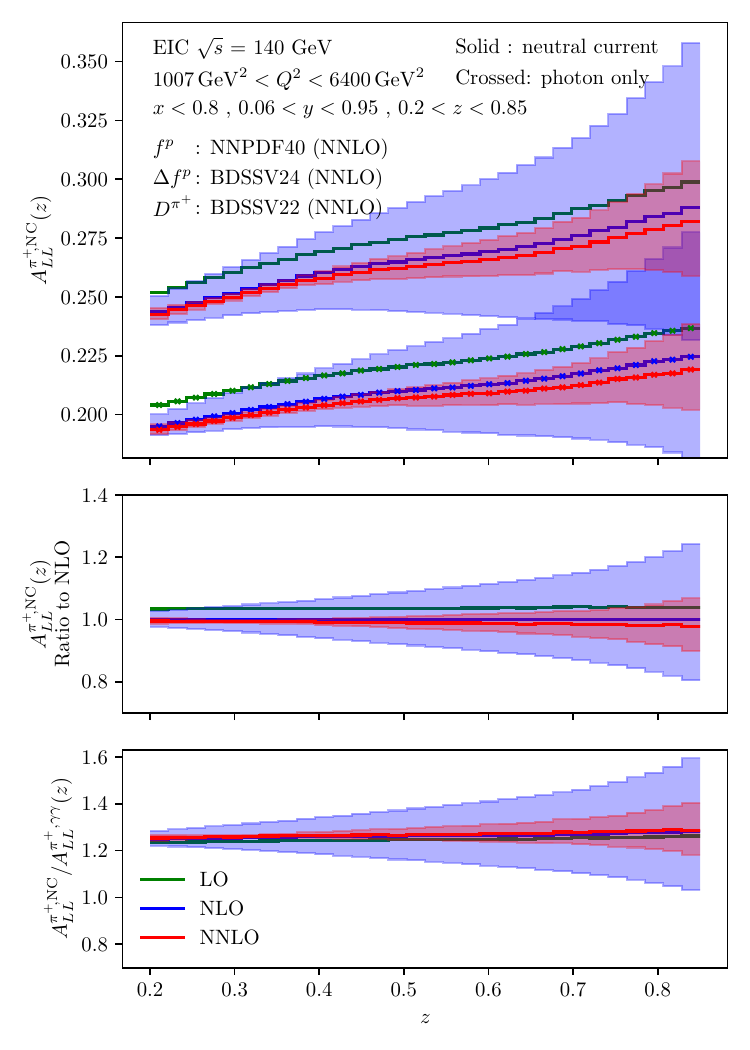}
\caption{High-$Q^2$}
\end{subfigure}
\begin{subfigure}{0.45\textwidth}
\centering
\includegraphics[width=1.0\linewidth]{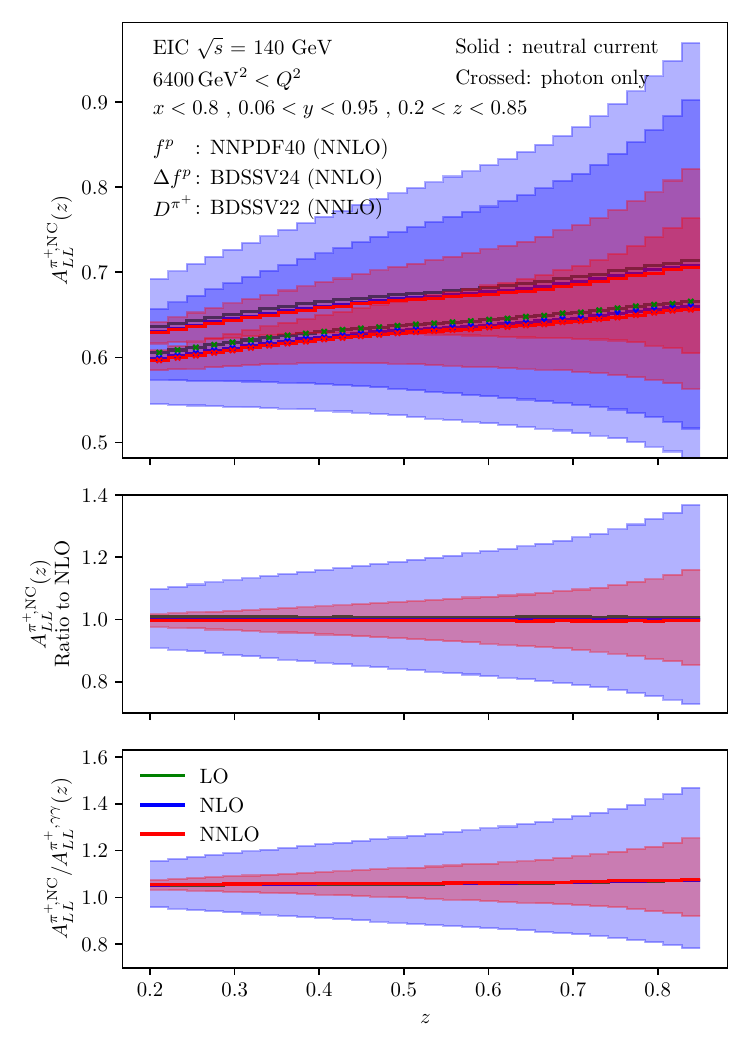}
\caption{Extreme-$Q^2$}
\end{subfigure}

\caption{
Double-longitudinal spin asymmetry $A_{LL}^{\pi^+}$ as a function of $z$ for NC exchange.
Top panels: Asymmetries for full NC (solid) and photon-only (crossed) exchanges.
Centre panels: NC Ratio to NLO.
Bottom panels: NC over photon-only exchange ratios.
}
\label{fig:Z_A_LL_NC_pip}

\end{figure}

In Fig.~\ref{fig:Z_A_LL_NC_pip} we study the double-longitudinal spin asymmetry~\eqref{eq:AhLLNC} for $\pi^+$ production in NC exchange as a function of $z$ in the four ranges of $Q^2$.
The asymmetry increases with $z$ in all ranges of $Q^2$. We also observe an increase of the asymmetry in average value from Low-$Q^2$ to Extreme-$Q^2$, from a minimum of $0.006$ to a maximum of $0.7$.
Electroweak effects enhance the asymmetry with respect to the photon-only exchange case.
Indeed $A^{\pi^+,\NC}_{LL}$ (solid in the top panel) is larger than $A^{\pi^+,\gamma\gamma}_{LL}$ (crossed) in all ranges of $Q^2$ and for all values of $z$.
This can be better seen in the ratio $(A^{\pi^+,\NC}_{LL}/A^{\pi^+,\gamma\gamma}_{LL})(z)$ shown in the bottom panels.
The ratio is flat in $z$ and grows with increasing $Q^2$ up to the High-$Q^2$ range to then decrease again at Extreme-$Q^2$, consistent with our observations on $(A^{\pi^+,\NC}_{LL}/A^{\pi^+,\gamma\gamma}_{LL})(Q^2)$ in Fig.~\ref{fig:Q2_A_LL_NC_pi}.
Electroweak effects are at few percent level at Low-$Q^2$; they grow to $20\%$ at Mid-$Q^2$ and up to $25\%$ at High-$Q^2$ to than decrease again to less than $10\%$ at Extreme-$Q^2$.
This is consistent with the behaviour of the asymmetry as a function of $Q^2$.
We observe good perturbative convergence and stability as shown in the centre panels for full NC exchange.
Perturbative corrections are at percent level at Low-$Q^2$ and they reduce
further to sub percent level with increasing average values of $Q^2$.
NNLO corrections are at sub percent level and lie within the NLO theory uncertainty band, with the exception of the Low-$Q^2$ region where at low $z$ they grow up to $10\%$.

\begin{figure}[p]
\centering
\begin{subfigure}{0.45\textwidth}
\centering
\includegraphics[width=1.0\linewidth]{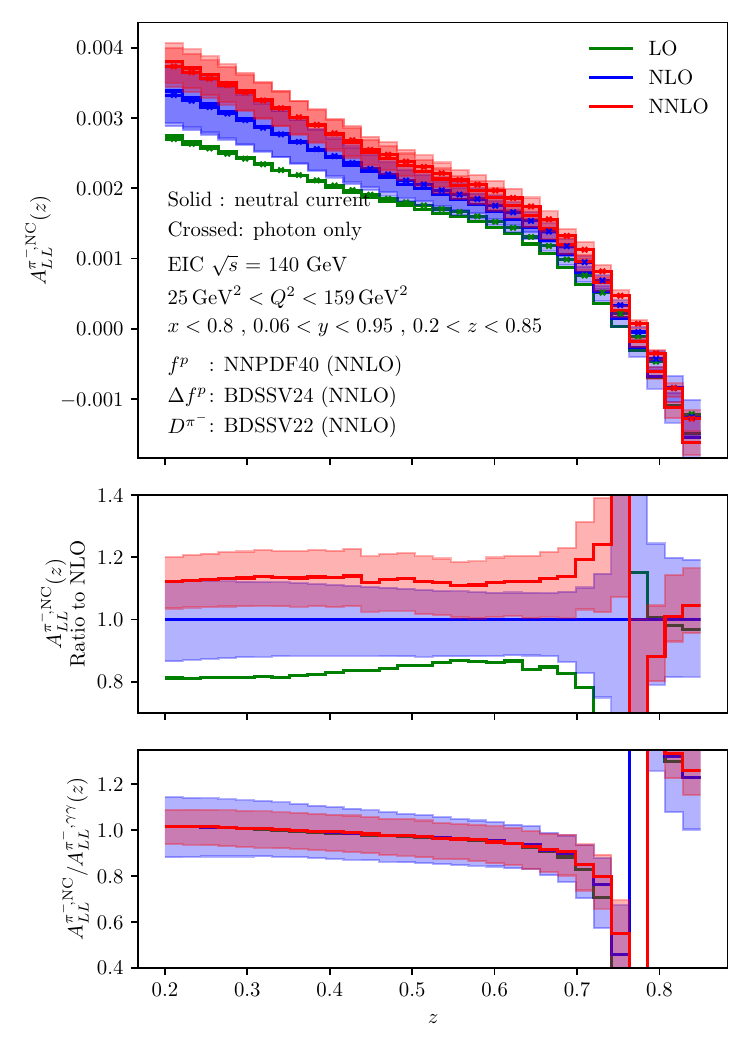}
\caption{Low-$Q^2$}
\end{subfigure}
\begin{subfigure}{0.45\textwidth}
\centering
\includegraphics[width=1.0\linewidth]{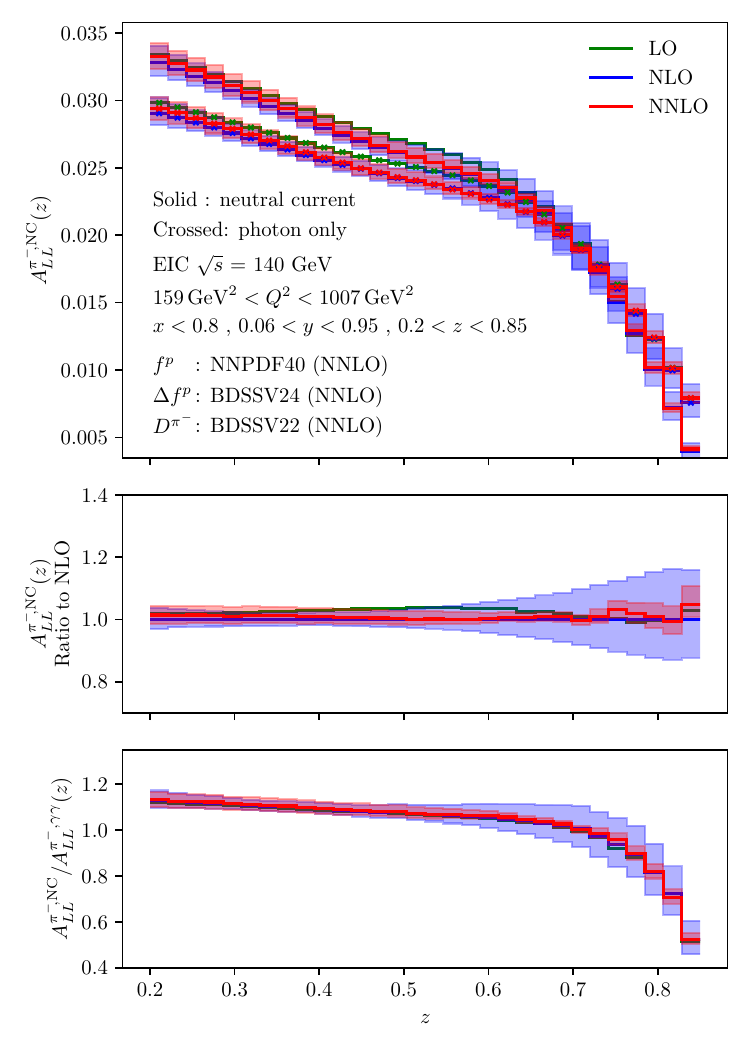}
\caption{Mid-$Q^2$}
\end{subfigure}
\begin{subfigure}{0.45\textwidth}
\centering
\includegraphics[width=1.0\linewidth]{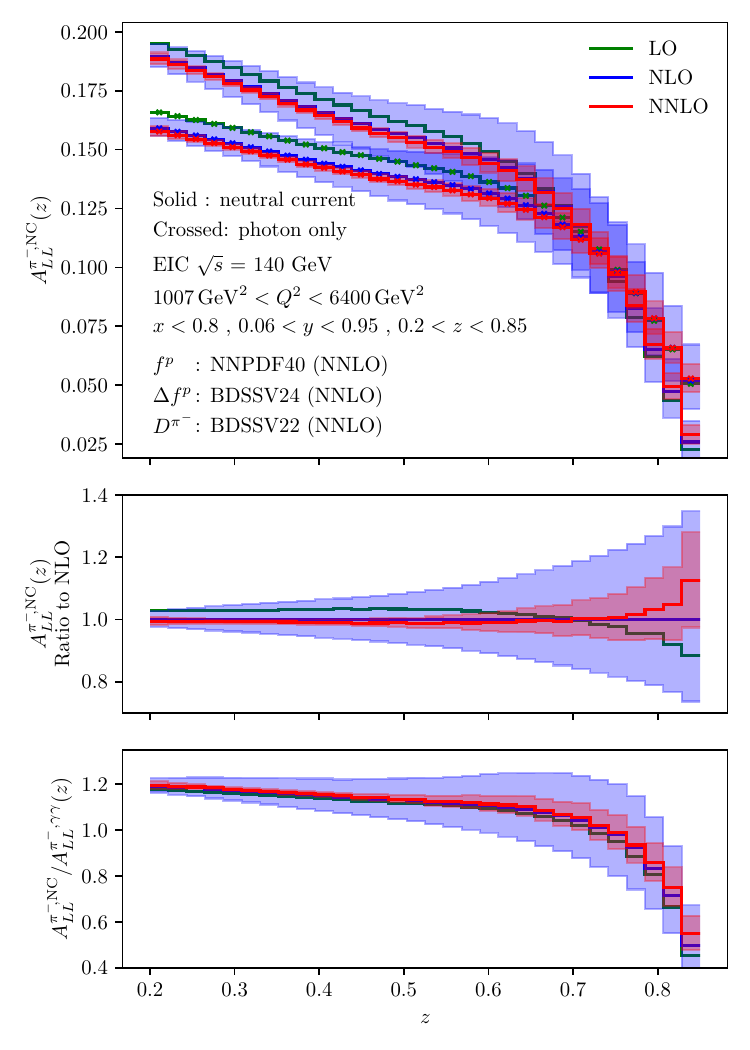}
\caption{High-$Q^2$}
\end{subfigure}
\begin{subfigure}{0.45\textwidth}
\centering
\includegraphics[width=1.0\linewidth]{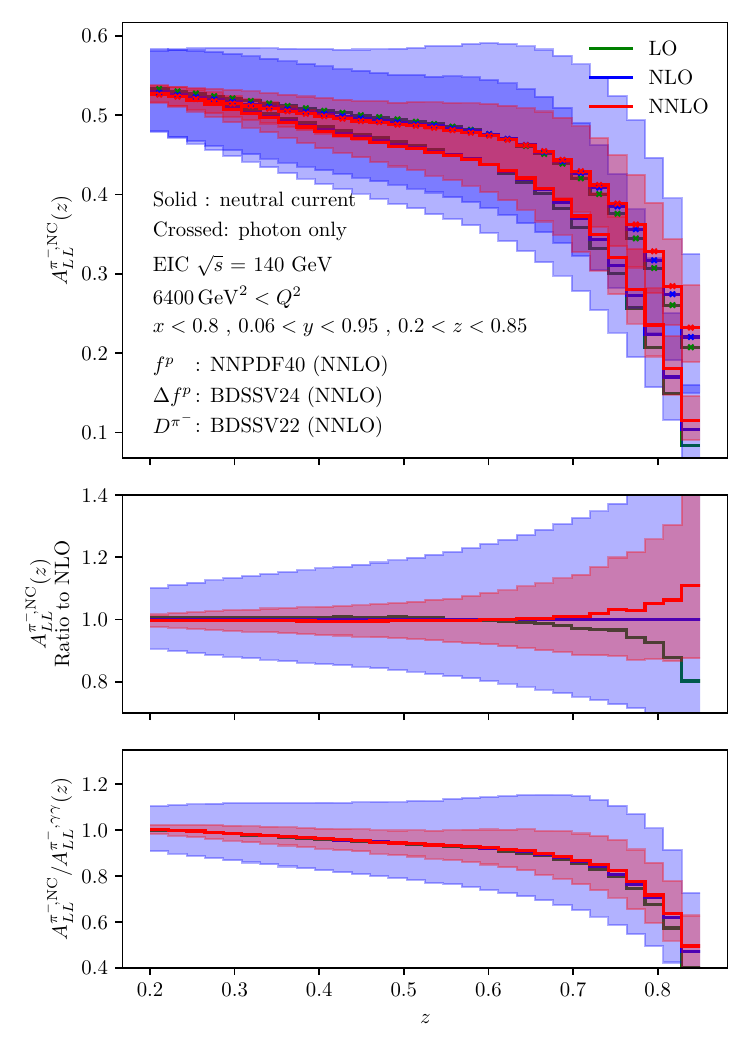}
\caption{Extreme-$Q^2$}
\end{subfigure}

\caption{
Double-longitudinal spin asymmetry $A_{LL}^{\pi^-}$ as a function of $z$ for NC interactions.
Top panels: Asymmetries for full NC (solid) and photon-only (crossed) exchanges.
Centre panels: NC Ratio to NLO.
Bottom panels: NC over photon-only exchange ratios.
}
\label{fig:Z_A_LL_NC_pim}

\end{figure}

In Fig.~\ref{fig:Z_A_LL_NC_pim} we study the double-longitudinal spin asymmetry~\eqref{eq:AhLLNC} for $\pi^-$ production in NC as a function of $z$.
The asymmetries are about $20\%$ smaller compared to $\pi^+$ production and they decrease in $z$ instead of increasing.
The behaviour of electroweak effects is very similar to the $\pi^+$ case: they grow from percent level at Low-$Q^2$ up to $20\%$ at High-$Q^2$ to than decrease to $10\%$ at Extreme-$Q^2$.
Perturbative corrections are large, up to $20\%$ at Low-$Q^2$. They decrease with the increase of the average value of $Q^2$ to few percent level.
Overall we observe good perturbative convergence and stability.

\begin{figure}[p]
\centering
\begin{subfigure}{0.457\textwidth}
\centering
\includegraphics[width=1.0\linewidth]{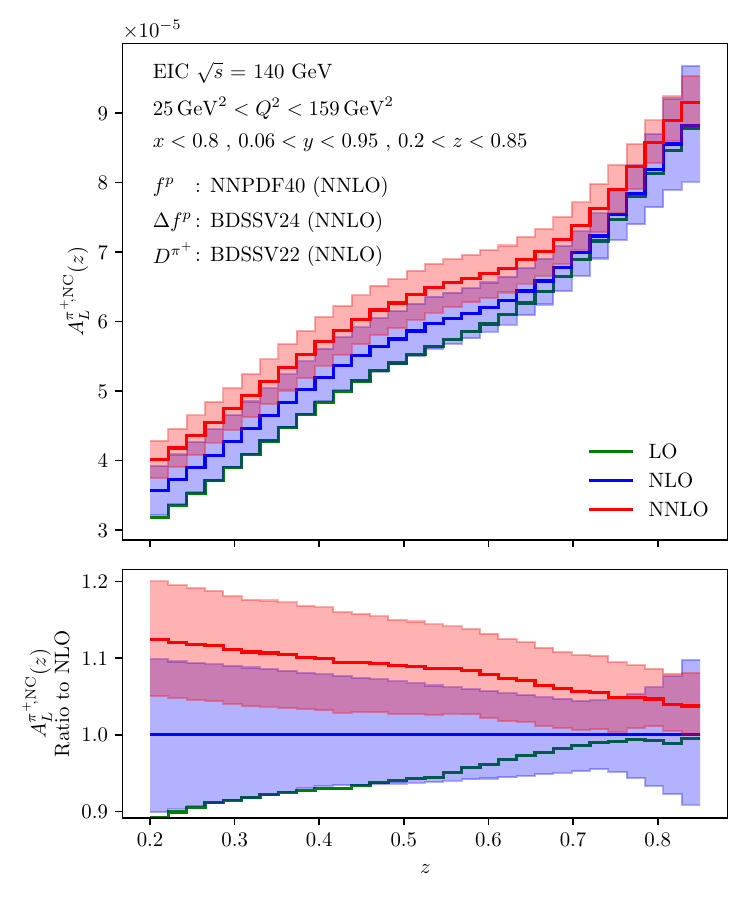}
\caption{Low-$Q^2$}
\end{subfigure}
\begin{subfigure}{0.45\textwidth}
\centering
\includegraphics[width=.993\linewidth]{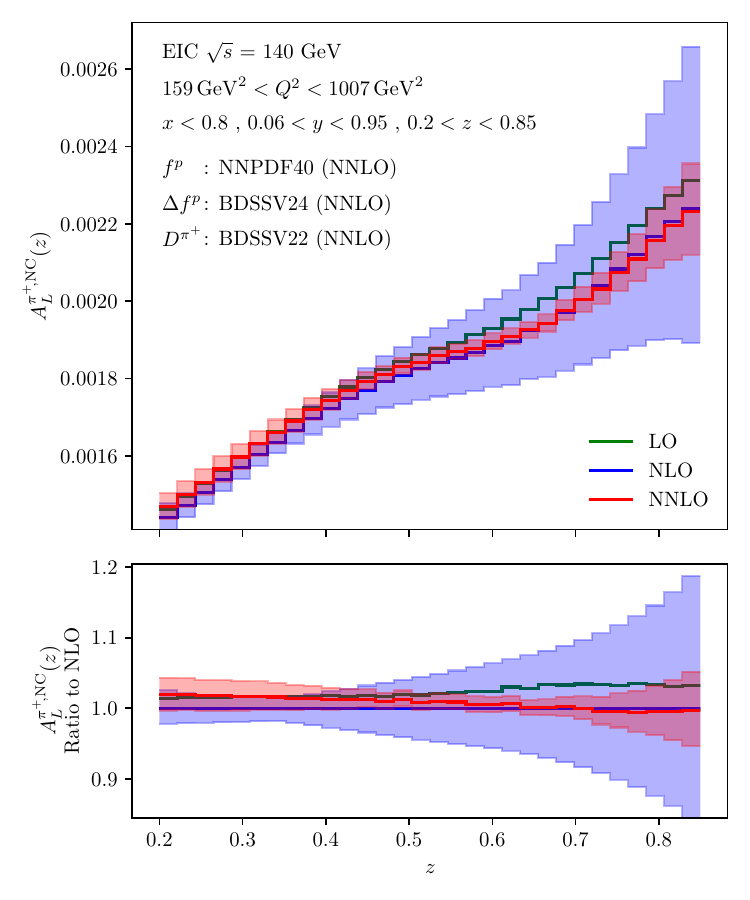}
\caption{Mid-$Q^2$}
\end{subfigure}
\begin{subfigure}{0.45\textwidth}
\centering
\includegraphics[width=1.0\linewidth]{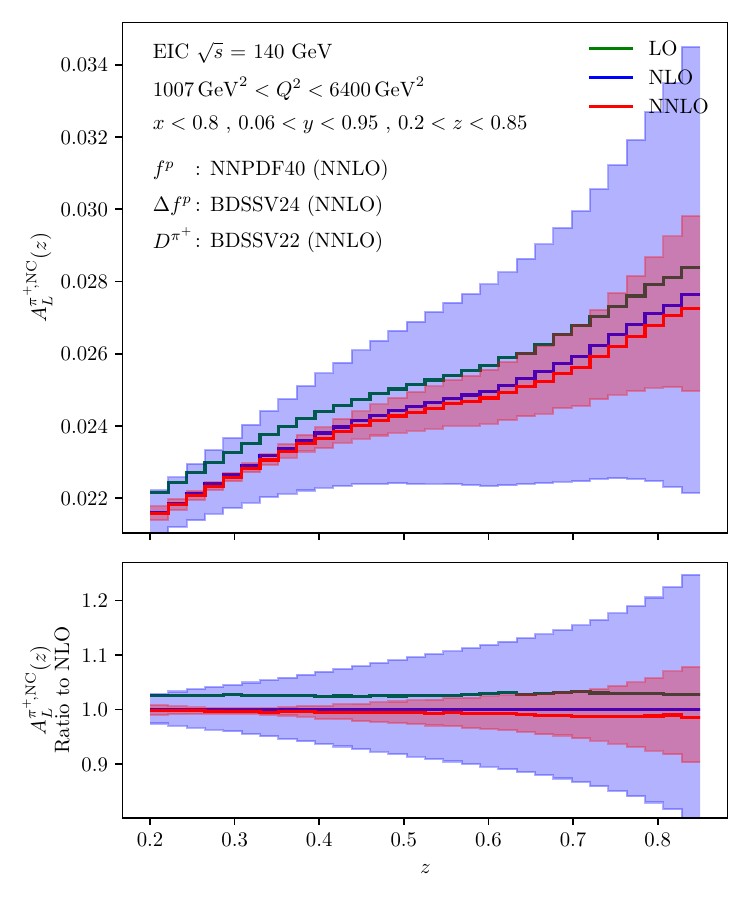}
\caption{High-$Q^2$}
\end{subfigure}
\begin{subfigure}{0.45\textwidth}
\centering
\includegraphics[width=1.0\linewidth]{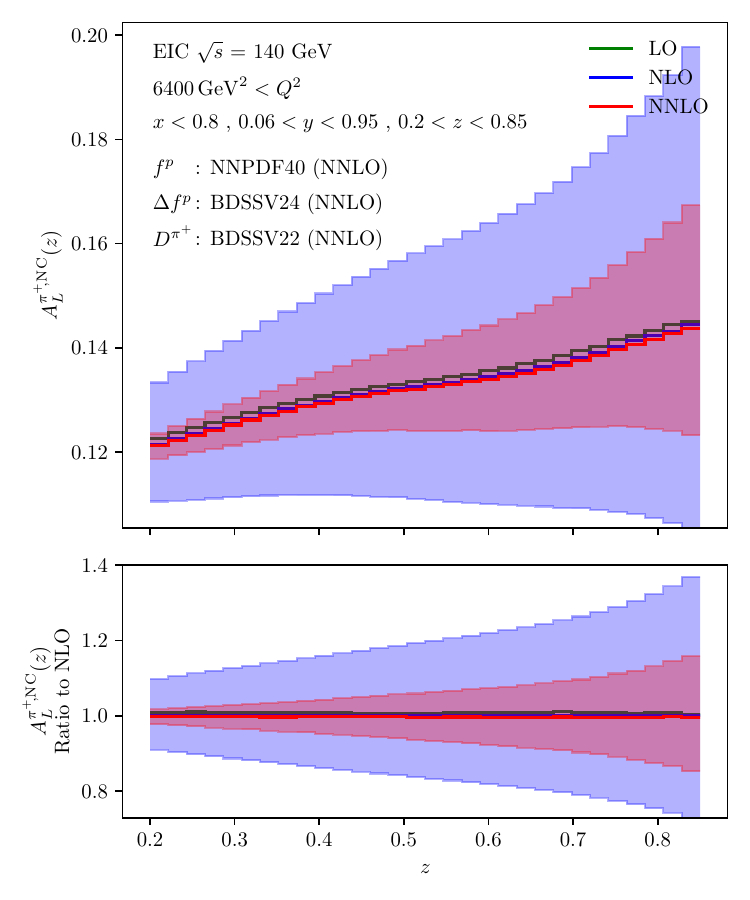}
\caption{Extreme-$Q^2$}
\end{subfigure}

\caption{
Single-longitudinal spin asymmetry $A_{L}^{\pi^+,\NC}$ as a function of $z$ for NC exchange.
Top panels: Asymmetries.
Bottom panels: Ratio to NLO.
}
\label{fig:Z_A_L_pip}

\end{figure}

\begin{figure}[p]
\centering
\begin{subfigure}{0.457\textwidth}
\centering
\includegraphics[width=1.0\linewidth]{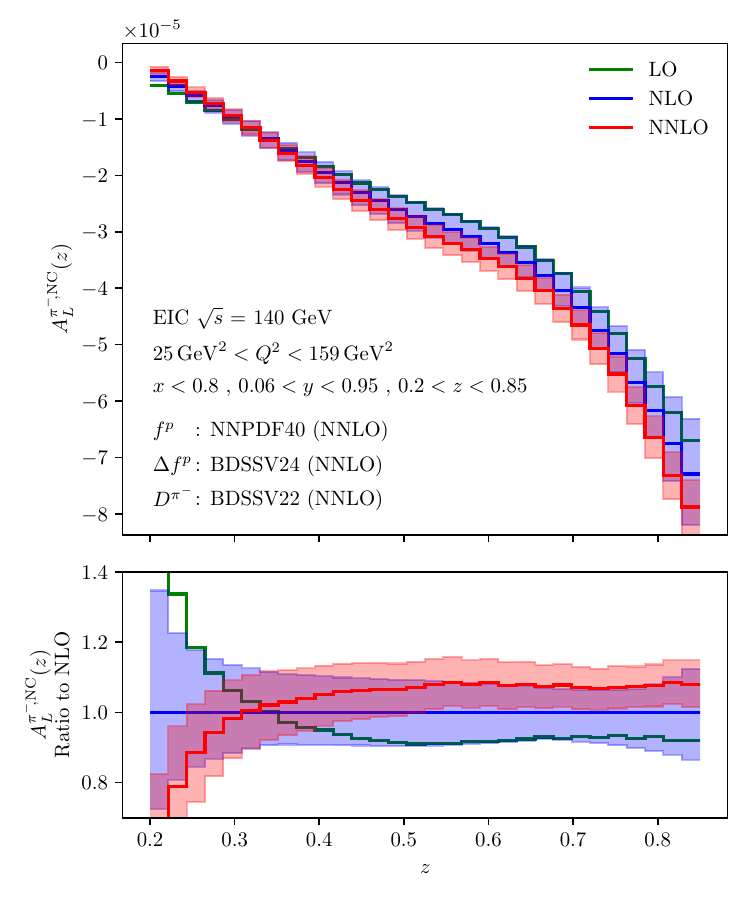}
\caption{Low-$Q^2$}
\end{subfigure}
\begin{subfigure}{0.45\textwidth}
\centering
\includegraphics[width=.994\linewidth]{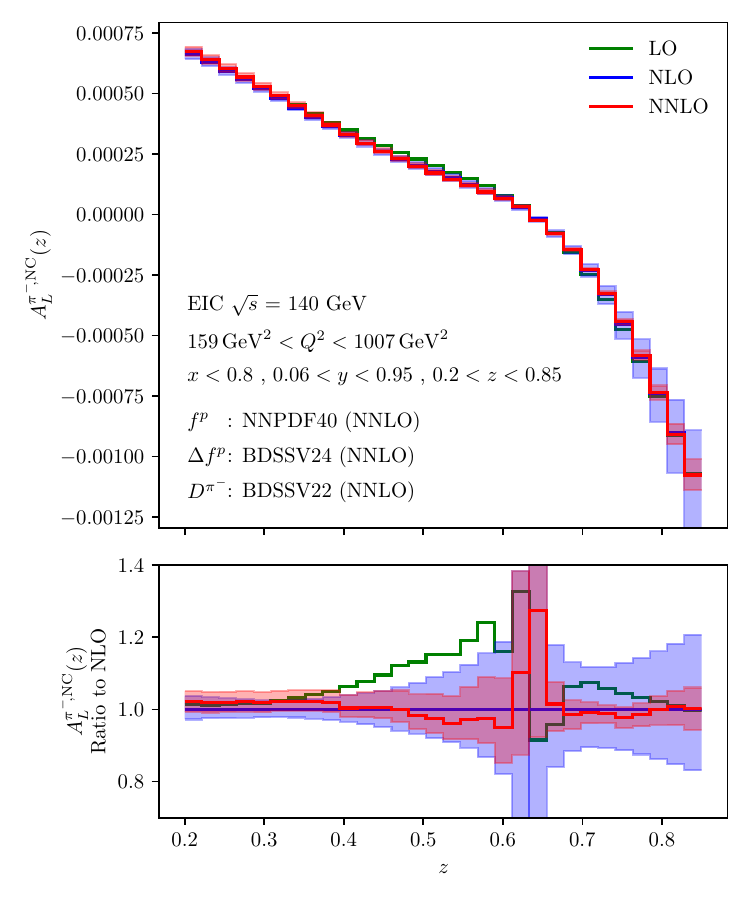}
\caption{Mid-$Q^2$}
\end{subfigure}
\begin{subfigure}{0.45\textwidth}
\centering
\includegraphics[width=1.0\linewidth]{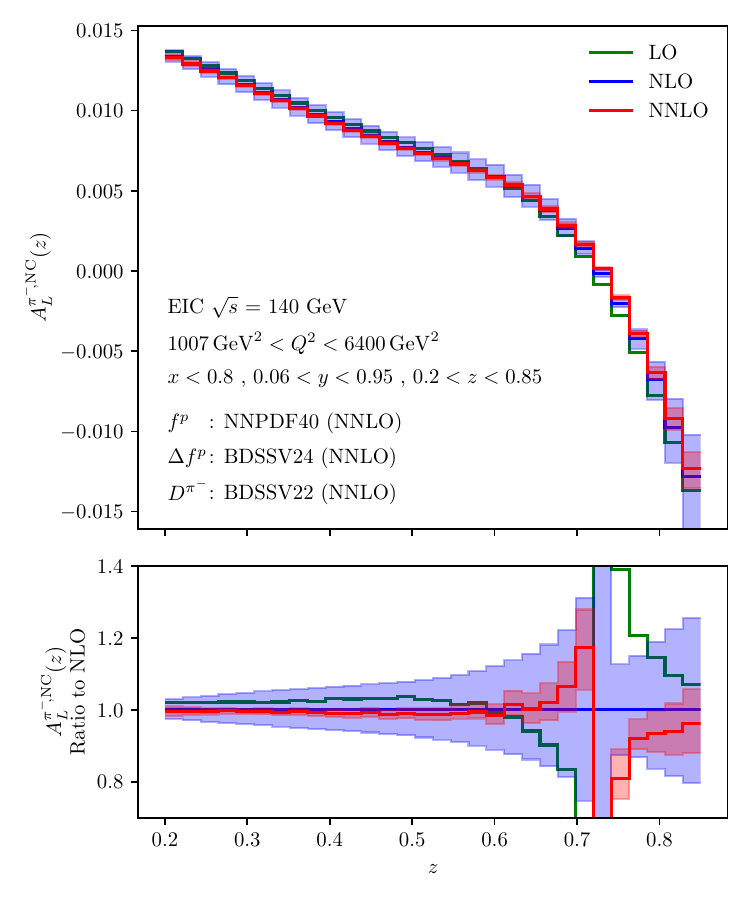}
\caption{High-$Q^2$}
\end{subfigure}
\begin{subfigure}{0.45\textwidth}
\centering
\includegraphics[width=.993\linewidth]{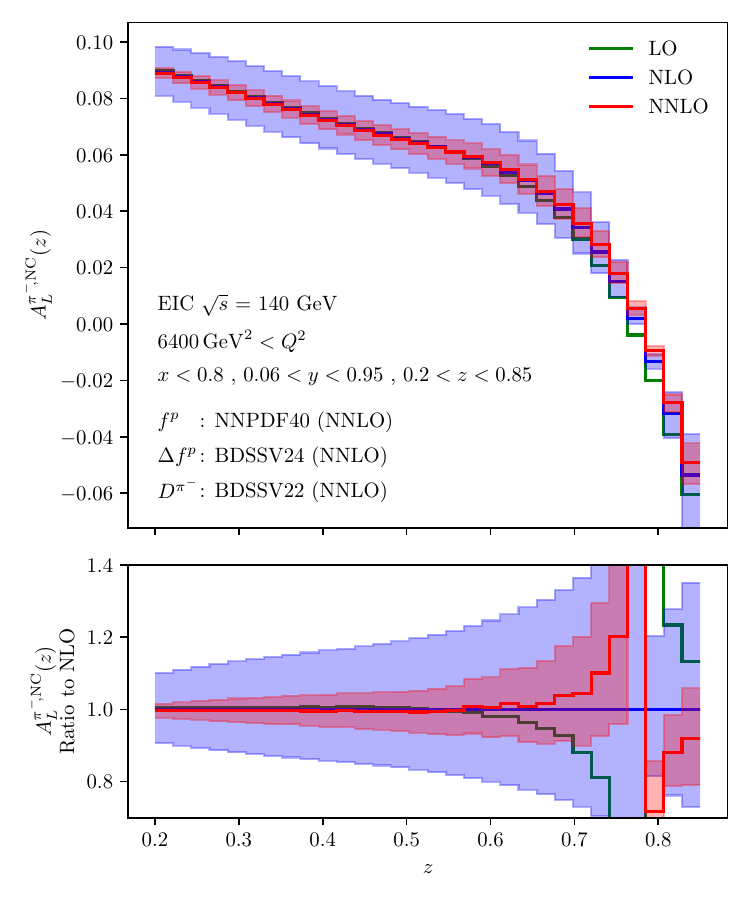}
\caption{Extreme-$Q^2$}
\end{subfigure}

\caption{
Single-longitudinal spin asymmetry $A_{L}^{\pi^-,\NC}$ as a function of $z$ for NC interactions.
Top panels: Asymmetries.
Bottom panels: Ratio to NLO.
}
\label{fig:Z_A_L_pim}

\end{figure}

In Fig.~\ref{fig:Z_A_L_pip} we study the single-longitudinal spin asymmetry~\eqref{eq:AhLNC} for $\pi^+$ production as a function of $z$ with the kinematic cuts of~\eqref{eq:cutsxyz} and~\eqref{eq:cutsQ2}.
The asymmetry grows with increasing $z$ in all ranges of $Q^2$ and it also increases in size with $Q^2$.
From Low-$Q^2$ to Mid-$Q^2$ the asymmetry is below percent level and it rises to few percent level in the High-$Q^2$ region and to $10-20\%$ in the Extreme-$Q^2$ region.
It reaches a maximum of $\sim 0.14$ at $z\sim 0.8$ in the Extreme-$Q^2$ region.
Perturbative corrections, as shown in the bottom panels in Fig.~\ref{fig:Z_A_L_pip},  are at about $10\%$  in the Low-$Q^2$ region.
They reduce to few percent level at Mid-$Q^2$ and High-$Q^2$ and to sub percent level at Extreme-$Q^2$.
We observe very good perturbative convergence and stability.

In Fig.~\ref{fig:Z_A_L_pim} we study the single-longitudinal spin asymmetry~\eqref{eq:AhLNC} for $\pi^-$ production as a function of $z$.
With the exception of the Low-$Q^2$ region the asymmetry crosses zero at $0.65<z<0.75$ depending on the average value of $Q^2$.
$A^{\pi^-,\NC}_{LL}$ grows in absolute value with the increase of $Q^2$.
It reaches percent level in the High-$Q^2$ region and grows up to almost $10\%$ level in the Extreme-$Q^2$ region.
We observe good perturbative convergence and stability, with visible large effects in the ratio to  NLO being an
artifact of the zero-crossing of the asymmetry at large $z$.

\subsubsection[\texorpdfstring{$x$-distributions}{x-distributions}]{\boldmath $x$-distributions}

\begin{figure}[p]
\centering

\begin{subfigure}{0.49\textwidth}
\centering
\includegraphics[width=\linewidth]{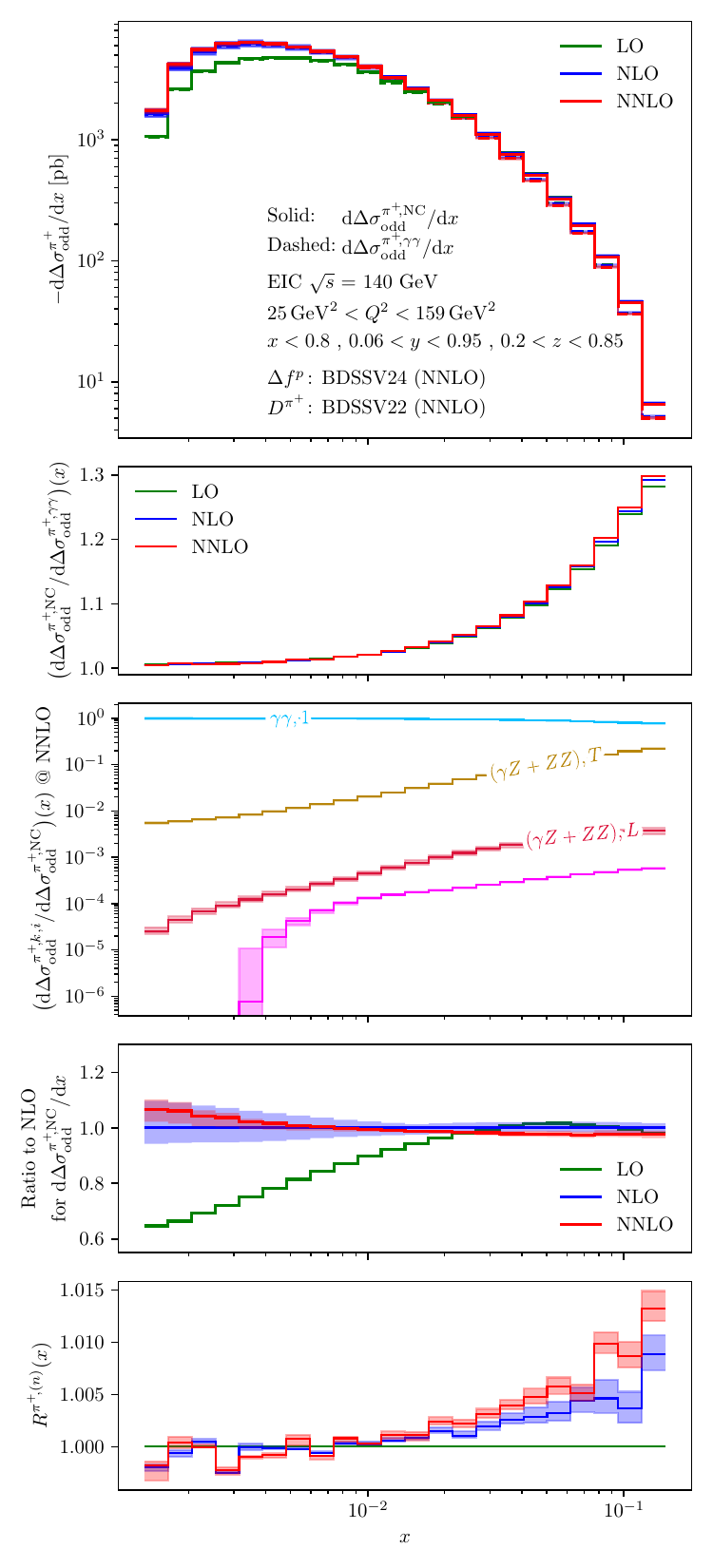}
\caption{Low-$Q^2$}
\end{subfigure}
\begin{subfigure}{0.49\textwidth}
\centering
\includegraphics[width=\linewidth]{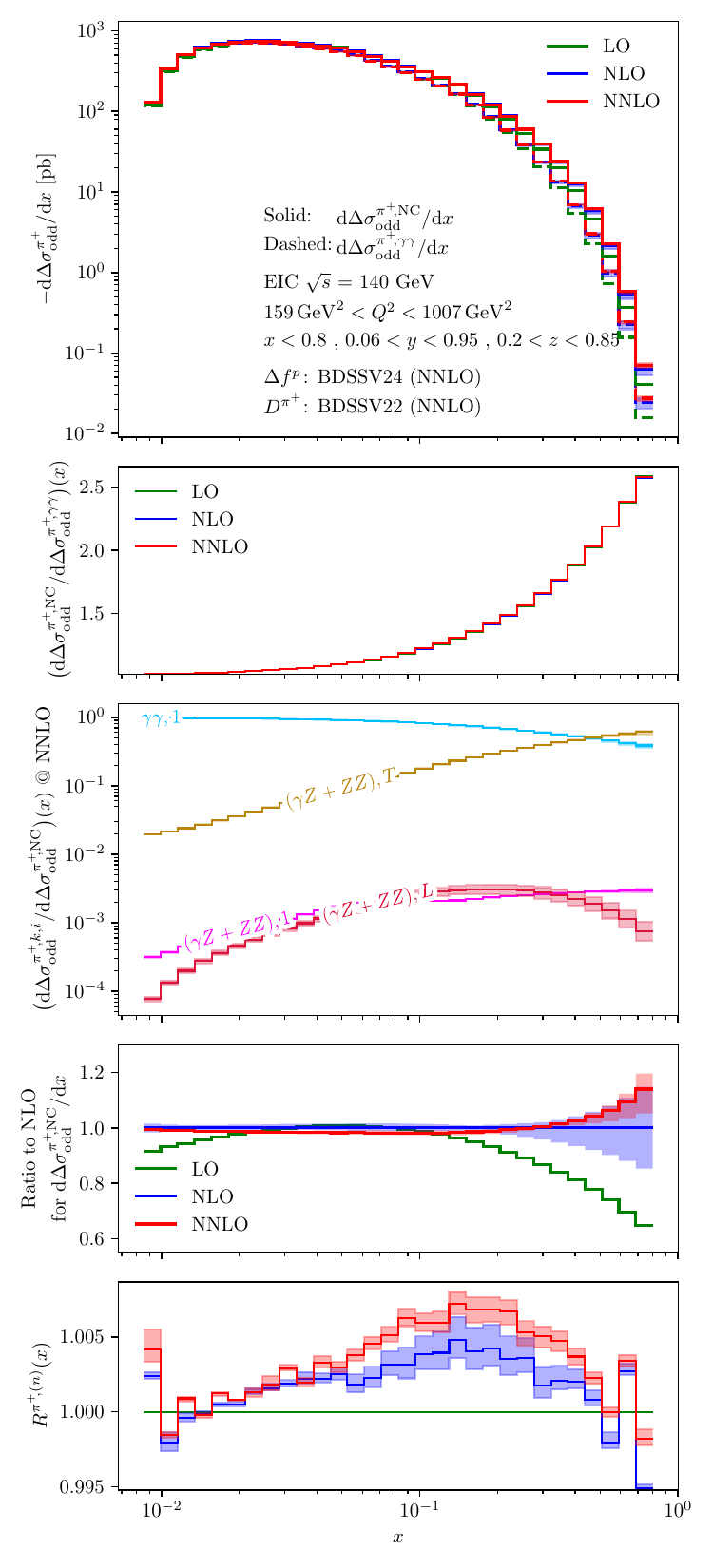}
\caption{Mid-$Q^2$}
\end{subfigure}

\caption{
NC cross section difference differential in $x$ for $\pi^+$ production at Low-$Q^2$ and Mid-$Q^2$.
First panels: total cross section $\dd\Delta\sigma^{\pi^+,\NC}_{\modd}/\dd x$.
Second panels: ratio $(\dd\Delta\sigma^{\pi^+,\NC}_{\modd}/\dd\Delta\sigma^{\pi^+,\gamma\gamma}_{\modd})(x)$.
Third panels: structure function contributions to the total cross section $\dd\Delta\sigma^{\pi^+,\NC}_{\modd}/\dd x$.
Fourth panels: $\dd\Delta\sigma^{\pi^+,\NC}_{\modd}/\dd x$ ratio to NLO.
Fifth panels: $R^{\pi^+,(n)}(x)$ ratio.
}
\label{fig:X_0.0_NC_lowQ2_pip}

\end{figure}

\begin{figure}[p]
\centering

\begin{subfigure}{0.49\textwidth}
\centering
\includegraphics[width=1.0\linewidth]{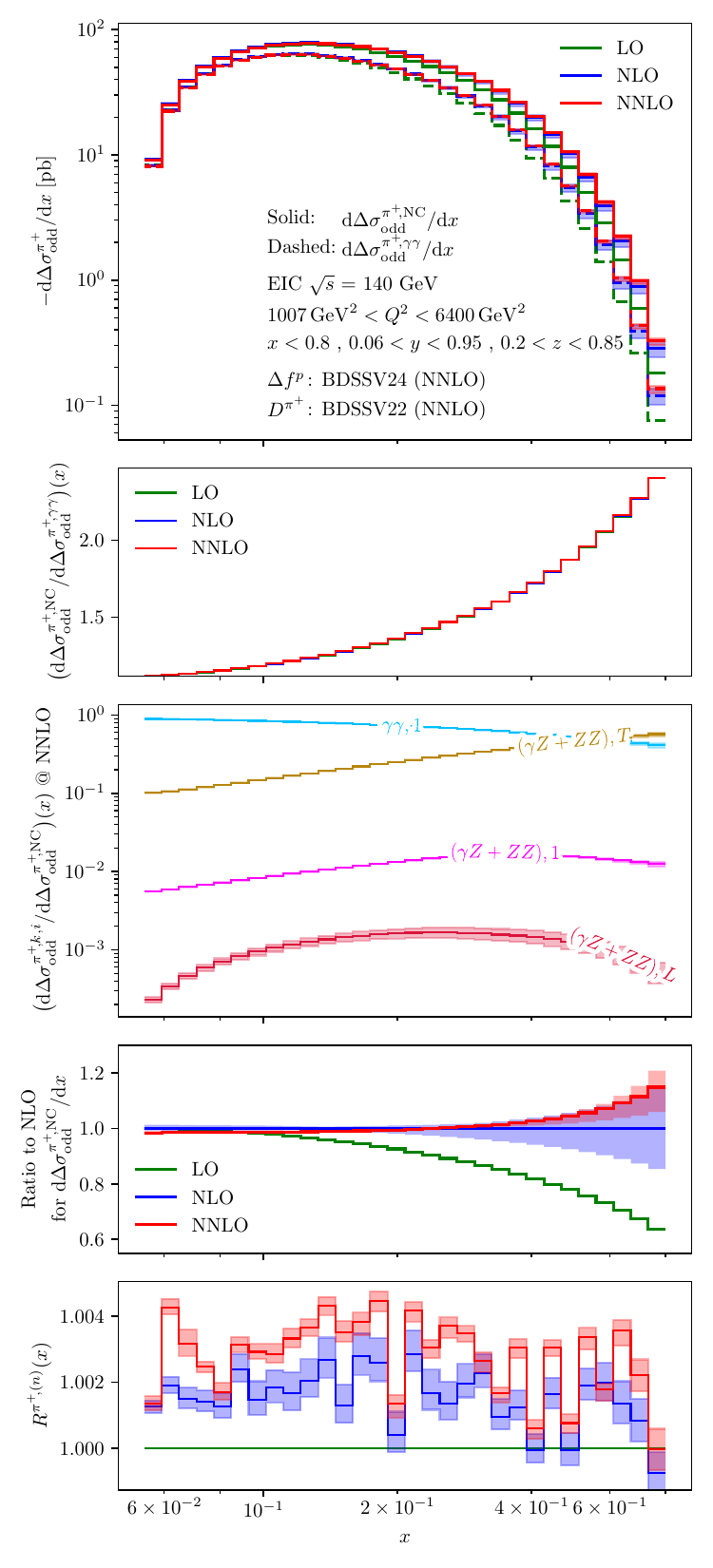}
\caption{High-$Q^2$}
\end{subfigure}
\begin{subfigure}{0.49\textwidth}
\centering
\includegraphics[width=.999\linewidth]{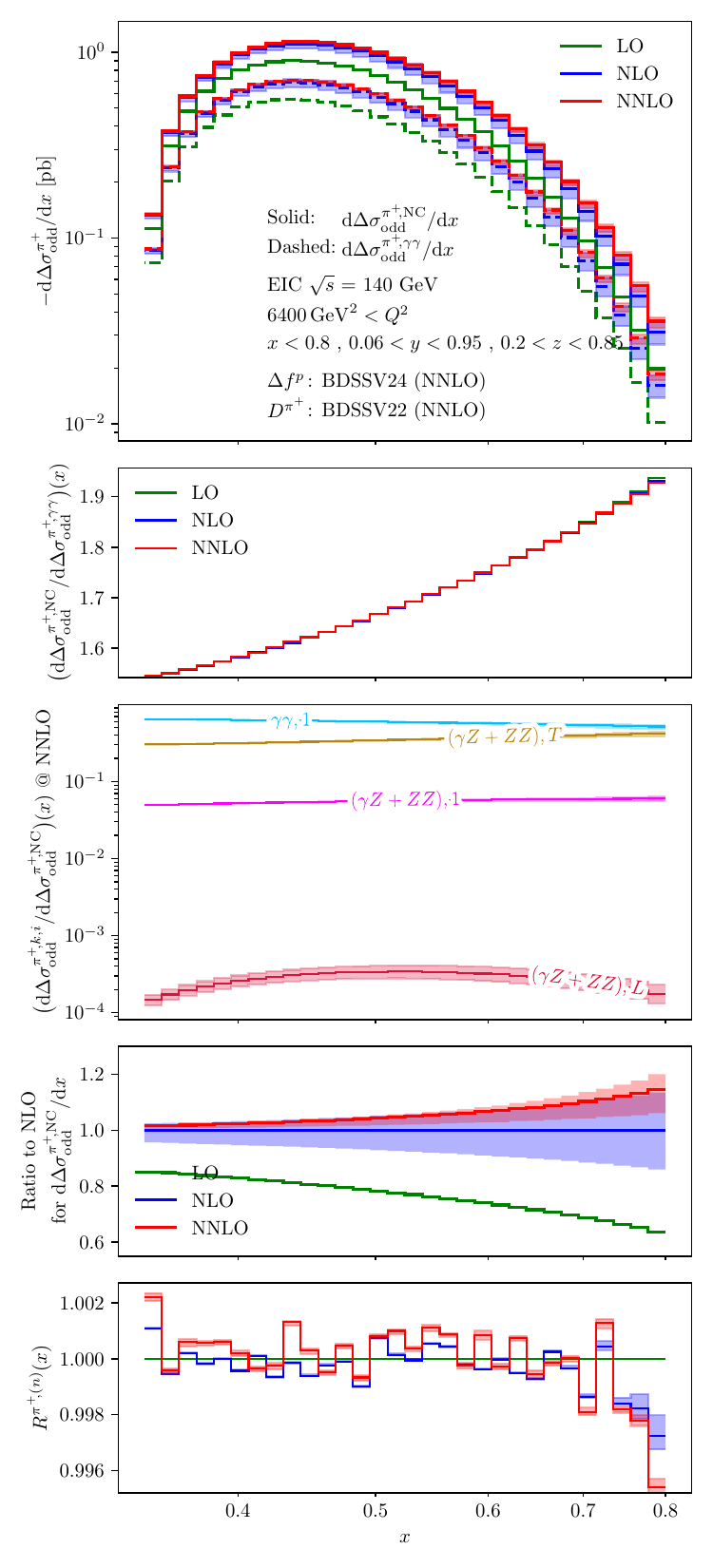}
\caption{Extreme-$Q^2$}
\end{subfigure}

\caption{
NC cross section difference differential in $x$ for $\lambda_\ell$-averaged $\pi^+$ production at High-$Q^2$ and Extreme-$Q^2$.
First panels: total cross section $\dd \Delta\sigma^{\pi^+,\NC}_{\modd}/\dd x$.
Second panels: ratio $(\dd \Delta\sigma^{\pi^+,\NC}_{\modd}/\dd \Delta\sigma^{\pi^+,\gamma\gamma}_{\modd})(x)$.
Third panels: structure function contributions to the total cross section $\dd \Delta\sigma^{\pi^+,\NC}_{\modd}$.
Fourth panels: $\dd \Delta\sigma^{\pi^+,\NC}_{\modd}$ ratio to NLO.
Fifth panels: $R^{\pi^+,(n)}(x)$ ratios.
}
\label{fig:X_0.0_NC_highQ2_pip}

\end{figure}

In Fig.~\ref{fig:X_0.0_NC_lowQ2_pip} and \ref{fig:X_0.0_NC_highQ2_pip} we show the $x$-differential distributions for the cross section difference $\dd\Delta\sigma^{\pi^+,\NC}_{\modd}/\dd x$ in $\pi^+$ production for the four ranges in $Q^2$ defined in~\eqref{eq:cutsQ2}.
The $x$-range is determined by $Q^2=xys$ with the restriction $x<0.8$.
As already observed for unpolarized SIDIS~\cite{Bonino:2025qta} where the same kinematic cuts were used, the shape in $x$ is largely affected by the cut on $y$.
As in unpolarized SIDIS the cross section difference increases at first with increasing $x$ to attain its maximum and then falling off rapidly.
This behaviour is present in all ranges of $Q^2$ and can be seen in the first panels in Fig.~\ref{fig:X_0.0_NC_lowQ2_pip} and \ref{fig:X_0.0_NC_highQ2_pip}.
The accelerated decline visible at larger average $Q^2$ values is due to the depletion of the polarized PDFs at high $x$.
Overall $\dd\Delta\sigma^{\pi^+,\NC}_{\modd}/\dd x$ decreases with increasing average value of $Q^2$ as predicted by the DIS scaling.

In the first panels the full NC exchange (solid) is compared to the photon only exchange (dashed).
The magnitude of the electroweak effects is studied in the second panels with the ratio $(\dd\Delta\sigma^{\pi^+,\NC}_{\modd}/\dd\Delta\sigma^{\pi^+,\gamma\gamma}_{\modd})(x)$.
The electroweak effects are very large, already at percent level at Low-$Q^2$.
The ratio $(\dd\Delta\sigma^{\pi^+,\NC}_{\modd}/\dd\Delta\sigma^{\pi^+,\gamma\gamma}_{\modd})(x)$ is maximal for Mid-$Q^2$ values where it reaches up to 2.5 at high $x$.
It then slowly decreases with further increase of $Q^2$.
The ratio overall grows with the increase of $x$.
This differs significantly with respect to the size of the electroweak effects observed in unpolarized SIDIS for the $x$ distributions, where a maximum of $60\%$ was reached at Extreme-$Q^2$ \cite{Bonino:2025qta}.
This stands as evidence for an increase in sensitivity of polarized cross sections to electroweak effects, as already observed in the $Q^2$ distributions.
The ratio $(\dd\Delta\sigma^{\pi^+,\NC}_{\modd}/\dd\Delta\sigma^{\pi^+,\gamma\gamma}_{\modd})(x)$ is very stable under QCD perturbative corrections as shown in the fifth panels.
With per-mille level accuracy the $n^{\mathrm{th}}$ perturbative order $\dd\Delta\sigma^{\pi^+,\NC,(n)}_{\modd}/\dd x$ can be predicted from the photon-only process $\dd\Delta\sigma^{\pi^+,\gamma\gamma,(n)}_{\modd}/\dd x$ multiplied by the LO ratio $\dd\Delta\sigma^{\pi^+,\NC,(0)}_{\modd}/\dd\Delta\sigma^{\pi^+,\gamma\gamma,(0)}_{\modd}$.

The large electroweak effects can be explained from the structure function decomposition of the cross section difference $\dd\Delta\sigma^{\pi^+,\NC}_{\modd}/\dd x$,
 shown in the third panel of each figure.
At Low-$Q^2$ the $\Gcal^{\pi^+,(\gamma Z+ZZ)}_T$ structure function already appears to be of the same order of magnitude as the photon-only contribution in $\Gcal^{\pi^+}_1$, for $x>0.03$.
At Mid-$Q^2$ and High-$Q^2$ $\Gcal^{\pi^+,(\gamma Z+ZZ)}_T$ even surpasses the photon-only contribution in $\Gcal^{\pi^+}_1$ at $x\sim 0.4$ and $x\sim 0.5$
respectively.
The $\Gcal^{\pi^+,(\gamma Z+ZZ)}_T$ structure function also carries a moderate $x$ dependence: it grows with $x$ at a rate similar to the decrease of $\Gcal^{\pi^+,\gamma\gamma}_1$.
The electroweak effects in $\Gcal^{\pi^+}_1$ are always small whilst growing in size with the average value of $Q^2$.
At Extreme-$Q^2$ $\Gcal^{\pi^+,(\gamma Z+ZZ)}_1$ is only one order of magnitude smaller than the photon-only contribution.
The longitudinal structure function is negligible in all $Q^2$ ranges.

In the fourth panels of Fig.~\ref{fig:X_0.0_NC_lowQ2_pip} and \ref{fig:X_0.0_NC_highQ2_pip} perturbative convergence and stability is studied for the perturbative corrections to $\dd\Delta\sigma^{\pi^+,\NC}_{\modd}/\dd x$, as ratio to NLO.
We observe good perturbative stability and convergence, with NNLO corrections always within the NLO envelope.
The NNLO corrections grow larger with the increase of $x$ for $x>0.1$.

\begin{figure}[p]
\centering

\begin{subfigure}{0.49\textwidth}
\centering
\includegraphics[width=\linewidth]{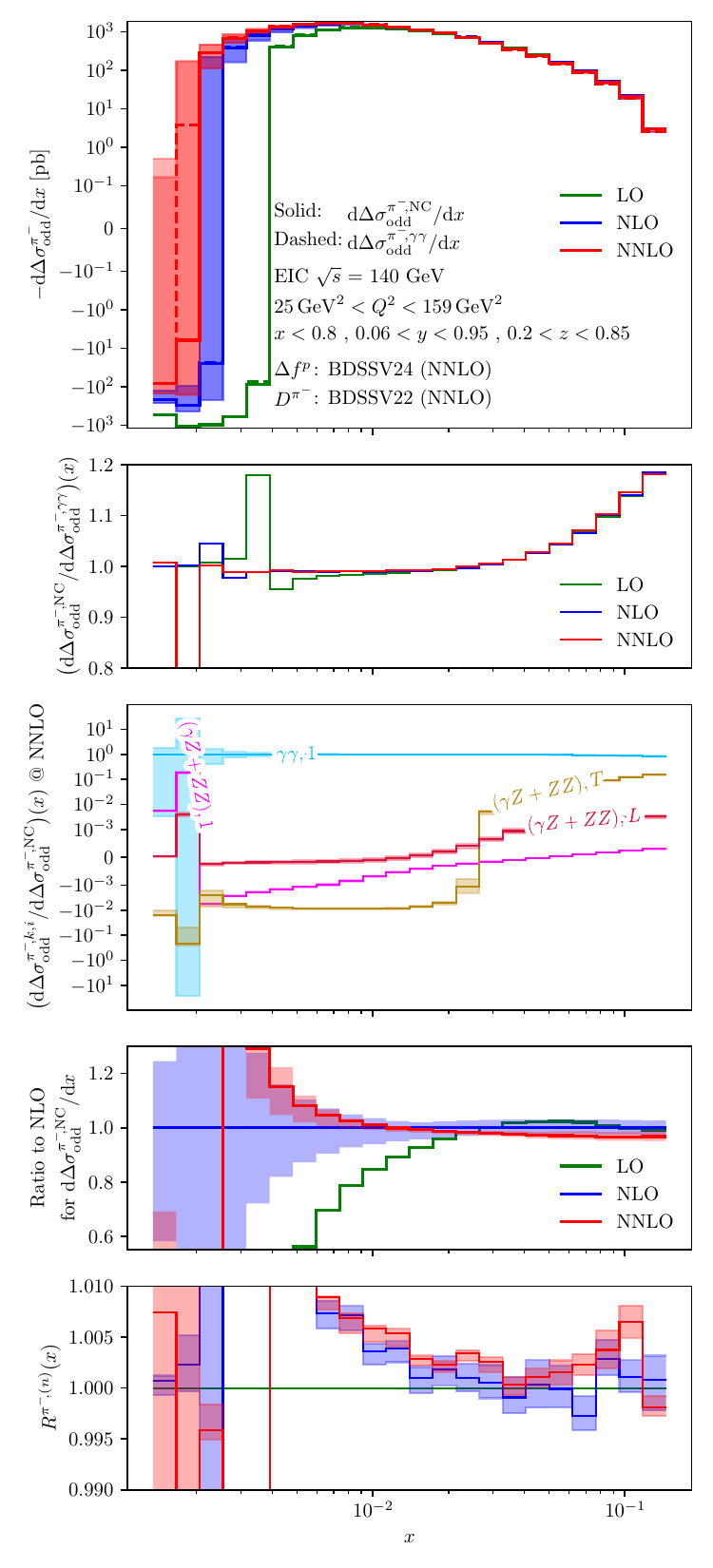}
\caption{Low-$Q^2$}
\end{subfigure}
\begin{subfigure}{0.49\textwidth}
\centering
\includegraphics[width=\linewidth]{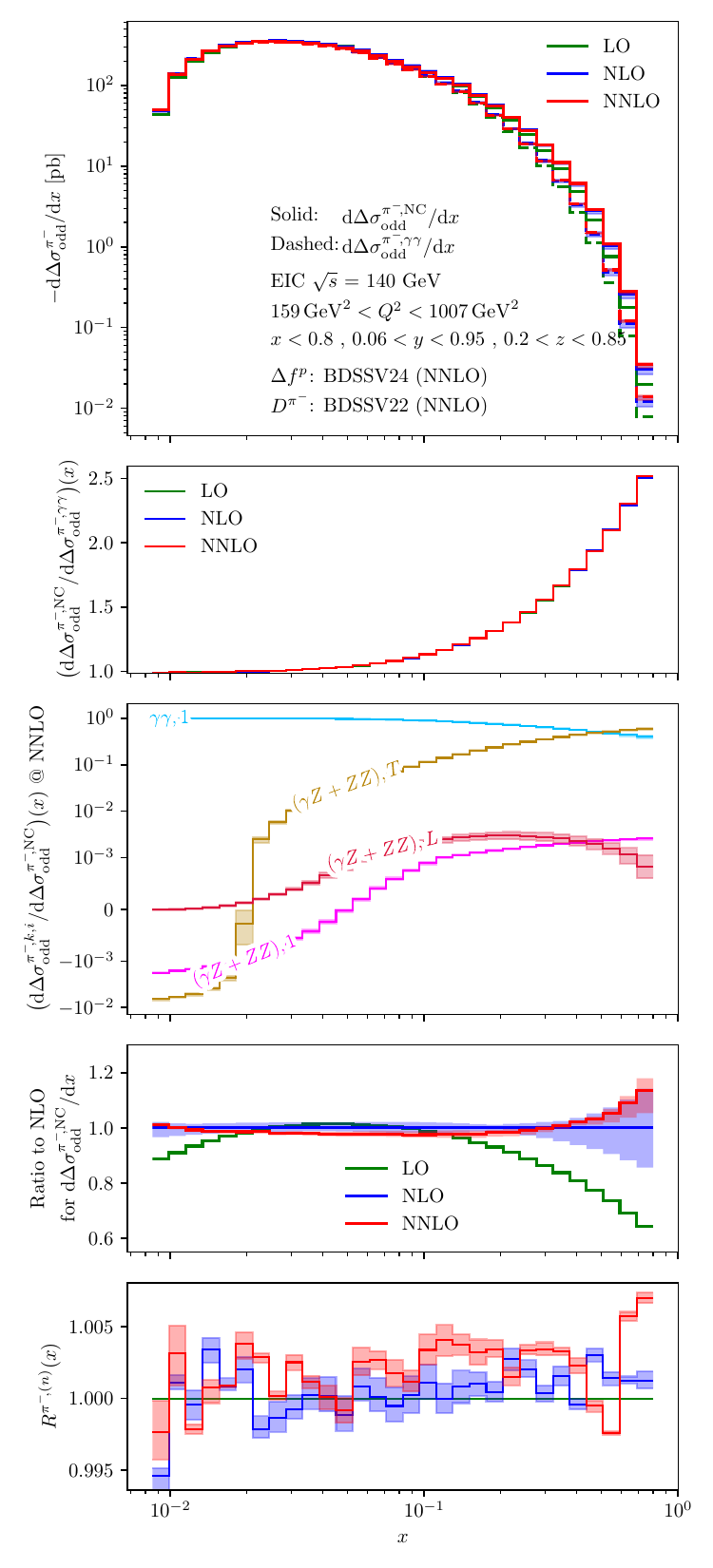}
\caption{Mid-$Q^2$}
\end{subfigure}

\caption{
NC cross section difference differential in $x$ for $\pi^-$ production at low and intermediate $Q^2$.
First panels: total cross section $\dd \Delta\sigma^{\pi^-}_{\modd}$
Second panels: ratio $(\dd \Delta\sigma^{\pi^-,\NC}_{\modd}/\dd \Delta\sigma^{\pi^-,\gamma\gamma}_{\modd})(x)$.
Third panels: structure function contributions to the total cross section.
Fourth panels: $\dd \Delta\sigma^{\pi^-,\NC}_{\modd}$ ratio to NLO.
Fifth panels: $R^{\pi^-,(n)}(x)$ ratio.
}
\label{fig:X_0.0_NC_lowQ2_pim}

\end{figure}

\begin{figure}[p]
\centering

\begin{subfigure}{0.49\textwidth}
\centering
\includegraphics[width=1.0\linewidth]{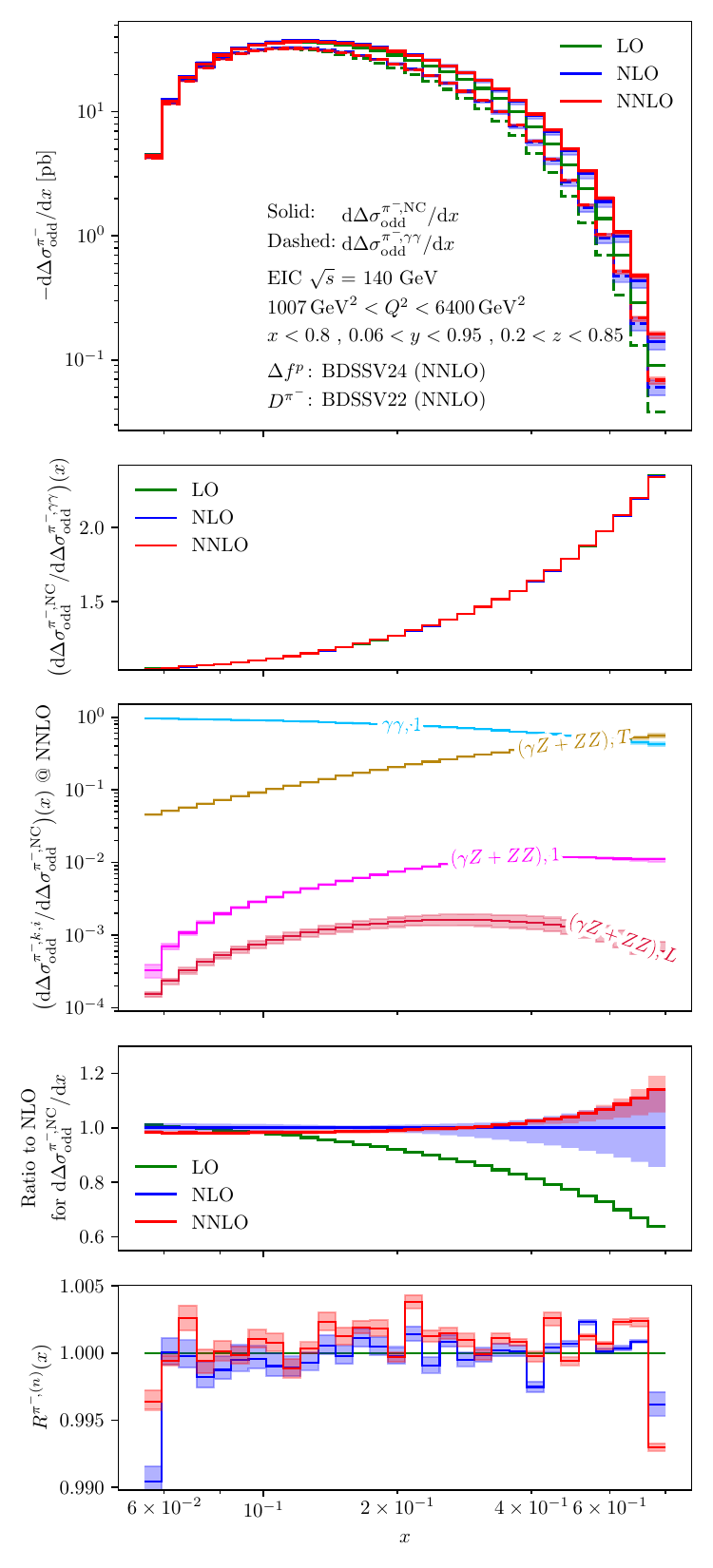}
\caption{High-$Q^2$}
\end{subfigure}
\begin{subfigure}{0.49\textwidth}
\centering
\includegraphics[width=1.0\linewidth]{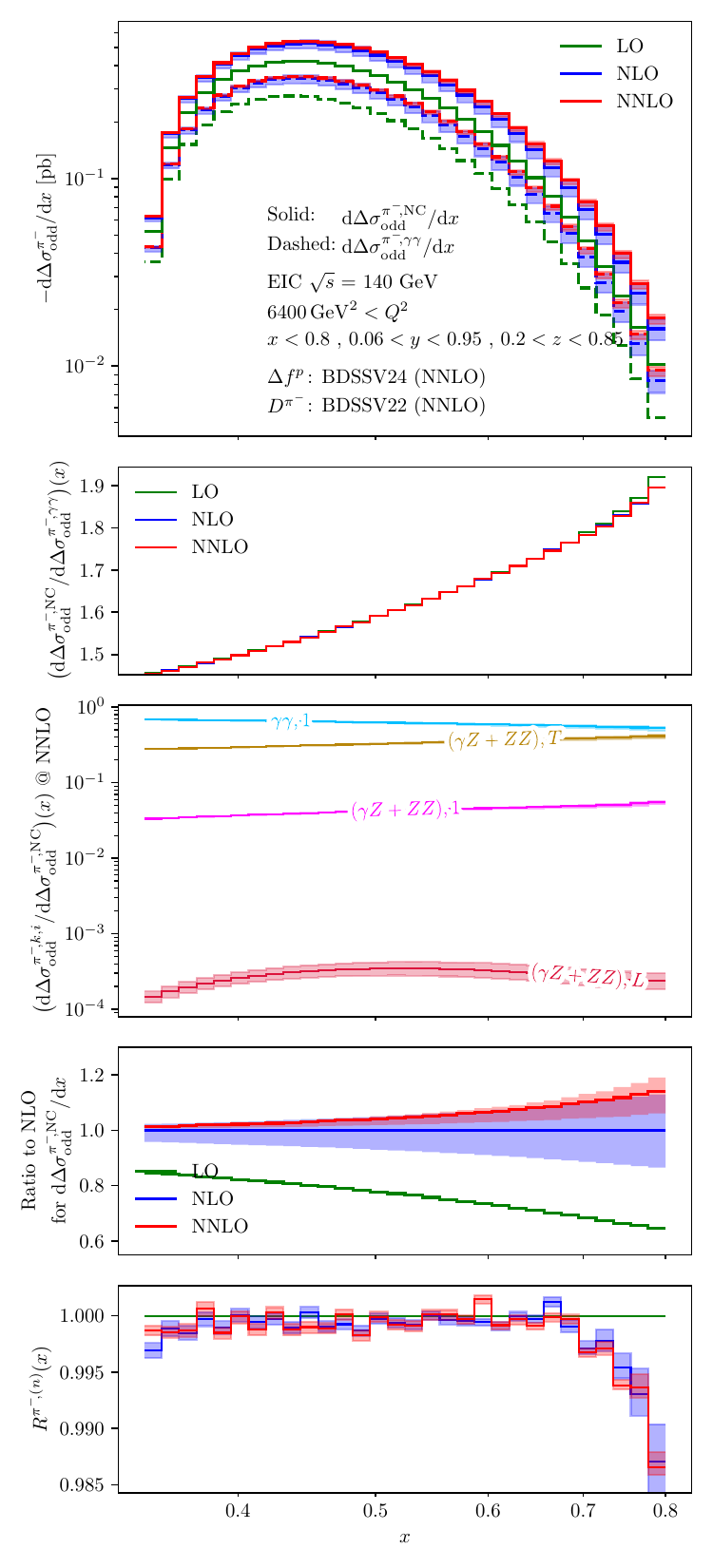}
\caption{Extreme-$Q^2$}
\end{subfigure}

\caption{
NC cross section difference differential in $x$ for $\pi^-$ production at high $Q^2$.
First panels: total cross section $\dd \Delta\sigma^{\pi^-}_{\modd}/\dd x$.
Second panels: ratio $(\dd \Delta\sigma^{\pi^-,\NC}_{\modd}/\dd \Delta\sigma^{\pi^-,\gamma\gamma}_{\modd})(x)$.
Third panels: structure function contributions to the total cross section $\dd \Delta\sigma^{\pi^-,\NC}_{\modd}/\dd x$.
Fourth panels: $\dd \Delta\sigma^{\pi^-,\NC}_{\modd}/\dd x$ ratio to NLO.
Fifth panels: $R^{\pi^+,(n)}(x)$ ratios.
}
\label{fig:X_0.0_NC_highQ2_pim}

\end{figure}

In Fig.~\ref{fig:X_0.0_NC_lowQ2_pim} and \ref{fig:X_0.0_NC_highQ2_pim} we show the NC cross section difference $\dd \Delta\sigma^{\pi^-,\NC}_{\modd}/\dd x$ for $\pi^-$ production  in each range in $Q^2$.
In the first panel the cross section difference is shown for both full NC (solid) and photon-only exchange (dashed).
The overall behaviour is very similar to the $\pi^+$ case, with the $x$-dependence mostly constrained by the kinematic cuts and polarized PDFs depletion at large $x$.
Due to the suppression of the $u\to u$ channel in the fragmentation process, the distributions are roughly an order of magnitude smaller for $\pi^-$ as compared to $\pi^+$.
At Low-$Q^2$ the cross section difference crosses zero for $x\sim 2\times 10^{-3}$.

The magnitude of the electroweak effects in $\dd \Delta\sigma^{\pi^-,\NC}_{\modd}/\dd x$ is shown in the second panels with the ratio $\left(\dd \Delta\sigma^{\pi^-,\NC}_{\modd}/\dd \Delta\sigma^{\pi^-,\gamma\gamma}_{\modd}\right)(x)$.
The electroweak effects are very large, already at $2\%$ level at Low-$Q^2$ and they peak in the Mid-$Q^2$ region where the ratio reaches $2.5$.
The size of the electroweak effects is comparable between $\pi^+$ and $\pi^-$.
This ratio is very stable under QCD corrections: it is possible to predict at per-mille level accuracy the $n^{\mathrm{th}}$ QCD order $\dd \Delta\sigma^{\pi^-,\NC,(n)}_{\modd}/\dd x$ by multiplying the corresponding photon-only exchange $\dd \Delta\sigma^{\pi^-,\gamma\gamma,(n)}_{\modd}/\dd x$ with the LO ratio $\dd \Delta\sigma^{\pi^-,\NC,(0)}_{\modd}/\dd \Delta\sigma^{\pi^-,\gamma\gamma,(0)}_{\modd}$.
This is shown in the fifth panels of Fig.~\ref{fig:X_0.0_NC_lowQ2_pim} and \ref{fig:X_0.0_NC_highQ2_pim}.

The very large electroweak effects are reflected in the structure function decomposition in the third panels.
As for $\pi^+$ the contribution from the transverse structure function grows quickly both in $x$ and $Q^2$ surpassing the photon-only contribution at Mid-$Q^2$ and High-$Q^2$.
At Extreme-$Q^2$ the electroweak and photon-only effects are comparable in size.
The only significant difference between $\pi^+$ and $\pi^-$ is visible in the Low-$Q^2$ region where the cross section difference crosses zero.
This behaviour is already present at LO for photon exchange only and is it due to large cancellation in the partonic channel contributions.
At small $x$ many sea quark and gluonic channels are sizeable and conspire to cancel the $u\to u$ channel contribution, see Fig.~\ref{fig:x_NC_AS_channel_pim} below.

\begin{figure}[tb]
\centering

\begin{subfigure}{0.45\textwidth}
\centering
\includegraphics[width=1.0\linewidth]{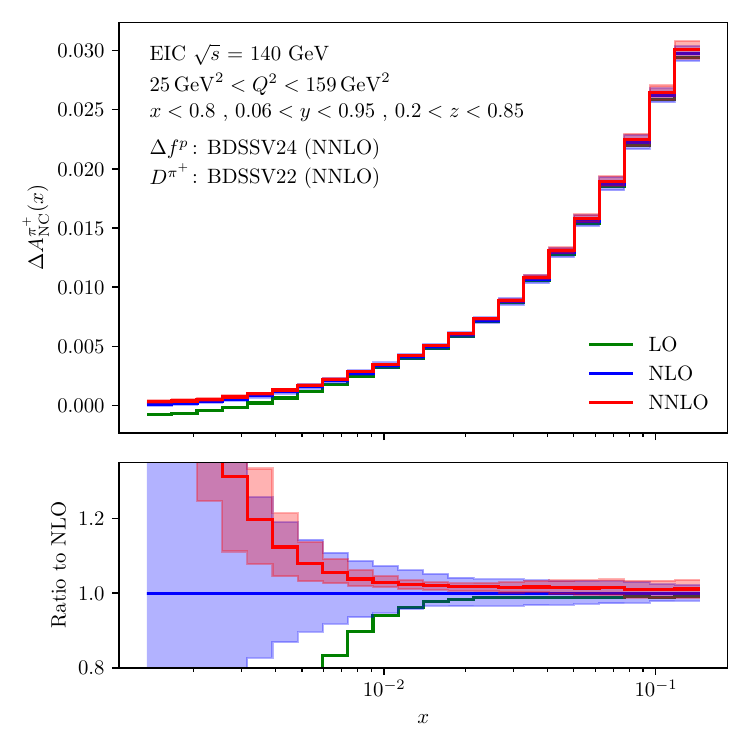}
\caption{Low-$Q^2$}
\end{subfigure}
\begin{subfigure}{0.45\textwidth}
\centering
\includegraphics[width=1.0\linewidth]{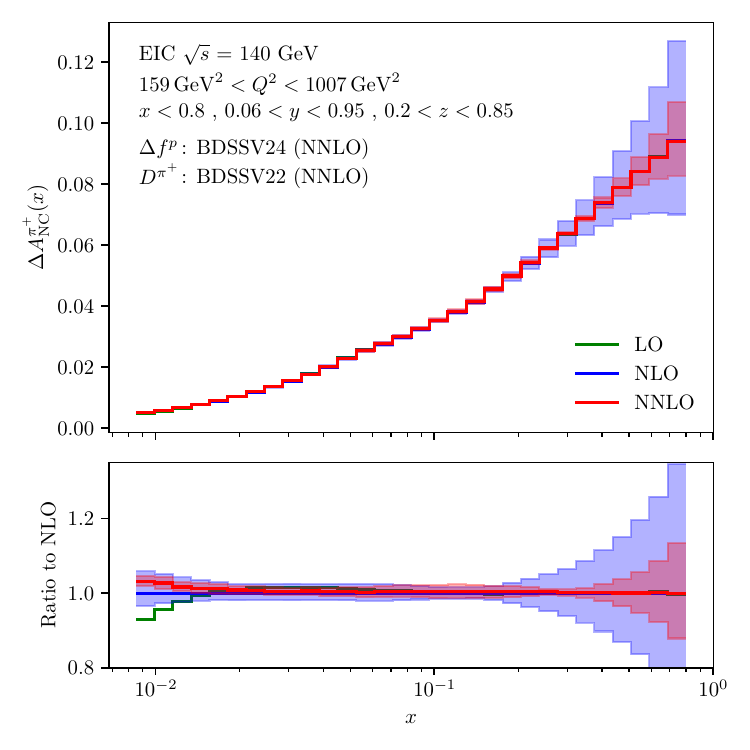}
\caption{Mid-$Q^2$}
\end{subfigure}
\begin{subfigure}{0.45\textwidth}
\centering
\includegraphics[width=1.0\linewidth]{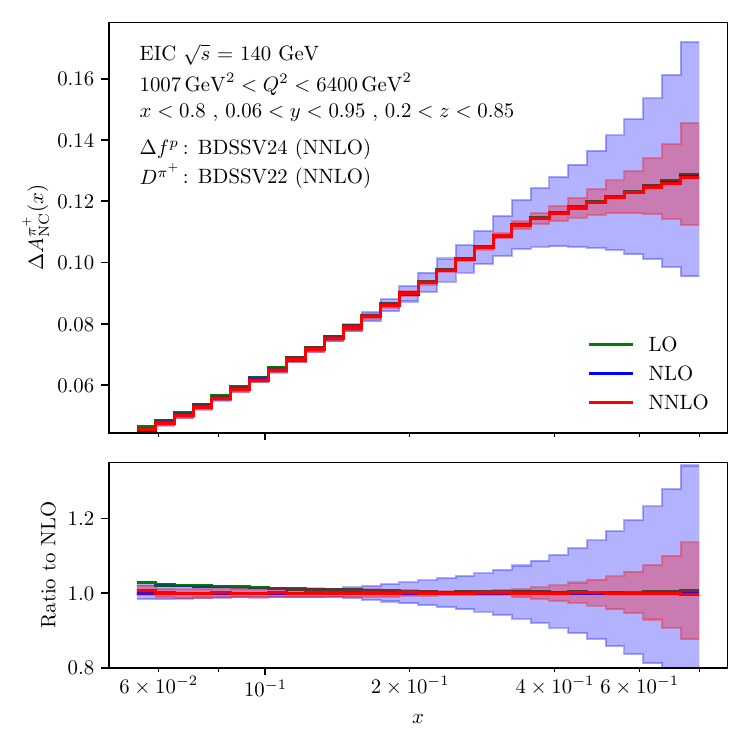}
\caption{High-$Q^2$}
\end{subfigure}
\begin{subfigure}{0.45\textwidth}
\centering
\includegraphics[width=1.0\linewidth]{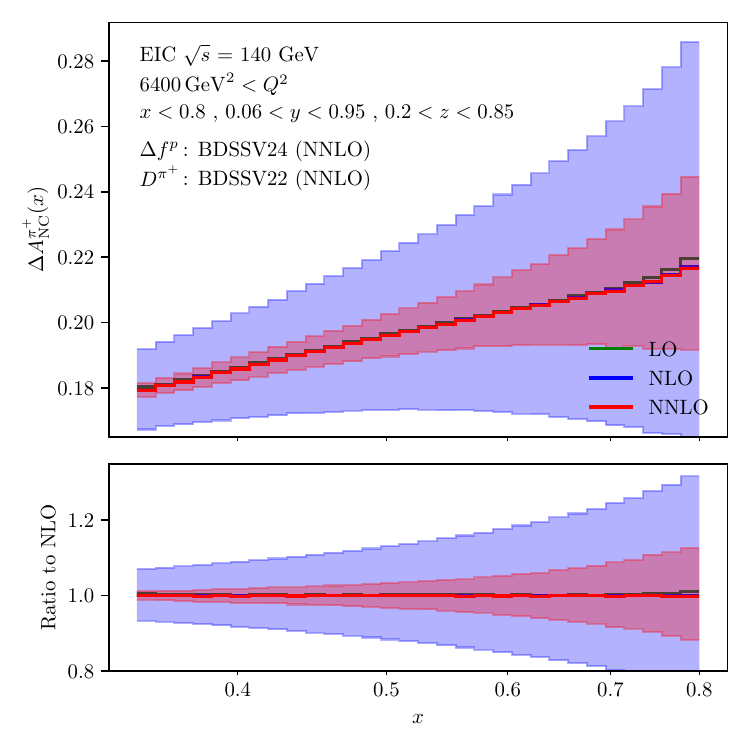}
\caption{Extreme-$Q^2$}
\end{subfigure}

\caption{
NC asymmetry $\Delta A_\mathrm{NC}^{\pi^+}$ in $\pi^+$ production as a function of $x$.
Top panels: Asymmetries.
Bottom panels: Ratio to NLO.
}
\label{fig:X_A_NC_pip}
\end{figure}

\begin{figure}[tb]
\centering

\begin{subfigure}{0.45\textwidth}
\centering
\includegraphics[width=1.0\linewidth]{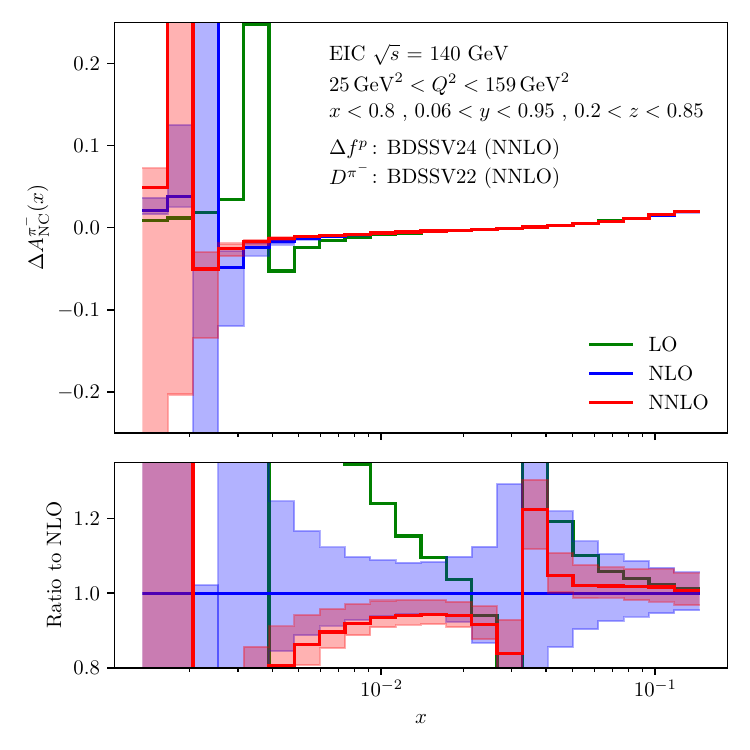}
\caption{Low-$Q^2$}
\end{subfigure}
\begin{subfigure}{0.45\textwidth}
\centering
\includegraphics[width=1.0\linewidth]{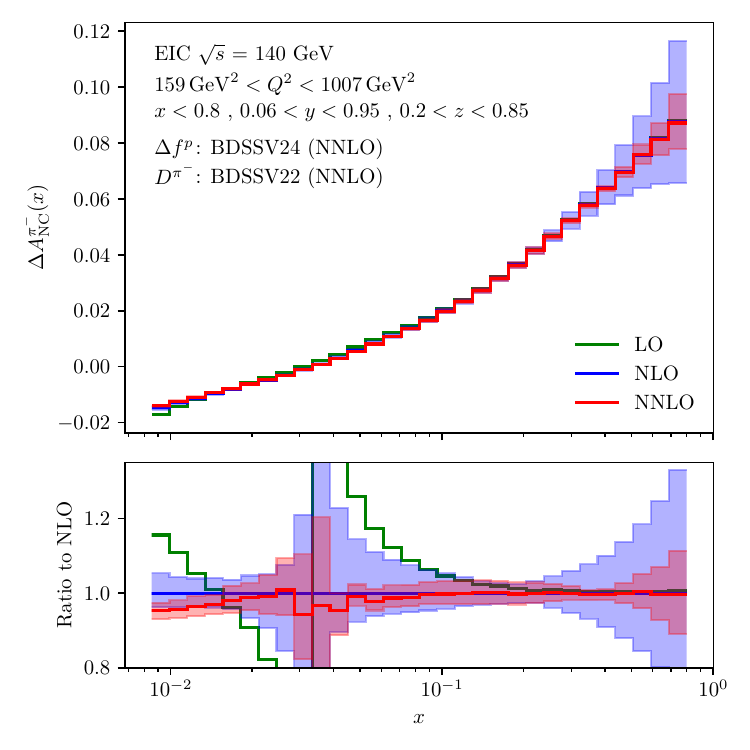}
\caption{Mid-$Q^2$}
\end{subfigure}
\begin{subfigure}{0.45\textwidth}
\centering
\includegraphics[width=1.0\linewidth]{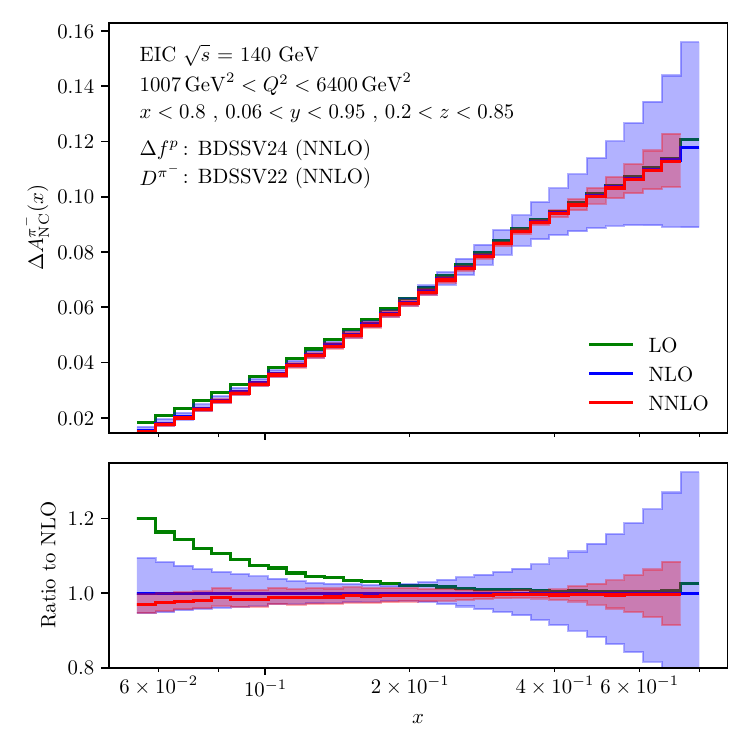}
\caption{High-$Q^2$}
\end{subfigure}
\begin{subfigure}{0.45\textwidth}
\centering
\includegraphics[width=1.0\linewidth]{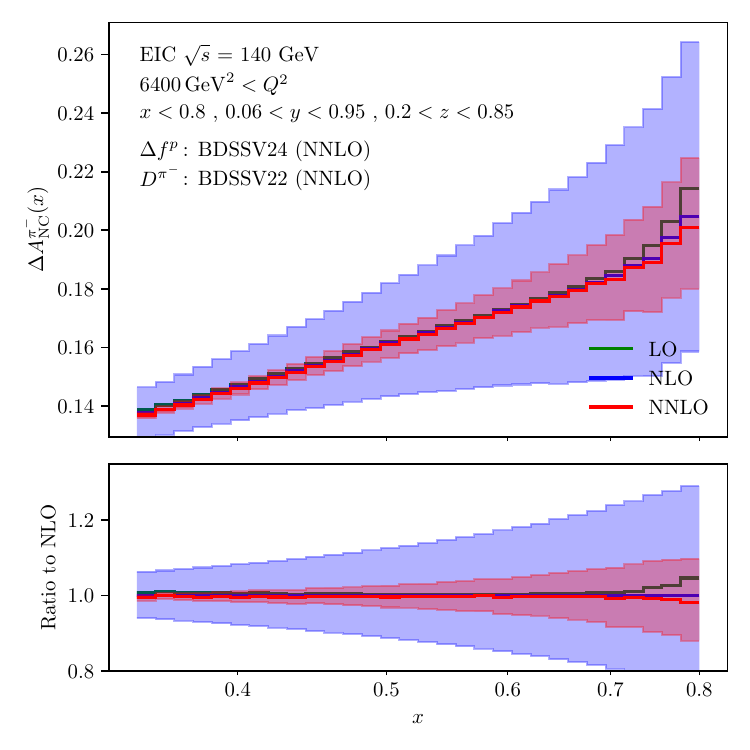}
\caption{Extreme-$Q^2$}
\end{subfigure}

\caption{
NC asymmetry $\Delta A_\mathrm{NC}^{\pi^-}$ in $\pi^-$ production as a function of $x$.
Top panels: Asymmetries.
Bottom panels: Ratio to NLO.
}
\label{fig:X_A_NC_pim}
\end{figure}

In Fig.~\ref{fig:X_A_NC_pip} we study the $x$ dependence of the polarized neutral current asymmetry $\Delta A^{\pi^+}_\NC$ in the four $Q^2$ ranges.
From the top panels of Fig.~\ref{fig:X_A_NC_pip} it is visible that the $\lambda_\ell$-even electroweak effects are strongly suppressed with respect to the $\lambda_\ell$-odd contributions. They grow in size with increasing average value of $Q^2$ from  $3\%$ to $25\%$.
The asymmetries also present a moderate growth in $x$ in all $Q^2$ ranges.
We observe a good perturbative convergence and stability throughout as shown in the bottom panels.
For $x>0.02$ the perturbative corrections are below percent level.
For $x<0.02$ the perturbative correction grow sizeable to more than $20\%$ in the lowest kinematic bins.

For $\pi^-$ production, Fig.~\ref{fig:X_A_NC_pim}, the asymmetry $\Delta A^{\pi^-}_\NC$ grows in $x$ and $Q^2$ with a similar trend and to a similar size as for $\pi^+$ production.
At Low-$Q^2$ large cancellations between the partonic channels spoil the perturbative convergence of the calculation.
A less severe crossing of zero is also present in the Mid-$Q^2$ region at $x\sim 0.03$.
Large perturbative corrections at percent level are present for $x<0.1$ in the High-$Q^2$ region.
For larger values of $x$ perturbative corrections are below percent level and good perturbative convergence is observed.

\begin{figure}[p]
\centering
\includegraphics[width=\textwidth]{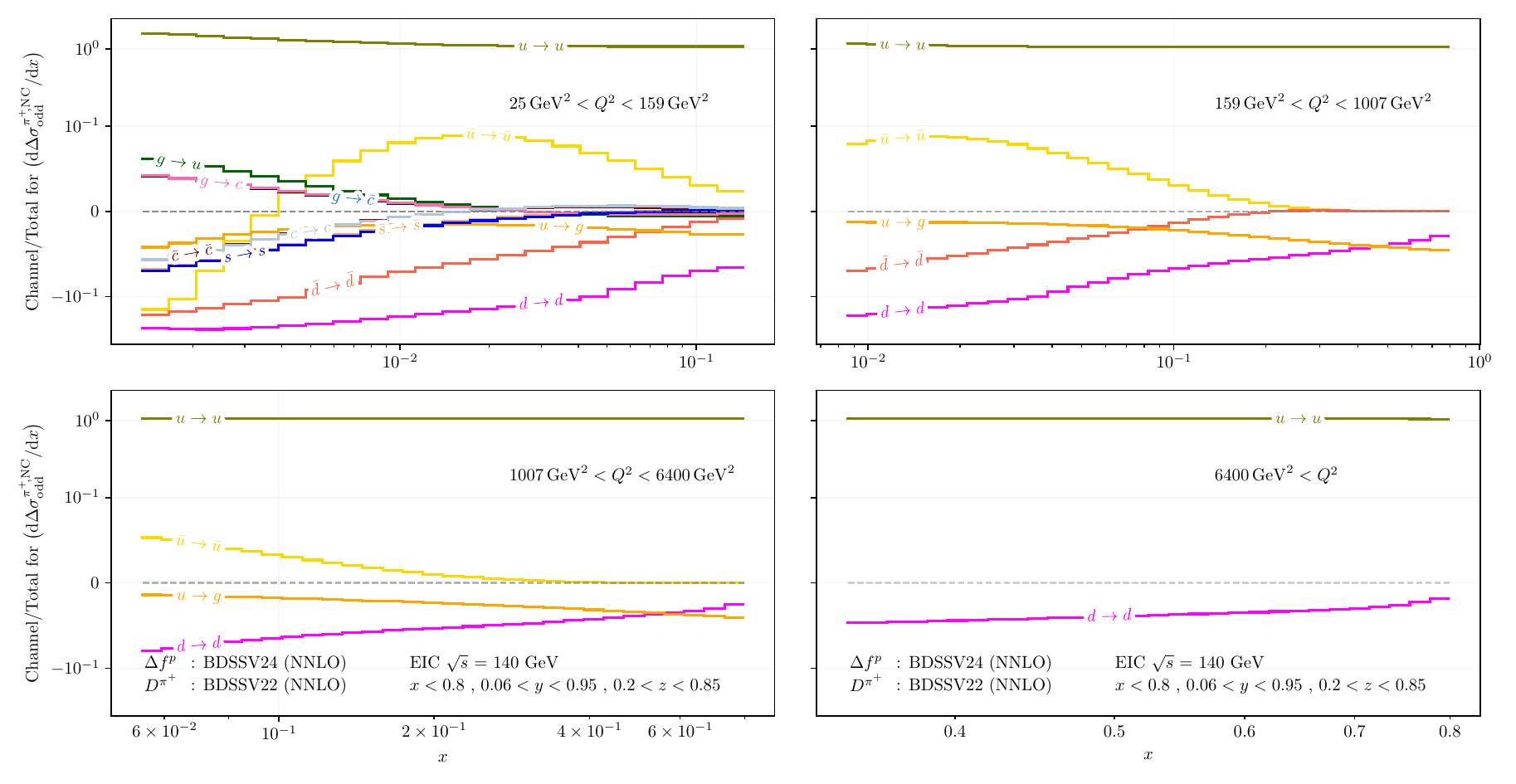}
\caption{
Channel decomposition for $\dd \Delta\sigma^{\pi^+,\NC}_{\modd}$ as a function of $x$.
Only channels contributing more than $4\,\%$ in any bin are displayed.
The range $(-0.1, 0.1)$ on the vertical axis is linear, the ranges above and below are plotted logarithmically.
}
\label{fig:x_NC_AS_channel_pip}

\bigskip

\includegraphics[width=\textwidth]{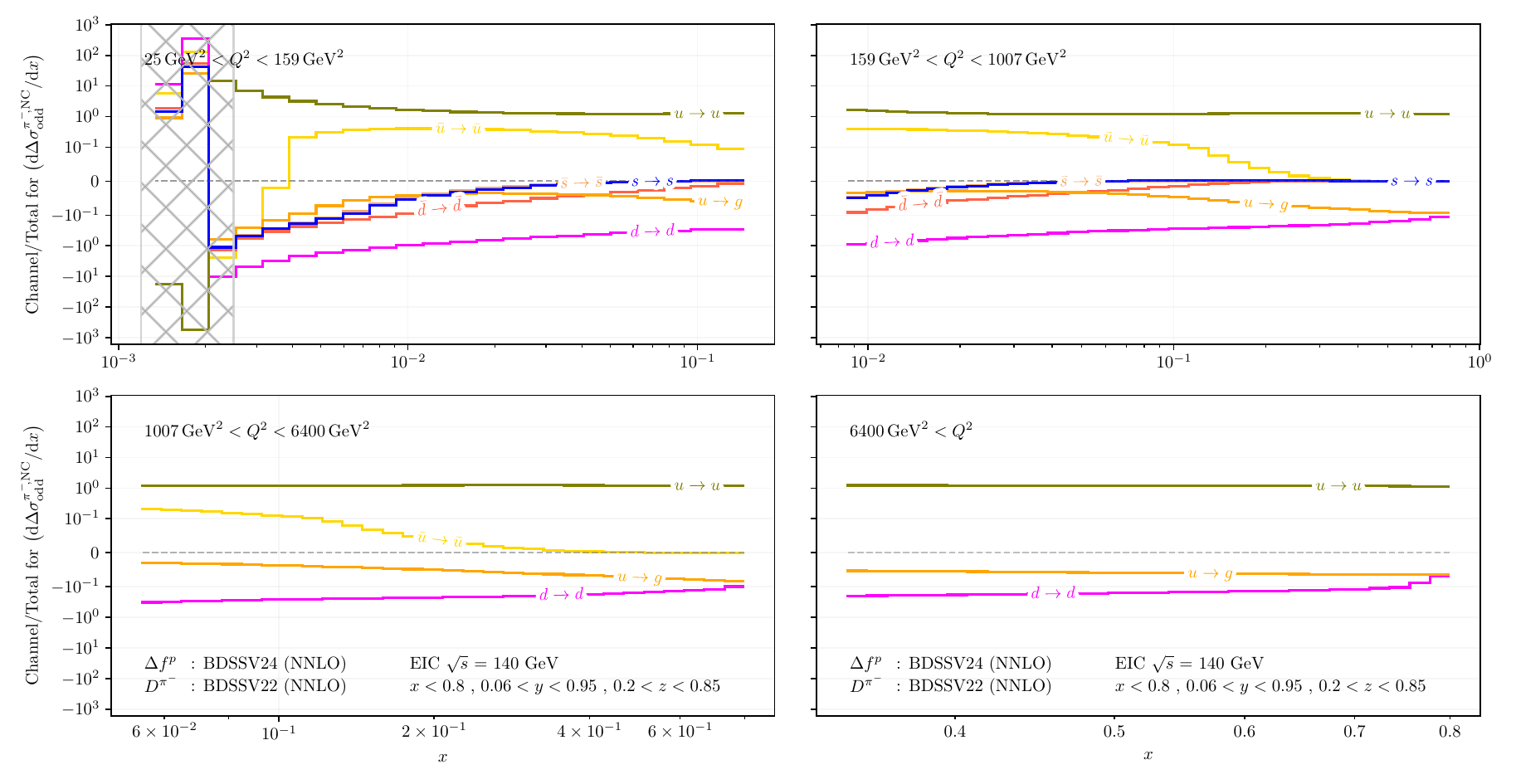}
\caption{
Channel decomposition for $\dd \Delta\sigma^{\pi^-,\NC}_{\modd}$ as a function of $x$.
Only channels contributing more than $4\,\%$ in any bin are displayed.
The bins where are division by zero occurs are not taken into account and are shaded in grey.
The range $(-0.1, 0.1)$ on the vertical axis is linear, the ranges above and below are plotted logarithmically.
}
\label{fig:x_NC_AS_channel_pim}
\end{figure}

In Fig.~\ref{fig:x_NC_AS_channel_pip} we show the channel decomposition for the NC cross section difference $\dd\Delta\sigma^{\pi^+,\NC}_{\modd}/\dd x$. We present $x$ distributions in the four ranges of $Q^2$.
The $u\to u$ channel is the most dominant due to its valence contribution in both initial and final state identified hadrons.
The second most dominant contribution is the negative $d\to d$ channel which is suppressed by the $D^{\pi^+}_d$ fragmentation functions and the electromagnetic coupling.
Its negativity is a consequence of the negative polarized PDF $\Delta f^p_d$.
At Low-$Q^2$ sea quarks and gluonic channels are sizeable because of the small values of $x$ probed.
The gluon-initiated channels are large and positive at small $x$.
At small $x$ polarized PDFs are small in size and tend to zero, resulting in larger contributions from sea quarks and gluons in the Low-$Q^2$ region.
With the increase of $Q^2$ the sea channels reduce in number and size.
Only channels that contribute to the valence of initial and/or final state identified hadrons survive at Mid-$Q^2$ and High-$Q^2$.

In the channel decomposition for $\pi^-$ production, the $u\to u$ is still dominant due its electromagnetic charge enhancement and the abundance of $u$ inside the proton.
Despite its valence role in $\pi^-$  the $d\to d$ channel is significantly smaller than the $u\to u$ channel and still negative.
Due to the suppression of the $D^{\pi^-}_u$ fragmentation function sea quarks and gluon channels play a significant role up to larger values of $x$ and $Q^2$.
For Low-$Q^2$ and $x<0.003$ large positive and negative contributions from sea and valence quarks and gluons conspire to cancel each other out and nullify the cross section difference (grey shaded region) in Fig.~\ref{fig:x_NC_AS_channel_pim}.
This is due to all polarized PDFs tending to very small values for $x\to 0$.
For both $\pi^+$ and $\pi^-$ production the only non trivial $x$ dependence is carried by the $\bar{u}\to\bar{u}$ channel.

\begin{figure}[p]
\centering
\includegraphics[width=\textwidth]{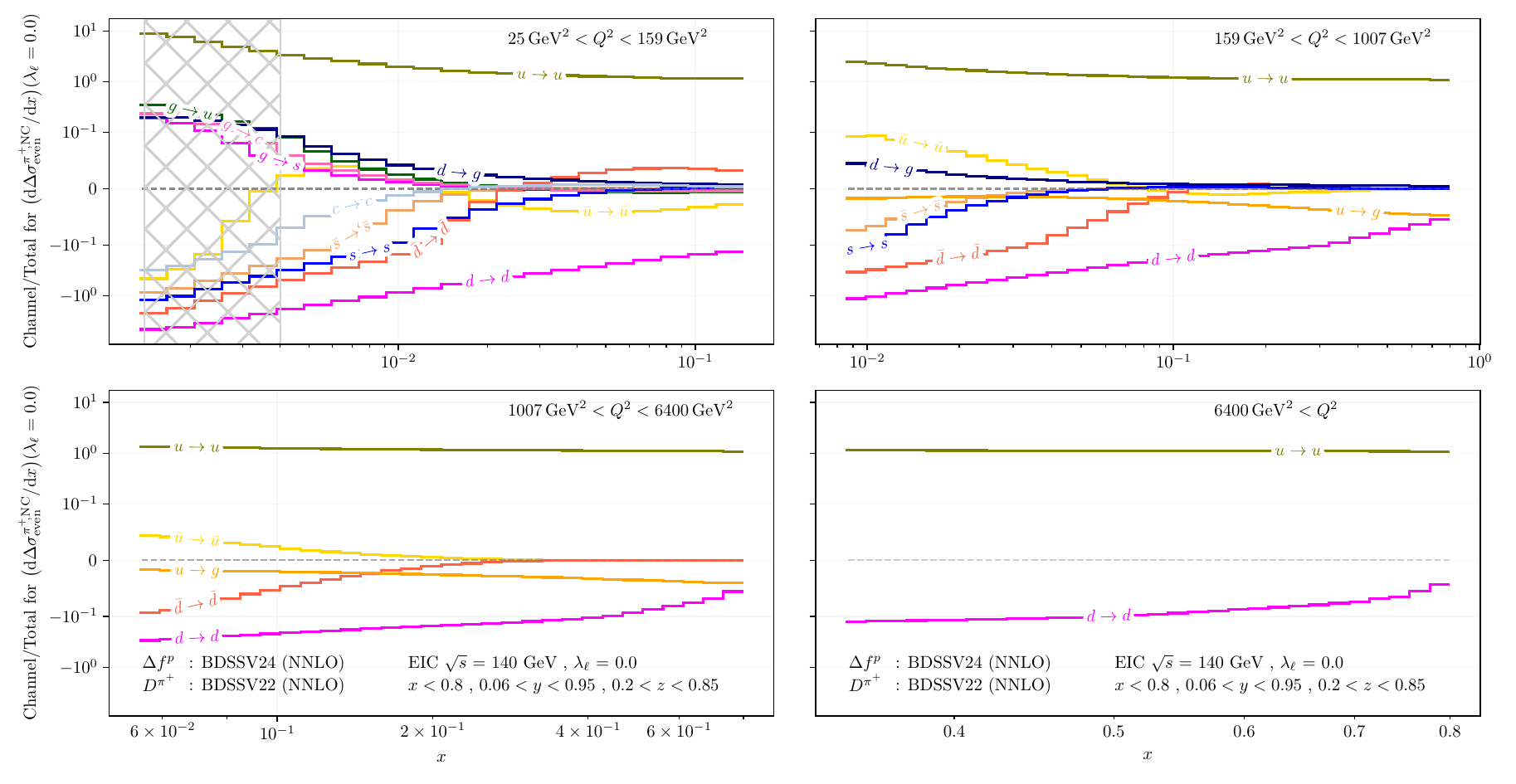}
\caption{
Channel decomposition for $\dd \Delta\sigma^{\pi^+,\NC}_{\meven}$ as a function of $x$.
Only channels contributing more than $4\,\%$ in any bin are displayed.
The range $(-0.1, 0.1)$ on the vertical axis is linear, the ranges above and below are plotted logarithmically.
}
\label{fig:x_NC_STD_channel_pip}

\bigskip

\includegraphics[width=\textwidth]{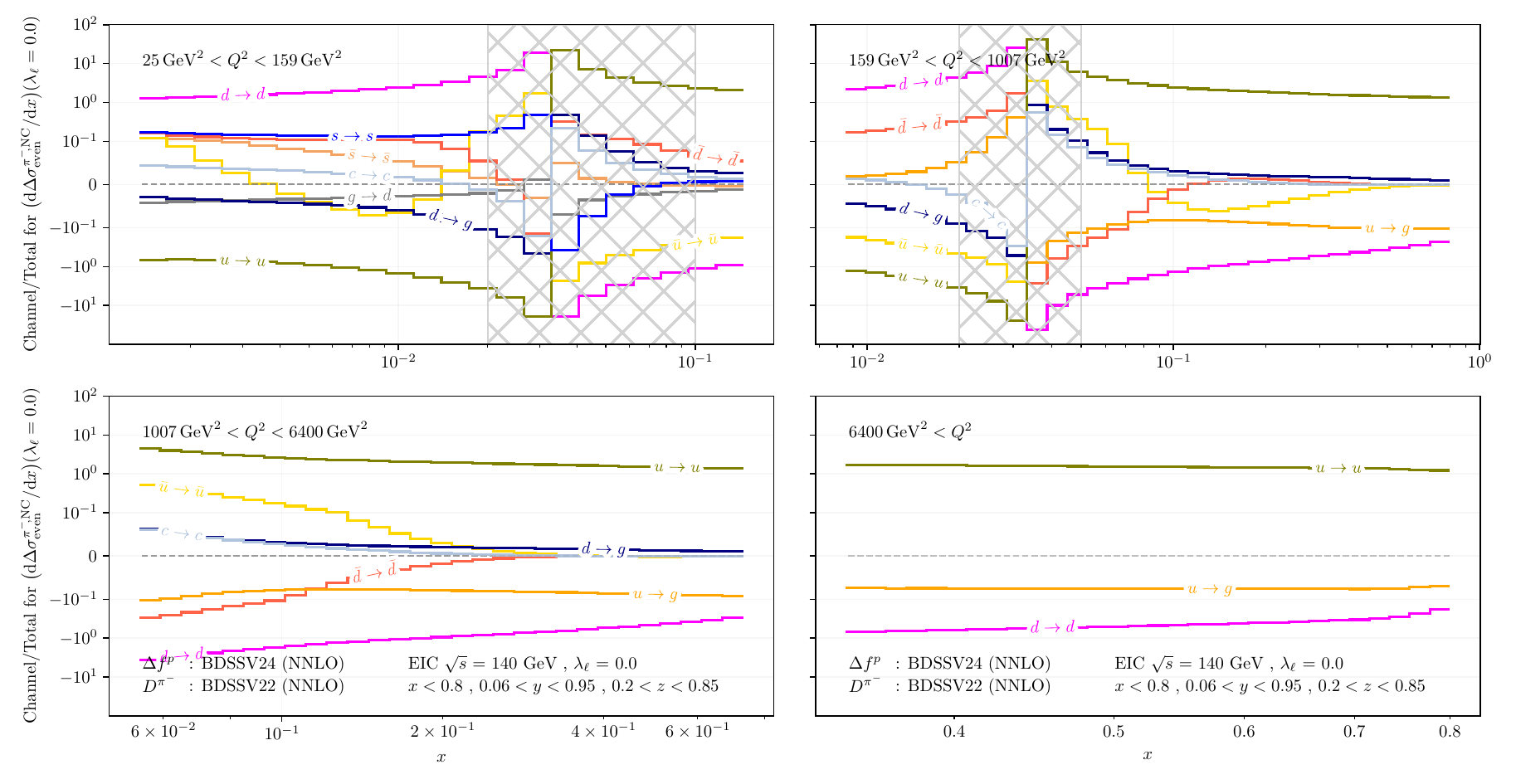}
\caption{
Channel decomposition for $\dd \Delta\sigma^{\pi^-,\NC}_{\meven}$ as a function of $x$.
Only channels contributing more than $4\,\%$ in any bin are displayed.
The bins where are division by zero occurs are not taken into account and are shaded in grey.
The range $(-0.1, 0.1)$ on the vertical axis is linear, the ranges above and below are plotted logarithmically.
}
\label{fig:x_NC_STD_channel_pim}
\end{figure}

In Fig.~\ref{fig:x_NC_STD_channel_pip} and \ref{fig:x_NC_STD_channel_pim} we show the channel decomposition of $\dd \Delta\sigma^{\pi^+,\NC}_{\meven}/\dd x$ and $\dd \Delta\sigma^{\pi^-,\NC}_{\meven}/\dd x$ respectively.
In both cases the hierarchy of the channels is the same as compared to the $\lambda_\ell$-odd channel decomposition, exception made for a few sea quark and gluonic channels.
As already observed for the $z$-distributions, the $d\to d$ channel is about a factor of five larger in the $\lambda_\ell$-even case with respect to the $\lambda_\ell$-odd one, for both $\pi^+$ and $\pi^-$.
This is again due to the electroweak couplings being of the same absolute magnitude for up and down quarks.
$\dd \Delta\sigma^{\pi^-,\NC}_{\meven}/\dd x$ is subject to large cancellation and is negative in both Low-$Q^2$ and Mid-$Q^2$ ranges where crosses zero at different values of $x$.
This results in a change of sign in the channel decomposition due to the normalisation with respect to $\dd \Delta\sigma^{\pi^-,\NC}_{\meven}/\dd x$.

\begin{figure}[p]
\centering
\begin{subfigure}{0.45\textwidth}
\centering
\includegraphics[width=1.0\linewidth]{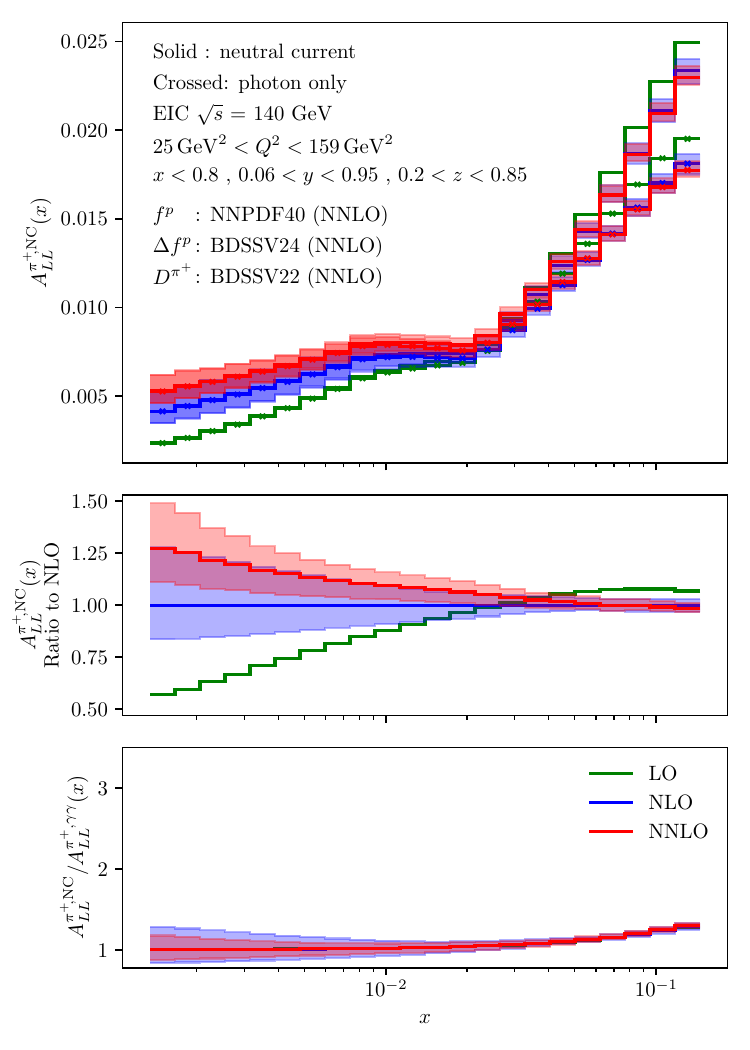}
\caption{Low-$Q^2$}
\end{subfigure}
\begin{subfigure}{0.45\textwidth}
\centering
\includegraphics[width=1.0\linewidth]{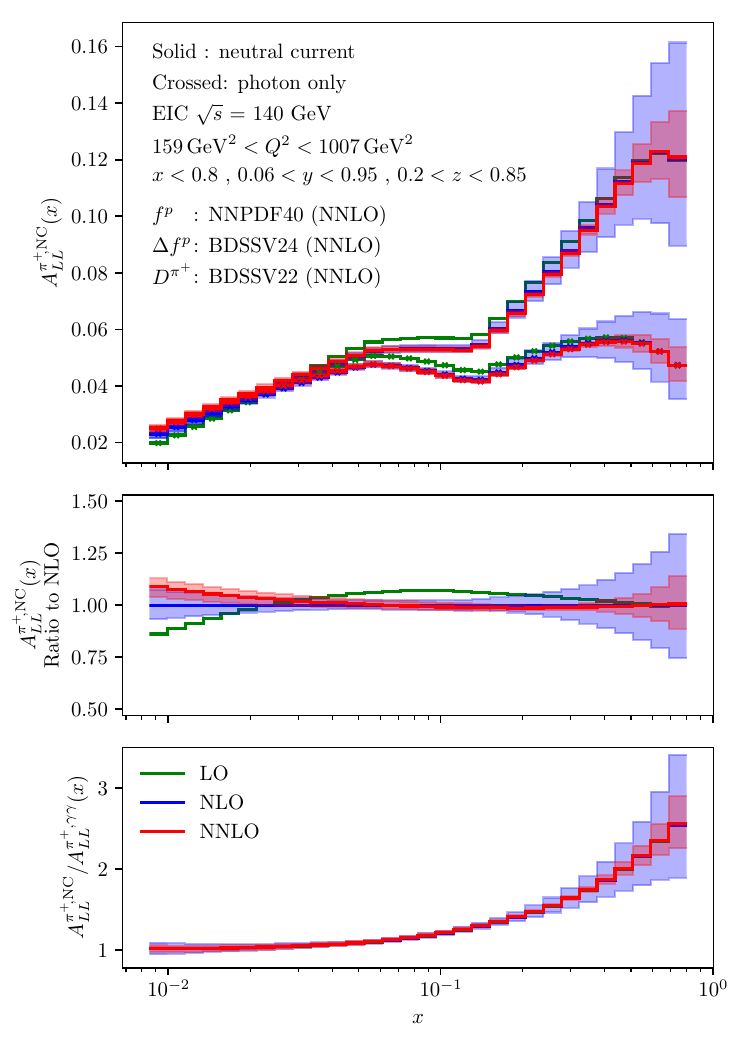}
\caption{Mid-$Q^2$}
\end{subfigure}
\begin{subfigure}{0.45\textwidth}
\centering
\includegraphics[width=1.0\linewidth]{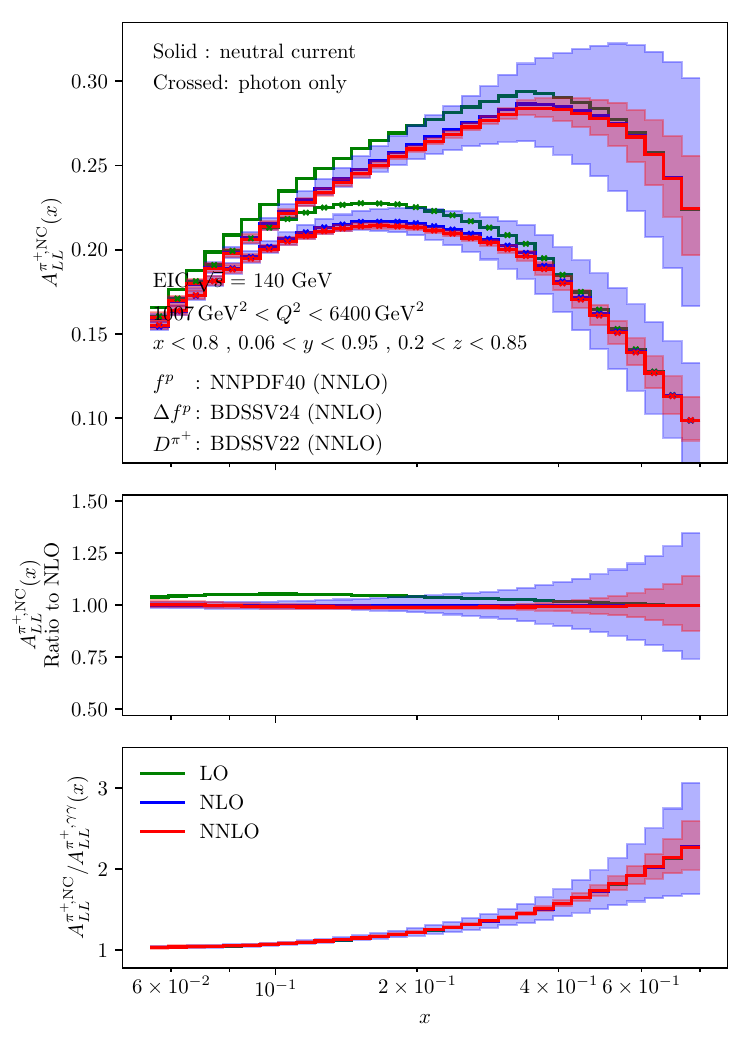}
\caption{High-$Q^2$}
\end{subfigure}
\begin{subfigure}{0.45\textwidth}
\centering
\includegraphics[width=1.0\linewidth]{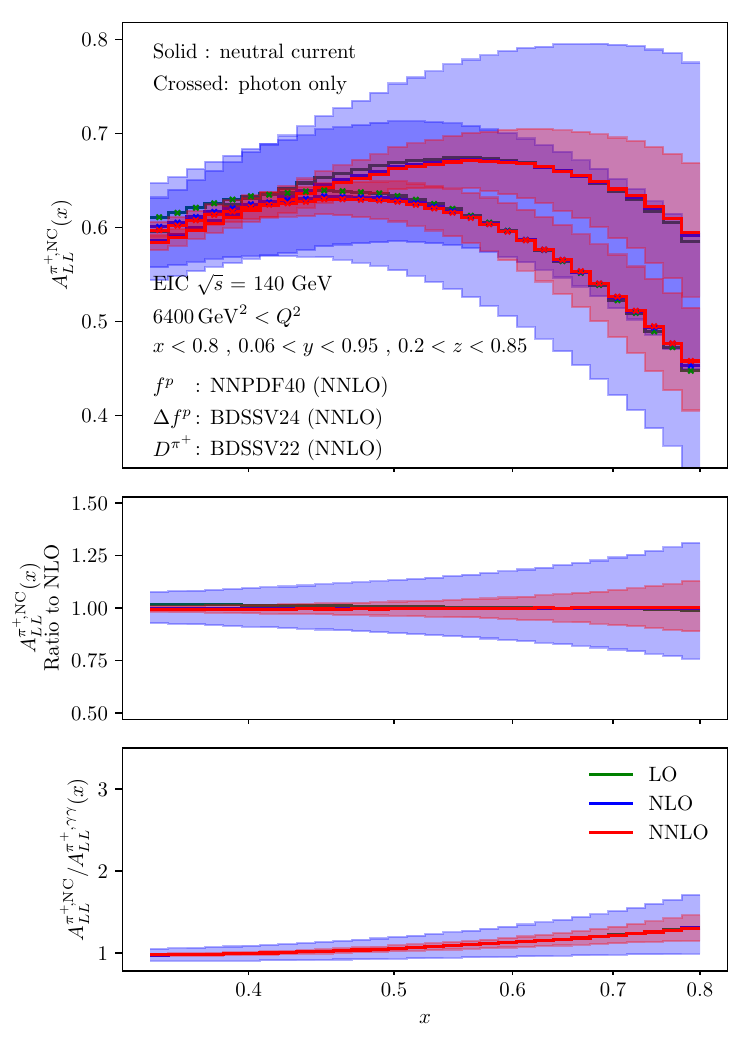}
\caption{Extreme-$Q^2$}
\end{subfigure}

\caption{
Double-longitudinal spin asymmetry $A_{LL}^{\pi^+,\NC}$ in NC exchange as a function of $x$.
Top panels: Asymmetries for full NC (solid) and photon-only (crossed) exchanges.
Centre panels: NC Ratio to NLO.
Bottom panels: NC over photon-only exchange ratios.
}
\label{fig:X_A_LL_NC_pip}

\end{figure}

\begin{figure}[p]
\centering
\begin{subfigure}{0.45\textwidth}
\centering
\includegraphics[width=1.0\linewidth]{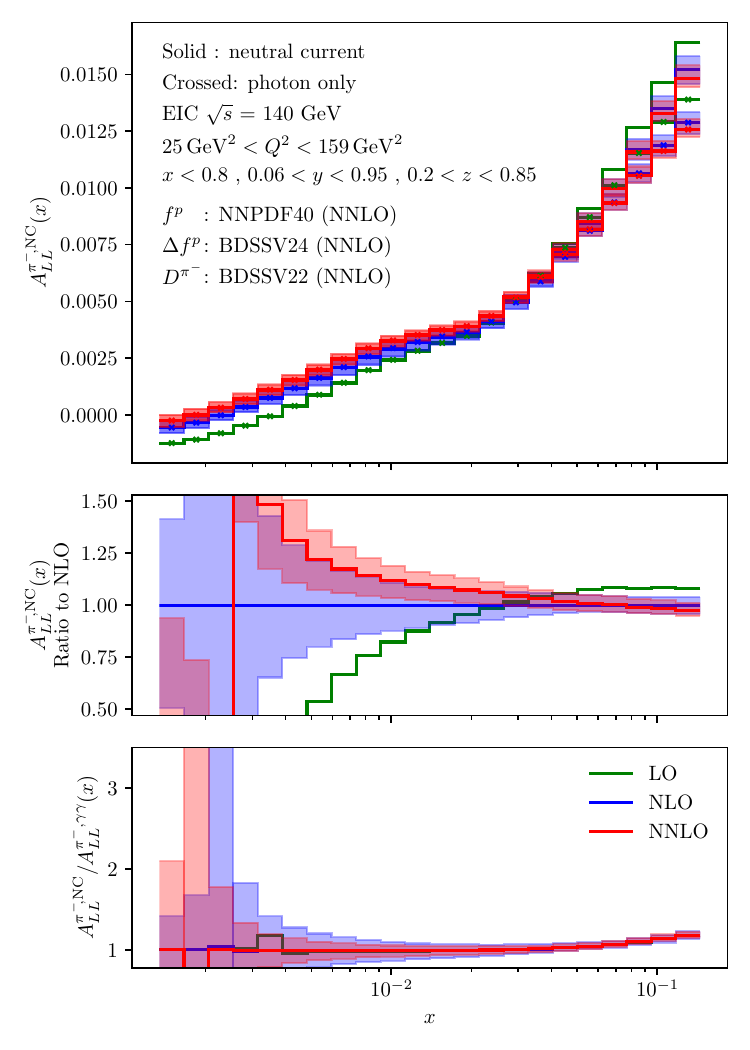}
\caption{Low-$Q^2$}
\end{subfigure}
\begin{subfigure}{0.45\textwidth}
\centering
\includegraphics[width=1.0\linewidth]{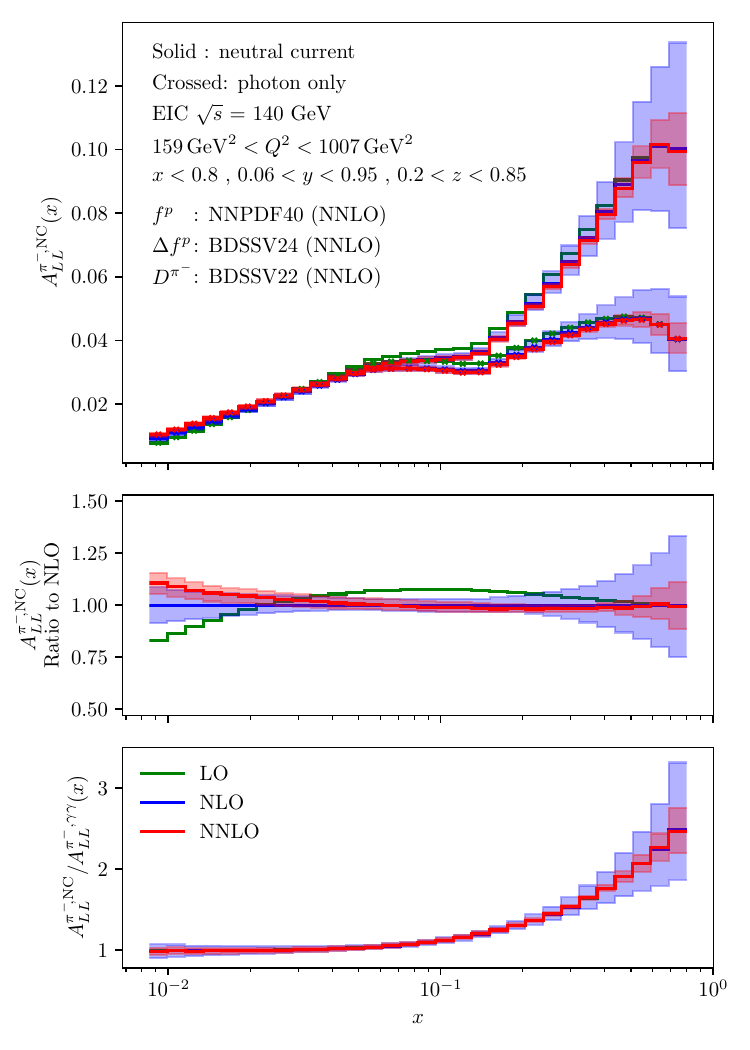}
\caption{Mid-$Q^2$}
\end{subfigure}
\begin{subfigure}{0.45\textwidth}
\centering
\includegraphics[width=1.0\linewidth]{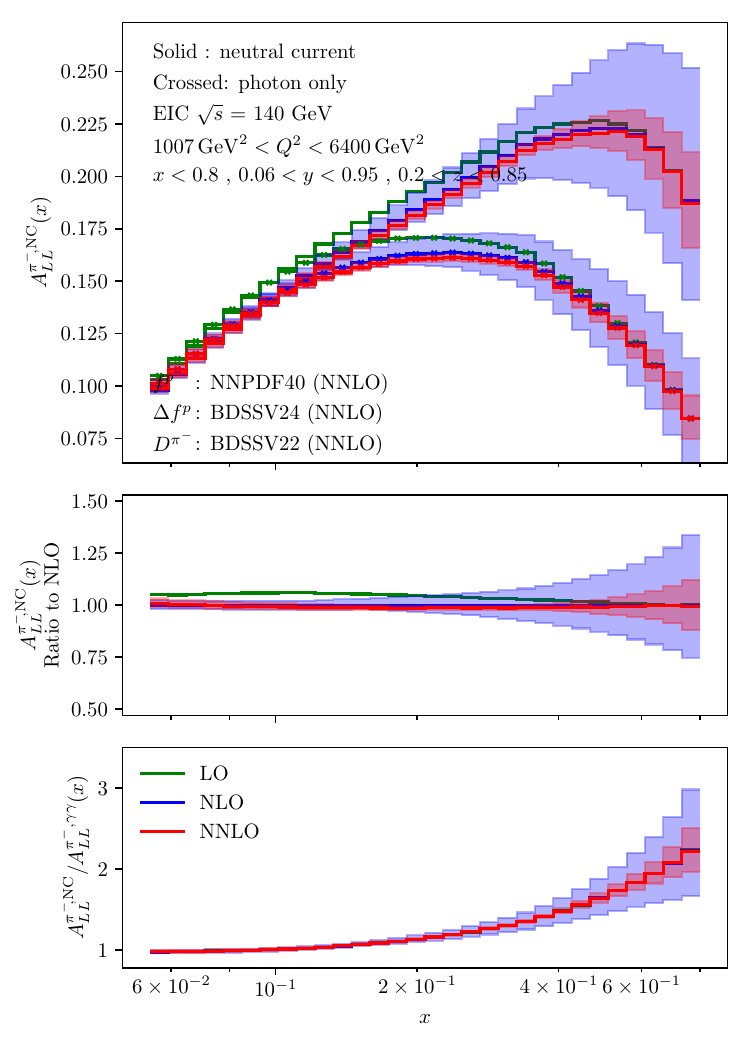}
\caption{High-$Q^2$}
\end{subfigure}
\begin{subfigure}{0.45\textwidth}
\centering
\includegraphics[width=1.0\linewidth]{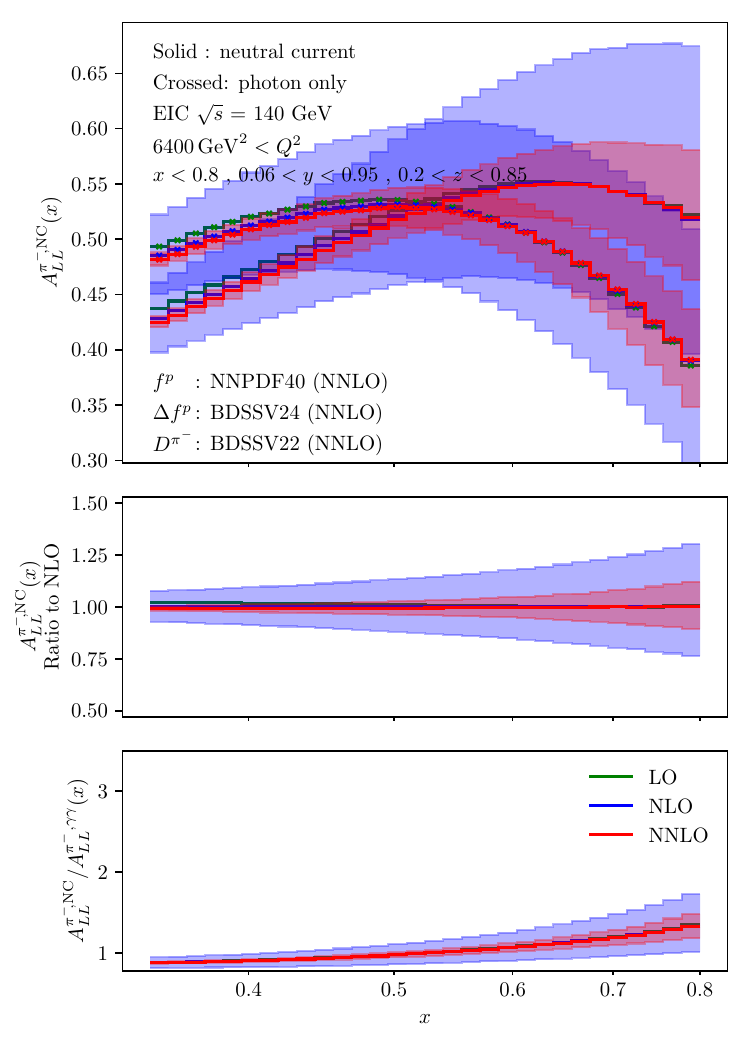}
\caption{Extreme-$Q^2$}
\end{subfigure}

\caption{
Double-longitudinal spin asymmetry $A_{LL}^{\pi^-,\NC}$ in NC exchange as a function of $x$.
Top panels: Asymmetries for full NC (solid) and photon-only (crossed) exchanges.
Centre panels: NC Ratio to NLO.
Bottom panels: NC over photon-only exchange ratios.
}
\label{fig:X_A_LL_NC_pim}

\end{figure}

In Fig.~\ref{fig:X_A_LL_NC_pip} we study the  double-longitudinal NC asymmetry~\eqref{eq:AhLLNC}
 for $\pi^+$ production as a function of $x$ in
 the four ranges of $Q^2$.
The average value of the asymmetry grows with increasing  average value of $Q^2$, from $0.015$ up to $0.65$.
The asymmetry also grows with $x$ up to a maximum that depends on the range of $Q^2$ probed.
The decrease observed for $x>0.4$ is due to the depletion of polarized PDFs at high $x$.
At Low-$Q^2$ and Mid-$Q^2$ the flat region appearing at moderate values of $x$ is due to the cut on $y$ restricting the available phase space.
In the top panels the asymmetry is shown for both NC (solid) and photon-only (crossed) exchanges.
The NC asymmetry is larger than the photon-only exchange one with the exception of a few bins at Extreme-$Q^2$ and low $x$.
The size of the electroweak effects is visible in the ratio $(A^{\pi^+,\NC}_{LL}/A^{\pi^+,\gamma\gamma}_{LL})(x)$ in the bottom panels.
As already observed as function of $z$ or $Q^2$, the electroweak effects grow with the increase of $Q^2$ up to the High-$Q^2$ region to then reduce again in size.
Since $Q^2$ and $x$ are not independent variables, the electroweak effects also grow significantly with
increasing $x$.
At Low-$Q^2$ they grow from sub-percent level up to almost $30\%$.
At Mid- and High-$Q^2$ they grow from percent level to a factor of $3$ the size of the photon-only contribution.
At Extreme-$Q^2$ they grow again only up to $50\%$.
Comparing the asymmetry as a function of $x$ with the asymmetry as a function of $Q^2$ or $z$ (Fig.~\ref{fig:Q2_A_LL_NC_pi} or Fig.~\ref{fig:Z_A_LL_NC_pip}) we observe that electroweak effects are much larger for $x$ distributions.
This is consistent with previous observations at the level of the cross section differences and is due to a non trivial $x$ dependence carried by the large transverse structure function.
Perturbative corrections are very large at Low-$Q^2$, where we observe almost $50\%$ NLO corrections and up to $25\%$ NNLO ones.
They reduce in size with increasing $x$ and average value of $Q^2$, decreasing from $10\%$ level to sub-percent level.
This hints towards a good perturbative convergence and stability of the QCD series expansion, as shown in the centre panels for full NC exchange.
The NNLO predictions typically lie within the NLO theory uncertainty band.

In Fig.~\ref{fig:X_A_LL_NC_pim} we study the double-longitudinal NC asymmetry~\eqref{eq:AhLLNC}
 for $\pi^-$ production as a function of $x$ in the same ranges of $Q^2$.
The overall behaviour in $x$ is similar to $\pi^+$ production, due to the cut on $y$ and the behaviour of the polarized PDFs.
The asymmetry is smaller in size of about $20\%$ due to the flavour-unfavoured transition being probed.
The electroweak effects are of very similar size and shape in the $\pi^+$ and $\pi^-$ processes.
The only visible differences appear in the Extreme-$Q^2$ region where $A^{\pi^-,\gamma\gamma}_{LL}$ stays larger than $A^{\pi^-,\NC}_{LL}$ up to $x\sim 0.5$.
Also the behaviour of QCD corrections is similar between $\pi^+$ and $\pi^-$ production.
NLO and NNLO corrections are very large, more than $20\%$, at Low-$Q^2$ and small $x$.
They reduce in size with increasing $x$ and $Q^2$ being at percent level at Mid-$Q^2$ and High-$Q^2$ and at sub percent level at Extreme-$Q^2$.
We observe good perturbative convergence and stability throughout.

\begin{figure}[p]
\centering
\begin{subfigure}{0.45\textwidth}
\centering
\includegraphics[width=1.0\linewidth]{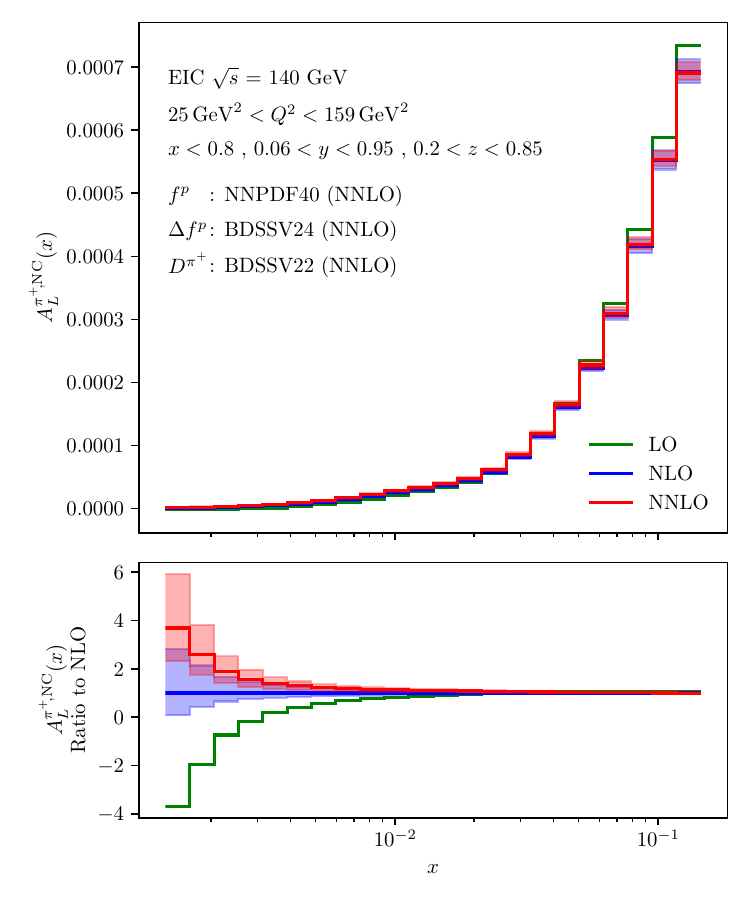}
\caption{Low-$Q^2$}
\end{subfigure}
\begin{subfigure}{0.45\textwidth}
\centering
\includegraphics[width=1.0\linewidth]{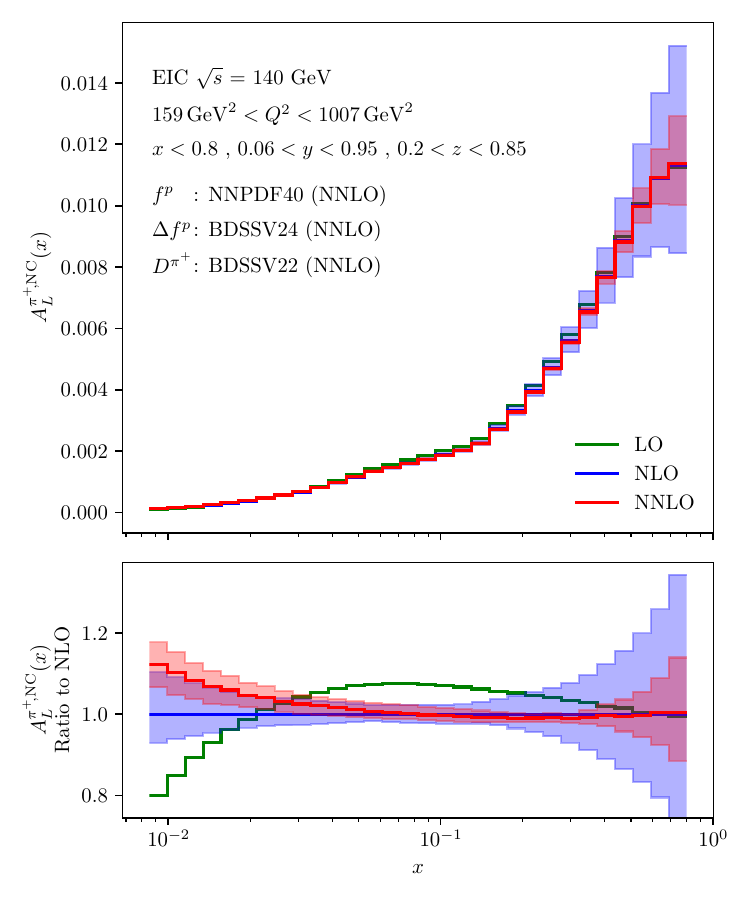}
\caption{Mid-$Q^2$}
\end{subfigure}
\begin{subfigure}{0.45\textwidth}
\centering
\includegraphics[width=1.0\linewidth]{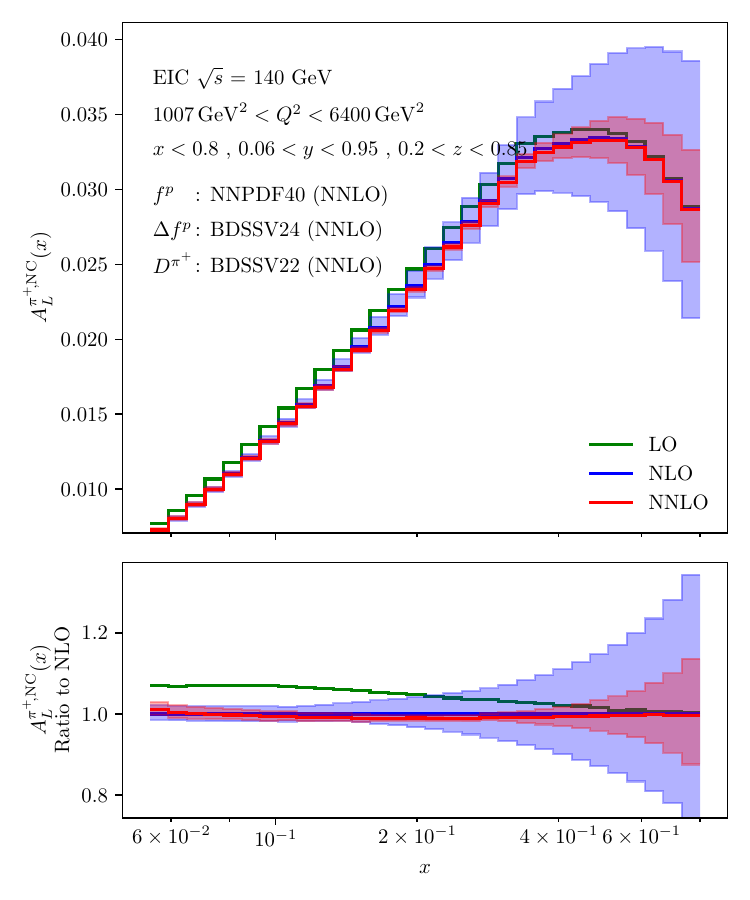}
\caption{High-$Q^2$}
\end{subfigure}
\begin{subfigure}{0.45\textwidth}
\centering
\includegraphics[width=1.0\linewidth]{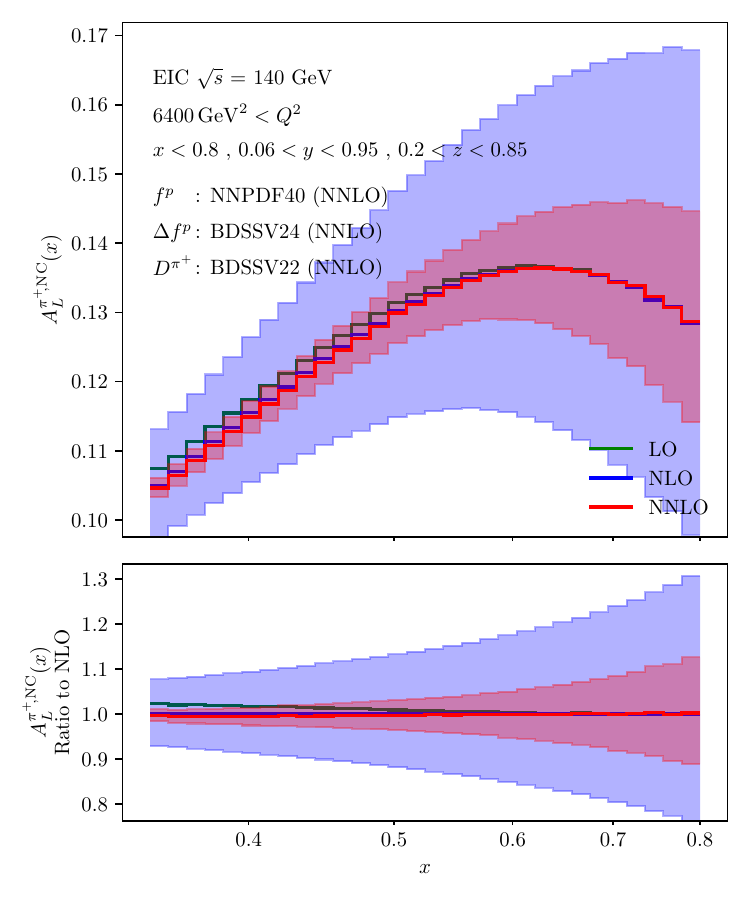}
\caption{Extreme-$Q^2$}
\end{subfigure}

\caption{
Single-longitudinal spin asymmetry $A_{L}^{\pi^+,\NC}$ in NC exchange as a function of $x$.
Top panels: Asymmetries.
Bottom panels: Ratio to NLO.
}
\label{fig:X_A_L_pip}

\end{figure}

\begin{figure}[p]
\centering
\begin{subfigure}{0.45\textwidth}
\centering
\includegraphics[width=1.0\linewidth]{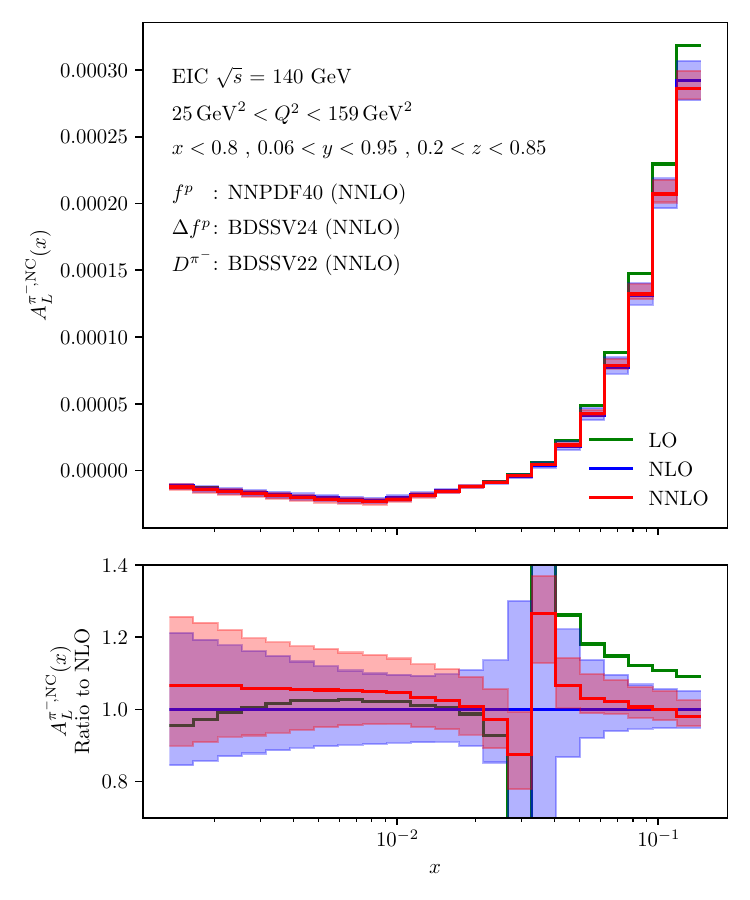}\caption{Low-$Q^2$}
\end{subfigure}
\begin{subfigure}{0.45\textwidth}
\centering
\includegraphics[width=1.0\linewidth]{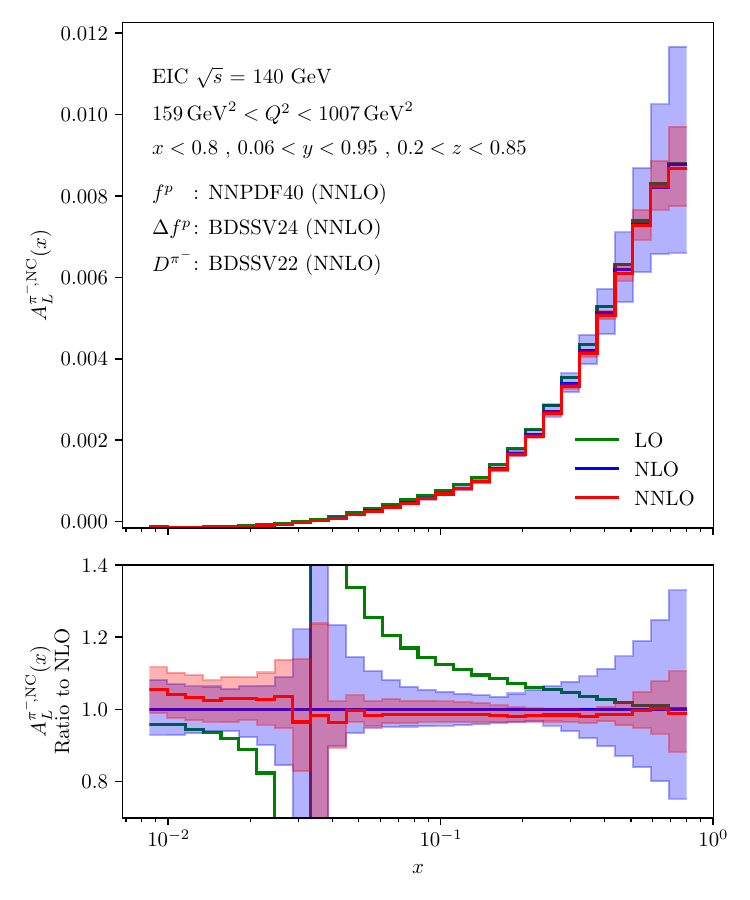}
\caption{Mid-$Q^2$}
\end{subfigure}
\begin{subfigure}{0.45\textwidth}
\centering
\includegraphics[width=1.0\linewidth]{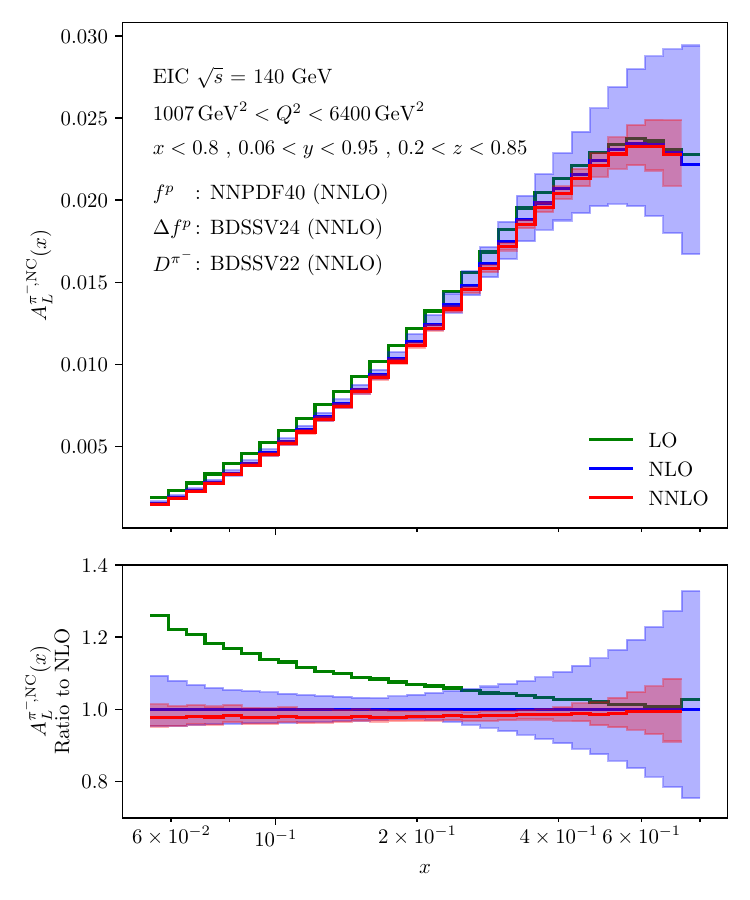}
\caption{High-$Q^2$}
\end{subfigure}
\begin{subfigure}{0.45\textwidth}
\centering
\includegraphics[width=1.0\linewidth]{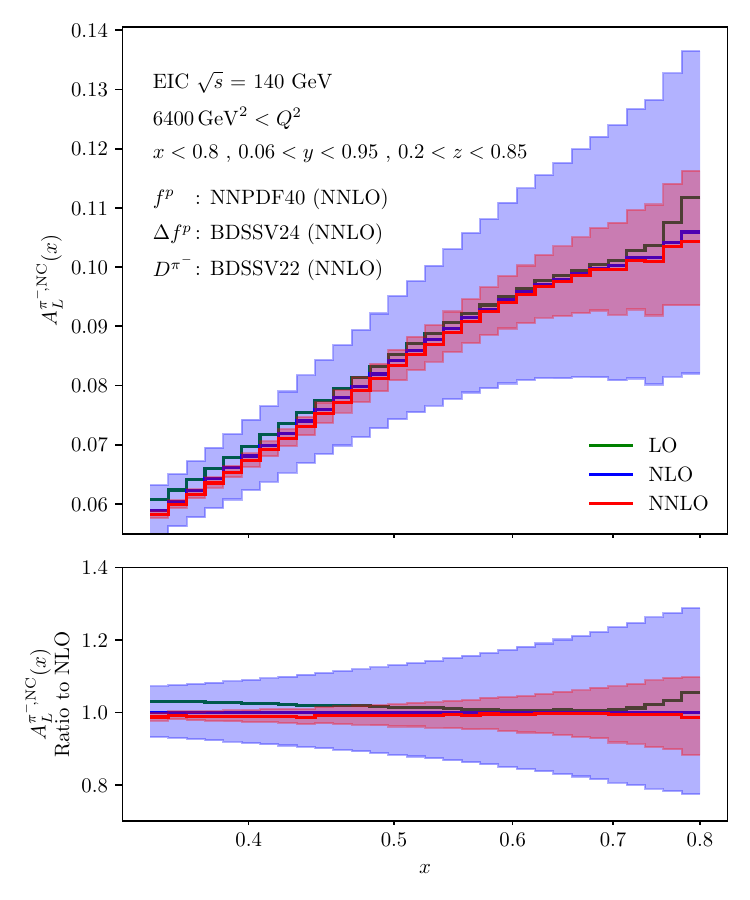}
\caption{Extreme-$Q^2$}
\end{subfigure}

\caption{
Single-longitudinal spin asymmetry $A_{L}^{\pi^-,\NC}$ in NC exchange as a function of $x$.
Top panels: Asymmetries.
Bottom panels: Ratio to NLO.
}
\label{fig:X_A_L_pim}

\end{figure}

In Fig.~\ref{fig:X_A_L_pip} we study the single-longitudinal spin asymmetry~\eqref{eq:AhLNC} for $\pi^+$ production as a function of $x$, employing the kinematics cuts~\eqref{eq:cutsxyz} and~\eqref{eq:cutsQ2}.
For Low-$Q^2$ and Mid-$Q^2$ the asymmetry grows with increasing $x$ reaching percent level already at Mid-$Q^2$ for $x>0.3$.
At High-$Q^2$ and Extreme-$Q^2$ the asymmetry grows up to a maximum to then decrease at the largest values of $x$ probed. Its maximum value is $0.032$ for High-$Q^2$ and  $0.135$ for Extreme-$Q^2$.
Perturbative convergence and stability increases with the increase of $Q^2$.
At Mid-$Q^2$ NLO  and NNLO corrections are well above percent level.
They decrease in relative size with the increase of $Q^2$ to reach few percent level at High-$Q^2$ and sub-percent level at Extreme-$Q^2$.

In Fig.~\ref{fig:X_A_L_pim} we study the single-longitudinal spin asymmetry~\eqref{eq:AhLNC} for $\pi^-$ production as a function of $x$.
We observe an similar overall behaviour as compared to $\pi^+$ production with the asymmetries being about $20\%$ smaller in all $Q^2$ ranges.
We observe good perturbative convergence and stability, with NLO and NNLO corrections reducing in size increasing the average value of $Q^2$ probed.
The single-longitudinal asymmetries as function of $x$ are of comparable size to the asymmetries in $Q^2$ while being significantly larger than the ones as a function of $z$, for  both $\pi^+$ and $\pi^-$.

\subsection{Charged Current}

In this Section we present phenomenological predictions for polarized charged current cross section asymmetries, both for $\pi^+$ and $\pi^-$ production.
We consider electron-initiated processes and consequently only processes with a $W^-$ exchange from the lepton to the quark line.
We apply the kinematical cuts of~\eqref{eq:cutsxyz} and~\eqref{eq:cutsQ2}  and show the results differential in the variables $Q^2$, $z$ or $x$.
We present results for the cross section difference~\eqref{eq:SFNCpol} as well as for the \textit{longitudinal charged-neutral current asymmetry} defined as
\begin{equation}\label{eq:ALCCNC}
A^{h,L}_{\CC/\NC} = \frac{\dd\Delta \sigma^{h,\CC}_{\meven}}{\dd\Delta \sigma^{h,\NC}_{\meven}} \, .
\end{equation}
The $\lambda_\ell$-even component of the CC cross section difference is defined as
\begin{equation}
\dd\Delta \sigma^{h,\CC}_{\meven}=\dd\Delta \sigma^{h,\CC}(\lambda_\ell=+1)+\dd \Delta\sigma^{h,\CC}(\lambda_\ell=-1)=
    \begin {cases}
         +\dd\Delta \sigma^{h,\CC}_{\modd}  & \text{for } e^{+} \,, \\
         -\dd\Delta \sigma^{h,\CC}_{\modd}  & \text{for } e^{-} \, .
    \end{cases}
\end{equation}
By projecting on the $\lambda_\ell$-even component, the dominant photon exchange from the
denominator of~\eqref{eq:ALCCNC} is removed, thereby enhancing the electroweak contributions.
$A^{h,L}_{\CC/\NC}$ is well suited for experimental measurements due to the cancellation of systematic uncertainties in both numerator and denominator, analogously to what has been discussed for the unpolarized $A^{h}_{\CC/\NC}$ asymmetry~\cite{Bonino:2025qta}.
Theory uncertainties in $A^{h,L}_{\CC/\NC}$ are obtained with a 31-point uncorrelated scale variation.

We also study the electron-initiated \textit{double-longitudinal spin asymmetry} for charged current exchange
\begin{equation}\label{eq:AhLLCC}
A^{h,\CC}_{LL}=\frac{1}{2}\frac{\dd\Delta\sigma^{h,\CC}_{\modd}}{\dd\sigma^{h,\CC}_{\meven}} \,  .
\end{equation}
Both numerator and denominator are obtained by setting $\lambda_\ell=-1$ in~\eqref{eq:SFNCpol}
and~\eqref{eq:xsNCCC} respectively.
The CC double-longitudinal spin asymmetry is useful to understand the relative size of CC events in polarized with respect to unpolarized collisions.

\subsubsection[\texorpdfstring{$Q^2$-distributions}{Q\^{}2-distributions}]{\boldmath $Q^2$-distributions}

\begin{figure}[tb]
\centering
\begin{subfigure}{.47\textwidth}
\centering
\includegraphics[width=\textwidth]{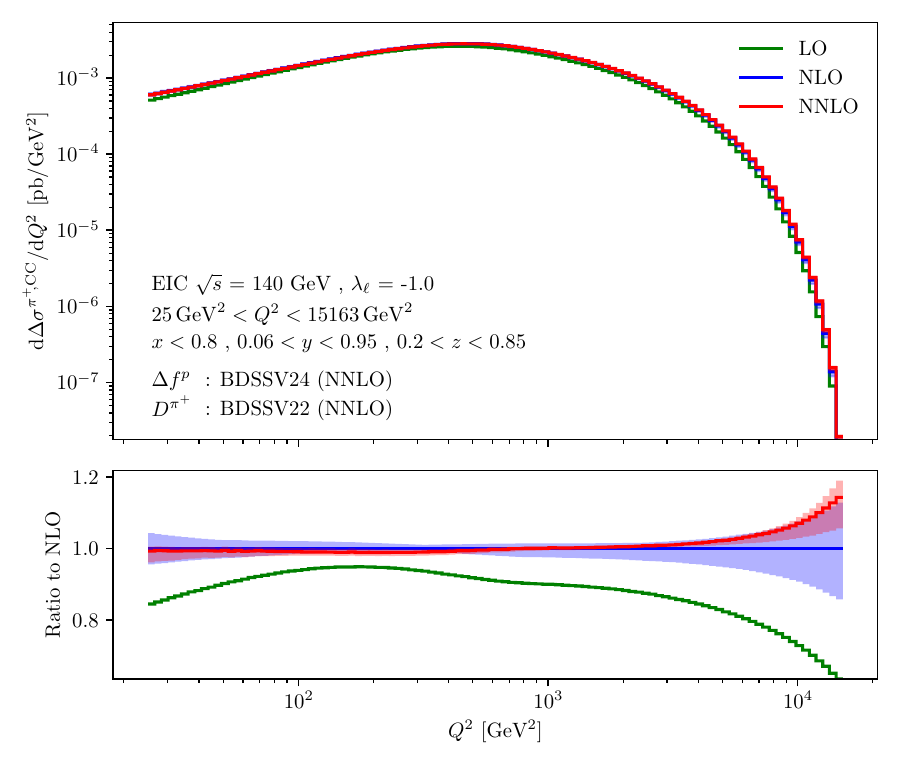}
\end{subfigure}
\begin{subfigure}{.47\textwidth}
\centering
\includegraphics[width=\textwidth]{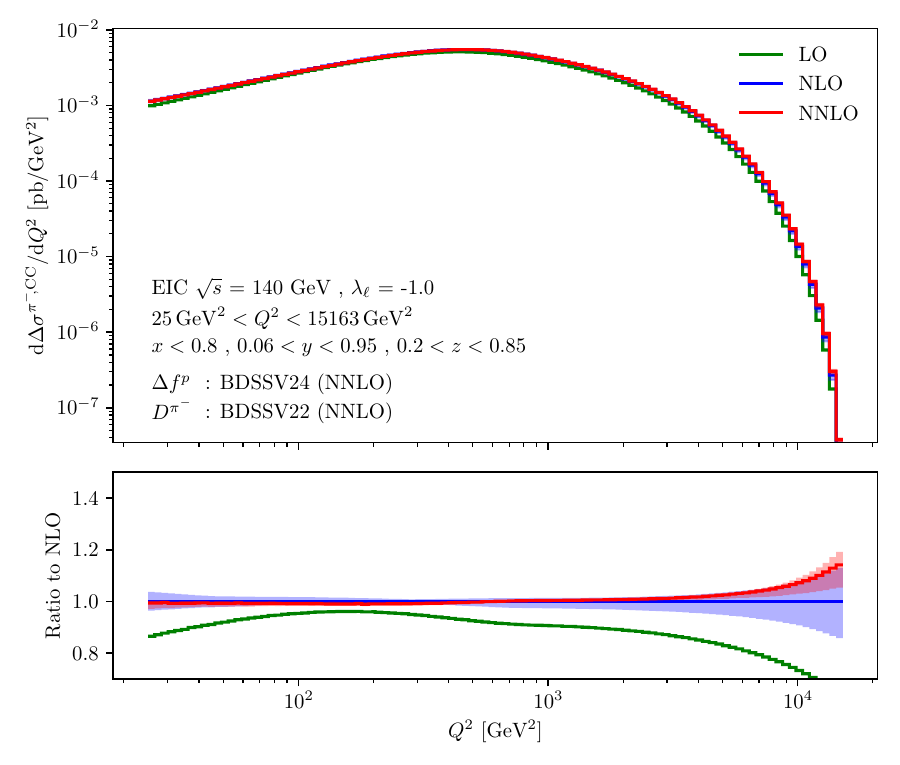}
\end{subfigure}
\caption{
Electron-initiated CC $Q^2$-distribution for $\pi^+$ and $\pi^-$ production.
Top panels: total cross section difference.
Bottom panels: Ratio to NLO.
}
\label{fig:Q2_CH}
\end{figure}

In Fig.~\ref{fig:Q2_CH} we show the $Q^2$-differential electron-initiated CC cross section difference~\eqref{eq:SFNCpol} for $\pi^+$ and $\pi^-$ production.
For both pion charges the cross section difference increases in absolute value with $Q^2$ peaking at $Q^2\sim 400 \, \GeV^2$.
It then sharply decreases due to the polarized PDFs depletion at the large values of $x$ probed.
The behaviour in $Q^2$ is very similar between $\pi^+$ (flavour-unfavoured) and $\pi^-$ (flavour-favoured) with the latter being approximately $60\,\%$ larger.
The enlargement in $\pi^-$ production in $W^-$ exchange is due to the size of the $\Delta f^p_u$ distribution and its LO transition into a $d$ quark due to the CC interaction.
The resulting dominant partonic channel is $u\to d$, which has valence content of both initial and the final state identified hadrons (see Fig.~\ref{fig:z_CH_channels} below).
This feature and the overall shape of the distribution is very similar between polarized and unpolarized SIDIS (see Fig.~25 in \cite{Bonino:2025qta}).
$\dd\Delta\sigma^{h,\CC}/\dd Q^2$ is approximately an order of magnitude smaller than $\dd\sigma^{h,\CC}/\dd Q^2$ and presents a depletion at small $Q^2$ due to the depletion of polarized PDFs at small $x$.
We observe good perturbative convergence and stability for both $\pi^+$ and $\pi^-$, as can be observed in the bottom panels of Fig.~\ref{fig:Q2_CH} from a ratio to NLO.
NLO corrections are at the level of $10\%$, with NNLO corrections at a few per-cent level increasing with $Q^2$.
The NNLO corrections lie within the NLO theory uncertainty band in all $Q^2$ ranges.

\begin{figure}[tb]
\centering
\begin{subfigure}{.47\textwidth}
\centering
\includegraphics[width=\textwidth]{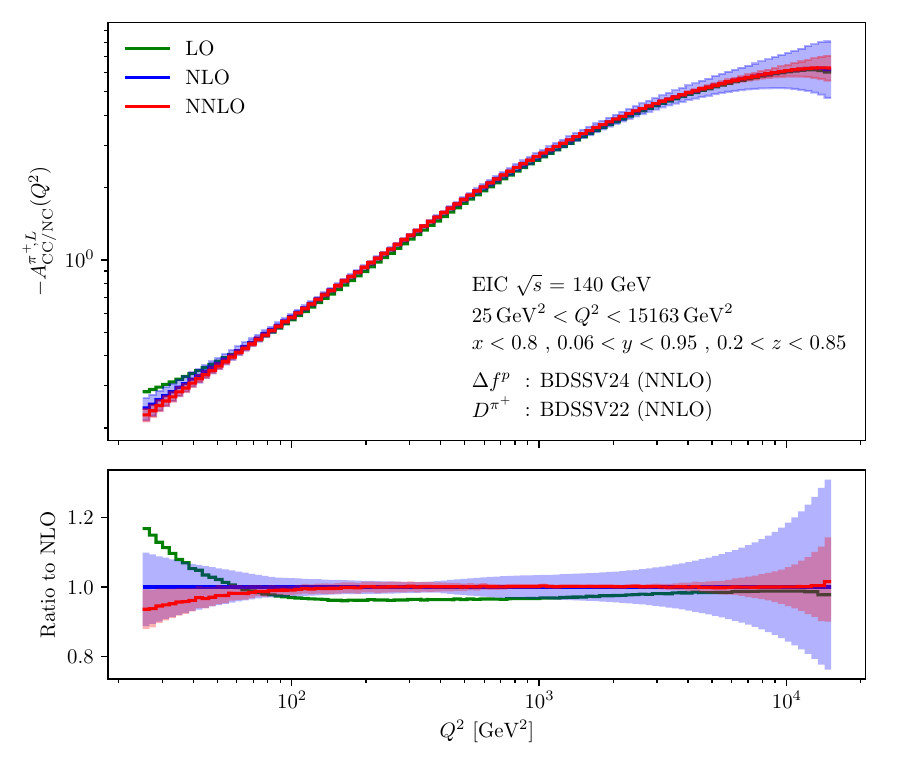}
\end{subfigure}
\begin{subfigure}{.47\textwidth}
\centering
\includegraphics[width=\textwidth]{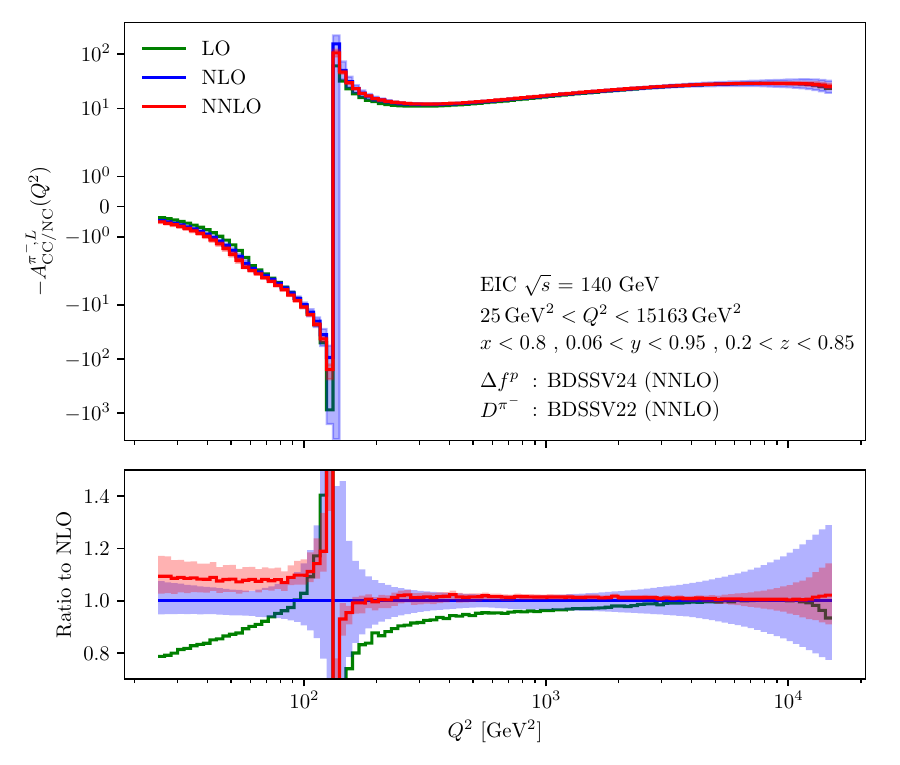}
\end{subfigure}

\caption{
Electron-initiated longitudinal charged-neutral current asymmetry $A^{\pi^\pm,L}_\mathrm{CC/NC}$ as a function of $Q^2$ in $\pi^+$ and $\pi^-$ production.
Top panels: total asymmetry.
Bottom panels: Ratio to NLO.
}
\label{fig:Q2_A_CCNC}
\end{figure}

In Fig.~\ref{fig:Q2_A_CCNC} we show the electron-initiated longitudinal charged-neutral current asymmetry of~\eqref{eq:ALCCNC} as a function of $Q^2$ in the entire kinematic range.
For $\pi^+$ (left) the absolute value of the asymmetry grows with $Q^2$. The CC contribution quickly surpasses in size the NC $\lambda_\ell$-even contribution already at $Q^2\sim 200\, \GeV^2$.
$A^{\pi^+,L}_{\CC/\NC}$ is approximately one order of magnitude larger than its unpolarized counterpart $A^{\pi^+}_{\CC/\NC}$ (see Fig.~26 in \cite{Bonino:2025qta}).
Due to the interplay of partonic channels, $A^{\pi^-,L}_{\CC/\NC}$ presents a pole at $Q^2\sim 150 \, \GeV^2$ where the denominator crosses zero.
For $Q^2>300\,\GeV^2$ the asymmetry  $A^{\pi^-,L}_{\CC/\NC}$ is very large, with the CC contribution growing to more than a factor of 10 in size with respect to the NC $\lambda_\ell$-even contribution.
Also for $\pi^-$ we observe a factor of 10 larger asymmetry in polarized SIDIS with respect to unpolarized SIDIS.
As can be seen from the bottom panels in Fig.~\ref{fig:Q2_A_CCNC}, perturbative corrections are at few percent level at most.
From NLO level we observe good perturbative convergence and stability with percent level NNLO corrections at small $Q^2$.

\begin{figure}[tb]
\centering
\begin{subfigure}{.45\textwidth}
\centering
\includegraphics[width=\textwidth]{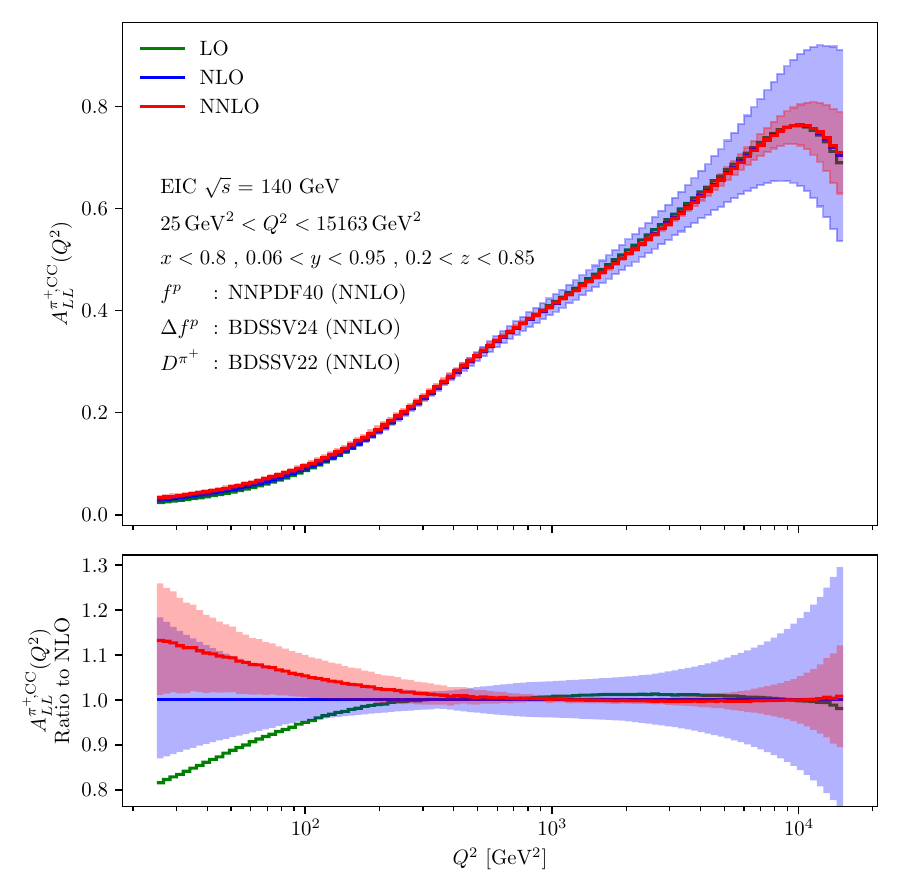}
\end{subfigure}
\begin{subfigure}{.45\textwidth}
\centering
\includegraphics[width=\textwidth]{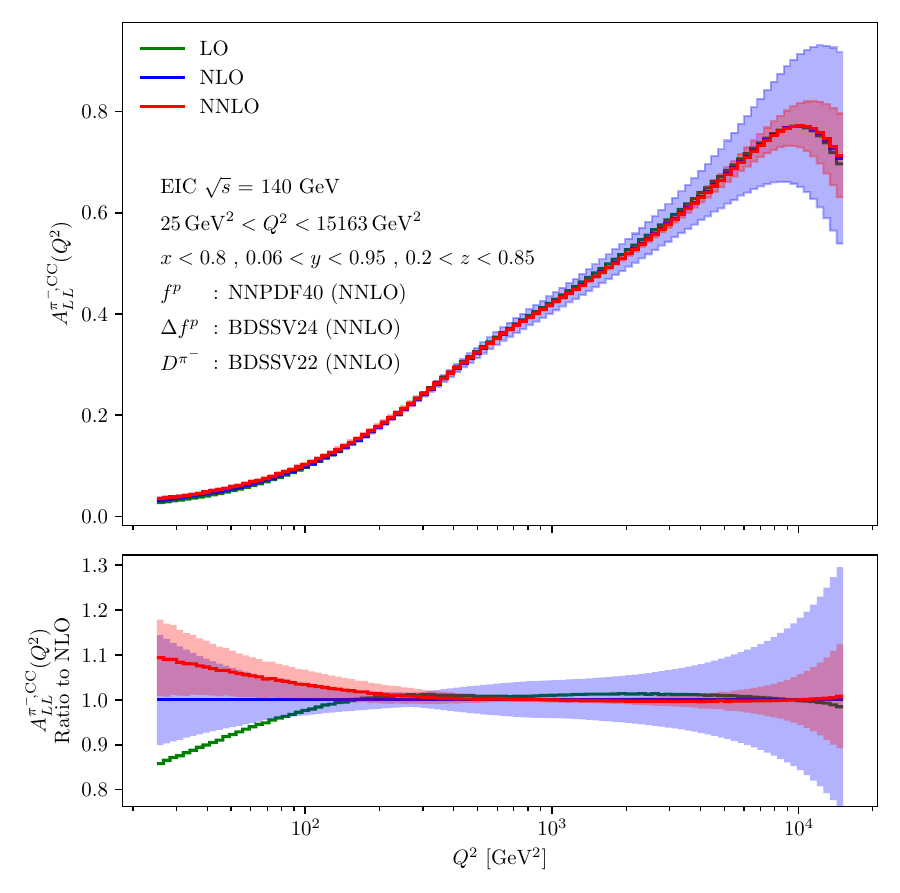}
\end{subfigure}

\caption{
Electron-initiated double-longitudinal charged current asymmetry $A^{\pi^\pm,\CC}_{LL}$ as a function of $Q^2$ in $\pi^+$ and $\pi^-$ production.
Top panels: total asymmetry.
Bottom panels: Ratio to NLO.
}
\label{fig:Q2_A_LL_CC}
\end{figure}

In Fig.~\ref{fig:Q2_A_LL_CC} we show the CC double-longitudinal spin asymmetry~\eqref{eq:AhLLCC} for $\pi^+$ (left) and $\pi^-$ production (right) as a function of $Q^2$.
The asymmetry is as large as the NC double-longitudinal spin asymmetry of Fig.~\ref{fig:Q2_A_LL_NC_pi} for both identified hadron charges.
It grows with increasing $Q^2$ reaching a maximum of $0.8$ at $Q^2\sim 9000 \, \GeV^2$.
The polarized CC cross section differences are of the same order of magnitude as the unpolarized CC cross sections for most of the values of virtuality probed at the EIC.
The shape and size of the asymmetry is very similar between $\pi^+$ and $\pi^-$ production.
For both $\pi^+$ and $\pi^-$ production we observe good perturbative convergence and stability.
Perturbative corrections are at $10\%$ level in the Low-$Q^2$ region and reduce to few percent level already in the Mid-$Q^2$ region.
For $Q^2>400\,\GeV^2$ perturbative corrections are at sub percent level.

\subsubsection[\texorpdfstring{$z$-distributions}{z-distributions}]{\boldmath $z$-distributions}

\begin{figure}[tb]
\centering

\begin{subfigure}{0.45\textwidth}
\centering
\includegraphics[width=1.0\linewidth]{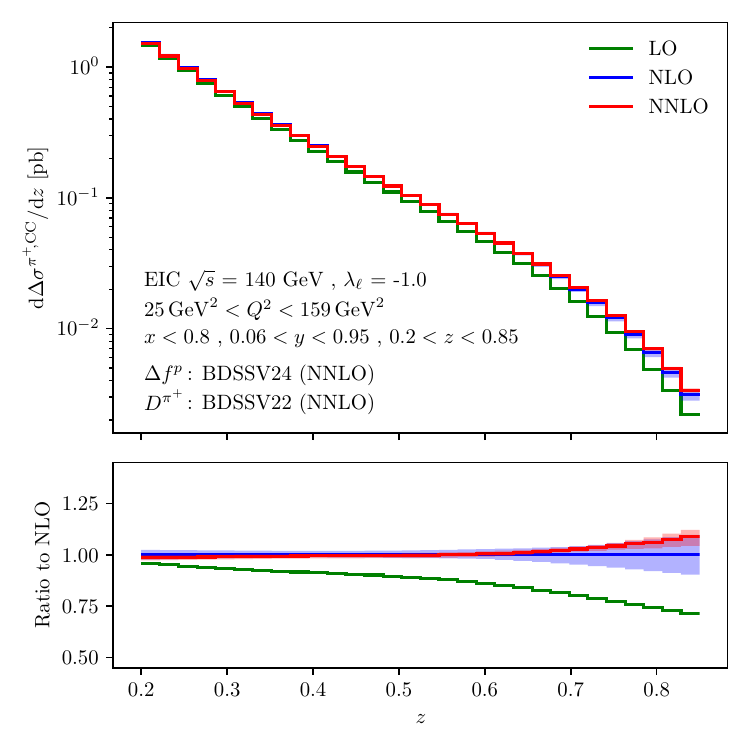}
\caption{Low-$Q^2$}
\end{subfigure}
\begin{subfigure}{0.45\textwidth}
\centering
\includegraphics[width=1.0\linewidth]{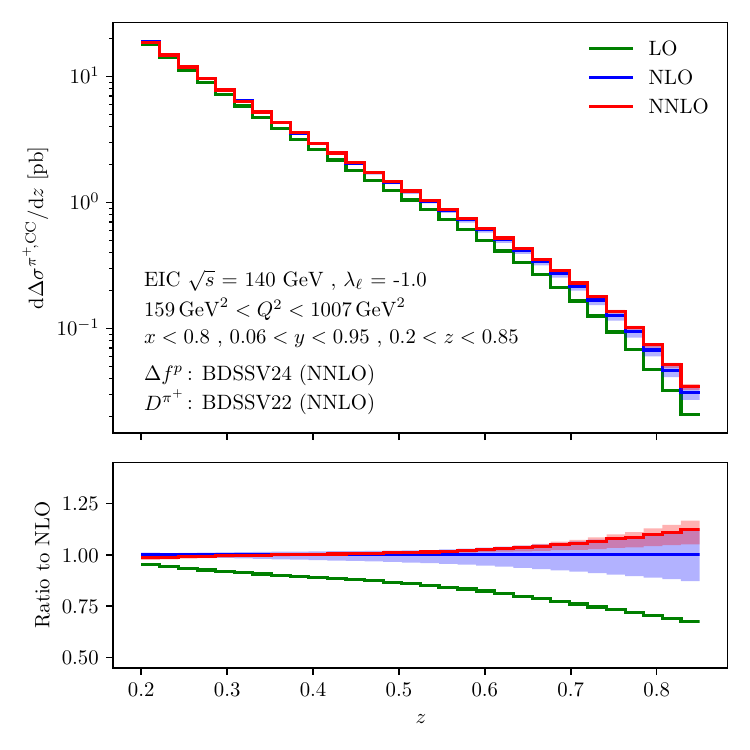}
\caption{Mid-$Q^2$}
\end{subfigure}
\begin{subfigure}{0.45\textwidth}
\centering
\includegraphics[width=1.0\linewidth]{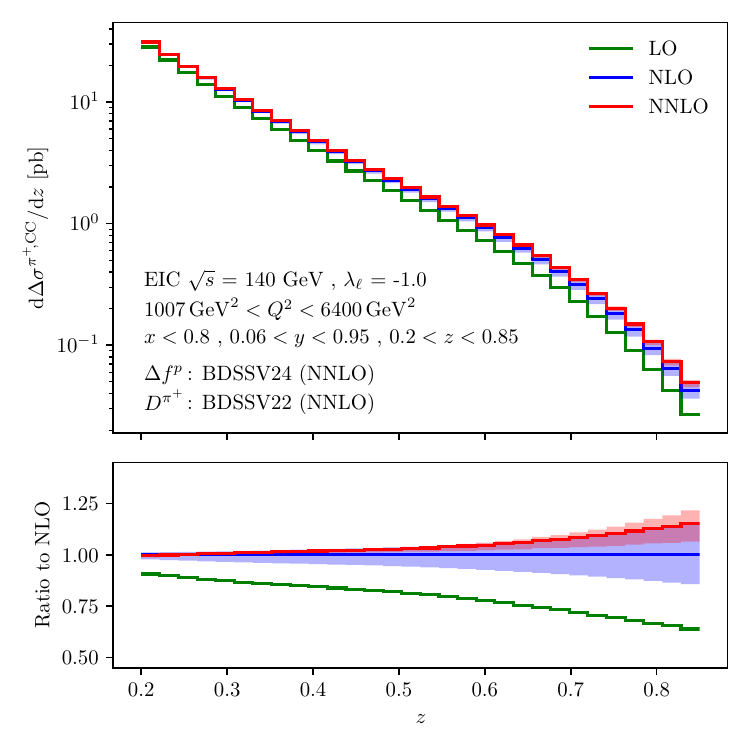}
\caption{High-$Q^2$}
\end{subfigure}
\begin{subfigure}{0.45\textwidth}
\centering
\includegraphics[width=1.0\linewidth]{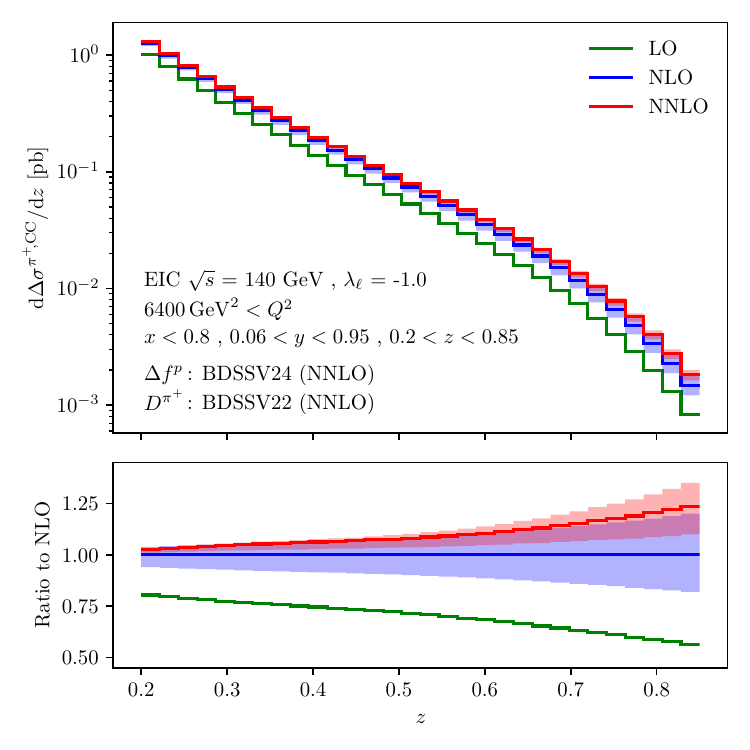}
\caption{Extreme-$Q^2$}
\end{subfigure}

\caption{
Electron-initiated CC $z$-distributions for $\pi^+$ production (unfavoured).
Top panels: total cross section difference.
Bottom panels: Ratio to NLO.
}
\label{fig:Z_CH_pip}

\end{figure}

\begin{figure}[tb]
\centering

\begin{subfigure}{0.45\textwidth}
\centering
\includegraphics[width=1.0\linewidth]{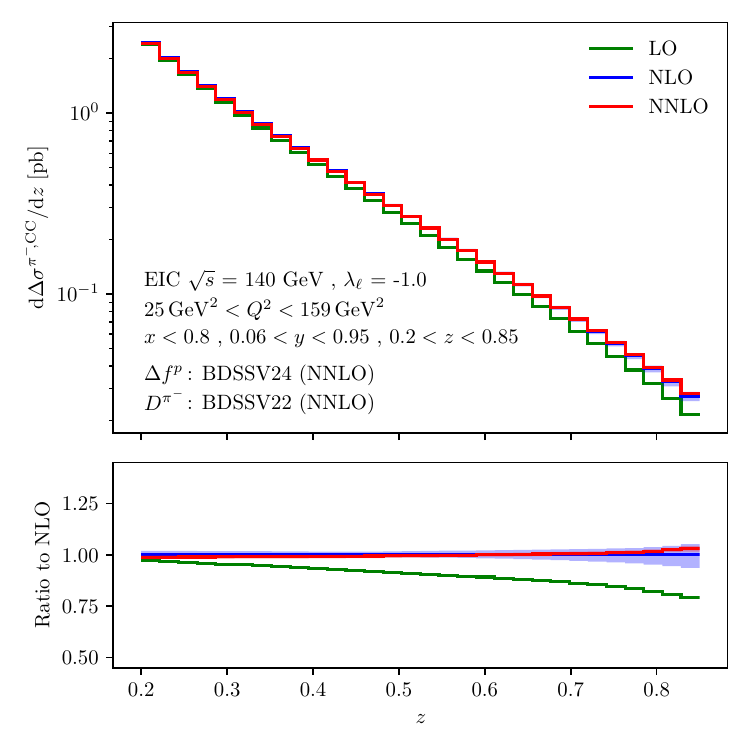}
\caption{Low-$Q^2$}
\end{subfigure}
\begin{subfigure}{0.45\textwidth}
\centering
\includegraphics[width=1.0\linewidth]{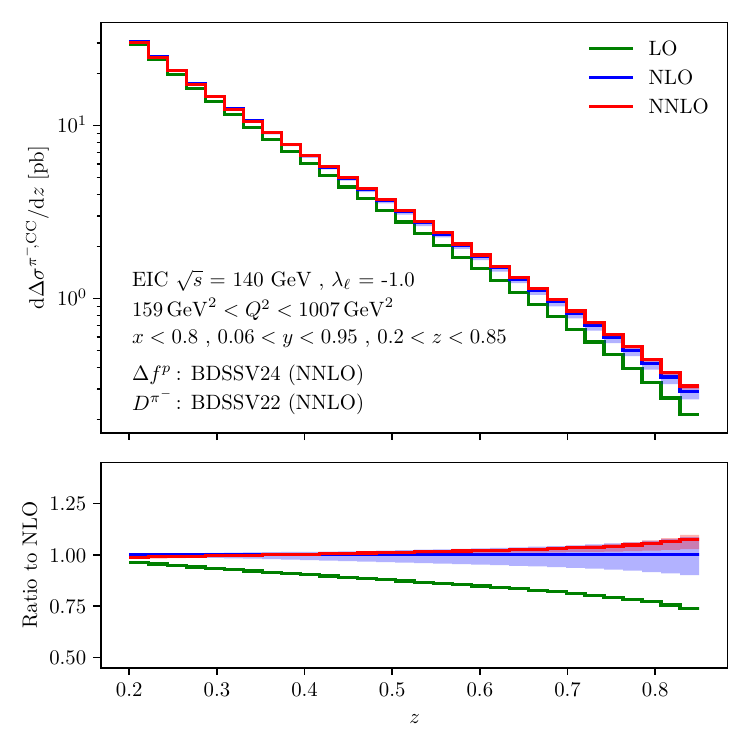}
\caption{Mid-$Q^2$}
\end{subfigure}
\begin{subfigure}{0.45\textwidth}
\centering
\includegraphics[width=1.0\linewidth]{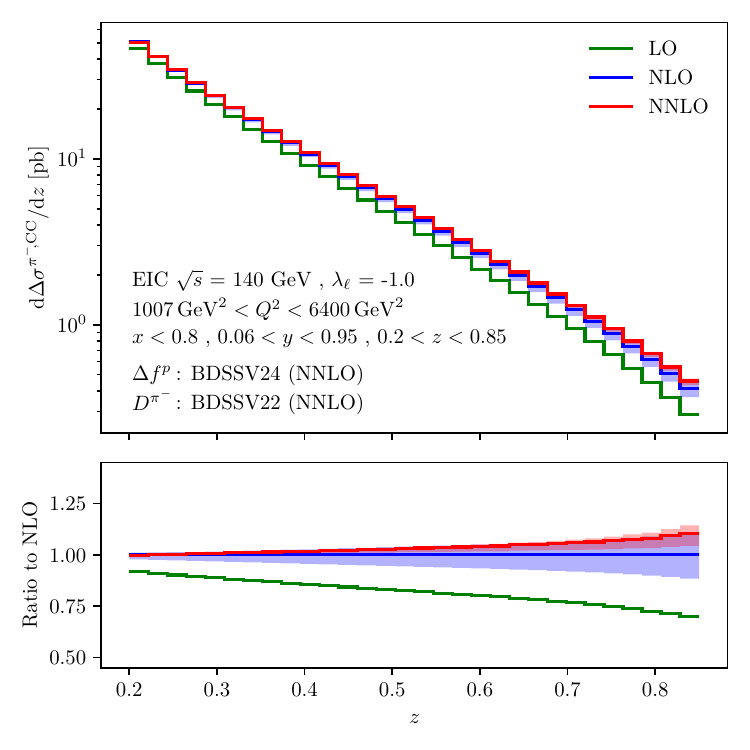}
\caption{High-$Q^2$}
\end{subfigure}
\begin{subfigure}{0.45\textwidth}
\centering
\includegraphics[width=1.0\linewidth]{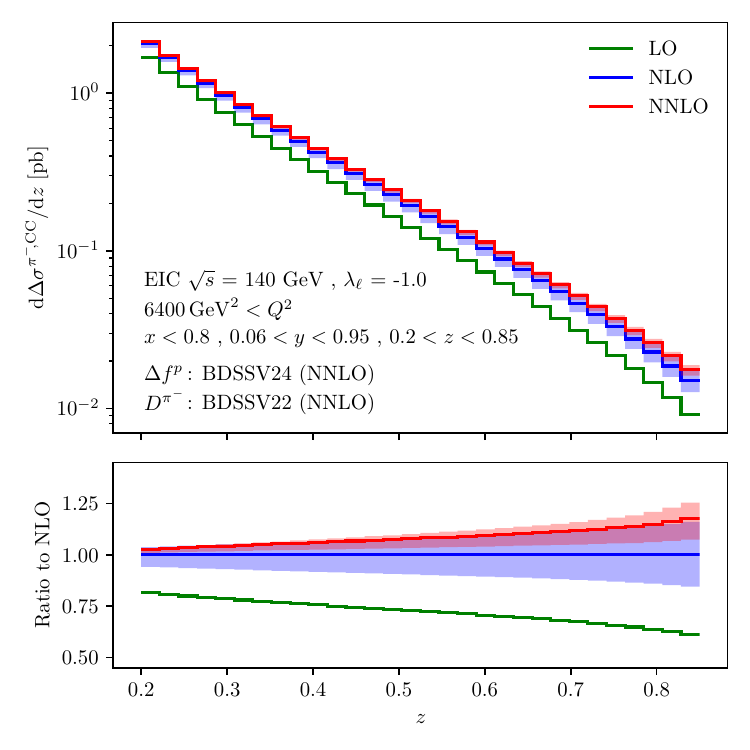}
\caption{Extreme-$Q^2$}
\end{subfigure}

\caption{
Electron-initiated CC $z$-distributions for $\pi^-$ production (favoured).
Top panels: total cross section difference.
Bottom panels: Ratio to NLO.
}
\label{fig:Z_CH_pim}

\end{figure}

In Fig.~\ref{fig:Z_CH_pip} we show the electron-initiated $z$-differential cross section difference $\dd\Delta \sigma ^{\pi^+,\CC}/\dd z$ for $\pi^+$ production in the four $Q^2$ ranges of~\eqref{eq:cutsQ2}.
The differential cross section decreases with increasing $z$ following the trend of the fragmentation functions as can be seen from the top panels.
$\dd\Delta \sigma ^{\pi^+,\CC}/\dd z$ overall increases from Low-$Q^2$ to High-$Q^2$.
From High-$Q^2$ to Extreme-$Q^2$ it decreases due to the strong PDF depletion at high $x$.
The polarized cross section difference $\dd\Delta \sigma ^{\pi^+,\CC}/\dd z$ is approximately one order of magnitude smaller than the unpolarized one $\dd \sigma ^{\pi^+,\CC}/\dd z$ at Low-$Q^2$, while being of comparable size in the other $Q^2$ ranges.

In the bottom panels we show the ratio to the NLO prediction where we observe good perturbative convergence and stability.
The NNLO corrections grow in size with the increase of $z$ and $Q^2$ from per-mille level to up to $20\%$.
They lie within the NLO theory uncertainty with the exception of $z>0.65$ at Extreme-$Q^2$.
The increase in relative size of the perturbative corrections at large-$z$ is an indication of large soft gluon emission effects.

In Fig.~\ref{fig:Z_CH_pim} we show the electron-initiated $z$-differential cross section difference for $\pi^-$ production
$\dd\Delta \sigma ^{\pi^-,\CC}/\dd z$ in the same four $Q^2$ ranges.
Due to the flavour-favoured production of $\pi^-$ in $W^-$ exchange, the differential cross sections are about a factor of two larger than the flavour-unfavoured $\pi^+$ ones.
The trend of the distributions and the behaviour of the perturbative corrections are very similar to the ones for $\pi^+$
discussed above.

\begin{figure}[p]
\centering
\begin{subfigure}{\textwidth}
\centering
\includegraphics[width=\textwidth]{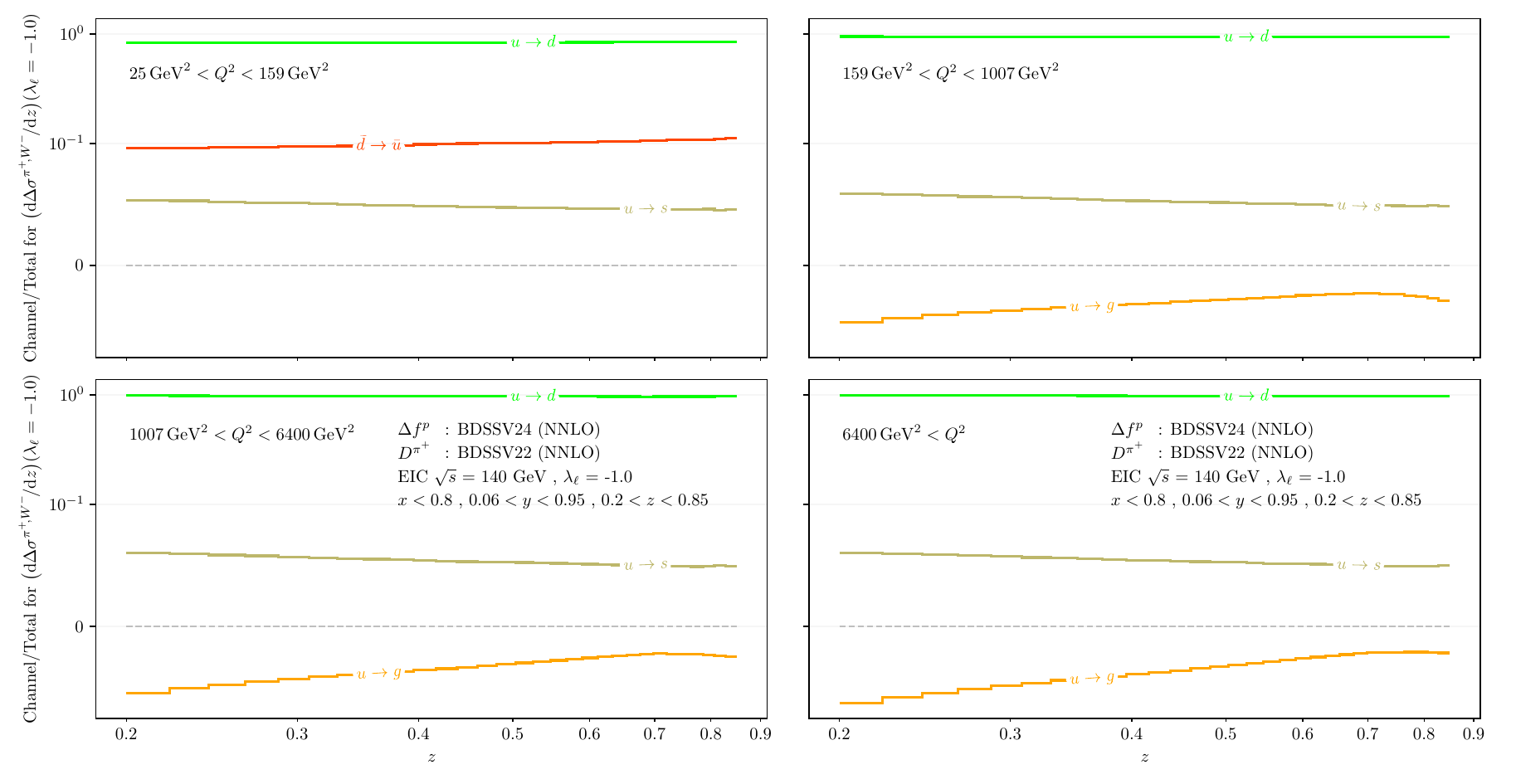}
\caption{Flavour-unfavoured transition $\dd \Delta \sigma^{\pi^+, W^-}/\dd z$}
\label{fig:z_CH_channels_pip}
\end{subfigure}
\begin{subfigure}{\textwidth}
\centering
\includegraphics[width=\textwidth]{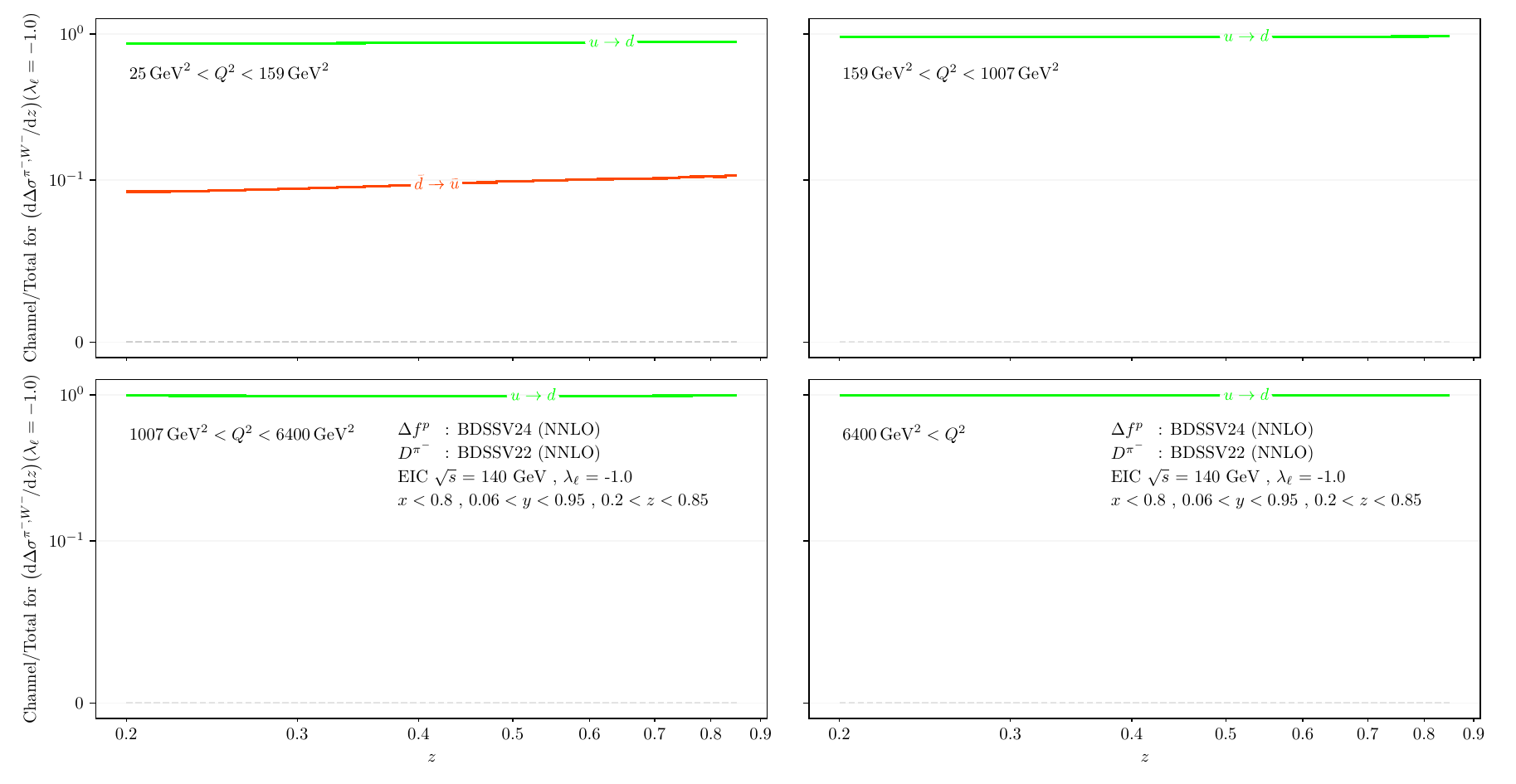}
\caption{Flavour-favoured transition $\dd \Delta \sigma^{\pi^-, W^-}/\dd z$}
\label{fig:z_CH_channels_pim}
\end{subfigure}
\caption{
Channel decomposition for electron-initiated $\CC$ $\pi^\pm$ production for $\dd \sigma^{\pi^\pm,W^-}$ as a function of $z$.
Only channels contributing more than $4 \,\%$ in any bin are displayed.
The range $(-0.1, 0.1)$ on the vertical axis is linear, the ranges above and below are plotted logarithmically.}
\label{fig:z_CH_channels}
\end{figure}

In Fig.~\ref{fig:z_CH_channels} we show the channel decomposition for the $z$-differential distributions of Fig.~\ref{fig:Z_CH_pip} and Fig.~\ref{fig:Z_CH_pim}.
The flavour-unfavoured transition $\dd \Delta\sigma^{\pi^+,W^-}/\dd z$ is given in Fig.~\ref{fig:z_CH_channels_pip}.
The $u\to d$ channel is the largest partonic channel due to the size of the $\Delta f^p_u$ polarized PDF.
This effect is particularly pronounced in the Mid-$Q^2$ to Extreme-$Q^2$ ranges, where
the only active channels are $u$-initiated.
The channel decomposition is similar to the unpolarized SIDIS case~\cite{Bonino:2025qta}
 for High-$Q^2$ and Extreme-$Q^2$.
The differences that can be observed at Low-$Q^2$ and Mid-$Q^2$ are due to the behaviour of the polarized PDFs.
At small $x$ the gluon and quark sea polarized PDFs are small in size and do not grow as the unpolarized PDFs.
As a result at Low-$Q^2$ and Mid-$Q^2$ in polarized SIDIS the $u$-initiated channels are the only ones contributing significantly.
The $\bar{d}\to\bar{u}$ channel is enhanced by the size of the CKM coefficient at LO and is sizeable at Low-$Q^2$.

The channel decomposition of the flavour-favoured transition $\dd \Delta\sigma^{\pi^-,W^-}/\dd z$ is shown in Fig.~\ref{fig:z_CH_channels_pim}.
As for unpolarized SIDIS, the $u\to d$ channel dominates the differential cross section due to its contribution to the valence content of both initial and final state identified hadrons.
The $\bar{d}\to \bar{u}$ channel is also sizeable at Low-$Q^2$ due to its valence role in the final state hadron and it being CKM-favoured at LO.
The channel decomposition of Fig.~\ref{fig:z_CH_channels_pim} is
 similar to the corresponding unpolarized SIDIS one~\cite{Bonino:2025qta}, with the exception of having
 considerably fewer sea quarks and gluonic channels at Low-$Q^2$ and Mid-$Q^2$.
This is again due to the vanishing of sea quark and gluon polarized PDFs at small $x$.

\begin{figure}[tb]
\centering

\begin{subfigure}{0.45\textwidth}
\centering
\includegraphics[width=1.0\linewidth]{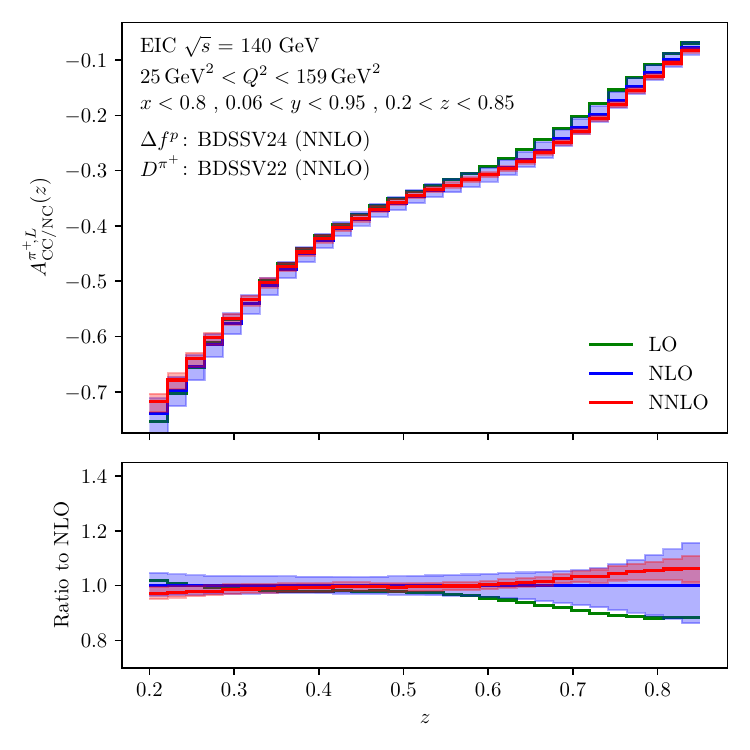}
\caption{Low-$Q^2$}
\end{subfigure}
\begin{subfigure}{0.45\textwidth}
\centering
\includegraphics[width=1.0\linewidth]{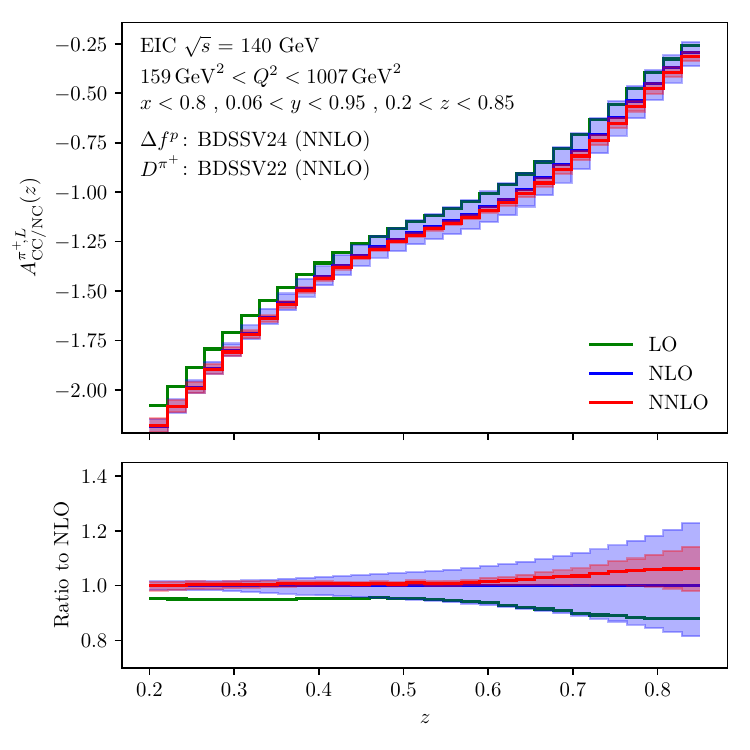}
\caption{Mid-$Q^2$}
\end{subfigure}
\begin{subfigure}{0.45\textwidth}
\centering
\includegraphics[width=1.0\linewidth]{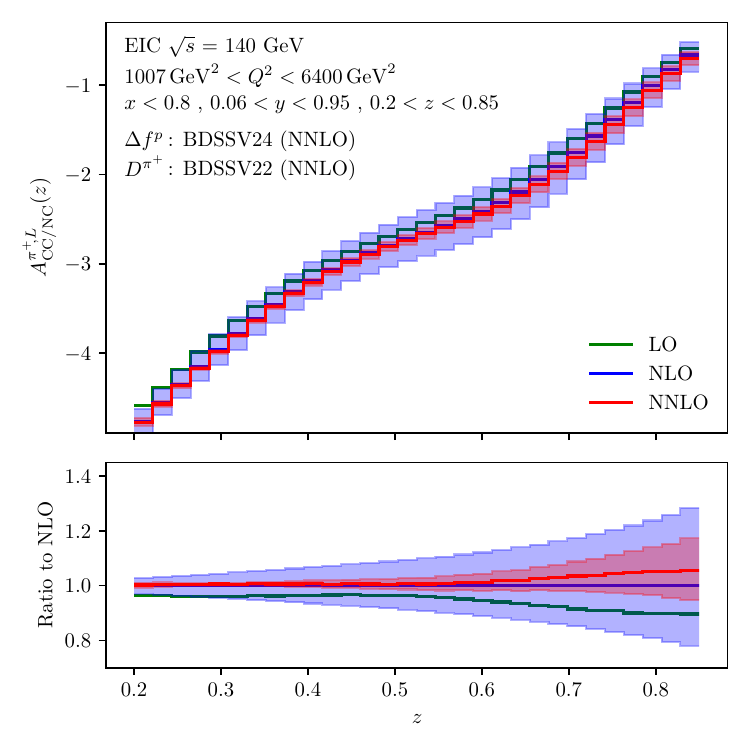}
\caption{High-$Q^2$}
\end{subfigure}
\begin{subfigure}{0.45\textwidth}
\centering
\includegraphics[width=1.0\linewidth]{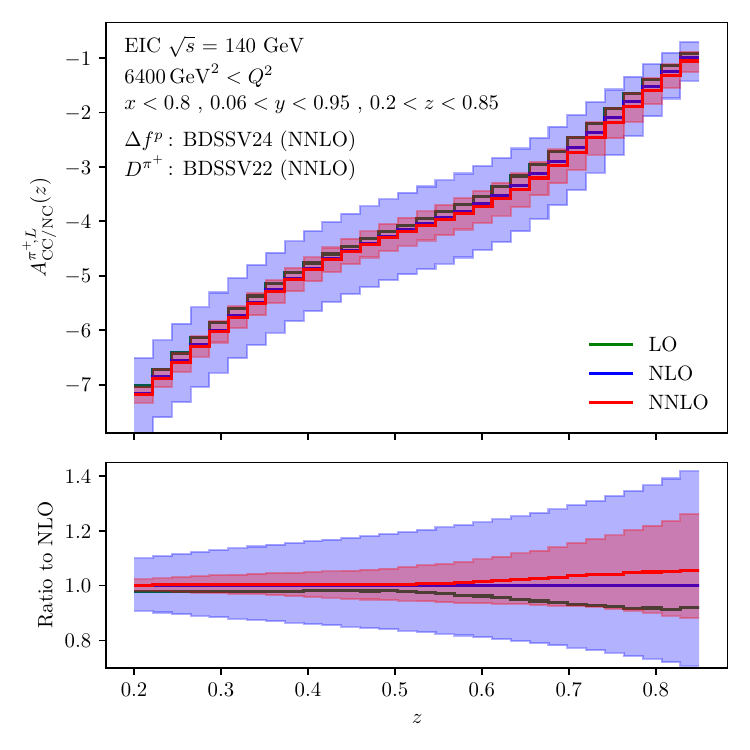}
\caption{Extreme-$Q^2$}
\end{subfigure}

\caption{
Electron-initiated longitudinal charged-neutral current asymmetry $A^{\pi^+,L}_\mathrm{CC/NC}$  in $\pi^+$ production as a function of $z$.
Top panels: total asymmetry.
Bottom panels: Ratio to NLO.
}
\label{fig:Z_A_CCNC_pip}

\end{figure}

\begin{figure}[tb]
\centering

\begin{subfigure}{0.45\textwidth}
\centering
\includegraphics[width=1.0\linewidth]{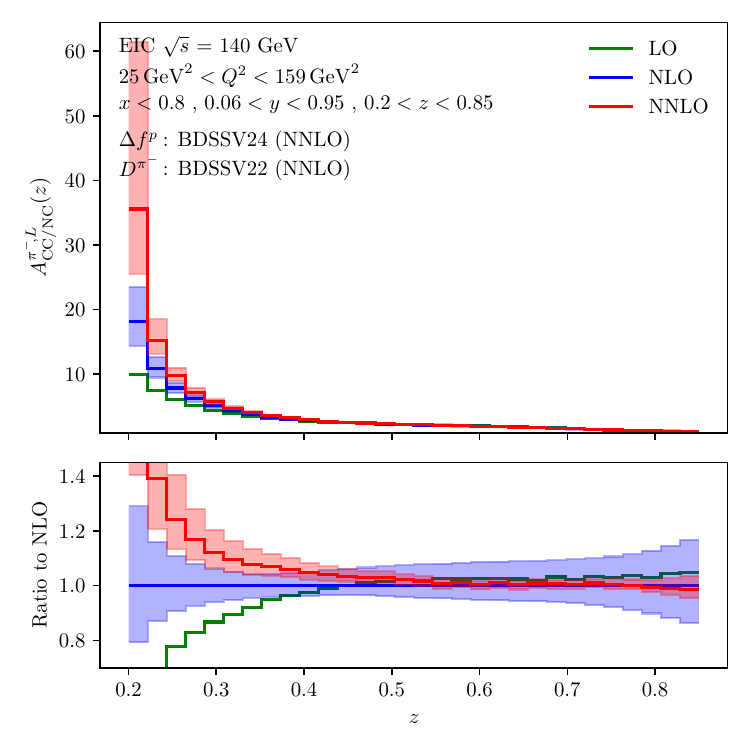}
\caption{Low-$Q^2$}
\end{subfigure}
\begin{subfigure}{0.45\textwidth}
\centering
\includegraphics[width=1.0\linewidth]{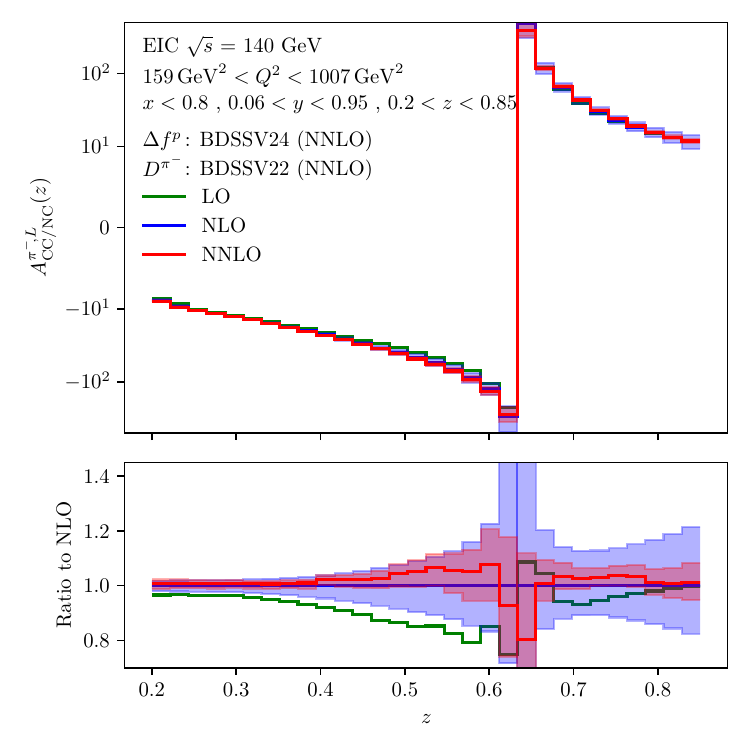}
\caption{Mid-$Q^2$}
\end{subfigure}
\begin{subfigure}{0.45\textwidth}
\centering
\includegraphics[width=1.0\linewidth]{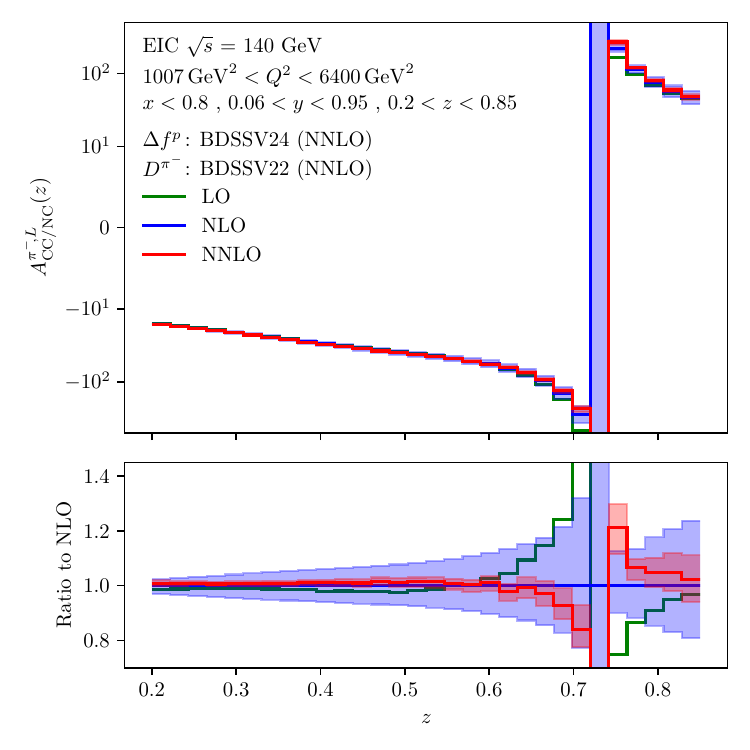}
\caption{High-$Q^2$}
\end{subfigure}
\begin{subfigure}{0.45\textwidth}
\centering
\includegraphics[width=1.0\linewidth]{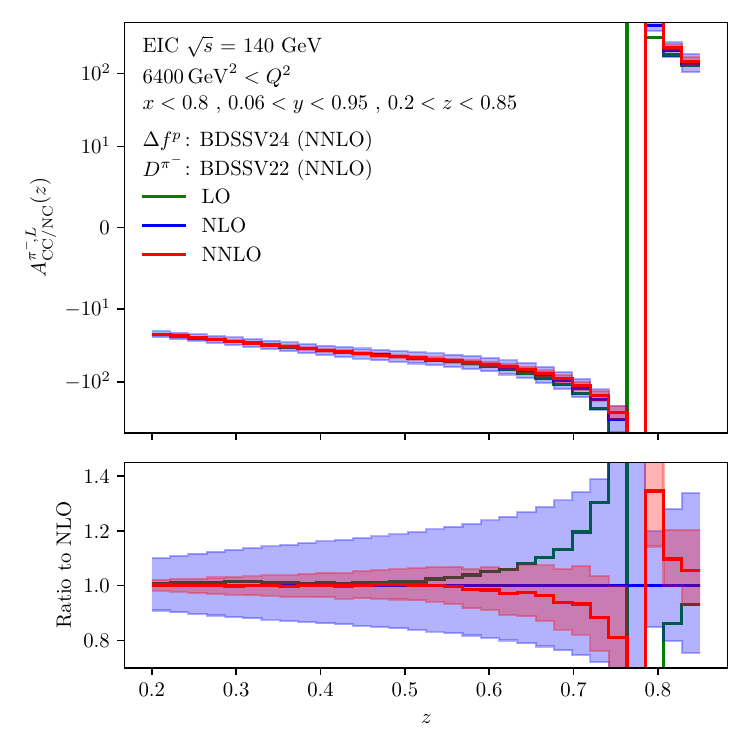}
\caption{Extreme-$Q^2$}
\end{subfigure}

\caption{
Electron-initiated longitudinal asymmetry $A^{\pi^-,L}_\mathrm{CC/NC}$ in $\pi^-$ production as a function of $z$.
Top panels: total asymmetry.
Bottom panels: Ratio to NLO.
}
\label{fig:Z_A_CCNC_pim}

\end{figure}

In Fig.~\ref{fig:Z_A_CCNC_pip} we show the electron-initiated longitudinal charged-neutral current asymmetry~\eqref{eq:ALCCNC} for $\pi^+$ as a function of $z$ in the usual $Q^2$ ranges.
The asymmetries decrease with increasing $z$ in all $Q^2$ ranges.
At Low-$Q^2$ the CC exchange is between $30\%$ to $90\%$ smaller than the NC $\lambda_\ell$-even contribution.
At Mid-$Q^2$, for $z\sim 0.5$ the CC exchange surpasses in size the NC $\lambda_\ell$-even contribution.
For Extreme-$Q^2$ and High-$Q^2$ the CC exchange is from the same size up to seven times larger than the NC $\lambda_\ell$-even contribution.
From the bottom panels in Fig.~\ref{fig:Z_A_CCNC_pip} we observe good perturbative convergence and stability.
Perturbative corrections grow sizeable to percent level at large $z$.
NNLO corrections are within the NLO theory uncertainty in the entire kinematic range.

In Fig.~\ref{fig:Z_A_CCNC_pim} we show the electron-initiated longitudinal charged-neutral current asymmetry for $\pi^-$ as a function of $z$.
The CC contribution surpasses in size the NC $\lambda_\ell$-even contribution in all $Q^2$ regions.
At different values of $z$ depending on the $Q^2$ range the NC asymmetry in the denominator crosses zero, leading
to the observed instabilities in the Mid-$Q^2$ to Extreme-$Q^2$ ranges. Outside these instabilities, a very good
perturbative convergence from NLO to NNLO is observed.
\begin{figure}[p]
\centering

\begin{subfigure}{0.45\textwidth}
\centering
\includegraphics[width=1.0\linewidth]{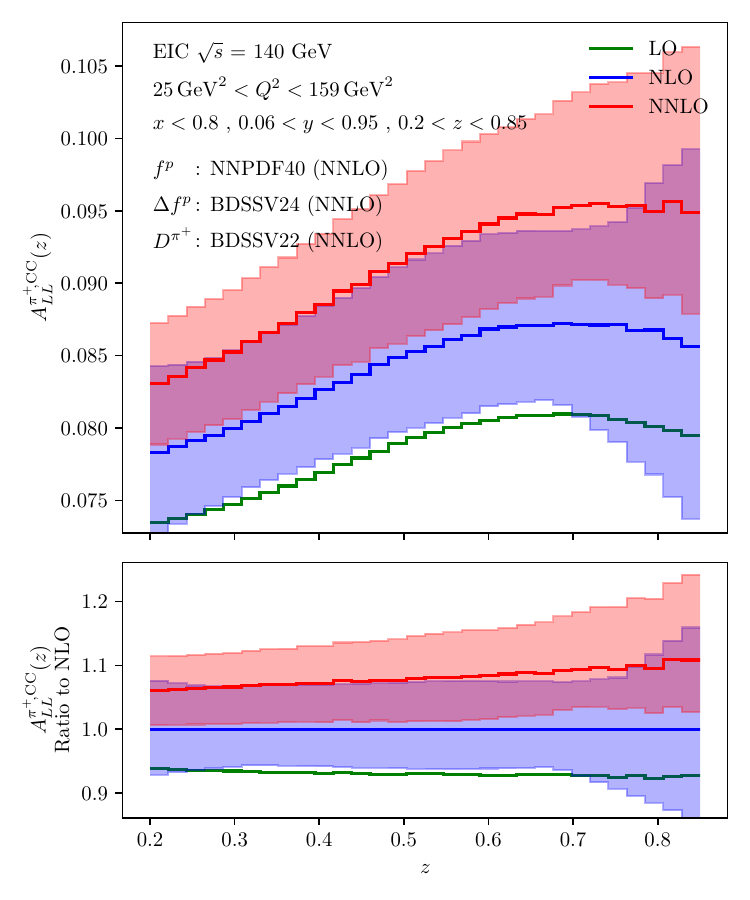}
\caption{Low-$Q^2$}
\end{subfigure}
\begin{subfigure}{0.45\textwidth}
\centering
\includegraphics[width=1.0\linewidth]{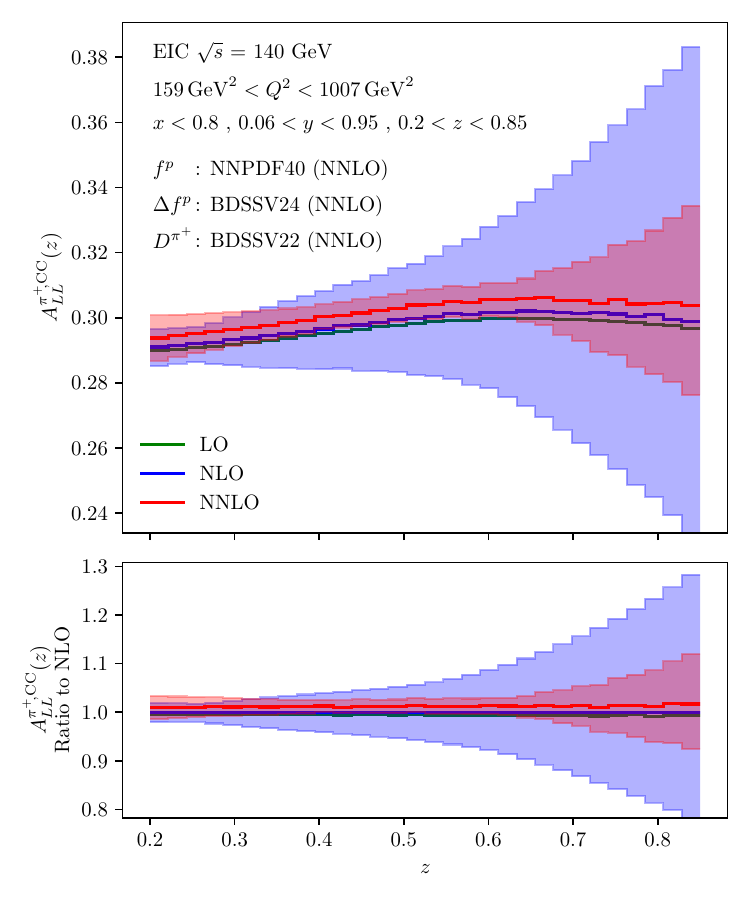}
\caption{Mid-$Q^2$}
\end{subfigure}
\begin{subfigure}{0.45\textwidth}
\centering
\includegraphics[width=1.0\linewidth]{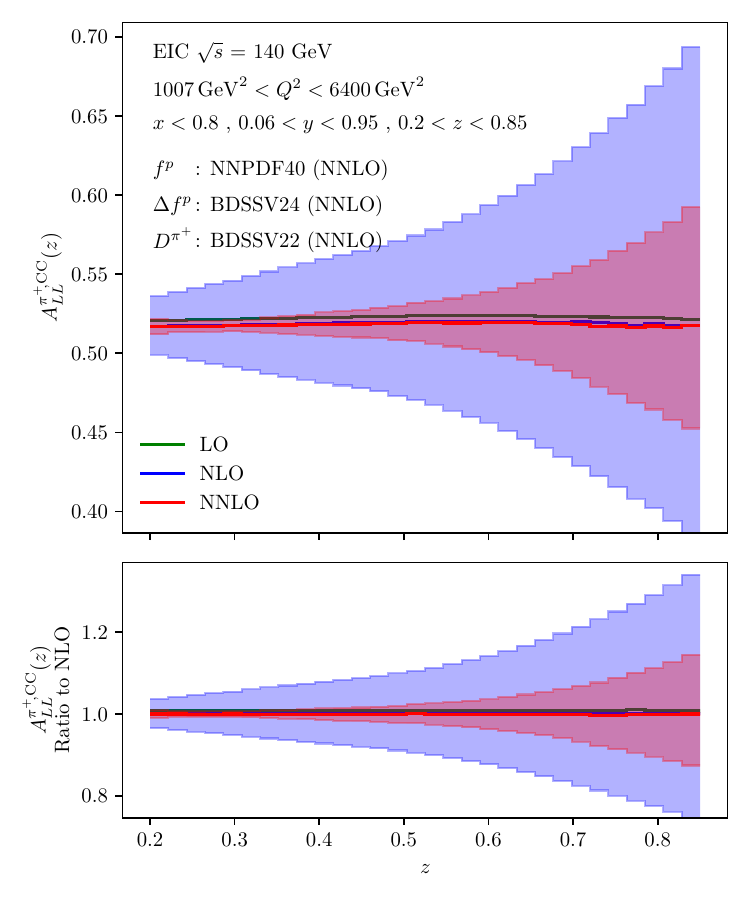}
\caption{High-$Q^2$}
\end{subfigure}
\begin{subfigure}{0.45\textwidth}
\centering
\includegraphics[width=1.0\linewidth]{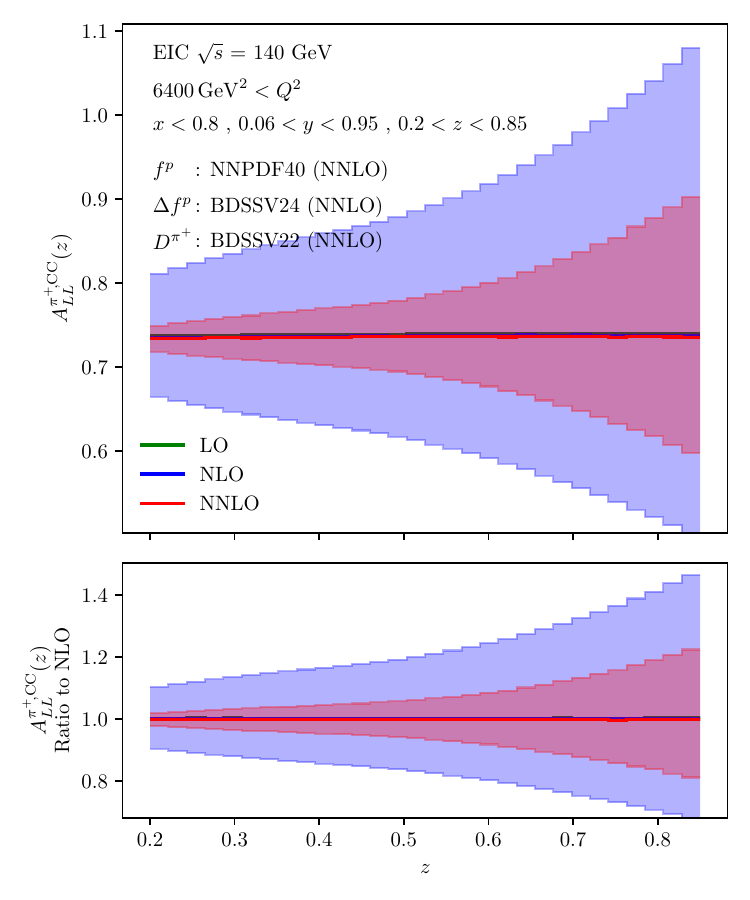}
\caption{Extreme-$Q^2$}
\end{subfigure}

\caption{
Electron-initiated double-longitudinal CC spin asymmetry $A^{\pi^+,\CC}_{LL}$  in $\pi^+$ production as a function of $z$.
Top panels: total asymmetry.
Bottom panels: Ratio to NLO.
}
\label{fig:Z_A_LL_CC_pip}

\end{figure}

\begin{figure}[p]
\centering

\begin{subfigure}{0.45\textwidth}
\centering
\includegraphics[width=1.0\linewidth]{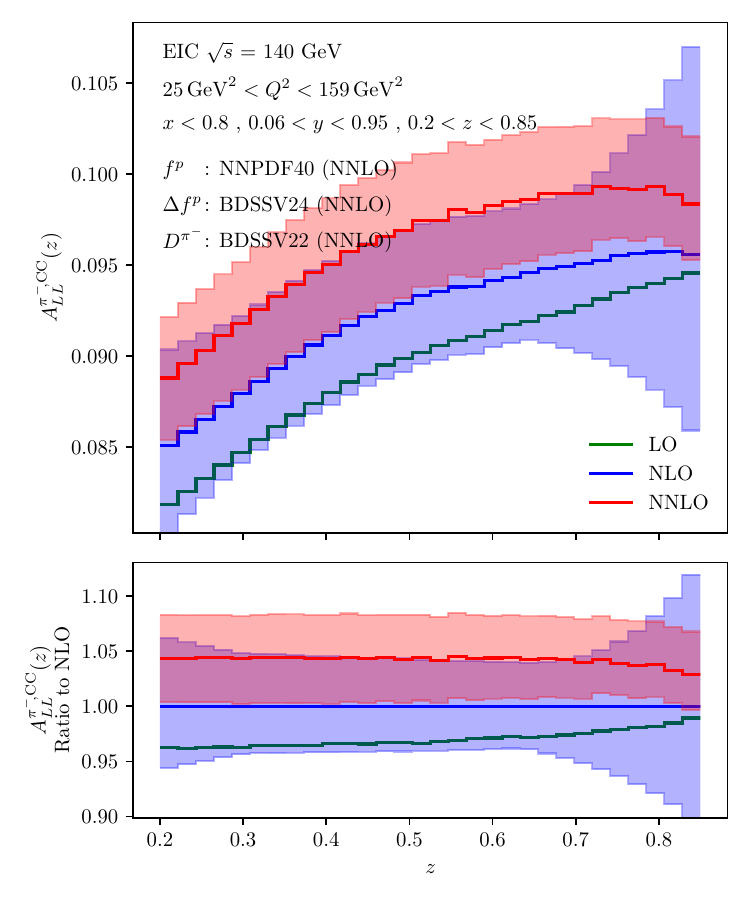}
\caption{Low-$Q^2$}
\end{subfigure}
\begin{subfigure}{0.45\textwidth}
\centering
\includegraphics[width=1.0\linewidth]{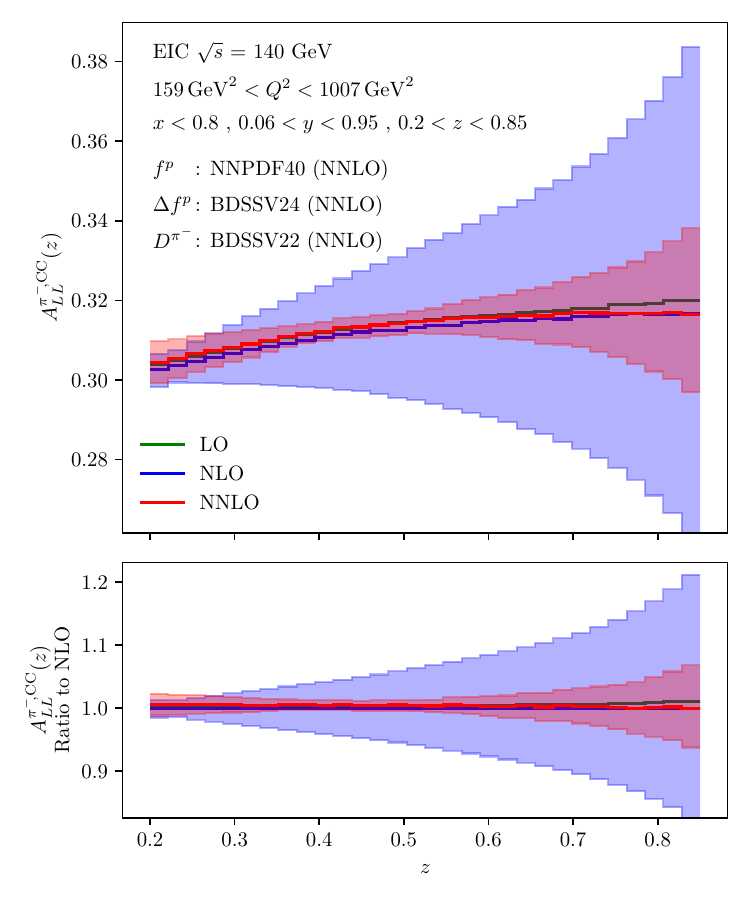}
\caption{Mid-$Q^2$}
\end{subfigure}
\begin{subfigure}{0.45\textwidth}
\centering
\includegraphics[width=1.0\linewidth]{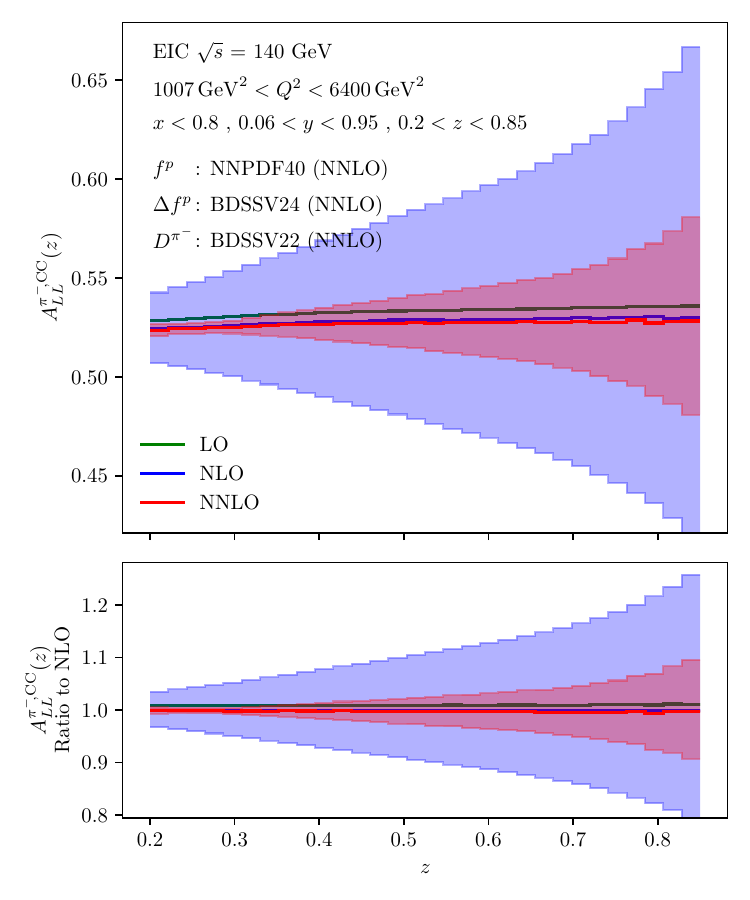}
\caption{High-$Q^2$}
\end{subfigure}
\begin{subfigure}{0.45\textwidth}
\centering
\includegraphics[width=1.0\linewidth]{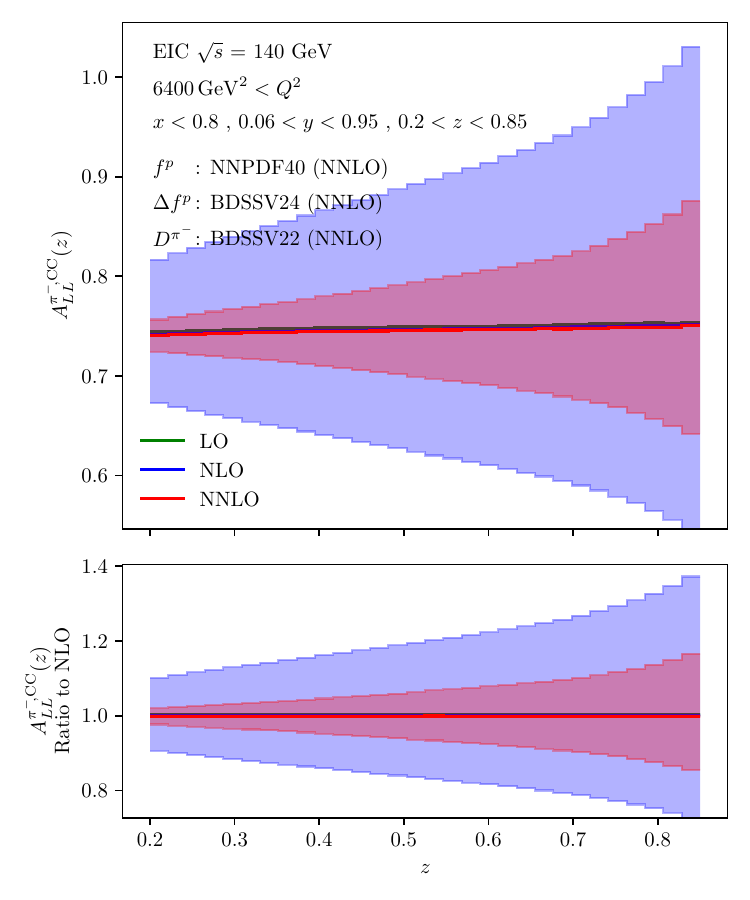}
\caption{Extreme-$Q^2$}
\end{subfigure}

\caption{
Electron-initiated double-longitudinal CC spin asymmetry $A^{\pi^-,\CC}_{LL}$ in $\pi^-$ production as a function of $z$.
Top panels: total asymmetry.
Bottom panels: Ratio to NLO.
}
\label{fig:Z_A_LL_CC_pim}

\end{figure}

In Fig.~\ref{fig:Z_A_LL_CC_pip} we study the CC electron-initiated double-longitudinal spin asymmetry~\eqref{eq:AhLLCC} for $\pi^+$ production as a function of $z$.
The asymmetry shows only a moderate $z$ dependence visible at Low-$Q^2$ and Mid-$Q^2$.
The asymmetry is below $0.1$ at Low-$Q^2$ and it grows with increasing $Q^2$.
For High-$Q^2$ and Extreme-$Q^2$ the polarized CC cross section difference is of equal size or even larger than the unpolarized CC cross section.
The asymmetry peaks at Extreme-$Q^2$ where it is approximately equal to $0.75$.
We observe that the CC double-longitudinal  asymmetry as a function of $z$ is larger that the NC double-longitudinal asymmetry as a function of $z$ in Fig.~\ref{fig:Z_A_LL_NC_pip}.
At Low-$Q^2$ we observe percent level perturbative corrections at both NLO and NNLO.
The ratio to NLO, as visible in the bottom panels of Fig.~\ref{fig:Z_A_LL_CC_pip}. appears to be independent on $z$.
The perturbative corrections reduce in size with the increase of the average value of $Q^2$ and are at subpercent level at High-$Q^2$ and Extreme-$Q^2$.

In Fig.~\ref{fig:Z_A_LL_CC_pim} we study the CC electron-initiated double-longitudinal spin asymmetry~\eqref{eq:AhLLCC} for $\pi^-$ production as a function of $z$.
The overall behaviour of the asymmetries is very similar as for $\pi^+$ and we refer to the above paragraph for a more detailed description.
$A^{\pi^-,\CC}_{LL}$ is slightly larger than $A^{\pi^+,\CC}_{LL}$ as can be seen at Low-$Q^2$ and Mid-$Q^2$ due to the flavour-favoured transition.
The double-longitudinal asymmetry is less sensitive to the type of flavour transition probed as compared to the cross sections differences (Fig.~\ref{fig:Z_CH_pip} and Fig.~\ref{fig:Z_CH_pim}).  This is due to the equal sensitivity to the flavour transition probed of both numerator and denominator in the ratio.

\subsubsection[\texorpdfstring{$x$-distributions}{x-distributions}]{\boldmath $x$-distributions}

\begin{figure}[tb]
\centering

\begin{subfigure}{0.45\textwidth}
\centering
\includegraphics[width=1.0\linewidth]{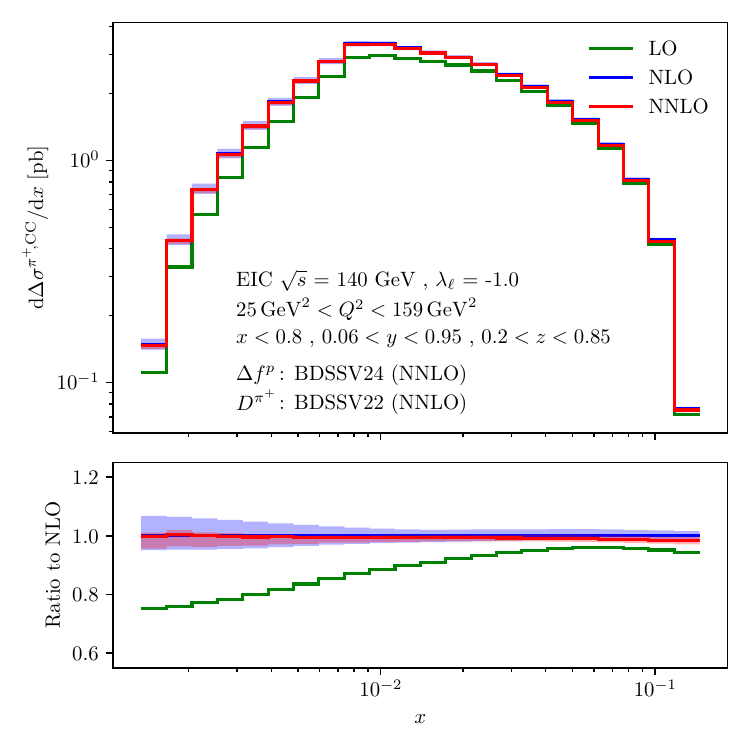}
\caption{Low-$Q^2$}
\end{subfigure}
\begin{subfigure}{0.45\textwidth}
\centering
\includegraphics[width=1.0\linewidth]{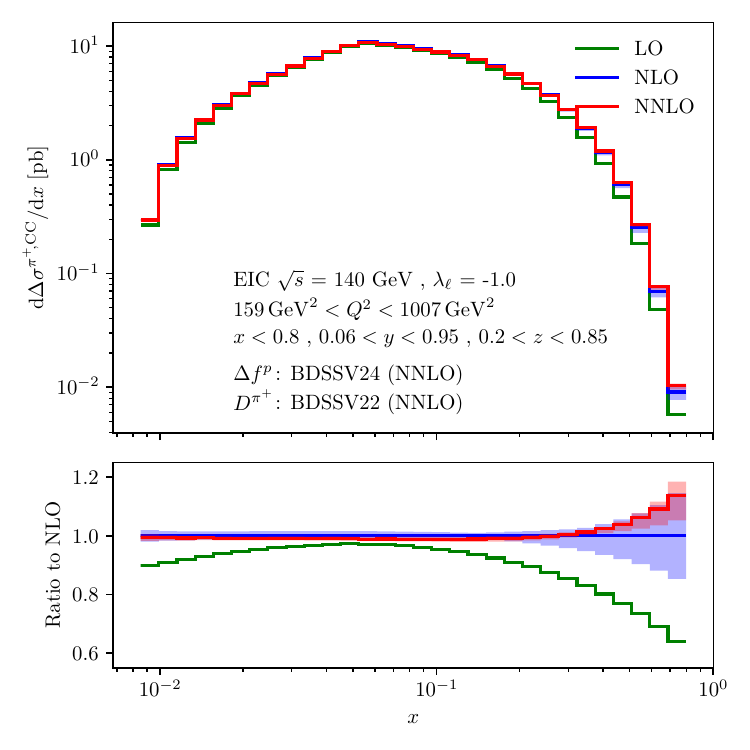}
\caption{Mid-$Q^2$}
\end{subfigure}
\begin{subfigure}{0.45\textwidth}
\centering
\includegraphics[width=1.0\linewidth]{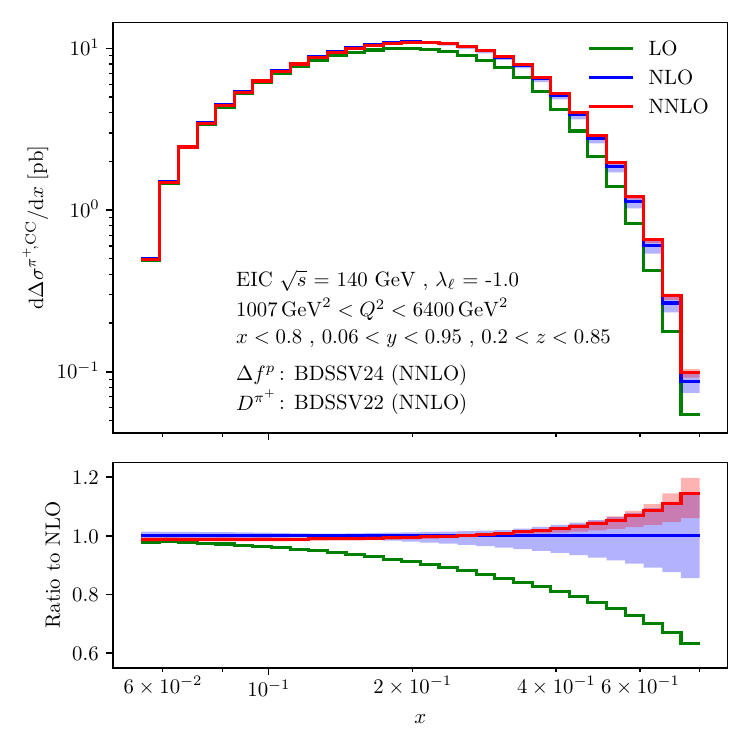}
\caption{High-$Q^2$}
\end{subfigure}
\begin{subfigure}{0.45\textwidth}
\centering
\includegraphics[width=1.0\linewidth]{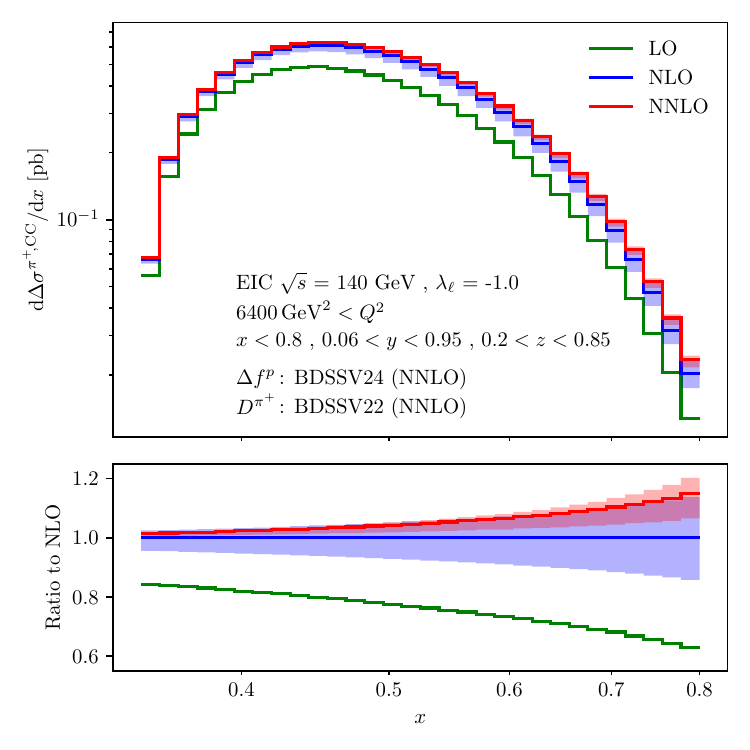}
\caption{Extreme-$Q^2$}
\end{subfigure}

\caption{
Electron-initiated CC $x$-distributions in $\pi^+$ production.
Top panels: total cross section difference.
Bottom panels: Ratio to NLO.
}
\label{fig:X_CH_pip}

\end{figure}

\begin{figure}[tb]
\centering

\begin{subfigure}{0.45\textwidth}
\centering
\includegraphics[width=1.0\linewidth]{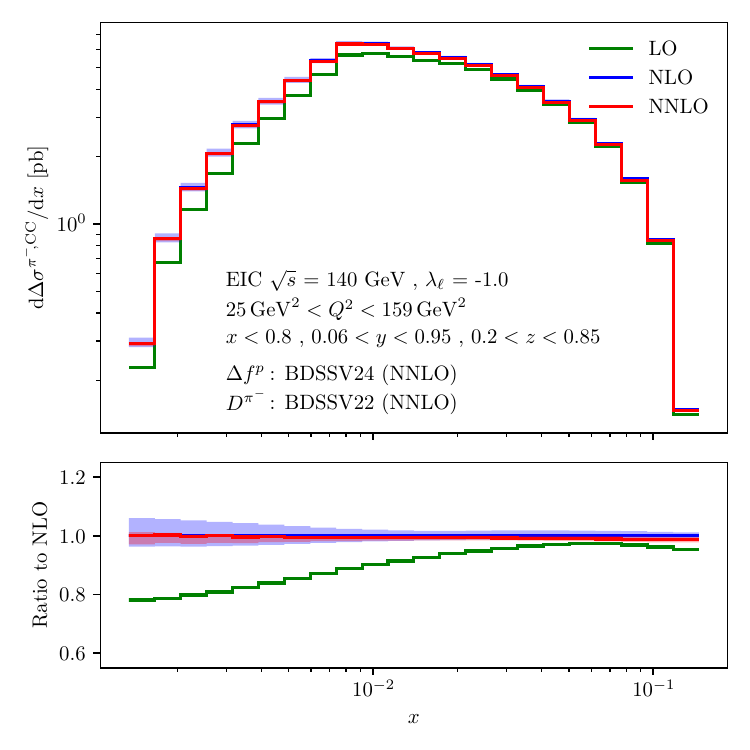}
\caption{Low-$Q^2$}
\end{subfigure}
\begin{subfigure}{0.45\textwidth}
\centering
\includegraphics[width=1.0\linewidth]{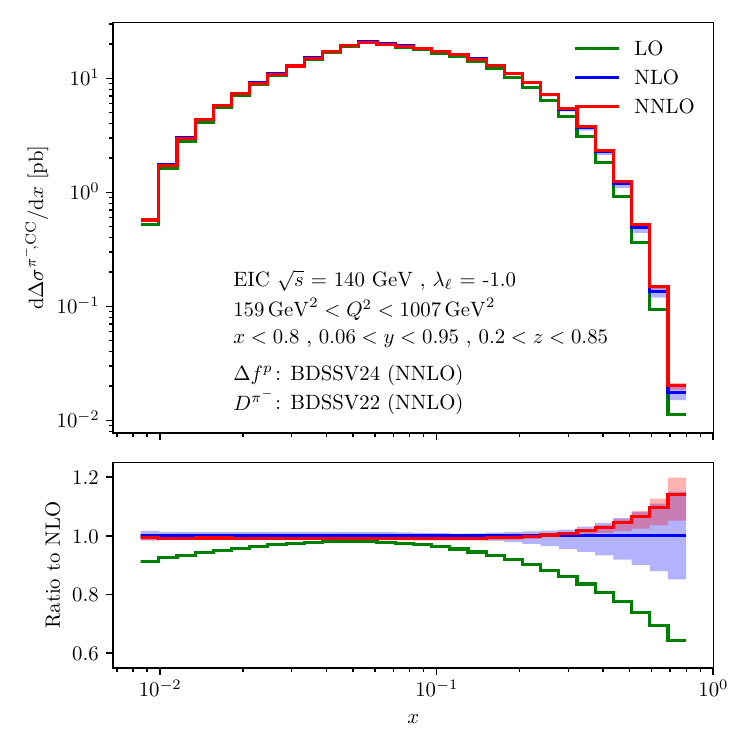}
\caption{Mid-$Q^2$}
\end{subfigure}
\begin{subfigure}{0.45\textwidth}
\centering
\includegraphics[width=1.0\linewidth]{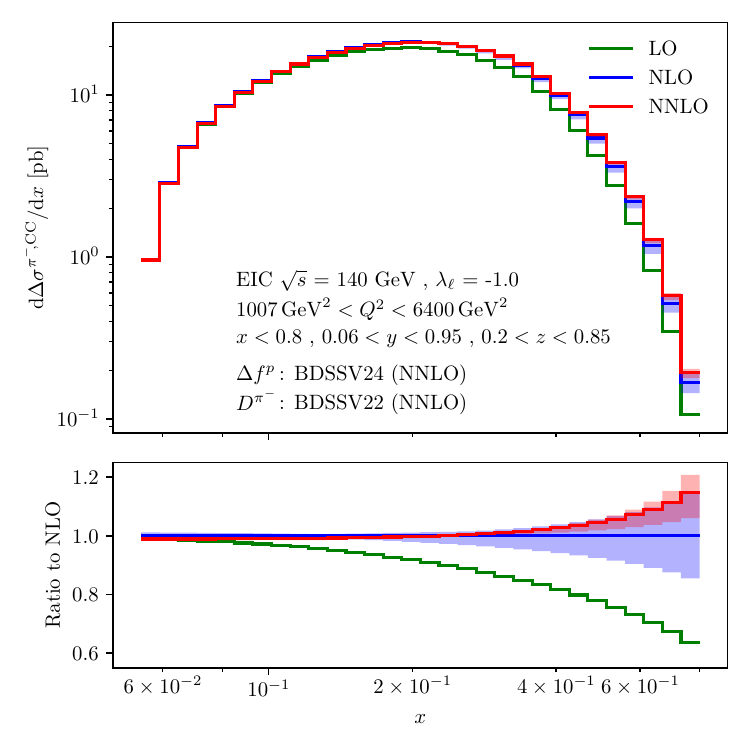}
\caption{High-$Q^2$}
\end{subfigure}
\begin{subfigure}{0.45\textwidth}
\centering
\includegraphics[width=1.0\linewidth]{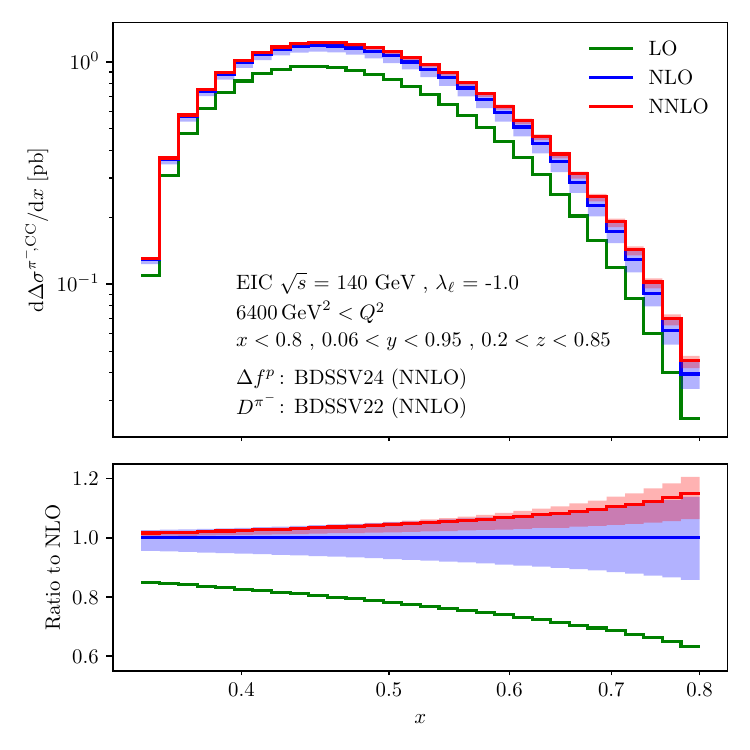}
\caption{Extreme-$Q^2$}
\end{subfigure}

\caption{
Electron-initiated CC $x$-distributions in $\pi^-$ production.
Top panels: total cross section difference.
Bottom panels: Ratio to NLO.
}
\label{fig:X_CH_pim}

\end{figure}

In Fig.~\ref{fig:X_CH_pip} we show the $x$-differential electron-initiated cross section difference $\dd \Delta \sigma^{\pi^+,\CC}/\dd x$ for $\pi^+$ production in the four kinematics ranges of $Q^2$.
At small $x$ the distributions increase up to a maximum which depends on the average value of $Q^2$.
This in due to the increase of the available phase space caused by the softening of the $y$ cut.
After reaching the maximum the distributions decrease due to the depletion of the polarized PDFs at high $x$.
From Low-$Q^2$ to High-$Q^2$ the average value of the cross section increases due to the increase in the polarized PDFs with the increase of the average $x$ probed.
The cross section instead decreases on average at Extreme-$Q^2$ due to the high values of $x$ probed.
The shape of the distributions is very similar in polarized and unpolarized SIDIS.
For Low-$Q^2$ and Mid-$Q^2$ the polarized distributions are about a factor of ten smaller than their unpolarized counterparts~\cite{Bonino:2025qta}.
For High-$Q^2$ and Extreme-$Q^2$ they are instead of comparable size.
This again indicates a faster growth of electroweak effects in polarized as compared to unpolarized SIDIS.
In the bottom panels of Fig.~\ref{fig:X_CH_pip} the ratio to NLO is shown.
Perturbative corrections are sizeable, especially from LO to NLO.
NNLO corrections are within the NLO theory uncertainty band and grow sizeable with the increase of $x$.

In Fig.~\ref{fig:X_CH_pim} we show the $x$-differential electron-initiated cross section difference for $\pi^-$ production $\dd \Delta \sigma^{\pi^-,\CC}/\dd x$.
The shape of the distributions and the behaviour of the perturbative corrections is very similar between $\pi^+$ and $\pi^-$.
The $\pi^-$ flavour-favoured transition is about a factor of two larger than the $\pi^+$ flavour-unfavoured transition, analogously to what observed for the $Q^2$ and $z$ distributions.

\begin{figure}[p]
\centering
\begin{subfigure}{\textwidth}
\centering
\includegraphics[width=\textwidth]{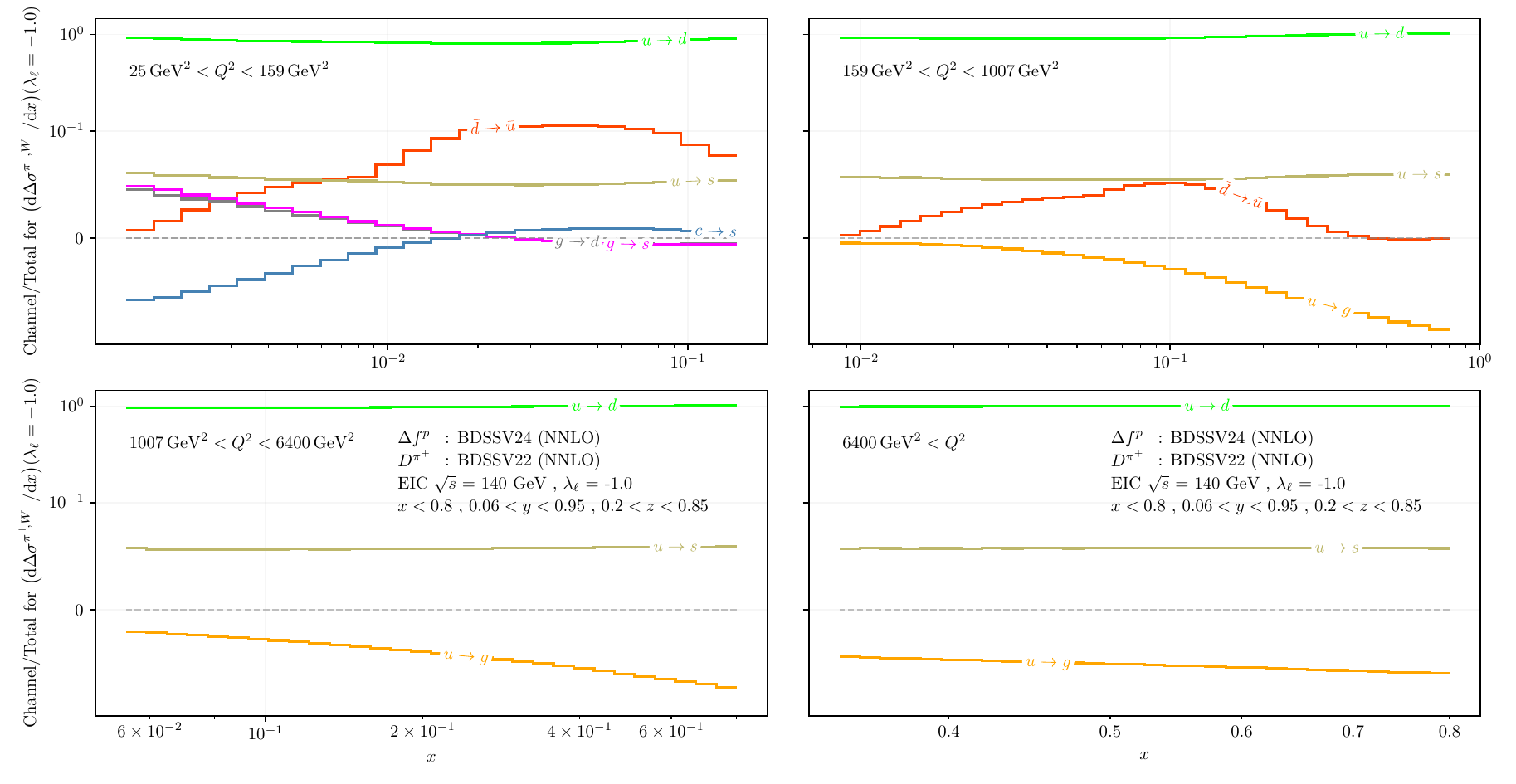}
\caption{Flavour-unfavoured transition $\dd \Delta \sigma^{\pi^+, W^-}/\dd x$}
\label{fig:x_CH_channels_pip}
\end{subfigure}
\begin{subfigure}{\textwidth}
\centering
\includegraphics[width=\textwidth]{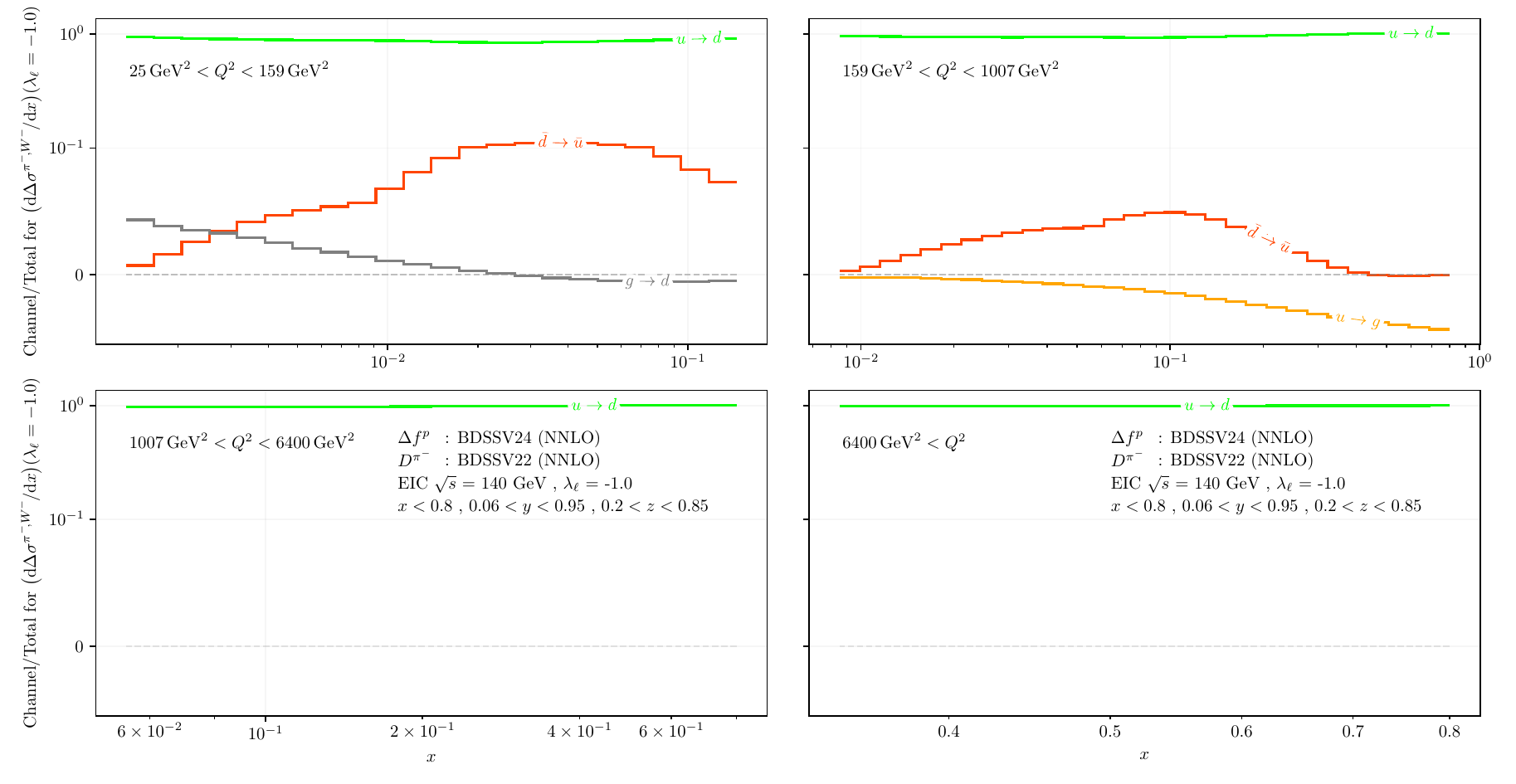}
\caption{Flavour-favoured transition $\dd \Delta \sigma^{\pi^-, W^-}/\dd x$}
\label{fig:x_CH_channels_pim}
\end{subfigure}
\caption{
Channel decomposition for electron-initiated $\CC$ $\pi^\pm$ production for $\dd \sigma^{\pi^\pm,W^-}$ as a function of $x$.
Only channels contributing more than $4 \,\%$ in any bin are displayed.
The range $(-0.1, 0.1)$ on the vertical axis is linear, the ranges above and below are plotted logarithmically.
}
\label{fig:x_CH_channels}
\end{figure}

In Fig.~\ref{fig:x_CH_channels} we show the partonic channel decomposition relative to the  cross sections presented in Fig.~\ref{fig:X_CH_pip} and Fig.~\ref{fig:X_CH_pim}.
The $u\to d$ channels dominates both flavour-unfavoured and flavour-favoured transitions, Fig.~\ref{fig:x_CH_channels_pip} and Fig.~\ref{fig:x_CH_channels_pim} respectively.
This is due to the size of the $\Delta f^p_u$ polarized PDF and the relative size of the CKM coefficient.
The resulting smaller overall size of $\dd \Delta\sigma^{\pi^+,W^-}/\dd x$ with respect to $\dd \Delta\sigma^{\pi^-,W^-}/\dd x$ is due to the suppression of the FF $D^{\pi^+}_d$ with respect to $D^{\pi^-}_d$ at large $z$.
At Low-$Q^2$ the second most important channel is the $\bar{u}\to \bar{d}$ channel, again due to its favoured contribution by the CKM flavour mixing, both in $\pi^+$ and $\pi^-$ production.
For $\pi^+$ production more sea quarks and gluonic channels contribute at Low-$Q^2$ and Mid-$Q^2$ due to the FF suppression of the flavour-unfavoured transition $u\to d$.
In $\pi^+$ production at Low-$Q^2$ and Mid-$Q^2$ only a handful of sea quark and gluonic channels contribute.
From the values of $x$ probed it is clear that in polarized SIDIS the smallness of the polarized PDFs at low $x$ results in a significant suppression of sea and gluonic channels.
With the increase of virtuality only $u$-initiated channels survive because of the size of $\Delta f^p_u$.
Since the partonic channel $u\to d$ is flavour-favoured,  $\pi^-$ production is even more dominated by this single channel in the majority of the $Q^2$ ranges.

\begin{figure}[tb]
\centering

\begin{subfigure}{0.45\textwidth}
\centering
\includegraphics[width=1.0\linewidth]{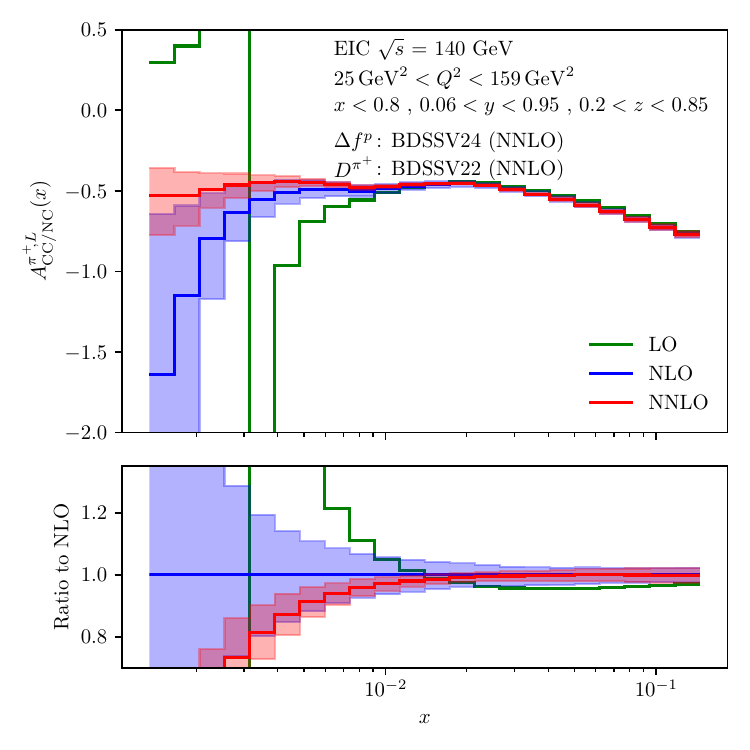}
\caption{Low-$Q^2$}
\end{subfigure}
\begin{subfigure}{0.45\textwidth}
\centering
\includegraphics[width=1.0\linewidth]{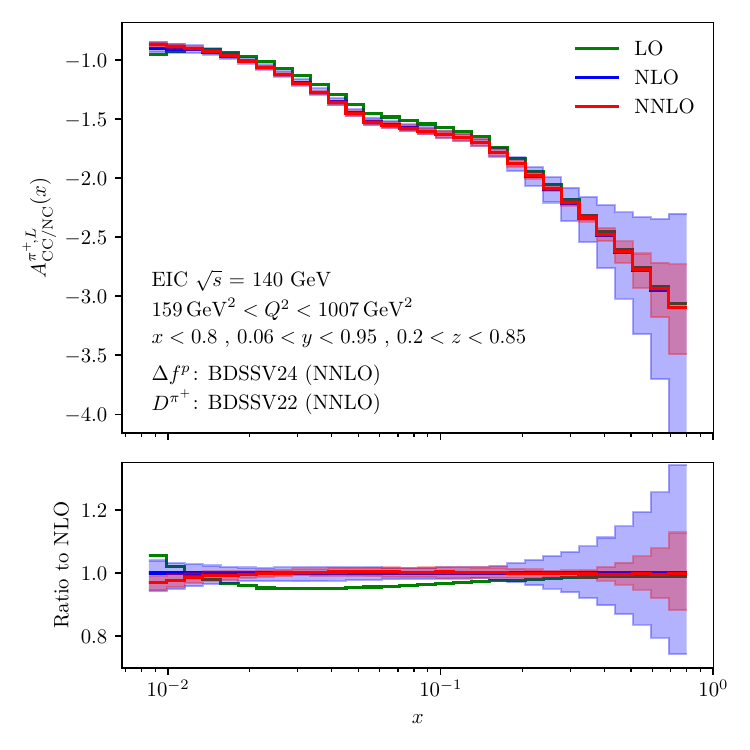}
\caption{Mid-$Q^2$}
\end{subfigure}
\begin{subfigure}{0.45\textwidth}
\centering
\includegraphics[width=1.0\linewidth]{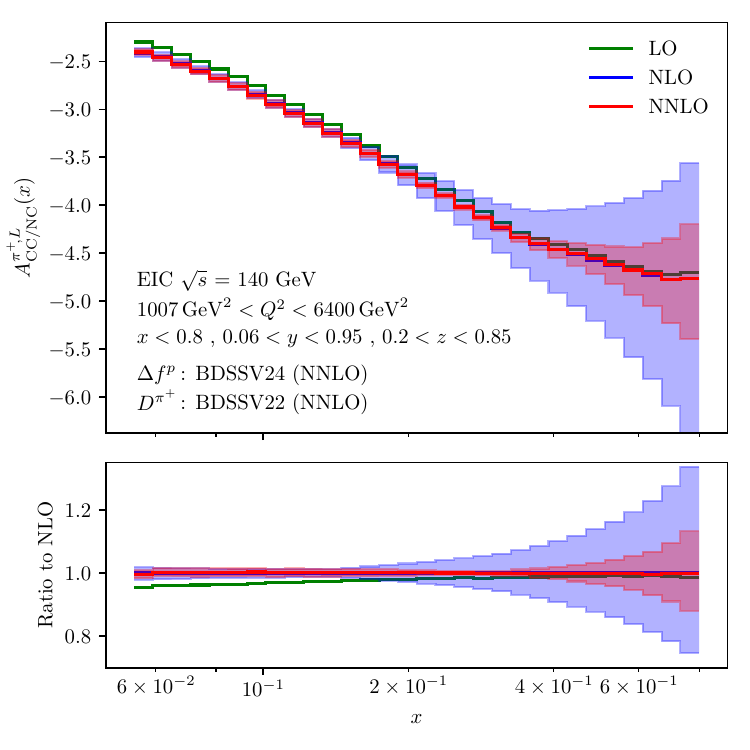}
\caption{High-$Q^2$}
\end{subfigure}
\begin{subfigure}{0.45\textwidth}
\centering
\includegraphics[width=1.0\linewidth]{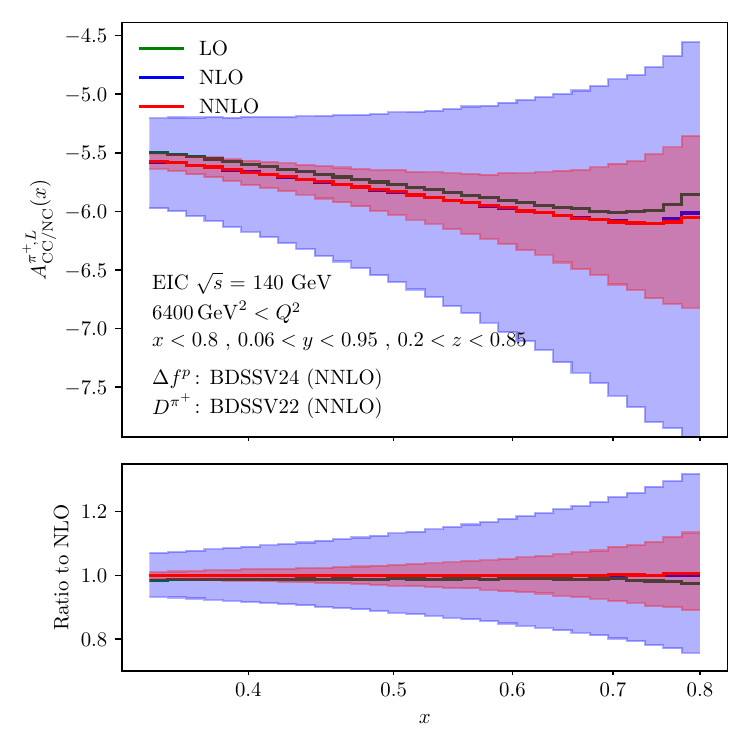}
\caption{Extreme-$Q^2$}
\end{subfigure}

\caption{
Electron-initiated longitudinal charge-neutral current asymmetry $A^{\pi^+,L}_\mathrm{CC/NC}$ as a function of $x$ in $\pi^+$ production.
Top panels: total asymmetry.
Bottom panels: Ratio to NLO.
}
\label{fig:X_A_CCNC_pip}

\end{figure}

\begin{figure}[tb]
\centering

\begin{subfigure}{0.45\textwidth}
\centering
\includegraphics[width=1.0\linewidth]{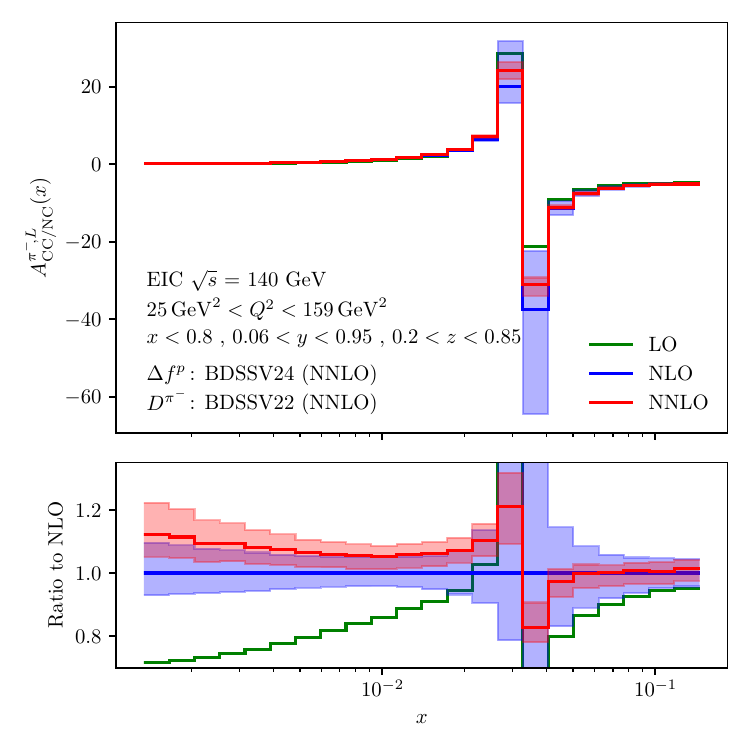}
\caption{Low-$Q^2$}
\end{subfigure}
\begin{subfigure}{0.45\textwidth}
\centering
\includegraphics[width=1.0\linewidth]{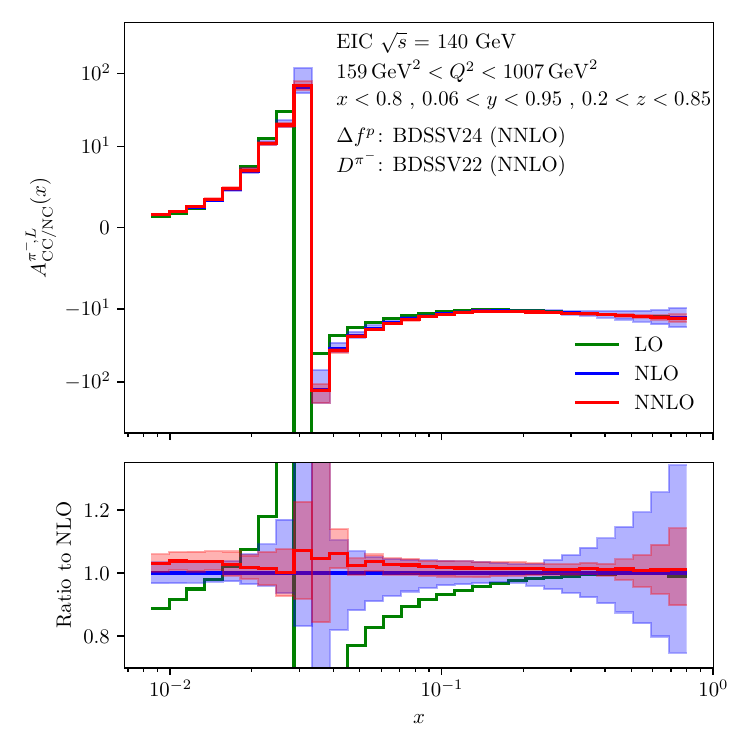}
\caption{Mid-$Q^2$}
\end{subfigure}
\begin{subfigure}{0.45\textwidth}
\centering
\includegraphics[width=1.0\linewidth]{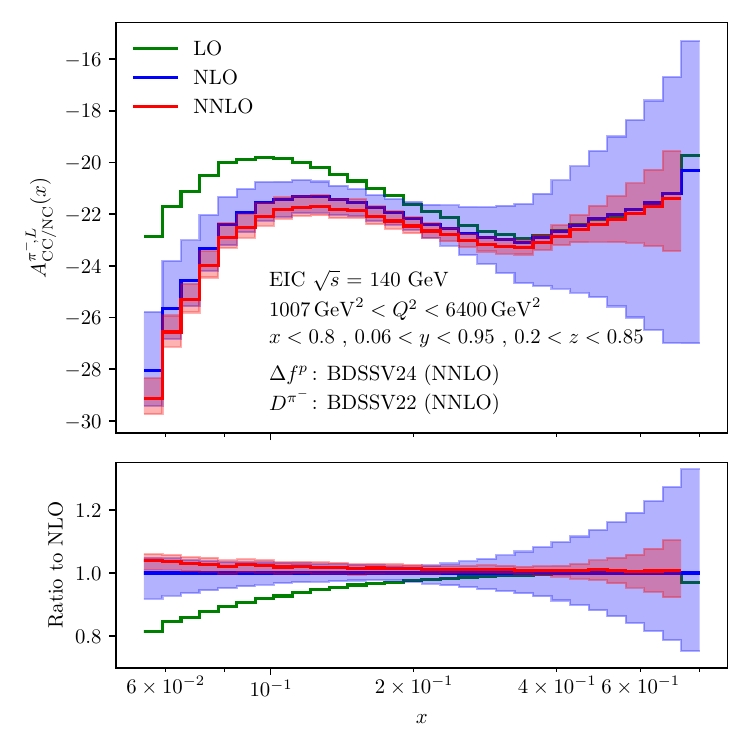}
\caption{High-$Q^2$}
\end{subfigure}
\begin{subfigure}{0.45\textwidth}
\centering
\includegraphics[width=1.0\linewidth]{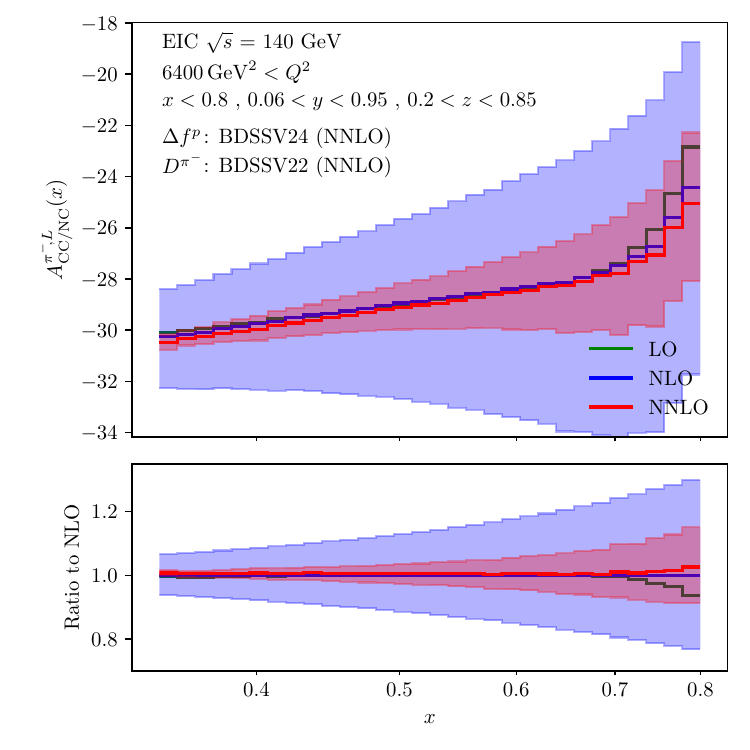}
\caption{Extreme-$Q^2$}
\end{subfigure}

\caption{
Electron-initiated longitudinal charged-neutral current asymmetry $A^{\pi^-,L}_\mathrm{CC/NC}$ as a function of $x$ in $\pi^-$ production.
Top panels: total asymmetry.
Bottom panels: Ratio to NLO.
}
\label{fig:X_A_CCNC_pim}
\end{figure}

In Fig.~\ref{fig:X_A_CCNC_pip} we show the electron-initiated longitudinal charged-neutral current asymmetry~\eqref{eq:ALCCNC} for $\pi^+$ production as a function of $x$.
Only at Low-$Q^2$ the absolute value of the asymmetry at NNLO is smaller than unity, corresponding to
 a smaller relative magnitude of the CC with respect to the NC $\lambda_\ell$-even contributions.
For Mid-$Q^2$, High-$Q^2$ and Extreme-$Q^2$ the asymmetry is always larger than one and increases in absolute value with increasing $x$.
At High-$Q^2$ and Extreme-$Q^2$ the CC contribution is about a factor of five larger than the NC $\lambda_\ell$-even contribution.
As shown in the bottom panels of Fig.~\ref{fig:X_A_CCNC_pip} perturbative corrections are at percent and sub-percent level for Mid-$Q^2$, High-$Q^2$ and Extreme-$Q^2$.
At Low-$Q^2$ we observe very large perturbative corrections for $x<0.01$.
This is due to all partonic channels becoming of similar (and small) size as an effect of the polarized PDFs at small $x$.

In Fig.~\ref{fig:X_A_CCNC_pim} the asymmetry~\eqref{eq:ALCCNC}
 is shown for $\pi^-$ production.
Overall the asymmetry is about a factor of 5 larger than for $\pi^+$ production and decreases with increasing $x$.
At Low-$Q^2$ and Mid-$Q^2$ the denominator crosses zero at values of $x$ that depend on the average virtuality,
leading to the observed instabilities in the asymmetry. Outside these instabilities, NNLO corrections are moderate,
and good convergence from NLO to NNLO is observed.

\begin{figure}[p]
\centering

\begin{subfigure}{0.45\textwidth}
\centering
\includegraphics[width=.99\linewidth]{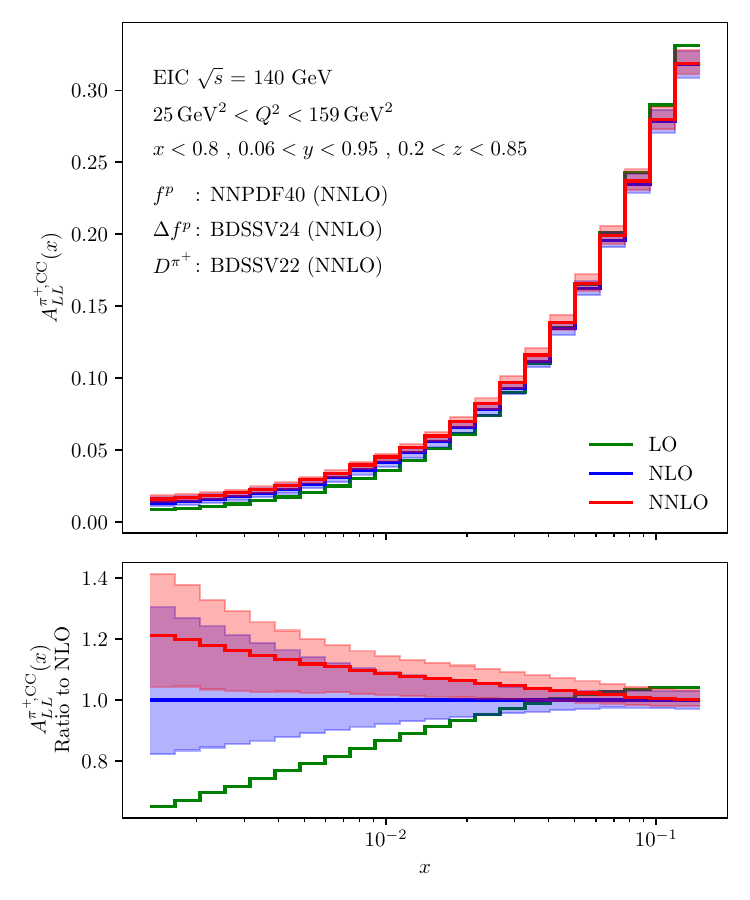}
\caption{Low-$Q^2$}
\end{subfigure}
\begin{subfigure}{0.45\textwidth}
\centering
\includegraphics[width=1.0\linewidth]{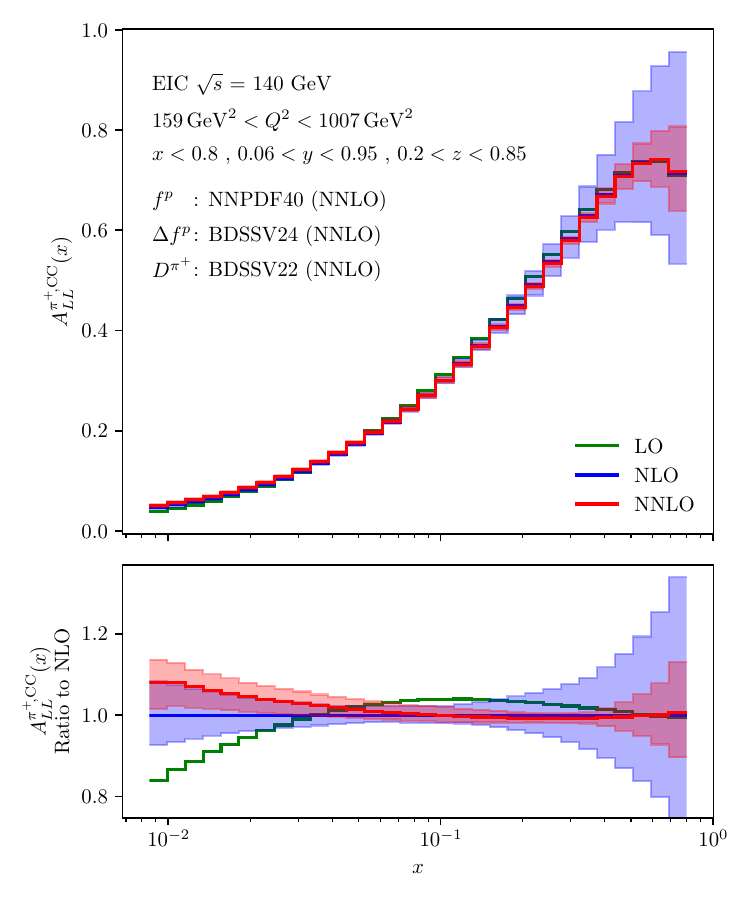}
\caption{Mid-$Q^2$}
\end{subfigure}
\begin{subfigure}{0.45\textwidth}
\centering
\includegraphics[width=1.0\linewidth]{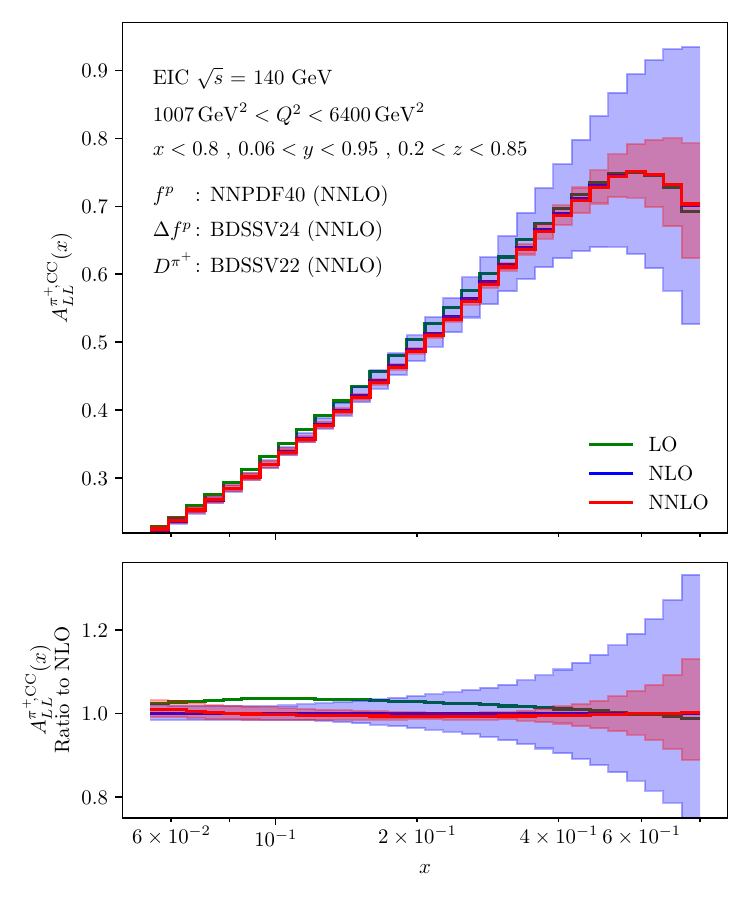}
\caption{High-$Q^2$}
\end{subfigure}
\begin{subfigure}{0.45\textwidth}
\centering
\includegraphics[width=1.0\linewidth]{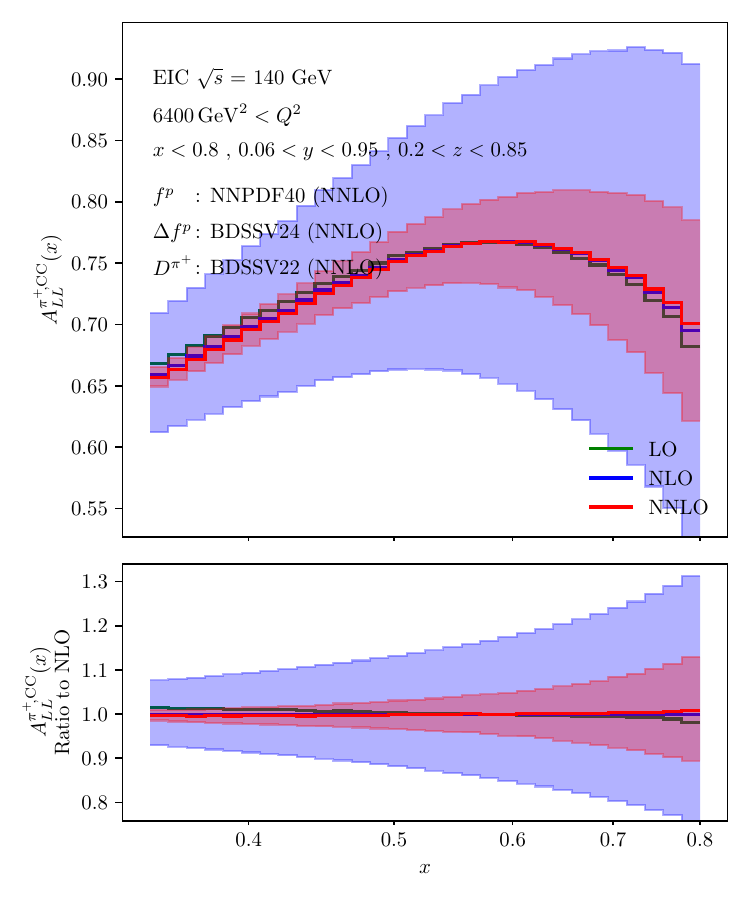}
\caption{Extreme-$Q^2$}
\end{subfigure}

\caption{
Electron-initiated CC double-longitudinal spin asymmetry $A^{\pi^+,\CC}_{LL}$ as a function of $x$ in $\pi^+$ production.
Top panels: total asymmetry.
Bottom panels: Ratio to NLO.
}
\label{fig:X_A_LL_CC_pip}

\end{figure}

\begin{figure}[p]
\centering

\begin{subfigure}{0.45\textwidth}
\centering
\includegraphics[width=1.0\linewidth]{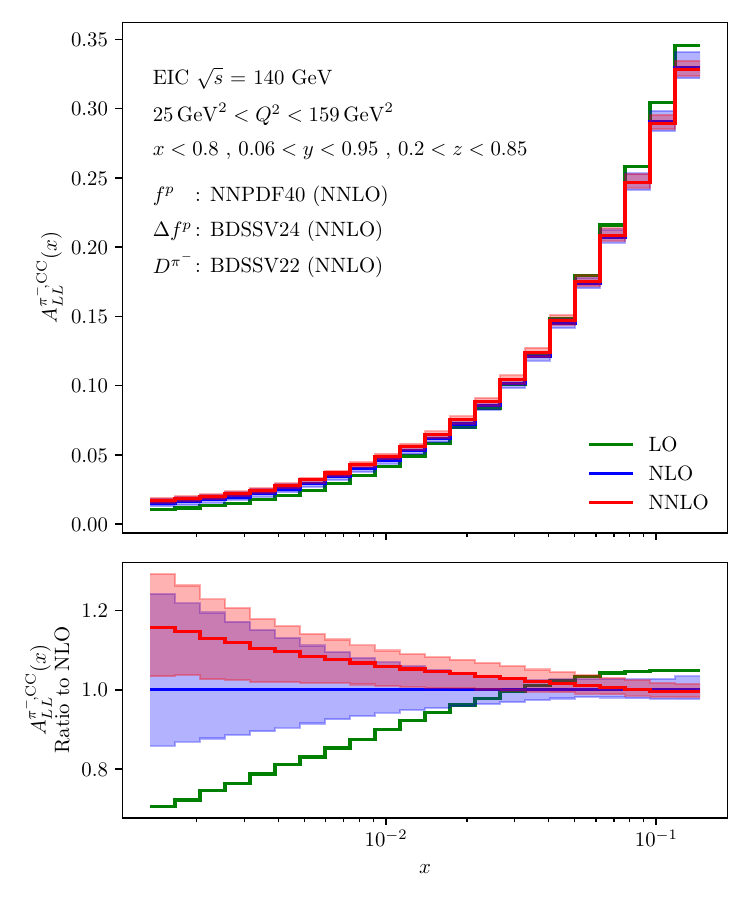}
\caption{Low-$Q^2$}
\end{subfigure}
\begin{subfigure}{0.45\textwidth}
\centering
\includegraphics[width=1.0\linewidth]{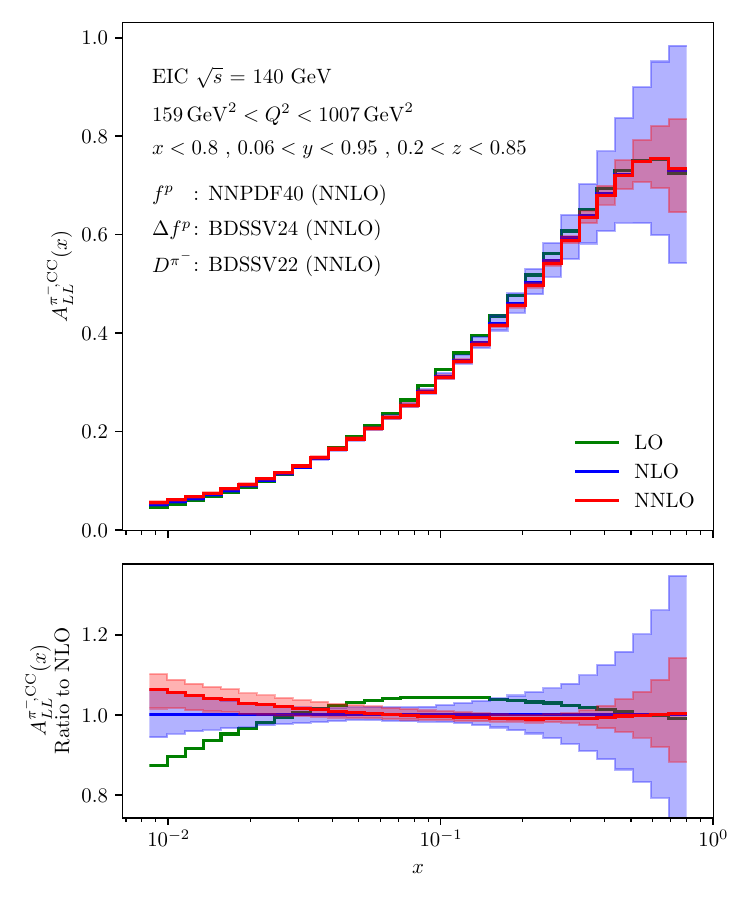}
\caption{Mid-$Q^2$}
\end{subfigure}
\begin{subfigure}{0.45\textwidth}
\centering
\includegraphics[width=1.0\linewidth]{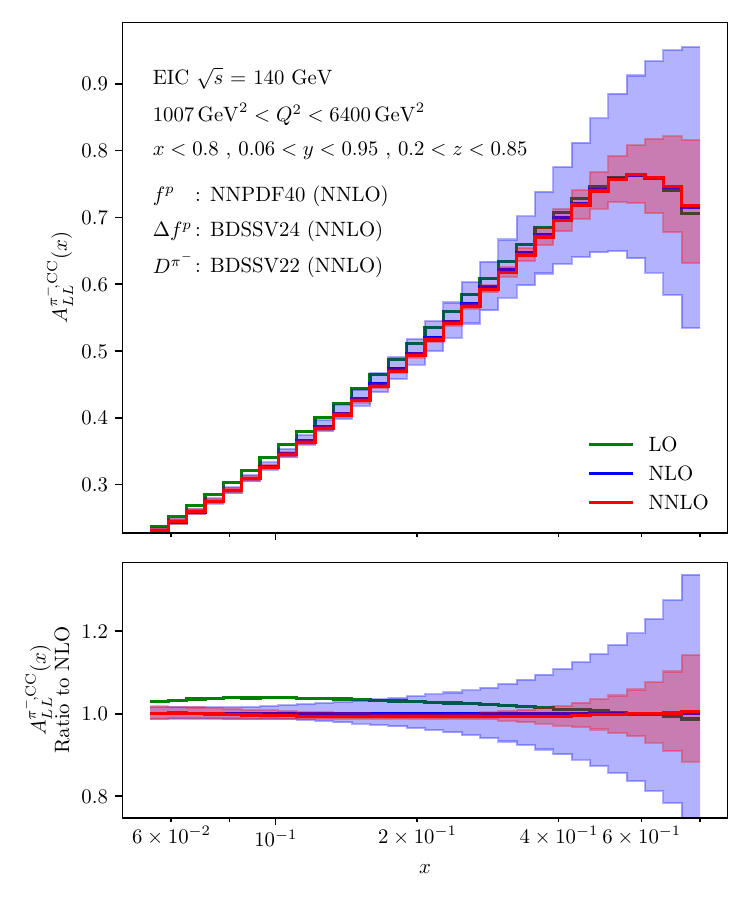}
\caption{High-$Q^2$}
\end{subfigure}
\begin{subfigure}{0.45\textwidth}
\centering
\includegraphics[width=1.0\linewidth]{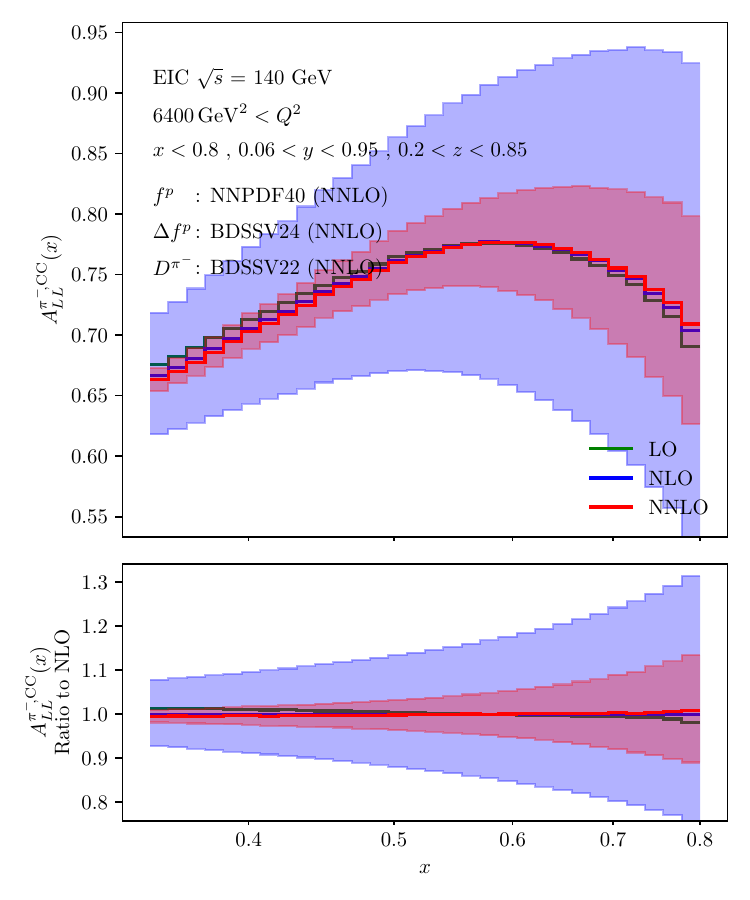}
\caption{Extreme-$Q^2$}
\end{subfigure}

\caption{
Electron-initiated CC double-longitudinal spin asymmetry $A^{\pi^-,\CC}_{LL}$ as a function of $x$ in $\pi^-$ production.
Top panels: total asymmetry.
Bottom panels: Ratio to NLO.
}
\label{fig:X_A_LL_CC_pim}
\end{figure}

In Fig.~\ref{fig:X_A_LL_CC_pip} we study the electron-initiated CC double-longitudinal spin
asymmetry~\eqref{eq:AhLLCC} for $\pi^+$ production as a function of $x$.
For Low-$Q^2$ the asymmetry grows with increasing $x$ up to a maximum of about $0.35$.
This is due to a quicker growth of the electroweak effects with the increase of $x$ (and therefore $Q^2$) in polarized SIDIS as compared to unpolarized SIDIS.
For Mid-$Q^2$ the asymmetry grows even faster with the increase of $Q^2$ reaching a maximum of almost $0.8$ at $x\sim 0.5$.
It then starts to decrease again due to the depletion of polarized PDFs at higher values of $x$.
A similar behaviour can be observed at High-$Q^2$ where the asymmetry grows from $\sim 0.3$ up to $\sim 0.75$ at $x\sim 0.5$.
Finally at Extreme-$Q^2$ the depletion of polarized PDFs is even more visible: the asymmetry has a moderate $x$ dependence and it averages around $0.7$.
Perturbative corrections become smaller with the increase of the average value of $Q^2$ as visible in the bottom panels of Fig.~\ref{fig:X_A_LL_CC_pip} from a ratio to NLO.
At Low-$Q^2$ and small $x$ NLO corrections are above $20\%$ of the LO prediction and NNLO corrections
yield another 20\%.
Perturbative corrections become smaller with the increase of $x$ and reduce to a few percent level.
At Mid-$Q^2$ perturbative corrections start around the $10\%$ level at lower values of $x$ to diminish to sub percent level at higher values of $x$.
At High-$Q^2$ and Extreme-$Q^2$ perturbative corrections are at few percent level and below for all values of $x$.
We observe overall good perturbative convergence and stability with NNLO corrections always within NLO theory uncertainties.

In Fig.~\ref{fig:X_A_LL_CC_pim} we study the electron-initiated CC double-longitudinal spin asymmetry~\eqref{eq:AhLLCC} for $\pi^-$ production as a function of $x$.
The overall behaviour of the asymmetry and of its perturbative corrections is very similar with respect to $\pi^+$ production and we refer to the above paragraph for a more detailed discussion.
The flavour-favoured $\pi^-$ asymmetry is only slightly larger than the $\pi^+$ flavour-unfavoured one due to the equal sensitivity to the flavour transition probed in numerator and denominator of~\eqref{eq:AhLLCC}.

We finally notice that both asymmetries as a function of $x$ of Fig.~\ref{fig:X_A_LL_CC_pip} and Fig.~\ref{fig:X_A_LL_CC_pim} are larger than the same asymmetries given as a function of $z$ in Fig.~\ref{fig:Z_A_LL_CC_pip} and Fig.~\ref{fig:Z_A_LL_CC_pim}.
This is due to a greater sensitivity of the $x$-distributions to the details of the polarized PDFs.

\section{Conclusion}
\label{sec:conc}

In this paper, we derived the NNLO QCD corrections to semi-inclusive hadron production cross sections in
deeply inelastic charged lepton scattering on longitudinally polarized nucleons. We extended previous NNLO results
for polarized SIDIS for photon exchange~\cite{Bonino:2024wgg,Goyal:2024tmo,Goyal:2024emo} by accounting for
electroweak gauge boson exchanges in neutral and charged current processes.
In rederiving these earlier
results~\cite{Bonino:2024wgg,Goyal:2024tmo,Goyal:2024emo}, we identified a previous misassignment of finite
renormalization contributions between flavour non-singlet and singlet channels, which we now overcome by consistently
working on the basis of individual flavour channels for initial and final state partons, as documented in detail
in Appendix~\ref{sec:Scheme_flavour_basis}.

We observe in our computation of the coefficient functions of the $\Gcal^h_T$ and $\Gcal^h_L$ structure functions that applying the polarized quark projector in a na\"ive Larin prescription yields incorrect results.
We resolve this issue by using a projector for polarized quarks,~\eqref{eq:quark_projector}, which was previously
introduced in the context of polarized OMEs in~\cite{Schonwald:2019gmn,Behring:2019tus} to eliminate evanescent
structures in the quark polarization.

Based on the newly derived NNLO corrections, we perform a detailed phenomenological study of longitudinally polarized
SIDIS neutral and charged current processes for the kinematics of the EIC, differentially in
$Q^2$, $z$ and $x$.
Several new polarized structure functions become accessible through the chiral couplings of the weak gauge bosons.
Starting from the cross section differences between the two nucleon polarizations, we investigate the role
of contributions that are even or odd in the incoming lepton helicity (allowing the construction of single or double spin
asymmetries). In the NC case, we observe that electroweak tree-level corrections
to the photon-exchange process in polarized SIDIS are typically one order of
magnitude larger than in the unpolarized case~\cite{Bonino:2025qta}.
Even at relatively low values of $Q^2 \sim 25 \, \GeV^2$ we observe percent-level corrections, which rapidly grow to around $ 10\,\%$ at $200\,\GeV^2$, cf.\ Fig.~\ref{fig:Q2_NC_pip}.
This feature can be largely understood from
the chiral coupling nature of the electroweak gauge bosons, whose contribution becomes enhanced by taking
nucleon polarization differences.
In contrast to the unpolarized case an estimate of electroweak effects obtained by rescaling the electroweak leading order by photon K-factors gives only gives percent-level agreement with the full calculation.

For all NC and CC SIDIS processes, we investigate the flavour-decomposition of the underlying partonic processes.
We observe a dominance of flavour combinations relating to the valence content of the nucleon target and the observed
hadron. This dominance is more pronounced than in unpolarized SIDIS~\cite{Bonino:2025qta}, and it persists at
small $x$, resulting from the suppression of gluon and sea quark polarized PDFs compared to their unpolarized
counterparts.

By projecting onto even and odd contributions in the lepton helicity, we construct single and double spin asymmetries.
In many cases these turn out to be sizeable at EIC kinematics and offer opportunities to probe the underlying polarized structure functions.

\acknowledgments
We would like to thank Ignacio Borsa and Daniel de Florian for useful discussions and the authors of~\cite{Goyal:2024tmo,Goyal:2024emo} for confirming the corrections to the scheme transformation previously applied in~\cite{Bonino:2024wgg,Goyal:2024tmo,Goyal:2024emo}, as described in Section \ref{sec:checks} and Appendix~\ref{sec:Scheme_transform}.
This work has received funding from the Swiss National Science Foundation (SNF) under contract 200020-204200 and from the European Research Council (ERC) under the European Union’s Horizon 2020 research and innovation program grant agreement 101019620 (ERC Advanced Grant TOPUP).
The work of KS was supported by the UZH Postdoc Grant, grant no.~[FK-24-115].

\appendix

\section{Mass Factorization and Finite Polarized Scheme Transformation}
\label{sec:Scheme_transform}
Apart from the divergent and multiplicative renormalization of the axial vector vertices, another finite renormalization is required to restore the conservation of the external polarized current, transforming it into the $\MSbar$ scheme.
In Appendix~\ref{sec:Scheme_NS_basis} we review the formalism in the non-singlet/singlet basis, and reformulate it in Appendix~\ref{sec:Scheme_flavour_basis} in the flavour basis.

\subsection{Finite Polarized Scheme Transformation in the Non-Singlet/Singlet Basis}
\label{sec:Scheme_NS_basis}
The finite polarized scheme transformation has been worked out within operator matrix element calculus~\cite{Matiounine:1998re} and has been successfully applied to the calculation of the polarized splitting functions~\cite{Moch:2014sna,Blumlein:2021enk,Blumlein:2021ryt}
and the \emph{inclusive} Wilson coefficients to the $g_1$ structure function~\cite{Blumlein:2022gpp} up to three loops in the non-singlet case.

Based on scheme invariance of the physical observable $\Gcal_1^h$ the scheme transformation is
schematically realised in form of a matrix equation,
\begin{align}
\Gcal_1^h &= D^h \otimes \left[\Delta C^{1,\MSbar} \otimes \Delta f^{\MSbar} \right]
\nonumber \\ &
= D^h \otimes \left[ (\Delta C^{1,\Larin} \otimes Z^{-1} ) \otimes ( Z \otimes \Delta f^{\Larin} )\right]
\nonumber \\ &
\equiv D^h \otimes \left[ \Delta C^{1,\Larin} \otimes \Delta f^{\Larin}\right] \, .
\end{align}
For the following discussion we omit the convolution with the fragmentation function, which is carried out in a different kinematic variable ($z$ instead of $x$) and remains unaffected by the initial state scheme transformation.

The finite polarized scheme transformation is formulated in the basis of non-singlet (NS) and singlet (S) splitting and coefficient functions~\cite{Matiounine:1998re,Moch:2014sna}. The non-singlet scheme transformation is given by
\begin{align}
\Delta C^{1,\MSbar}_\text{NS} \otimes \Delta f_{q}^{\mathrm{NS}}
=  (\Delta C^{1,\Larin}_\mathrm{NS} \otimes z^{-1}_{qq,\mathrm{NS}} ) \otimes ( z_{qq,\mathrm{NS}} \otimes \Delta f_{q}^{\mathrm{NS},\Larin} ) \,.
\end{align}
For the singlet the scheme transformation is given by the matrix equation
\begin{align*}
& \begin{pmatrix}
\Delta C^{1,\MSbar}_{q,\mathrm{S}} & \Delta C_g^{1,\MSbar}
\end{pmatrix}
\otimes
\begin{pmatrix}
\Delta \Sigma^{\MSbar} \\ \Delta f_{g}^{\MSbar}
\end{pmatrix} \\
& =
\begin{pmatrix}
\Delta C^{1,\Larin}_{q,\mathrm{S}} & \Delta C_g^{1,\Larin}
\end{pmatrix}
\otimes
\begin{pmatrix}
z_{qq,\mathrm{S}} & z_{qg,\mathrm{S}} \\
z_{gq,\mathrm{S}} & z_{gg,\mathrm{S}}
\end{pmatrix}^{-1}
\otimes
\begin{pmatrix}
z_{qq,\mathrm{S}} & z_{qg,\mathrm{S}} \\
z_{gq,\mathrm{S}} & z_{gg,\mathrm{S}}
\end{pmatrix}
\otimes
\begin{pmatrix}
\Delta \Sigma^{\Larin} \\ \Delta f_{g}^{\Larin}
\end{pmatrix}
\, , \numberthis
\end{align*}
with the polarized singlet distribution $\Delta \Sigma = \sum_q \left(\Delta f_{q} + \Delta f_{\bar{q}}\right)$.

The scheme transformation coefficients admit a series expansion in terms of the strong coupling constant,
\begin{align}
z_{ij,\mathrm{(N)S}} = \delta(1-x) \, \delta_{ij} + \sum_{k=1}^\infty \left( \frac{\alpha_s}{2\pi} \right)^k z_{ij,\mathrm{(N)S}}^{(k)}
\end{align}
The expressions for the series coefficients of $z_{qq,\mathrm{NS}}$ and $z_{qq,\mathrm{PS}} = z_{qq,\mathrm{S}} - z_{qq,\mathrm{NS}}$ up to NNLO are given in~\cite{Matiounine:1998re,Moch:2014sna}.
In the conventional transformation to the $\MSbar$ scheme the following relations are imposed:
$z_{qg} = z_{gq} = 0$, and $z_{gg} = \delta(1-x)$.

\subsection{Finite Polarized Scheme Transformation in the Flavour Basis}
\label{sec:Scheme_flavour_basis}
Unlike in inclusive DIS a final state particle is identified in SIDIS.
As the finite scheme transformations renormalize the polarized splitting functions, they inherit the particle-identity- or flavour-changing nature of the splittings.
Explicit formulae for the mass factorization and the finite renormalization of the individual channels in polarized SIDIS~\cite{Bonino:2024wgg,Goyal:2024tmo} are presented in~\cite{Goyal:2024emo}.

Expressing the finite scheme transformation in a flavour basis, we are able to uniquely associate the contributions to the SIDIS channels.
In doing so we find differences in the implementation of the finite renormalization.
The differences originate from misallocating the finite renormalization counter terms with the appropriate SIDIS channels in~\cite{Bonino:2024wgg,Goyal:2024tmo,Goyal:2024emo}.

Working in the basis of partonic channels the finite renormalization can be written as
\begin{align}
	\Delta C_{p^\prime p}^{i,k} &= \Delta C_{p^\prime m}^{i,k,\Larin} \otimes \left( Z^{-1} \right)_{m p} \, ,
\end{align}
where $p,m,p^\prime \in \left\{ q, \qb, \qp, \qbp, g \right\}$ with $\qp\neq q$ and a summation over $m$ is implied.
The coefficient functions with the superscript $\Larin$ are the UV-renormalized and mass factorized
coefficients obtained using Larin's prescription for $\gamma_5$, while the ones without additional superscript
are the ones in the $\MSbar$-scheme.
The matrix $Z$ is given by
\begin{align}
	Z &=
	\begin{pmatrix}
		Z_{qq} & Z_{q\qb} & Z_{q \qp} & Z_{q \qbp} & Z_{q g} \\
		Z_{\qb q} & Z_{\qb\qb} & Z_{\qb \qp} & Z_{\qb \qbp} & Z_{\qb g} \\
		Z_{\qp q} & Z_{\qp \qb} & Z_{\qp \qp} & Z_{\qp \qbp} & Z_{\qp g} \\
		Z_{\qbp q} & Z_{\qp \qb} & Z_{\qbp \qp} & Z_{\qbp \qbp} & Z_{\qbp g} \\
		Z_{gq} & Z_{g\qb} & Z_{g \qp} & Z_{g \qbp} & Z_{g g}
	\end{pmatrix} \, .
\end{align}
Due to the symmetries of QCD $\left(Z\right)_{ij} = \left(Z\right)_{ji}$ is imposed for
$i,j \in \left\{ q, \qb, \qp, \qbp \right\}$ and
\begin{alignat}{4}
&	Z_{qq} &&= Z_{\qb\qb} &&= Z_{\qp\qp} &&= Z_{\qbp \qbp}   \, , \\
&	Z_{q \qp} &&= Z_{q \qbp} &&= Z_{\qb \qbp} &&= Z_{\qb \qp} \, , \\
&	Z_{q g} &&= Z_{\qb g} &&= Z_{\qp g} &&= Z_{\qbp g} \, , \\
&	Z_{g q} &&= Z_{g \qb} &&= Z_{g \qp} &&= Z_{g \qbp} \, .
\end{alignat}

\subsubsection*{The non-singlet flavour decomposition}
In order to find the relation between the independent terms in the flavour basis and the renormalization terms in the non-singlet/singlet basis, it is sufficient to find the equivalent relation for the polarized splitting functions.
The polarized splitting functions can be decomposed into
\begin{align}
\Delta P_{q_i q_k} &= \delta_{ik} \Delta P_{qq}^v + \Delta P_{qq}^s \, , &
\Delta P_{q_i \qb_k} &= \delta_{ik} \Delta P_{q\qb}^v + \Delta P_{q\qb}^s \, .
\end{align}
It is instructive to look into the decomposition of the flavour asymmetry
\begin{align}
\Delta f_{q_i q_k}^{\mathrm{NS},\pm} = (\Delta f_{q_i} \pm \Delta f_{\qb_i}) - (\Delta f_{q_k} \pm \Delta f_{\qb_k})\, .
\end{align}
The flavour asymmetry by the above definitions evolves as
\begin{align*}
\frac{\dd \Delta f_{q_i q_k}^{\mathrm{NS},\pm}}{\dd \log \mu^2} &=
\sum_a \biggl[
	\biggl(
		\Delta P_{q_i q_a}
		\pm  \Delta P_{\qb_i q_a}
	\biggr) \otimes \Delta f_{q_a}
	+ \biggl(
		\Delta P_{q_i  \qb_a}
		\pm \Delta P_{\qb_i \qb_a}
	\biggr) \otimes \Delta f_{\qb_a}
	\biggr]
	- \left(i\leftrightarrow k\right) \\
&= \left( \Delta P_{qq}^v \pm \Delta P_{q\qb}^v \right) \otimes  \Delta f_{q_i q_k}^{\mathrm{NS},^\pm} \\
&= \Delta P_\mathrm{NS}^\pm \otimes \Delta f_{q_i q_k}^{\mathrm{NS},^\pm} \, . \numberthis
\end{align*}
Therefore,
\begin{align}
\Delta P_{qq}^v &= \frac{1}{2} \left[ \Delta P_\mathrm{NS}^+ + \Delta P_\mathrm{NS}^- \right] \, , &
\Delta P_{q\qb}^v &= \frac{1}{2} \left[ \Delta P_\mathrm{NS}^+ - \Delta P_\mathrm{NS}^- \right] \, .
\end{align}

We now separate the even and odd Mellin moments of the polarized splitting functions.
We observe that the structure function $\Gcal^h_1 \propto (\Delta q + \Delta \qb) $ only yields odd moments in Mellin-$N$ space. Therefore,
\begin{align}
\Delta P_\mathrm{NS}^{+} &= \frac{1-(-1)^N}{2} \Delta P_\mathrm{NS}^{+} \, , &
\Delta P_\mathrm{NS}^{-} &= \frac{1+(-1)^N}{2} \Delta P_\mathrm{NS}^{-} \, ,
\end{align}
using that $[1\pm(-1)^N]/2$ projects onto the even/odd moments.

We can therefore define
\begin{align}
	\Delta P_{qq}^{\mathrm{NS}}(N) = \Delta P_\mathrm{NS}^{+}(N) + \Delta P_\mathrm{NS}^{-}(N) &= \int\limits_{0}^{1} \dd x \, x^{N-1} \biggl(
		\Delta P_{qq}^{v} - (-1)^N \Delta P_{\bar{q}q}^{v}
		\biggr) \, .
\end{align}
In this way odd moments ($(-1)^N \to -1$) give the combination $\Delta P_{qq}^v + \Delta P_{\qb q}^v$ and
even moments ($(-1)^N \to 1$) give the combination $\Delta P_{qq}^v - \Delta P_{\qb q}^v$.
This is in contrast to the unpolarized case, where even moments describe the
combination $P_{qq}^v + P_{\qb q}^v$ and for which we therefore have to define
\begin{align}
	P_{qq}^{\text{NS}}(N) &= \int\limits_{0}^{1} \dd x \, x^{N-1} \biggl(
		P_{qq}^{v} + (-1)^N P_{\bar{q}q}^{v}
		\biggr) \, .
\end{align}
The same discussion applies to the additional finite renormalization, for which we find
\begin{align}
	z_{qq}^{\text{NS}}(N) &= \int\limits_{0}^{1} {\rm d}x \, x^{N-1} \biggl(
		z_{qq}^{v} - (-1)^N z_{\bar{q}q}^{v}
	\biggr) \, .
\end{align}

\subsubsection*{The pure-singlet flavour decomposition}
For the singlet distribution the evolution is given by
\begin{align}
\frac{\dd \Delta \Sigma}{\dd \log \mu^2}  = \left[ \Delta P_\mathrm{NS}^+ + \nf \left( \Delta P_{qq}^s + \Delta P_{q\qb}^s \right) \right] \otimes \Delta \Sigma + 2 \nf \,  \Delta P_{qg} \otimes \Delta f_{g} \,.
\end{align}
The pure-singlet piece of the evolution is $\Delta P_{qq}^\mathrm{PS} = \nf \left( \Delta P_{qq}^s + \Delta P_{q\qb}^s \right)$.
Since the pure-singlet diagrams are insensitive to the fermion flow of the external quark, we find
\begin{align}
\Delta P_{qq}^s= \Delta P_{q\qb}^s = \frac{1}{2 \nf}\, \Delta P_{qq}^\mathrm{PS} \, .
\end{align}
As a consequence the finite pure-singlet scheme transformation in the flavour basis is given by
\begin{align}
z_{qq}^{s} = z_{q\qb}^{s} = \frac{1}{2\nf} z_{qq}^\mathrm{PS} \, .
\end{align}

\subsubsection*{Flavour decomposition for the finite scheme transformation}

The independent terms of the finite scheme transformation were calculated in \cite{Matiounine:1998re}
up to $\mathcal{O}(\alpha_s^2)$ and read
\begin{alignat}{2}
&	Z_{qq} &&=
	\delta(1-x)
	+ \frac{\alpha_s}{2\pi} z_{qq}^{v,(0)}
	+ \left(\frac{\alpha_s}{2\pi} \right)^2 \left( z_{qq}^{v,(1)} + z_{qq}^{s,(1)} \right)\,, \\
&	Z_{\bar{q}q} &&= \left(\frac{\alpha_s}{2\pi} \right)^2 \left(
		z_{\bar{q}q}^{v,(1)} + z_{qq}^{s,(1)}
	\right) \,, \\
&	Z_{q^\prime q} &&= \left(\frac{\alpha_s}{2\pi} \right)^2 z_{qq}^{s,(1)} \,, \\
&	Z_{q g} &&= 0 \,, \\
&	Z_{g q} &&= 0 \,, \\
&	Z_{g g} &&= 0 \,,
\end{alignat}
with
\begin{align}
	z_{qq}^{v,(0)} &= z_{qq}^{(0)} = -4 \CF (1 - x)  \,, \\
	z_{qq}^{v,(1)} &=
	\CF \TR \nf (1-x)\biggl[
		\frac{20}{9} + \frac{4}{3} \ln(x)
	\biggr]
	- \CF^2 \biggl[
		4 (1 - x)
		+ \bigl(
			2 (2 + x)
			\nonumber \\ &
			- 4 (1 - x) \ln(1 - x)
		\bigr) \ln(x)
	\biggr]
	- \CF \CA  \biggl[
		 \frac{148}{9} (1 - x)
		- \frac{1}{3} \pi^2 (1 - x)
		\nonumber \\ &
		+ \frac{2}{3} (10 - x) \ln(x)
		+ (1 - x) \ln(x)^2
	\biggr] \,, \\
	z_{\qb q}^{v,(1)} &= -\CF \left( \CF - \frac{\CA}{2} \right)
	\biggl[
		- 14 (1-x)
		+\frac{2}{3} \pi^2 (1 + x)
		- 2 (1 + x) \ln(x)^2
		\nonumber \\ &
		- (1 + x) \left(
			  6
			- 8 \ln(1 + x)
		\right) \ln(x)
		+ 8 (1+x) \text{Li}_{2}(-x)
	\biggr] \,, \\
	z_{qq}^{s,(1)} &= z_{\qb q}^{s,(1)} = \frac{1}{2\nf} z_{qq}^{\text{PS},(1)} =
	\CF \TR \biggl(
		2(1 - x)
		+ (3 - x) \ln(x)
		+ \frac{1}{2}(2 + x) \ln^2(x)
	\biggr) \,,
\end{align}
according to our flavour decomposition.
Here $\tr[T^a T^b] = \TR \delta_{ab}$ for the colour group generators and $\CA = N_c$ and $\CF=\TR(N_c^2-1)/N_c$ are the colour coefficients.
Note that we assume $\qp \neq q$ in this discussion.

\subsection{The Finite Polarized Scheme Transformation in Polarized SIDIS}

Up to $\mathcal{O}(\alpha_s^2)$ we find the following expressions for the finite renormalization of the polarized splitting functions,
\begin{align*}
	\Delta \Ccal_{q q}^{i,k} &= \Delta \Ccal_{q q}^{i,k,\Larin}
	- \frac{\alpha_s}{2 \pi}
	\biggl[
		\Delta \Ccal_{q q}^{i,k,(0),\Larin} \otimes z_{qq}^{v,(0)}
	\biggr]
			\\ &
	+ \left(\frac{\alpha_s}{2 \pi}\right)^2
	\biggl[
		\Delta \Ccal_{q q}^{i,k,(0),\Larin} \otimes\left(z_{qq}^{v,(0)} \otimes z_{qq}^{v,(0)} - z_{qq}^{v,(1)} - z_{qq}^{s,(1)} \right)
		- \Delta \Ccal_{q q}^{i,k,(1),\Larin} \otimes z_{qq}^{v,(0)}
	\biggr] \hspace{-0.404pt} \, ,\\
	\Delta \Ccal_{\qb q}^{i,k} &= \Delta \Ccal_{\qb q}^{i,k,\Larin}
	- \left(\frac{\alpha_s}{2 \pi}\right)^2
	\biggl[
		\Delta \Ccal_{\qb \qb}^{i,k,(0),\Larin} \otimes
		\left( z_{\qb q}^{v,(1)} + z_{\qb q}^{s,(1)} \right)
	\biggr] \\
	&= \Delta \Ccal_{\qb q}^{i,k,\Larin}
	- \left(\frac{\alpha_s}{2 \pi}\right)^2
	\biggl[
		\Delta \Ccal_{\qb \qb}^{i,k,(0),\Larin} \otimes
		\left( z_{\qb q}^{v,(1)} + z_{q q}^{s,(1)} \right)
	\biggr] 	\, ,\\
	\Delta \Ccal_{\qp q}^{i,k} &= \Delta \Ccal_{\qp q}^{i,k,\Larin}
	- \left(\frac{\alpha_s}{2 \pi}\right)^2
	\biggl[
		\Delta \Ccal_{\qp \qp}^{i,k,(0),\Larin} \otimes z_{\qp q}^{s,(1)}
	\biggr]
	\\
	&= \Delta \Ccal_{\qp q}^{i,k,\Larin}
	- \left(\frac{\alpha_s}{2 \pi}\right)^2
	\biggl[
		\Delta \Ccal_{\qp \qp}^{i,k,(0),\Larin} \otimes z_{q q}^{s,(1)}
	\biggr]
	\, ,\\
	\Delta \Ccal_{\qbp q}^{i,k} &= \Delta \Ccal_{\qbp q}^{i,k,\Larin}
	- \left(\frac{\alpha_s}{2 \pi}\right)^2
	\biggl[
		\Delta \Ccal_{\qbp \qbp}^{i,k,(0),\Larin} \otimes z_{\qbp q}^{s,(1)}
	\biggr]
	\\
	&= \Delta \Ccal_{\qbp q}^{i,k,\Larin}
	- \left(\frac{\alpha_s}{2 \pi}\right)^2
	\biggl[
		\Delta \Ccal_{\qbp \qbp}^{i,k,(0),\Larin} \otimes z_{q q}^{s,(1)}
	\biggr]
	\, ,\\
	\Delta \Ccal_{g q}^{i,k} &= \Delta \Ccal_{g q}^{i,k,\Larin}
	- \frac{\alpha_s}{2 \pi}
	\biggl[
		\Delta \Ccal_{g q}^{i,k,(1),\Larin} \otimes z_{qq}^{v,(0)}
	\biggr]
	\, ,\\
	\Delta \Ccal_{q g}^{i,k} &= \Delta \Ccal_{q g}^{i,k,\Larin}
	\, ,\\
	\Delta \Ccal_{g g}^{i,k} &= \Delta \Ccal_{g g}^{i,k,\Larin} \, , \numberthis
\end{align*}
where we still demand $q^\prime \neq q$.
The other partonic channels can be
obtained by symmetries.
Note that the couplings of the lower order coefficient functions need to be adapted to the partonic content, e.g.\ $\Delta\Ccal_{\qbp\qbp}^{1,k,(0)} = (\vh_{\qp\qp}^{k} + \ah_{\qp\qp}^k)/(\vh_{qq}^{k} + \ah_{qq}^k) \, \Delta\Ccal_{\qb\qb}^{1,k,(0)}$ for NC exchange.
At the level of the building blocks, which are stripped of their couplings, the scheme transformation reads
\begin{align*}
\DC_{qq}^{i,(1)} &=
	\DC_{qq}^{i,(1),\Larin}
	- \frac{\alpha_s}{2 \pi}
	\biggl[
		\DC_{q q}^{i,(0),\Larin} \otimes z_{qq}^{v,(0)}
	\biggr]
	\, , \\
\DC_{qq}^{i,(2)} &=
	\DC_{qq}^{i,(2),\Larin}
	\\&
	+ \left(\frac{\alpha_s}{2 \pi}\right)^2
	\biggl[
		\DC_{q q}^{i,(0),\Larin} \otimes\left(z_{qq}^{v,(0)} \otimes z_{qq}^{v,(0)} - z_{qq}^{v,(1)} \right)
		- \DC_{q q}^{i,(1),\Larin} \otimes z_{qq}^{v,(0)}
	\biggr] \, ,\\
	\DC_{\qb q}^{i,(2)} &= \DC_{\qb q}^{i,(2),\Larin}
	- \left(\frac{\alpha_s}{2 \pi}\right)^2
	\biggl[
		\DC_{\qb\qb}^{i,(0),\Larin} \otimes
		z_{\qb q}^{v,(1)}
	\biggr] 	\, ,\\
	\DC_{\qp q,2}^{i,(2)} &= \DC_{\qp q,2}^{i,(2),\Larin}
	- \left(\frac{\alpha_s}{2 \pi}\right)^2
	\biggl[
		\DC_{qq}^{i,(0),\Larin} \otimes z_{q q}^{s,(1)}
	\biggr]
	\, ,\\
	\DC_{g q}^{i,(2)} &= \DC_{g q}^{i,(2),\Larin}
	- \frac{\alpha_s}{2 \pi}
	\biggl[
		\DC_{g q}^{i,(1),\Larin} \otimes z_{qq}^{v,(0)}
	\biggr]
	\, .
\end{align*}

\addtocontents{toc}{\protect\enlargethispage{\baselineskip}}
\section{Ancillary files with analytical expressions for coefficient functions}
\label{sec:files}

The ancillary file contains all coefficient functions up to NNLO as listed in Sections \ref{sec:polNCAntisymmCF}, \ref{sec:polNCSymmCF},  \ref{sec:polCCAntisymmCF} and \ref{sec:polCCSymmCF} in \texttt{FORM} readable format.
The notation used is explained in the header of the ancillary file and is consistent with the one adopted in the sections mentioned above.
The coefficient functions contained in the ancillary file are stripped of their couplings.
They are written in terms of plus distributions and contain the full dependence on the renormalization scale and the initial-state and final-state factorization scale.
Special functions appearing in the coefficient functions are also specified in the header of the ancillary file.
The following naming scheme is adopted in the ancillary file
\begin{align}
  \texttt{\textcolor{MidnightBlue}{[{Structure function}]}\textcolor{Green}{C}\textcolor{BrickRed}{[order]}\textcolor{Fuchsia}{[{PDF}]}2\textcolor{Red}{[{FF}]}\textcolor{BlueViolet}{[labels]}} \,,
\end{align}
e.g.
\begin{align}
  \texttt{\textcolor{MidnightBlue}{G1}\textcolor{Green}{C}\textcolor{BrickRed}{0}\textcolor{Fuchsia}{q}2\textcolor{Red}{q}\textcolor{BlueViolet}{M}}
  \leftrightarrow
  \textcolor{MidnightBlue}\Delta\textcolor{Green}{C}^{\textcolor{MidnightBlue}{1},\textcolor{BrickRed}{(0)}}_{\textcolor{Red}{q}\textcolor{Fuchsia}{q}}
\end{align}
with the label \textcolor{BlueViolet}{\texttt{M}} denoting that the coefficient function is in the $\overline{\mathrm{MS}}$ scheme.
Like throughout the paper, the label \texttt{NoW} denotes that the coefficient is only present if the flavour is conserved at the electroweak vertex.
The label \texttt{Fcon} denotes flavour-conserving contributions if the flavour is not conserved at the electroweak vertex, and the labels \texttt{A} and \texttt{AA} denote single and double anomaly insertions.
The names of the coefficient functions in this paper and their notation in the ancillary file are listed in Table~\ref{tab:ancillary}.

\renewcommand{\arraystretch}{1.3}
\begin{table}[p]
\centering
\resizebox{\textwidth}{!}{%
\begin{tabular}{c|c|c|c|c|c}
\toprule
  \multicolumn{2}{c|}{$\Gcal^h_1$ } &
  \multicolumn{2}{c|}{$\Gcal^h_T$} &
  \multicolumn{2}{c}{$\Gcal^h_L$} \\
\midrule
 paper & ancillary & paper & ancillary & paper & ancillary \\
 \midrule
 $\Delta C^{1,(0)}_{qq}$  & \texttt{G1C0q2qM} & $\Delta C^{T,(0)}_{qq}$ & \texttt{GTC0q2qM} &  & \\
 \midrule
 $\Delta C^{1,(1)}_{qq}$ & \texttt{G1C1q2qM} & $\Delta C^{L,(1)}_{qq}$ & \texttt{GTC1q2qM}  &  $\Delta C^{L,(1)}_{qq}$ & \texttt{GLC1q2qM} \\
 $\Delta C^{1,(1)}_{gq}$ & \texttt{G1C1q2gM} & $\Delta C^{L,(1)}_{gq}$ & \texttt{GTC1q2gM}  &  $\Delta C^{L,(1)}_{gq}$ & \texttt{GLC1q2gM} \\
 $\Delta C^{1,(1)}_{qg}$ & \texttt{G1C1g2qM} & $\Delta C^{L,(1)}_{qg}$ & \texttt{GTC1g2qM}  &  $\Delta C^{L,(1)}_{qg}$ & \texttt{GLC1g2qM} \\
  \midrule
 $\Delta C^{1,(2)}_{qq}$                                   & \texttt{G1C2q2qM}              &  $\Delta C^{T,(2)}_{qq}$                                   & \texttt{GTC2q2qM}             &  $\Delta C^{L,(2)}_{qq}$                                &  \texttt{GLC2q2qM} \\
 $\Delta C^{1,(2)}_{qq,\Fcon,1}$                      & \texttt{G1C2q2qMFcon1}    & $\Delta C^{T,(2)}_{qq,\Fcon,1}$                       & \texttt{GTC2q2qMFcon1}   & $\Delta C^{L,(2)}_{qq,\Fcon,1}$                        & \texttt{GLC2q2qMFcon1} \\
 $\Delta C^{1,(2)}_{qq,\Fcon,2}$                     & \texttt{G1C2q2qMFcon2}    &                                                                              &      &                                                                    & \\
 $\Delta C^{1,(2)}_{qq,\mathrm{A},\NoW}$   & \texttt{G1C2q2qMANoW}   & $\Delta C^{T,(2)}_{qq,\mathrm{A},\NoW}$   & \texttt{GTC2q2qMANoW}   & $\Delta C^{L,(2)}_{qq,\mathrm{A},\NoW}$     & \texttt{GLC2q2qMANoW} \\
 $\Delta C^{1,(2)}_{gq}$                                  & \texttt{G1C2q2gM}              & $\Delta C^{T,(2)}_{gq}$                                   & \texttt{GTC2q2gM}              & $\Delta C^{L,(2)}_{gq}$                                     & \texttt{GLC2q2gM} \\
 $\Delta C^{1,(2)}_{gq,\mathrm{A},\NoW}$   & \texttt{G1C2q2gMANoW}   & $\Delta C^{T,(2)}_{gq,\mathrm{A},\NoW}$   & \texttt{GTC2q2gMANoW}   & $\Delta C^{L,(2)}_{gq,\mathrm{A},\NoW}$      & \texttt{GLC2q2gMANoW} \\
 $\Delta C^{1,(2)}_{qg}$                                   & \texttt{G1C2g2qM}              & $\Delta C^{T,(2)}_{qg}$                                   & \texttt{GTC2g2qM}              & $\Delta C^{L,(2)}_{qg}$                                       & \texttt{GLC2g2qM} \\
 $\Delta C^{1,(2)}_{qg,\mathrm{A},\NoW}$   & \texttt{G1C2g2qMANoW}   & $\Delta C^{T,(2)}_{qg,\mathrm{A},\NoW}$   & \texttt{GTC2g2qMANoW}   & $\Delta C^{L,(2)}_{qg,\mathrm{A},\NoW}$        & \texttt{GLC2g2qMANoW} \\
 $\Delta C^{1,(2)}_{gg}$                                   & \texttt{G1C2g2gM}              &                                    &              &                                                                      & \\
 $\Delta C^{1,(2)}_{gg,\mathrm{AA},\NoW}$ & \texttt{G1C2g2gMAANoW} &                                   &              &                                                                      & \\
 $\Delta C^{1,(2)}_{\qp q,1}$                            & \texttt{G1C2q2qpM1}          & $\Delta C^{T,(2)}_{\qp q,1}$                             & \texttt{GTC2q2qpM1}           & $\Delta C^{L,(2)}_{\qp q,1}$                                & \texttt{GLC2q2qpM1} \\
 $\Delta C^{1,(2)}_{\qp q,2}$                            & \texttt{G1C2q2qpM2}         & $\Delta C^{T,(2)}_{\qp q,2}$                            & \texttt{GTC2q2qpM2}           & $\Delta C^{L,(2)}_{\qp q,2}$                               & \texttt{GLC2q2qpM2} \\
 $\Delta C^{1,(2)}_{\qp q,\NoW,3}$                 & \texttt{G1C2q2qpMNoW3} & $\Delta C^{T,(2)}_{\qp q,\NoW,3}$                 & \texttt{GTC2q2qpMNoW3}  & $\Delta C^{L,(2)}_{\qp q,\NoW,3}$                    & \texttt{GLC2q2qpMNoW3} \\
 $\Delta C^{1,(2)}_{\qp q,\NoW,4}$                 & \texttt{G1C2q2qpMNoW4} & $\Delta C^{T,(2)}_{\qp q,\NoW,4}$                 & \texttt{GTC2q2qpMNoW4}  & $\Delta C^{L,(2)}_{\qp q,\NoW,4}$                     & \texttt{GLC2q2qpMNoW4} \\
 $\Delta C^{1,(2)}_{\qb q}$                               & \texttt{G1C2q2qbM}            & $\Delta C^{T,(2)}_{\qb q}$                               & \texttt{GTC2q2qbM}            & $\Delta C^{L,(2)}_{\qb q}$                              & \texttt{GLC2q2qbM} \\
 $\Delta C^{1,(2)}_{\qb q,\Fcon}$                     & \texttt{G1C2q2qbMFcon}    & $\Delta C^{T,(2)}_{\qb q,\Fcon}$                    & \texttt{GTC2q2qbMFcon}    & $\Delta C^{L,(2)}_{\qb q,\Fcon}$                     & \texttt{GLC2q2qbMFcon} \\
  \bottomrule
\end{tabular}
}
\caption{List of coefficient functions and ancillary file notation grouped by structure function and perturbative order.}
\label{tab:ancillary}
\end{table}

\clearpage

\bibliographystyle{JHEP}
\bibliography{EW_SIDIS}

\providecommand{\href}[2]{#2}\begingroup\raggedright\begin{thebibliography}{10}

\bibitem{Aidala:2012mv}
C.A.~Aidala, S.D.~Bass, D.~Hasch and G.K.~Mallot, \emph{{The Spin Structure of
  the Nucleon}}, \href{https://doi.org/10.1103/RevModPhys.85.655}{\emph{Rev.
  Mod. Phys.} {\bfseries 85} (2013) 655}
  [\href{https://arxiv.org/abs/1209.2803}{{\ttfamily 1209.2803}}].

\bibitem{Accardi:2012qut}
A.~Accardi et~al., \emph{{Electron Ion Collider: The Next QCD Frontier}:
  {Understanding the glue that binds us all}},
  \href{https://doi.org/10.1140/epja/i2016-16268-9}{\emph{Eur. Phys. J. A}
  {\bfseries 52} (2016) 268} [\href{https://arxiv.org/abs/1212.1701}{{\ttfamily
  1212.1701}}].

\bibitem{AbdulKhalek:2021gbh}
R.~Abdul~Khalek et~al., \emph{{Science Requirements and Detector Concepts for
  the Electron-Ion Collider}: {EIC Yellow Report}},
  \href{https://doi.org/10.1016/j.nuclphysa.2022.122447}{\emph{Nucl. Phys. A}
  {\bfseries 1026} (2022) 122447}
  [\href{https://arxiv.org/abs/2103.05419}{{\ttfamily 2103.05419}}].

\bibitem{Aschenauer:2019kzf}
E.C.~Aschenauer, I.~Borsa, R.~Sassot and C.~Van~Hulse, \emph{{Semi-inclusive
  Deep-Inelastic Scattering, Parton Distributions and Fragmentation Functions
  at a Future Electron-Ion Collider}},
  \href{https://doi.org/10.1103/PhysRevD.99.094004}{\emph{Phys. Rev. D}
  {\bfseries 99} (2019) 094004}
  [\href{https://arxiv.org/abs/1902.10663}{{\ttfamily 1902.10663}}].

\bibitem{Moch:2014sna}
S.~Moch, J.A.M.~Vermaseren and A.~Vogt, \emph{{The Three-Loop Splitting
  Functions in QCD: The Helicity-Dependent Case}},
  \href{https://doi.org/10.1016/j.nuclphysb.2014.10.016}{\emph{Nucl. Phys. B}
  {\bfseries 889} (2014) 351}
  [\href{https://arxiv.org/abs/1409.5131}{{\ttfamily 1409.5131}}].

\bibitem{Blumlein:2021enk}
J.~Bl\"umlein, P.~Marquard, C.~Schneider and K.~Sch\"onwald, \emph{{The
  three-loop unpolarized and polarized non-singlet anomalous dimensions from
  off shell operator matrix elements}},
  \href{https://doi.org/10.1016/j.nuclphysb.2021.115542}{\emph{Nucl. Phys. B}
  {\bfseries 971} (2021) 115542}
  [\href{https://arxiv.org/abs/2107.06267}{{\ttfamily 2107.06267}}].

\bibitem{Blumlein:2021ryt}
J.~Bl\"umlein, P.~Marquard, C.~Schneider and K.~Sch\"onwald, \emph{{The
  three-loop polarized singlet anomalous dimensions from off-shell operator
  matrix elements}}, \href{https://doi.org/10.1007/JHEP01(2022)193}{\emph{JHEP}
  {\bfseries 01} (2022) 193}
  [\href{https://arxiv.org/abs/2111.12401}{{\ttfamily 2111.12401}}].

\bibitem{Zijlstra:1993sh}
E.B.~Zijlstra and W.L.~van Neerven, \emph{{Order-$\alpha_s^2$ corrections to
  the polarized structure function $g_1 (x,Q^2)$}},
  \href{https://doi.org/10.1016/0550-3213(94)90538-X}{\emph{Nucl. Phys. B}
  {\bfseries 417} (1994) 61}.

\bibitem{Blumlein:2022gpp}
J.~{Bl\"umlein}, P.~Marquard, C.~Schneider and K.~{Sch\"onwald}, \emph{{The
  massless three-loop Wilson coefficients for the deep-inelastic structure
  functions F$_{2}$, F$_{L}$, xF$_{3}$ and g$_{1}$}},
  \href{https://doi.org/10.1007/JHEP11(2022)156}{\emph{JHEP} {\bfseries 11}
  (2022) 156} [\href{https://arxiv.org/abs/2208.14325}{{\ttfamily
  2208.14325}}].

\bibitem{Bonino:2024wgg}
L.~Bonino, T.~Gehrmann, M.~L\"ochner, K.~Sch\"onwald and G.~Stagnitto,
  \emph{{Polarized Semi-Inclusive Deep-Inelastic Scattering at
  Next-to-Next-to-Leading Order in QCD}},
  \href{https://doi.org/10.1103/PhysRevLett.133.211904}{\emph{Phys. Rev. Lett.}
  {\bfseries 133} (2024) 211904}
  [\href{https://arxiv.org/abs/2404.08597}{{\ttfamily 2404.08597}}].

\bibitem{Goyal:2024tmo}
S.~Goyal, R.N.~Lee, S.-O.~Moch, V.~Pathak, N.~Rana and V.~Ravindran,
  \emph{{Next-to-Next-to-Leading Order QCD Corrections to Polarized
  Semi-Inclusive Deep-Inelastic Scattering}},
  \href{https://doi.org/10.1103/PhysRevLett.133.211905}{\emph{Phys. Rev. Lett.}
  {\bfseries 133} (2024) 211905}
  [\href{https://arxiv.org/abs/2404.09959}{{\ttfamily 2404.09959}}].

\bibitem{Goyal:2024emo}
S.~Goyal, R.N.~Lee, S.-O.~Moch, V.~Pathak, N.~Rana and V.~Ravindran,
  \emph{{NNLO QCD corrections to unpolarized and polarized SIDIS}},
  \href{https://doi.org/10.1103/PhysRevD.111.094007}{\emph{Phys. Rev. D}
  {\bfseries 111} (2025) 094007}
  [\href{https://arxiv.org/abs/2412.19309}{{\ttfamily 2412.19309}}].

\bibitem{Bertone:2024taw}
{\scshape MAP (Multi-dimensional Analyses of Partonic distributions)}
  collaboration, \emph{{Helicity-dependent parton distribution functions at
  next-to-next-to-leading order accuracy from inclusive and semi-inclusive
  deep-inelastic scattering data}},
  \href{https://doi.org/10.1016/j.physletb.2025.139497}{\emph{Phys. Lett. B}
  {\bfseries 865} (2025) 139497}
  [\href{https://arxiv.org/abs/2404.04712}{{\ttfamily 2404.04712}}].

\bibitem{Borsa:2024mss}
I.~Borsa, M.~Stratmann, W.~Vogelsang, D.~de~Florian and R.~Sassot,
  \emph{{Next-to-Next-to-Leading Order Global Analysis of Polarized Parton
  Distribution Functions}},
  \href{https://doi.org/10.1103/PhysRevLett.133.151901}{\emph{Phys. Rev. Lett.}
  {\bfseries 133} (2024) 151901}
  [\href{https://arxiv.org/abs/2407.11635}{{\ttfamily 2407.11635}}].

\bibitem{Bonino:2025qta}
L.~Bonino, T.~Gehrmann, M.~L{\"o}chner, K.~Sch{\"o}nwald and G.~Stagnitto,
  \emph{{Neutral and Charged Current Semi-Inclusive Deep-Inelastic Scattering
  at NNLO QCD}},  \href{https://arxiv.org/abs/2506.19926}{{\ttfamily
  2506.19926}}.

\bibitem{Bonino:2024qbh}
L.~Bonino, T.~Gehrmann and G.~Stagnitto, \emph{{Semi-Inclusive Deep-Inelastic
  Scattering at Next-to-Next-to-Leading Order in QCD}},
  \href{https://doi.org/10.1103/PhysRevLett.132.251901}{\emph{Phys. Rev. Lett.}
  {\bfseries 132} (2024) 251901}
  [\href{https://arxiv.org/abs/2401.16281}{{\ttfamily 2401.16281}}].

\bibitem{Goyal:2023zdi}
S.~Goyal, S.-O.~Moch, V.~Pathak, N.~Rana and V.~Ravindran,
  \emph{{Next-to-Next-to-Leading Order QCD Corrections to Semi-Inclusive
  Deep-Inelastic Scattering}},
  \href{https://doi.org/10.1103/PhysRevLett.132.251902}{\emph{Phys. Rev. Lett.}
  {\bfseries 132} (2024) 251902}
  [\href{https://arxiv.org/abs/2312.17711}{{\ttfamily 2312.17711}}].

\bibitem{ParticleDataGroup:2024cfk}
{\scshape Particle Data Group} collaboration, \emph{{Review of particle
  physics}}, \href{https://doi.org/10.1103/PhysRevD.110.030001}{\emph{Phys.
  Rev. D} {\bfseries 110} (2024) 030001}.

\bibitem{deFlorian:2012wk}
D.~de~Florian and Y.~Rotstein~Habarnau, \emph{{Polarized semi-inclusive
  electroweak structure functions at next-to-leading-order}},
  \href{https://doi.org/10.1140/epjc/s10052-013-2356-3}{\emph{Eur. Phys. J. C}
  {\bfseries 73} (2013) 2356}
  [\href{https://arxiv.org/abs/1210.7203}{{\ttfamily 1210.7203}}].

\bibitem{Nogueira:1991ex}
P.~Nogueira, \emph{{Automatic Feynman Graph Generation}},
  \href{https://doi.org/10.1006/jcph.1993.1074}{\emph{J. Comput. Phys.}
  {\bfseries 105} (1993) 279}.

\bibitem{Vermaseren:2000nd}
J.A.M.~Vermaseren, \emph{{New features of FORM}},
  \href{https://arxiv.org/abs/math-ph/0010025}{{\ttfamily math-ph/0010025}}.

\bibitem{Larin:1993tq}
S.A.~Larin, \emph{{The Renormalization of the axial anomaly in dimensional
  regularization}},
  \href{https://doi.org/10.1016/0370-2693(93)90053-K}{\emph{Phys. Lett. B}
  {\bfseries 303} (1993) 113}
  [\href{https://arxiv.org/abs/hep-ph/9302240}{{\ttfamily hep-ph/9302240}}].

\bibitem{deFlorian:1994wp}
D.~de~Florian and R.~Sassot, \emph{{O (alpha-s) spin dependent weak structure
  functions}}, \href{https://doi.org/10.1103/PhysRevD.51.6052}{\emph{Phys. Rev.
  D} {\bfseries 51} (1995) 6052}
  [\href{https://arxiv.org/abs/hep-ph/9412255}{{\ttfamily hep-ph/9412255}}].

\bibitem{Forte:2001ph}
S.~Forte, M.L.~Mangano and G.~Ridolfi, \emph{{Polarized parton distributions
  from charged current deep inelastic scattering and future neutrino
  factories}}, \href{https://doi.org/10.1016/S0550-3213(01)00101-8}{\emph{Nucl.
  Phys. B} {\bfseries 602} (2001) 585}
  [\href{https://arxiv.org/abs/hep-ph/0101192}{{\ttfamily hep-ph/0101192}}].

\bibitem{tHooft:1972tcz}
G.~'t~Hooft and M.J.G.~Veltman, \emph{{Regularization and Renormalization of
  Gauge Fields}},
  \href{https://doi.org/10.1016/0550-3213(72)90279-9}{\emph{Nucl. Phys. B}
  {\bfseries 44} (1972) 189}.

\bibitem{Breitenlohner:1977hr}
P.~Breitenlohner and D.~Maison, \emph{{Dimensional Renormalization and the
  Action Principle}}, \href{https://doi.org/10.1007/BF01609069}{\emph{Commun.
  Math. Phys.} {\bfseries 52} (1977) 11}.

\bibitem{Vogelsang:1996im}
W.~Vogelsang, \emph{{The Spin dependent two loop splitting functions}},
  \href{https://doi.org/10.1016/0550-3213(96)00306-9}{\emph{Nucl. Phys. B}
  {\bfseries 475} (1996) 47}
  [\href{https://arxiv.org/abs/hep-ph/9603366}{{\ttfamily hep-ph/9603366}}].

\bibitem{Mertig:1995ny}
R.~Mertig and W.L.~van Neerven, \emph{{The Calculation of the two loop spin
  splitting functions $\Delta P_{ij}^{(1)}(x)$}},
  \href{https://doi.org/10.1007/s002880050138}{\emph{Z. Phys. C} {\bfseries 70}
  (1996) 637} [\href{https://arxiv.org/abs/hep-ph/9506451}{{\ttfamily
  hep-ph/9506451}}].

\bibitem{Matiounine:1998re}
Y.~Matiounine, J.~Smith and W.L.~van Neerven, \emph{{Two loop operator matrix
  elements calculated up to finite terms for polarized deep inelastic lepton -
  hadron scattering}},
  \href{https://doi.org/10.1103/PhysRevD.58.076002}{\emph{Phys. Rev. D}
  {\bfseries 58} (1998) 076002}
  [\href{https://arxiv.org/abs/hep-ph/9803439}{{\ttfamily hep-ph/9803439}}].

\bibitem{Klein:2009ig}
S.W.G.~Klein, \emph{{Mellin Moments of Heavy Flavor Contributions to
  $F_2(x,Q^2)$ at NNLO}}, Ph.D. thesis, Dortmund U., Berlin, 2009.
\newblock \href{https://arxiv.org/abs/0910.3101}{{\ttfamily 0910.3101}}.
\newblock 10.1007/978-3-642-23286-2.

\bibitem{Schonwald:2019gmn}
K.~Sch\"onwald, \emph{{Massive two- and three-loop calculations in QED and
  QCD}}, Ph.D. thesis, Dortmund U., 2019.
\newblock 10.17877/DE290R-20310.

\bibitem{Behring:2019tus}
A.~Behring, J.~Bl\"umlein, A.~De~Freitas, A.~Goedicke, S.~Klein, A.~von
  Manteuffel et~al., \emph{{The Polarized Three-Loop Anomalous Dimensions from
  On-Shell Massive Operator Matrix Elements}},
  \href{https://doi.org/10.1016/j.nuclphysb.2019.114753}{\emph{Nucl. Phys. B}
  {\bfseries 948} (2019) 114753}
  [\href{https://arxiv.org/abs/1908.03779}{{\ttfamily 1908.03779}}].

\bibitem{Belusca-Maito:2023wah}
H.~B{\'e}lusca-Ma{\"\i}to, A.~Ilakovac, P.~K{\"u}hler,
  M.~Ma{\dj}or-Bo{\v{z}}inovi{\'c}, D.~St{\"o}ckinger and M.~Wei{\ss}wange,
  \emph{{Introduction to Renormalization Theory and Chiral Gauge Theories in
  Dimensional Regularization with Non-Anticommuting
  {\ensuremath{\gamma}}$_{5}$}},
  \href{https://doi.org/10.3390/sym15030622}{\emph{Symmetry} {\bfseries 15}
  (2023) 622} [\href{https://arxiv.org/abs/2303.09120}{{\ttfamily
  2303.09120}}].

\bibitem{vonManteuffel:2012np}
A.~von Manteuffel and C.~Studerus, \emph{{Reduze 2 - Distributed Feynman
  Integral Reduction}},  \href{https://arxiv.org/abs/1201.4330}{{\ttfamily
  1201.4330}}.

\bibitem{Chetyrkin:1981qh}
K.G.~Chetyrkin and F.V.~Tkachov, \emph{{Integration by parts: The algorithm to
  calculate $\beta$-functions in 4 loops}},
  \href{https://doi.org/10.1016/0550-3213(81)90199-1}{\emph{Nucl. Phys. B}
  {\bfseries 192} (1981) 159}.

\bibitem{Laporta:2000dsw}
S.~Laporta, \emph{{High-precision calculation of multiloop Feynman integrals by
  difference equations}},
  \href{https://doi.org/10.1142/S0217751X00002159}{\emph{Int. J. Mod. Phys. A}
  {\bfseries 15} (2000) 5087}
  [\href{https://arxiv.org/abs/hep-ph/0102033}{{\ttfamily hep-ph/0102033}}].

\bibitem{Gehrmann:1999as}
T.~Gehrmann and E.~Remiddi, \emph{{Differential equations for two loop four
  point functions}},
  \href{https://doi.org/10.1016/S0550-3213(00)00223-6}{\emph{Nucl. Phys. B}
  {\bfseries 580} (2000) 485}
  [\href{https://arxiv.org/abs/hep-ph/9912329}{{\ttfamily hep-ph/9912329}}].

\bibitem{Gehrmann:2005pd}
T.~Gehrmann, T.~Huber and D.~Maitre, \emph{{Two-loop quark and gluon
  form-factors in dimensional regularisation}},
  \href{https://doi.org/10.1016/j.physletb.2005.07.019}{\emph{Phys. Lett. B}
  {\bfseries 622} (2005) 295}
  [\href{https://arxiv.org/abs/hep-ph/0507061}{{\ttfamily hep-ph/0507061}}].

\bibitem{Gehrmann:2022cih}
T.~Gehrmann and R.~Sch\"urmann, \emph{{Photon fragmentation in the antenna
  subtraction formalism}},
  \href{https://doi.org/10.1007/JHEP04(2022)031}{\emph{JHEP} {\bfseries 04}
  (2022) 031} [\href{https://arxiv.org/abs/2201.06982}{{\ttfamily
  2201.06982}}].

\bibitem{Haug:2022hkr}
J.~Haug and F.~Wunder, \emph{{The massless single off-shell scalar box integral
  \textemdash{} branch cut structure and all-order epsilon expansion}},
  \href{https://doi.org/10.1007/JHEP02(2023)177}{\emph{JHEP} {\bfseries 02}
  (2023) 177} [\href{https://arxiv.org/abs/2211.14110}{{\ttfamily
  2211.14110}}].

\bibitem{Anastasiou:2002yz}
C.~Anastasiou and K.~Melnikov, \emph{{Higgs boson production at hadron
  colliders in NNLO QCD}},
  \href{https://doi.org/10.1016/S0550-3213(02)00837-4}{\emph{Nucl. Phys. B}
  {\bfseries 646} (2002) 220}
  [\href{https://arxiv.org/abs/hep-ph/0207004}{{\ttfamily hep-ph/0207004}}].

\bibitem{Bonino:2024adk}
L.~Bonino, T.~Gehrmann, M.~Marcoli, R.~Sch\"urmann and G.~Stagnitto,
  \emph{{Antenna subtraction for processes with identified particles at hadron
  colliders}}, \href{https://doi.org/10.1007/JHEP08(2024)073}{\emph{JHEP}
  {\bfseries 08} (2024) 073}
  [\href{https://arxiv.org/abs/2406.09925}{{\ttfamily 2406.09925}}].

\bibitem{Ahmed:2024owh}
T.~Ahmed, S.~Goyal, S.M.~Hasan, R.N.~Lee, S.-O.~Moch, V.~Pathak et~al.,
  \emph{{NNLO phase-space integrals for semi-inclusive deep-inelastic
  scattering}}, \href{https://doi.org/10.1103/22x4-dmmp}{\emph{Phys. Rev. D}
  {\bfseries 112} (2025) 014020}
  [\href{https://arxiv.org/abs/2412.16509}{{\ttfamily 2412.16509}}].

\bibitem{Remiddi:1999ew}
E.~Remiddi and J.A.M.~Vermaseren, \emph{{Harmonic polylogarithms}},
  \href{https://doi.org/10.1142/S0217751X00000367}{\emph{Int. J. Mod. Phys. A}
  {\bfseries 15} (2000) 725}
  [\href{https://arxiv.org/abs/hep-ph/9905237}{{\ttfamily hep-ph/9905237}}].

\bibitem{Maitre:2005uu}
D.~Maitre, \emph{{HPL, a mathematica implementation of the harmonic
  polylogarithms}},
  \href{https://doi.org/10.1016/j.cpc.2005.10.008}{\emph{Comput. Phys. Commun.}
  {\bfseries 174} (2006) 222}
  [\href{https://arxiv.org/abs/hep-ph/0507152}{{\ttfamily hep-ph/0507152}}].

\bibitem{Duhr:2019tlz}
C.~Duhr and F.~Dulat, \emph{{PolyLogTools \textemdash{} polylogs for the
  masses}}, \href{https://doi.org/10.1007/JHEP08(2019)135}{\emph{JHEP}
  {\bfseries 08} (2019) 135}
  [\href{https://arxiv.org/abs/1904.07279}{{\ttfamily 1904.07279}}].

\bibitem{Borsa:2022vvp}
I.~Borsa, R.~Sassot, D.~de~Florian, M.~Stratmann and W.~Vogelsang,
  \emph{{Towards a Global QCD Analysis of Fragmentation Functions at
  Next-to-Next-to-Leading Order Accuracy}},
  \href{https://doi.org/10.1103/PhysRevLett.129.012002}{\emph{Phys. Rev. Lett.}
  {\bfseries 129} (2022) 012002}
  [\href{https://arxiv.org/abs/2202.05060}{{\ttfamily 2202.05060}}].

\bibitem{NNPDF:2021njg}
{\scshape NNPDF} collaboration, \emph{{The path to proton structure at 1\%
  accuracy}}, \href{https://doi.org/10.1140/epjc/s10052-022-10328-7}{\emph{Eur.
  Phys. J. C} {\bfseries 82} (2022) 428}
  [\href{https://arxiv.org/abs/2109.02653}{{\ttfamily 2109.02653}}].

\bibitem{Buckley:2014ana}
A.~Buckley, J.~Ferrando, S.~Lloyd, K.~Nordstr\"om, B.~Page, M.~R\"ufenacht
  et~al., \emph{{LHAPDF6: parton density access in the LHC precision era}},
  \href{https://doi.org/10.1140/epjc/s10052-015-3318-8}{\emph{Eur. Phys. J. C}
  {\bfseries 75} (2015) 132} [\href{https://arxiv.org/abs/1412.7420}{{\ttfamily
  1412.7420}}].

\bibitem{COMPASS:2010hwr}
{\scshape COMPASS} collaboration, \emph{{Quark helicity distributions from
  longitudinal spin asymmetries in muon-proton and muon-deuteron scattering}},
  \href{https://doi.org/10.1016/j.physletb.2010.08.034}{\emph{Phys. Lett. B}
  {\bfseries 693} (2010) 227}
  [\href{https://arxiv.org/abs/1007.4061}{{\ttfamily 1007.4061}}].

\bibitem{COMPASS:2015mhb}
{\scshape COMPASS} collaboration, \emph{{The spin structure function $g_1^{\rm
  p}$ of the proton and a test of the Bjorken sum rule}},
  \href{https://doi.org/10.1016/j.physletb.2015.11.064}{\emph{Phys. Lett. B}
  {\bfseries 753} (2016) 18}
  [\href{https://arxiv.org/abs/1503.08935}{{\ttfamily 1503.08935}}].

\bibitem{HERMES:2004zsh}
{\scshape HERMES} collaboration, \emph{{Quark helicity distributions in the
  nucleon for up, down, and strange quarks from semi-inclusive deep-inelastic
  scattering}}, \href{https://doi.org/10.1103/PhysRevD.71.012003}{\emph{Phys.
  Rev. D} {\bfseries 71} (2005) 012003}
  [\href{https://arxiv.org/abs/hep-ex/0407032}{{\ttfamily hep-ex/0407032}}].

\bibitem{HERMES:2006jyl}
{\scshape HERMES} collaboration, \emph{{Precise determination of the spin
  structure function g(1) of the proton, deuteron and neutron}},
  \href{https://doi.org/10.1103/PhysRevD.75.012007}{\emph{Phys. Rev. D}
  {\bfseries 75} (2007) 012007}
  [\href{https://arxiv.org/abs/hep-ex/0609039}{{\ttfamily hep-ex/0609039}}].

\bibitem{HERMES:2018awh}
{\scshape HERMES} collaboration, \emph{{Longitudinal double-spin asymmetries in
  semi-inclusive deep-inelastic scattering of electrons and positrons by
  protons and deuterons}},
  \href{https://doi.org/10.1103/PhysRevD.99.112001}{\emph{Phys. Rev. D}
  {\bfseries 99} (2019) 112001}
  [\href{https://arxiv.org/abs/1810.07054}{{\ttfamily 1810.07054}}].

\bibitem{Borsa:2022cap}
I.~Borsa, D.~de~Florian and I.~Pedron, \emph{{NNLO jet production in neutral
  and charged current polarized deep inelastic scattering}},
  \href{https://doi.org/10.1103/PhysRevD.107.054027}{\emph{Phys. Rev. D}
  {\bfseries 107} (2023) 054027}
  [\href{https://arxiv.org/abs/2212.06625}{{\ttfamily 2212.06625}}].

\bibitem{Goyal:2025bzf}
S.~Goyal, S.-O.~Moch, V.~Pathak, N.~Rana and V.~Ravindran, \emph{{Soft and
  virtual corrections to semi-inclusive DIS up to four loops in QCD}},
  \href{https://arxiv.org/abs/2506.24078}{{\ttfamily 2506.24078}}.

\end{thebibliography}\endgroup

\end{document}